\documentclass[iop,numberedappendix,appendixfloats]{emulateapj}
\usepackage{apjfonts}
\usepackage{amssymb}
\usepackage{natbib}
\usepackage{amsmath}
\usepackage{xcolor}
\usepackage{amssymb}
\usepackage{natbib}
\usepackage[colorlinks,citecolor=blue]{hyperref}
\usepackage{booktabs}
\shorttitle{Keck Lyman Continuum Spectroscopic Survey}
\shortauthors{Steidel et al.}
\newcommand{\lya}{\ensuremath{\rm Ly\alpha}}

\newcommand{\zla}{\ensuremath{z_{\rm Ly\alpha}}}
\newcommand{\zis}{\ensuremath{z_{\rm IS}}}

\newcommand{\kms}{\rm km~s\ensuremath{^{-1}\,}}
\newcommand{\dvla}{\ensuremath{\rm \Delta v_{\lya}}}
\newcommand{\dvis}{\ensuremath{\rm \Delta v_{IS}}}

\newcommand{\lyb}{Ly$\beta$}

\newcommand{\nhi}{\ensuremath{N_{\rm HI}}}
\newcommand{\fobs}{\mbox{$\langle f_{900}/f_{1500}\rangle_{\rm obs}$}}
\newcommand{\fout}{\mbox{$\langle f_{900}/f_{1500}\rangle_{\rm out}$}}
\newcommand{\fabs}{\mbox{$\langle f_{\rm esc,abs} \rangle$}}
\newcommand{\wlya}{\ensuremath{W_{\lambda}(\lya)}}
\newcommand{\nhimin}{\ensuremath{N_{\rm HI,min}}}
\newcommand{\nhimax}{\ensuremath{N_{\rm HI,max}}}

\newcommand{\secpoint}{\mbox{$''\mskip-7.6mu.\,$}}

\def\ltsima{$\; \buildrel < \over \sim \;$}
\def\simlt{\lower.5ex\hbox{\ltsima}}
\def\gtsima{$\; \buildrel > \over \sim \;$}
\def\simgt{\lower.5ex\hbox{\gtsima}}
\def\arcs{$''~$}
\def\arcm{$'~$}

\begin{document}
\title{The Keck Lyman Continuum Spectroscopic Survey (KLCS): \\ the Emergent Ionizing Spectrum of Galaxies at $z\sim 3$\altaffilmark{1}}
\slugcomment{DRAFT: \today}
\author{\sc Charles C. Steidel\altaffilmark{2}, Milan Bogosavljevi\'c\altaffilmark{2,9}, Alice E. Shapley\altaffilmark{3}, 
Naveen A. Reddy\altaffilmark{4,5}, Gwen C. Rudie\altaffilmark{6}, \\  Max Pettini\altaffilmark{8}, Ryan F. Trainor\altaffilmark{7},
%Dawn K. Erb\altaffilmark{10}, 
Allison L. Strom\altaffilmark{2,6} }

\altaffiltext{1}{Based on data obtained at the 
W.M. Keck Observatory, which 
is operated as a scientific partnership among the California Institute of 
Technology, the
University of California, and NASA, and was made possible by the generous 
financial
support of the W.M. Keck Foundation. 
}
\altaffiltext{2}{Cahill Center for Astronomy and Astrophysics, California Institute of Technology, MS 249-17, Pasadena, CA 91125.}
\altaffiltext{3}{University of California, Los Angeles, Deparment of Physics and Astronomy, 430 Portola Plaza, Los Angeles, CA 90095.}
\altaffiltext{4}{University of California, Riverside, Deparment of Physics and Astronomy, 400 University Ave., Riverside, CA 92521.}
\altaffiltext{5}{Alfred P. Sloan Foundation Fellow}
\altaffiltext{6}{Carnegie Institute for Science, The Observatories, 813 Santa Barbara Street, Pasadena, CA 91107.}
\altaffiltext{7}{Franklin \& Marshall College, Department of Physics and Astronomy, 415 Harrisburg Pike, Lancaster, PA 17603.} 
\altaffiltext{8}{Institute of Astronomy, Madingley Road, Cambridge CB3 OHA, United Kingdom}
\altaffiltext{9}{Division of Science, New York University Abu Dhabi, P.O. Box 129188, Abu Dhabi, UAE.}
%\altaffiltext{10}{Center for Gravitation and Cosmology, Department of Physics, University of Wisconsin, Milwaukee, 3135 N. Maryland Ave., Milwaukee, WI 53211.}

\begin{abstract}
We present results of a deep spectroscopic survey 
quantifying the statistics of the escape of hydrogen-ionizing photons from star-forming galaxies at $z \sim 3$. 
The Keck Lyman Continuum Spectroscopic Survey (KLCS) 
includes spectra of 
124 galaxies with $\langle z \rangle = 3.05\pm0.18$ and $-22.1 \le M_{\rm uv} \le -19.5$, observed in 9 independent fields, 
covering a common
rest-wavelength range $880 \le \lambda_0/{\rm \textrm \AA} \simlt 1750$. 
We measure the ratio of ionizing to non-ionizing UV flux density \fobs, where 
$f_{900}$ is the mean flux density evaluated over the range $\lambda_0=[880,910]$ \AA.
To quantify  
\fout -- the {\it emergent} ratio of ionizing to non-ionizing UV flux density 
-- we use detailed Monte Carlo modeling of the opacity   
of \ion{H}{1} in the intergalactic (IGM) and circumgalactic (CGM) medium as a function of source redshift. 
By analyzing   
high-S/N composite spectra formed from sub-samples exhibiting common observed properties and numbers 
sufficient to reduce the uncertainty in the IGM+CGM correction, 
we obtain precise values of \fout, including a full-sample average $\fout = 0.057\pm0.006$. 
We further show that \fout\ increases monotonically with \lya\ rest equivalent
width \wlya, inducing an inverse correlation with UV luminosity as a by-product.  
To connect LyC leakage to intrinsic galaxy properties, 
we fit the composite spectra using stellar population synthesis (SPS) 
together with simple models of the ISM in which a fraction $f_{\rm c}$ of the stellar continuum
is covered by optically-thick gas with column density \nhi. We show that the composite spectra 
simultaneously constrain the intrinsic properties of the ionizing
stars $(L_{900}/L_{1500})_{\rm int}$ along with $f_{\rm c}$, \nhi, $E(B-V)$, and \fabs, the escape fraction of ionizing photons. We find
a sample-averaged $\fabs = 0.09\pm0.01$, and that subsamples fall along a linear relation $\fabs \simeq 0.75\langle \wlya/(110~\textrm{\AA})\rangle$ for
$0 \le \langle \wlya \rangle \simlt 60$; subsamples with $\langle \wlya \rangle < 0$ have \fabs\ consistent with zero.
We use the FUV luminosity function, the distribution function $n[\wlya]$, and   
the relationship between \wlya\ and \fout\ to estimate
the total ionizing emissivity of $z \sim 3$
star-forming galaxies with $M_{\rm UV} \le -19.5$: $\epsilon_{\rm LyC} \simeq 6\times10^{24}$ ergs s$^{-1}$ Hz$^{-1}$ Mpc$^{-3}$.   
This value exceeds the contribution of QSOs by a factor of $\sim 1.2-3.7$, and accounts for $\simgt 50$\%
of the {\it total} $\epsilon_{\rm LyC}$ estimated using indirect methods at $z \sim 3$.

\end{abstract}
\keywords{cosmology: observations --- galaxies: evolution --- galaxies: high-redshift }

\section{Introduction}
\label{sec:klcs_intro}

Substantial recent efforts have focused on establishing the demographics of star-forming galaxies in the redshift
range $6 \simlt z \simlt 10$ now believed to be most relevant for cosmic reionization \citep{planck2016}. Nevertheless, a detailed physical understanding 
of the reionization process remains elusive, due in large part to uncertainties that cannot be reduced simply by identifying
a larger number of potential sources of ionizing photons.  
Crucial missing ingredients include knowledge of the {\it intrinsic} ionizing spectra of sources, and, more importantly,
the net ionizing spectrum presented to the intergalactic medium (IGM) after passing through layers of gas and dust in the galaxy interstellar
medium (ISM.) 
Unfortunately, these will be impossible
to measure from direct photometric or spectroscopic study
of reionization-era galaxies, even when the {\it James Webb Space Telescope} ({\it JWST}) comes on line, due to the rapidly 
increasing \ion{H}{1} opacity with
redshift along extended lines of sight to high redshifts, even post-reionization.   

However, one can explore the likely behavior of Lyman continuum-producing objects, and perhaps
make testable predictions, using observations of analogous sources at lower redshifts, where the opacity of
intervening neutral H (\ion{H}{1}) is less limiting, and where ancillary multiwavelength observations
are more easily obtained.  One avenue that has enjoyed recent success is ultraviolet (UV) observations conducted
using {\it Hubble Space Telescope} ({\it HST}) of low-redshift galaxies that have been identified as likely
analogs of the high redshift objects (e.g., \citealt{borthakur14,izotov16a,izotov16b,leitherer16}), 
such as 
`Lyman break analogs'' (LBAs; \citealt{overzier09}), and ``Green Peas'' (\citealt{cardamone09}.) These objects, 
although rare in the present-day
universe, have many of the same properties typical of high-redshift star-forming galaxies-- e.g., high
UV luminosity, strong nebular emission lines, strong galaxy-scale outflows, compact sizes, and high specific
star formation rates (sSFRs).  Alternatively, one can obtain very deep observations of larger samples of
more distant galaxies, where direct observations of the Lyman continuum (LyC) are possible without necessarily observing
from space (e.g., \citealt{steidel01,shapley06,iwata09,nestor11,vanzella12,nestor13,mostardi13,mostardi15,grazian16}.) 
Redshifts $z \simgt 3$   
bring the rest-frame Lyman limit of hydrogen at 911.75 \AA\ (13.6 eV) to observed wavelengths well above the atmospheric 
cutoff near 3100-3200 \AA,  where the atmosphere is transparent, the terrestrial background is darkest, and 
the instrumental sensitivity of spectrometers on large ground-based telescopes is high. 

In general, ionizing photons produced by massive stars in \ion{H}{2} regions are a local phenomenon, with a sphere of influence 
measured in pc;     
the ``escape fraction'' of ionizing photons from an isolated ionization-bounded \ion{H}{2} region  
(i.e., where the ionized region is entirely embedded within a predominantly neutral region, and the extent of the \ion{H}{2}
region is determined by the production rate of ionizing photons) is zero, by definition. 
However, when the density of star formation is very high, as is often the case for high redshift star-forming systems, intense
episodes of star formation and frequent supernovae can in principle produce ionized bubbles that carve channels through which Lyman continuum photons might 
escape unimpeded. The net escaping ionizing radiation depends on the geometry of the sites of massive star formation
and the surrounding ISM, the lifetimes of the stars that produce the bulk of the ionizing photons (e.g., \citealt{ma16}), and the probability
that during that lifetime favorable conditions for LyC photon escape will occur. 

Once H-ionizing photons escape the ISM of a parent galaxy, the probability of detection   
by an observer at $z=0$ is governed by the effective opacity of intervening \ion{H}{1} along the line of sight. 
This opacity increases steeply with redshift (e.g., \citealt{madau95,steidel01,vanzella10,vanzella12,becker15}), and for
the wavelength range most relevant to LyC measurement (i.e., $\lambda_0 \simlt 912(1+z_{\rm s})$ for a source with redshift $z_{\rm s}$) 
it is also subject to large fluctuations from sightline to sightline, since it is dominated by the incidence of relatively small numbers of 
intervening \ion{H}{1} systems with $16 \simlt {\rm log}~(\nhi/{\rm cm}^{-2}) \simlt 18$ (e.g., \citealt{rudie13}).  
As we discuss in more detail below (\S\ref{sec:igm_trans}), 
these competing factors strongly favor the range $2.75 \simlt z \simlt 3.50$ for 
a ground-based survey. 

Even within the optimal redshift range for ground-based observations, practical sensitivity 
limits impose
severe restrictions on the dynamic range available for possible detections. 
The combination of limited sensitivity and detectability dominated
by the stochastic behavior of the IGM foreground means that  
individual detections of LyC signal i.e., those for which significant LyC flux is detected without stacking) 
are almost guaranteed to be unusual either in their intrinsic properties, in having a fortuitously transparent line of sight
through the IGM, or a combination of both.
Direct detections of individual sources at high redshift (\citealt{vanzella15,shapley16,vanzella17}) are valuable for demonstrating 
that at least {\it some} galaxies produce ionizing radiation that propagates beyond their own ISM,  
but they do not place strong constraints on more typical galaxies. 

In view of these challenges, successful characterization of the propensity for galaxies with particular common
properties to ``leak'' LyC radiation requires observations of an {\it ensemble}, in order
to marginalize over the fluctuations in the intervening IGM opacity.  It should also include
1) the most sensitive possible measurements of individual sources, made as close as possible
to the rest-frame Lyman limit of each; 2) a very accurate characterization of the statistics of intervening
\ion{H}{1} as a function of column density and redshift;  3) control over systematics -- those affecting measurement of
individual sources (e.g, background subtraction, contamination) and those that would invalidate the statistical
IGM correction. The latter suggests observing sources in several independent survey fields and avoiding regions
known to harbor unusual large-scale structures, if possible. 

Although spectroscopic surveys at $z \sim 3$ using 8m-class telescopes first became feasible in the mid-1990s (e.g., \citealt{steidel96,steidel03}),
the initially-available instruments were not optimized for high near-UV/blue sensitivity as required for the most effective observations
of LyC emission \citep{steidel01}. 
The situation changed substantially for the better  
with the commissioning on the Keck 1 telescope of the blue channel of the LRIS spectrograph (LRIS-B,
\citealt{steidel04}) in 2002, followed by the installation of the Keck 1 Cassegrain Atmospheric 
Dispersion Corrector (\citealt{phillips06}) in 2007; projects
demanding high efficiency in the wavelength range ($3100 - 4500$\,\AA) became much more feasible.
\cite{shapley06} [S06], using pilot data obtained immediately following LRIS-B commissioning, 
presented what were apparently the first direct detections of LyC emission in the spectra of individual
star-forming galaxies at $z\sim3$. 
S06 observed a sample of 14 LBGs with LRIS 
in multi-slit mode for a total of $\sim 8$ hrs (in the mode sensitive to LyC light), 
reaching an unprecedented depth for individual galaxy spectra of
$3 \times 10^{-31}$\,ergs\,s$^{-1}$\,cm$^{-2}$\,Hz$^{-1}$ at 
$\sim3600$\,\AA\space (3$\sigma$ detection limit), 
corresponding to $AB \simeq 27.6$, or $\sim 20$ times fainter 
than the (non-ionizing) continuum flux density of L$_{\ast}$ galaxies at $z\sim 3$. 
While the spectra presented by S06 were far superior for LyC detection compared to what had been available, 
it later turned out that 2 of the 3 putative detections of residual
LyC flux were due to contamination of the LyC rest-frame spectral region by faint, unrelated foreground
galaxies, based on subsequent near-IR spectra and HST imaging (\citealt{siana15}). 
In the years since, it has transpired that {\it most} apparent detections of significant 
LyC emission from $z\simgt 3$ galaxies have, on further inspection -- particularly using high spatial
resolution images obtained with {\it HST} -- been attributed to similar 
foreground contamination (\citealt{vanzella12,vanzella15,siana15,mostardi15}). 

Most of the more recent observational effort toward detecting LyC emission from intermediate and high redshift
galaxies has been invested in imaging surveys, which have an obvious multiplex advantage, particularly when aimed 
at fields containing known galaxy over-densities. In such fields, narrow or intermediate band filters can be
fine-tuned to lie just below the rest-frame Lyman limit at the redshift of interest (\citealt{inoue05,iwata09,nestor11,mostardi13}). 
Alternatively, one can use extremely deep broad-band UV images to search for LyC emission from galaxies having
known spectroscopic redshifts that ensure the band lies entirely shortward of the rest-frame Lyman limit (\citealt{malkan03,cowie09,bridge10,siana10,rutkowski16,grazian16,grazian17}). 
While imaging surveys obtain LyC measurements for every galaxy known or suspected to lie at high enough redshift
in the field of view, putative detections (and the quantification of non-detections) requires both follow-up spectroscopy
and/or high-resolution {\it HST} imaging (e.g., \citealt{vanzella10,nestor13,mostardi15}). 

In the present work, we return to using very deep spectroscopic observations, similar to those
presented by S06.  
The Keck Lyman Continuum Spectroscopic Survey (KLCS) 
expands and improves on the S06 study in several
respects: first, the sample is larger by an order
of magnitude, with a total of  
136 $z \simeq 3$ galaxies observed. 
Second, the observations were conducted
in 9 independent survey fields, which should drastically reduce the
sample variance of the results, particularly if there are large-scale
correlations in intergalactic LyC opacity that could have a very strong
effect on results based on a single field (see discussion in S06\footnote{Several of the initial surveys, including S06, \citet{iwata09,nestor11} 
were conducted in a single field (SSA22), focusing on a known proto-cluster at $z=3.09$ (\citealt{steidel98,steidel00,hayashino04,matsuda04}.)}.)
Third, KLCS covers a broader range in both redshift ($2.72 \le z \le 3.54$),
and galaxy luminosity ($0.2 < (L_{\rm uv}/L^{\ast}_{\rm uv}) < 3$) compared to S06.
Most importantly, however, KLCS has benefited from the accumulated insight and lessons
learned through experience -- e.g., the importance of false positive detections due to foreground contamination and the sensitivity
required for plausible detections -- as well as from advances in the physical interpretation
of the far-UV spectra of high redshift galaxies (e.g., \citealt{steidel16,eldridge17}) and in
the precision of our statistical knowledge of the foreground IGM+CGM opacity (e.g., \citealt{rudie13}.)

In this paper, we show that, through careful control of systematics and concerted efforts to eliminate contamination, 
ensembles of deep rest-UV spectra can be used to measure the ratio of LyC flux density to non-ionizing UV flux density 
(hereafter $(f_{900}/f_{1500})_{\rm obs}$) with high precision.  
The KLCS observations provide individual galaxy
spectra of unprecedented quality; composite spectra formed from 
substantial subsets provide templates 
that are the most sensitive ever obtained for similar high redshift objects,
enabling access to a remarkable range of stellar, interstellar, and nebular spectral
features, many of which have not been observed previously beyond the local
universe. 
As well as direct constraints on the leakage of ionizing photons from galaxies, high quality rest-frame  
far-UV spectra encode the ancillary information on the massive star populations, the geometry and porosity of the
ISM, the kinematics, physics, and chemistry of galaxy-scale outflows, and
stellar and ionized gas-phase metallicities of the same galaxies -- all of which are needed to place LyC leakage
within the broader context of galaxies and the diffuse IGM.  

The paper is organized as follows: \S\ref{sec:klcs_sample} describes the selection of the KLCS sample; \S\ref{sec:klcs_specobs} 
details the spectroscopic observations, while \S\ref{sec:cal_and_redux} describes the data reduction, including the steps taken
to minimize residual systematic errors in the sample. \S\ref{sec:final_sample} defines the final KLCS sample used for
subsequent analysis. LyC measurements from individual KLCS spectra are covered in \S\ref{sec:measure}.  \S\ref{sec:igm_trans}
describes modeling of the IGM and CGM transmission used to correct the KLCS LyC observations. In \S\ref{sec:composites}, we
form a number of KLCS sub-samples and discuss the construction of their stacked (composite) spectra; \S\ref{sec:galaxy_properties} 
relates the composite spectra of the sub-samples to the corresponding LyC measurements, while \S\ref{sec:implications}
discusses the implications of the measurements. In \S\ref{sec:lyc_escape} we evaluate the spectroscopic results in the
context of a simple model for LyC escape and its connection to other observed galaxy properties, and propose
the most appropriate method for calculating the total ionizing emissivity of the galaxy population at $z \sim 3$.  
Finally, \S\ref{sec:discussion} summarizes the principal results
and discusses their broader implications and suggestions for future work. Appendix A summarizes data reduction steps taken to minimize residual
systematic errors in KLCS as well as tests of their efficacy; Appendix B 
contains details of the IGM+CGM Monte Carlo transmission model used for
the analysis. 

Readers interested primarily in the results of the analysis, but not the details of the methods used to
obtain them, may wish to focus on the final 4 sections (\S\ref{sec:galaxy_properties} - \S\ref{sec:discussion}). 
 
Where relevant, we assume a CDM cosmology with $\Omega_{\rm m} = 0.3$, $\Omega_{\Lambda} = 0.7$, and $h=0.7$. 
All spectroscopic flux density measurements used in the paper are expressed as flux per unit frequency ($f_{\nu}$), and
are generally plotted as
a function of wavelength, so that a spectrum with constant $f_{\nu}$, constant $m_{\rm AB}$, or $f_{\lambda} \propto \lambda^{-\beta}$ with $\beta = 2$,
appears ``flat''. 

\section{The KLCS Sample }
\label{sec:klcs_sample}

\subsection{Target Redshift Range }

Ozone in the Earth's atmosphere efficiently blocks ultraviolet (UV) 
radiation with a sharp transparency cutoff preventing photons
with $\lambda \le 3100$\,\AA\space (at elevation of 4200m, as on Mauna Kea) from reaching the surface.
A consequence is that ground-based observations of rest-frame
LyC photons from celestial objects require observing them
at redshifts $z \simgt 2.725$, where the rest-frame Lyman 
limit of \ion{H}{1} ($\lambda_0 = 911.75$ \AA) 
falls at an observed wavelength of $\simeq 3400$ \AA. 
At higher redshifts, observations of the rest-frame LyC 
benefit from the generally higher instrumental throughput and atmospheric transmission at longer
wavelengths, but the sensitivity for detection of
LyC flux escaping from galaxies actually declines precipitously with increasing redshift beyond
$z \simeq 3.5$ (e.g., \citealt{madau95,steidel01,vanzella10}; see \S\ref{sec:igm_trans})

The decreasing sensitivity with increasing redshift is due to a combination of 
several effects: 
first and most obviously, galaxies of a given intrinsic UV luminosity 
($L_{\rm uv}$)
become apparently fainter with redshift; in addition 
the characteristic $L^{\ast}_{\rm uv}(z)$ itself dims 
as redshift increases beyond $z \sim 3$ (e.g., \citealt{reddy09,bouwens10,finkelstein15}). 
Although the net instrumental sensitivity at wavelengths $912(1+z_{\rm s})$ \AA\ increases with 
redshift from $z\sim 2.7$ to $z \sim 3.5$, at redshifts $z \simgt 3.5$ the throughput gains 
are more than offset by the increasing intensity of
the sky background against which any faint LyC signal must be measured.  

Most importantly, line and continuum
opacity of neutral
hydrogen (\ion{H}{1}) in the IGM along the line of sight increases steeply with
$z_{\rm s}$ (e.g., \citealt{madau95}).  Even
if intergalactic \ion{H}{1} contributed no net continuum opacity for ionizing photons emitted
from a source with $z_{\rm s}$, the LyC region will be blanketed by the effective opacity 
caused by Lyman series lines with $z \simlt 911.8/1215.7 \times (1+z_{\rm s})^{-1}$, reducing
the dynamic range accessible to LyC detection\footnote{According to \citet{becker11}, continuum blanketing 
from the 
Lyman $\alpha$ forest increases $\propto (1+z)^{2.8}$ over the redshift range $2.1 \simlt z_{\rm s} \simlt 5.5$.}.  
When one includes the net LyC opacity contributed by gas outside of the galaxy, but at redshifts near enough
to $z_{\rm s}$ to impact the net transmission averaged over the LyC detection band, 
the median transmission in the rest-wavelength
interval 880--910 \AA\ decreases by a factor of $\simeq 7$ between $z=3$ and $z=5$, 
(see discussion in \S\ref{sec:igm_trans}, and Table~\ref{tab:igm_models}.) 

Tallying all of the exacerbating factors, 
the overall difficulty of a (ground-based) detection of LyC signal from a $L_{\rm uv} = L_{\rm uv}^{\ast}(z_{\rm s})$ 
increases
by a factor of $\sim 35$ as one moves from $z \sim 3$ to $z \sim 5$.
Thus, there is strong impetus for a ground-based LyC survey to focus on sources 
with $2.75 \simlt z_{\rm s} \simlt 3.5$, as we have done for KLCS.

\subsection{Survey Design}
\label{sec:klcs_design}

 Targets for KLCS were selected in 9 separate fields on the
 sky (Table~\ref{tab:klcs_fields}), chosen from among
$\simeq 25$ high latitude fields in which we have obtained deep $U_{\rm n}G{\cal R}$ photometry
suitable for selecting LBG candidates at $z \sim 3$ (see \citealt{reddy12} for a nearly-complete
list) as well as spectroscopic follow-up observations.  
The final field selection was based on a combination of visibility time at low airmass  
during scheduled observing runs and the number of 
galaxies with previously-obtained spectroscopic redshifts in
the range $2.9 \simlt z \simlt 3.2$ that could be accommodated
on a single slit mask of the Low Resolution Imaging Spectrometer (LRIS; \citealt{oke95,steidel04}) 
on the Keck 1 telescope.   
For six of the nine KLCS fields (see Table~\ref{tab:klcs_fields}), selection of star-forming galaxy candidates
 using rest-UV continuum photometry was performed during the course of
 a survey for  $z \sim 3$ Lyman-break galaxies (LBGs; see \citealt{steidel03}). 
 Three additional fields (Q0100$+$1300, Q1009$+$2956 and HS1549$+$1919)
including $z\sim 3$ LBGs were observed during the period 2003 to 2009; these 
comprise part of the Keck Baryonic Structure Survey (KBSS; \citealt{steidel04,steidel2010,rudie12a,steidel14,strom17}.)

Selection of $z \sim 3$ LBGs in all 9 of the KLCS fields was based on photometric selection using the
3-band ($U_nG{\cal R}$) photometric system described in detail by \citet{steidel03}; photometric
and spectroscopic catalogs for 6 of these fields (as of 2003) were also presented in that work.
Full photometric and spectroscopic survey catalogs for the 3 KBSS fields will be presented elsewhere. 

\begin{deluxetable*}{lllllllcll}
\tabletypesize{\scriptsize}
\tablewidth{0pt}
\tablecaption{Keck Lyman Continuum Spectroscopic Survey: Observations}
\tablehead{\colhead{Field}  & \colhead{RA (J2000)\tablenotemark{a}} & \colhead{Dec (J2000)\tablenotemark{a}} & \colhead{Mask} & \colhead{$N_{\rm obs}$\tablenotemark{b}}  
& \colhead{$N$\tablenotemark{c}} & \colhead{$N_{\rm det}$\tablenotemark{d}} & \colhead{Date Observed}  & \colhead{ADC} & \colhead{$t_{\rm exp}$\tablenotemark{i}} } 
\startdata
Q0100$+$1300\tablenotemark{g}   &  01:03:11.27 & $+$13:16:18.0  & q0100\_L1 & ~15     & ~12  & ~0    & 2006 Dec  & no   &  ~5.1 \\
                                &   &   & q0100\_L2\tablenotemark{e} & \nodata     & \nodata  & \nodata    & 2007 Sep  & yes   &  ~5.2 \\
Q0256$-$000\tablenotemark{f}          &  02:59:05.13 & $+$00:11:06.8  & q0256\_L1 & ~15  & ~11   & ~0  & 2007 Nov         & yes       &  ~8.5  \\
B20902$+$34\tablenotemark{f}          &  09:05:31.23 & $+$34:08:01.7  & b20902\_L1 & ~14  & ~11  & ~0   & 2007 Nov   & yes       &  ~5.0  \\
          &   &   & b20902\_L1 & \nodata  & \nodata  & \nodata   & 2008 Apr   & yes       &  ~3.4  \\
Q0933$+$2854\tablenotemark{f}         &  09:33:36.09 & $+$28:45:34.8  & q0933\_L2 &   ~13     & ~10  & ~1    & 2007 Mar  & yes   &  ~9.2\tablenotemark{h}  \\
                                      &   &   & q0933\_L3 &   ~15     & ~14  & ~3    &  2008 Apr  &  yes   &  ~8.2\tablenotemark{h}  \\
Q1009$+$2956\tablenotemark{g}         &  10:11:54.49 & $+$29:41:33.5  & q1009\_L1 & ~12     & ~11  & ~0    & 2006 Dec         & no        &  ~7.0   \\
Westphal\tablenotemark{f,g}           &  14:17:43.21 & $+$52:28:48.5  & gws\_L1 & ~15     & ~15  & ~3    & 2008 Jun         & yes       &  ~8.5   \\
Q1422$+$2309\tablenotemark{f}         &  14:24:36.98 & $+$22:53:49.6  & q1422\_L2 & ~16 & ~15  & ~3 & 2008 Apr         & yes       &  ~8.2   \\
HS1549$+$1919\tablenotemark{g}        &  15:51:54.75 & $+$19:10:48.0  & q1549\_L2 & ~13 & ~13  & ~2  & 2008 Apr  & yes       &  ~4.3   \\
       &   &  & q1549\_L2 & \nodata & \nodata  & \nodata & 2008 Jun  & yes       &  ~4.2   \\
DSF2237b\tablenotemark{f}             &  22:39:34.10 & $+$11:51:38.8  & dsf2237b\_L1 & ~14~~~    & ~14  & ~3    & 2007 Sep  & yes       &   ~7.3  \\
             &   &   & dsf2237b\_L1 & \nodata   & \nodata   & \nodata   & 2007 Nov  & yes       &   ~5.5  \\
\hline \\
TOTAL  &   &   & 10  & 136    & 124  & 15    &  &    & 89.6    
\enddata
\tablenotetext{a}{Positions of the field centers. }
\tablenotetext{b}{Number of $z>2.7$ galaxies observed.} 
\tablenotetext{c}{Number of galaxies included in KLCS sample.}
\tablenotetext{d}{Number of galaxies with $>3\sigma$ detections of residual LyC flux. }
\tablenotetext{e}{Mask q0100\_L2 includes the same KLCS targets as q0100\_L1, but eliminated lower-priority objects from the mask design.}
\tablenotetext{f}{Fields from $z\sim 3.0$ LBG survey \citep{steidel03}. }
\tablenotetext{g}{Fields from KBSS (see also \citealt{rudie12a,steidel14}).} 
\tablenotetext{h}{Two objects are common to masks q0933\_L2 and q0933\_L3.}
\tablenotetext{i}{Total exposure time, in hours.}
\label{tab:klcs_fields}
\end{deluxetable*}

We used a slitmask design strategy that assigned highest priority to comparatively
brighter (${\cal R} \le 24.5$) LBGs known to have redshifts in the interval $2.9 \le z \le 3.3$.
Other star-forming galaxies in a broader redshift 
range $2.7 < z < 3.6$ were assigned somewhat lower priority.
If space on a mask was still available, we included additional
candidates that were identically selected but had not been previously observed spectroscopically\footnote{As discussed
in \cite{steidel03}, the contamination of the photometric selection windows used for
$z \sim 3$ galaxies is $< 5$\%, so that most of the new targets observed yielded redshifts
in the desired range.}. 
Objects that had already been classified as AGN based on their existing survey
spectra were deliberately given relatively high priority in the KLCS mask
design, as little is known about whether ionizing radiation escapes from low-luminosity
AGN or QSOs\footnote{Generally, only bright QSOs have been observed shortward of
the rest-frame Lyman limit.}. This small sub-sample will be addressed in future work. 

\begin{figure}[tp]
    \centerline{\includegraphics[width=8.5cm]{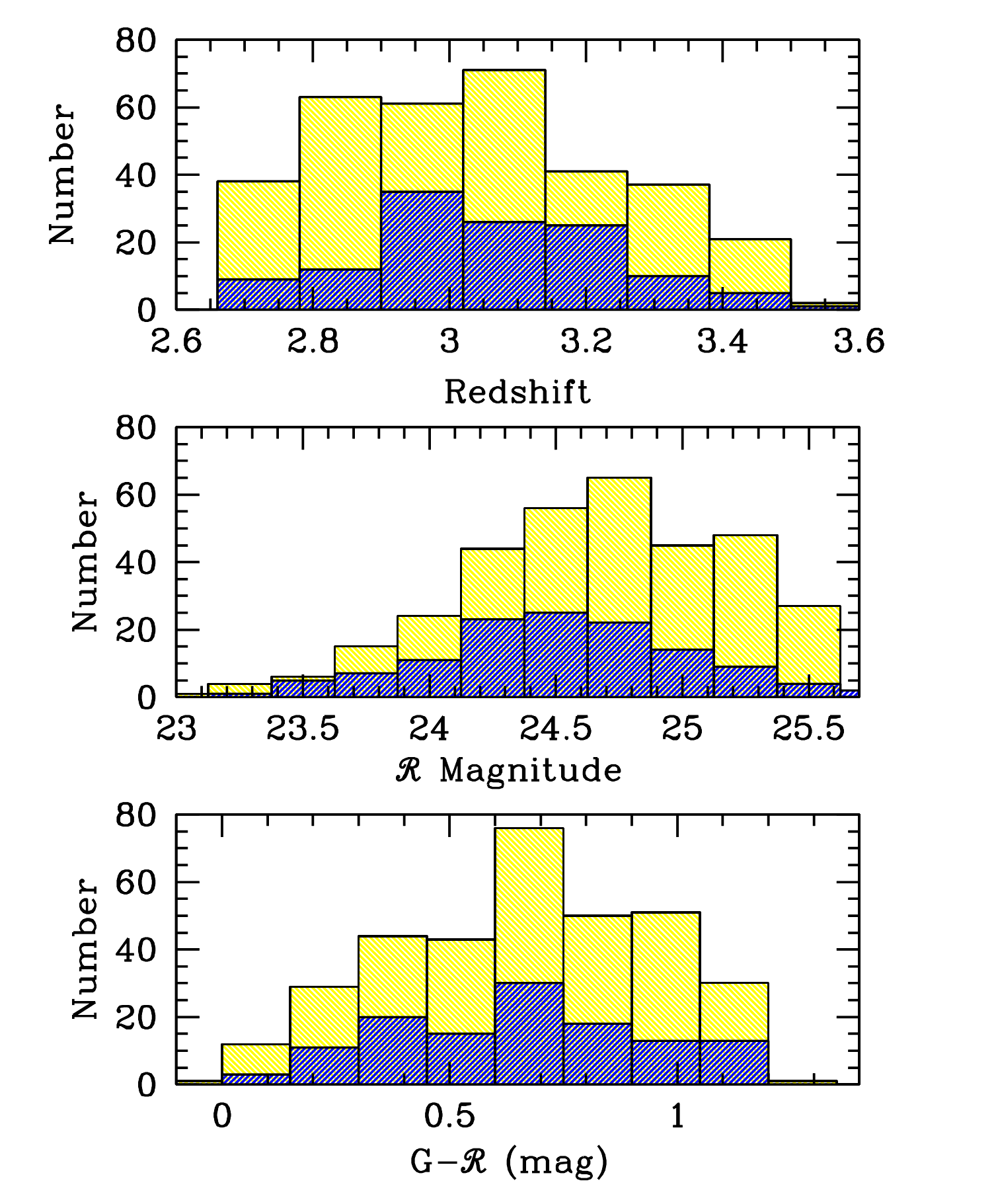}}
    \caption{
    The distribution of redshift (top), ${\cal R}$ magnitude (center), and
$G-{\cal R}$ color for the KLCS sample (dark shaded histograms) as compared
with those of all confirmed 
    LBGs in the 9 survey fields used (light shaded histograms). 
    Apart from the slight preference for brighter objects (${\cal R} \simlt 25$)
    and for objects at redshift $z \sim 3.0$, the KLCS sample is not significantly
    biased with respect to the LBG population.
  \label{fig:klcs_sample}
}
  \end{figure}

Figure\,\ref{fig:klcs_sample} presents comparisons between the parent
sample from which targets were selected (light histograms), and those that were successfully
observed (dark histograms), in terms of redshift, 
 apparent magnitude ${\cal R}$, and color $G-{\cal R}$. 
The ``parent'' sample in this case includes all galaxies with
redshifts $2.7 < z < 3.6$
 located in the same set of fields used in the KLCS.   
 The middle panel of Figure\,\ref{fig:klcs_sample} 
 shows that the KLCS sample has a moderate excess of sources with
 ${\cal R} < 24.5$ and a related deficiency of galaxies with
   $25.0 \le {\cal R} \le 25.5$,
 which is an expected consequence of our observing strategy.
 The redshift distribution of KLCS galaxies also demonstrates our slight
 preference for targets in $2.9\le z \le 3.2$ redshift ``window''.
 We find no difference in the distribution of $G-{\cal R}$
 color between KLCS and all spectroscopically identified galaxies in these fields
 (Figure \ref{fig:klcs_sample}, bottom panel).
Thus, the sub-sample of galaxies observed for the KLCS is slightly
brighter,  and has a slightly tighter redshift distribution, than
the LBGs in the same fields, but is otherwise representative. 

\section{Spectroscopic Observations}
\label{sec:klcs_specobs}

As summarized in Table~\ref{tab:klcs_fields}, a total of 136 galaxies 
was observed, in 9 independent fields using 10 different
slitmasks.  
Of the galaxies observed, 13 were later 
excluded from the LyC analysis because of uncalibrated 
slit defects, close companions on the slit, scattered light, or other potential sources
of systematic error that would make LyC flux measurements less secure
(see the detailed discussion in \S\ref{sec:final_sample} below).

All 10 slitmasks were designed with $1\secpoint2$ slit widths and individual
slit lengths between 10\arcsec and 30\arcsec; the median slit length for high priority KLCS
targets was 20\arcsec.
With the exception of the Q0933$+$2854 field, a single slitmask was observed in each field, containing 
between 8 and 16 objects known to lie at redshifts $2.7 < z < 3.6$. 
The observations were conducted over the course of 6 separate
observing runs using LRIS on the Keck 1 10m telescope between 
2006 December and 2008 June. As summarized in Table~\ref{tab:klcs_fields}, 
the total integration time per slitmask was between 8.2 and 12.8 hours.  Fields were generally
observed within 2 hours of the meridian in order to minimize attenuation by atmospheric extinction 
and (in the case of the observations made prior to 2007 August) slit losses due to 
differential atmospheric refraction. 

All observations were made using the same configuration
of the LRIS double-beamed spectrograph \citep{oke95,steidel04}, with the incoming
beam divided near 5000 \AA\ using the ``d500'' dichroic beamsplitter. Wavelengths shortward
of 5000 \AA\ were recorded by the ``blue'' spectrograph channel (LRIS-B) 
and those longward of $\simeq 5000$ \AA\ by the red channel (LRIS-R). 
LRIS-B was configured with a  
400 line/mm grism with first-order blaze at 3400~\AA, providing wavelength coverage from 
3100 \AA\ to beyond the dichroic split near 5000\AA. 
The LRIS-B detector was binned 1x2 (binned in the dispersion direction) 
at readout in order to minimize the effect of read noise, which for these devices is $\simeq 3.8$e$^{-}$ pix$^{-1}$, 
resulting in pixels that project to 0\secpoint135 on the sky in the spatial direction and 
$\simeq 2.18$ \AA\ per pixel in the dispersion direction. 
LRIS-R was configured with
a 600 line/mm grating
with first-order blaze at 5000\,\AA, with spectra recorded using the (pre-2009) Tektronix 2k x 2k (monolithic) 
detector with 24 $\mu$m pixels. With no on-chip binning, 
the scale at the detector was 0\secpoint211 per pixel spatially and 1.28 \AA\ pix$^{-1}$ in the dispersion
direction.  The LRIS-R grating was tilted such that all KLCS slits would have wavelength
coverage from shortward of the dichroic split to $\simgt 7000$ \AA\ depending
on the spatial position of the slit within the LRIS field of view.  

Individual exposure times were $1800$s 
and all LRIS-B and LRIS-R exposures 
were obtained simultaneously, resulting in identical
observing conditions and integration times for a given mask. 
Data were collected under mostly photometric observing conditions,
and all data used in the KLCS were obtained with seeing $< 1$\secpoint0,
and typically $\simlt 0$\secpoint7. 

Prior to the commissioning of the Keck 1 Cassegrain Atmospheric Dispersion Corrector
(ADC; \citealt{phillips06}) in the summer of 2007,  all KLCS observations were obtained at
elevations within $30^\circ$ of zenith and position angle close
to the parallactic in order to minimize effects of differential
refraction.  Once commissioned, the ADC was used for all observations, 
greatly enhancing the efficiency of the survey by allowing position angle to
be unconstrained during mask design, thus allowing inclusion of a larger number
of high priority targets. 
By correcting differential refraction before the slitmask, the ADC improved the data quality
particularly for LRIS-B, and enables the use of simpler (and more robust) data reduction 
and extraction techniques (see \S\ref{sec:redux}).

The spectral resolution achieved varied slightly depending on observing conditions
and the angular size of objects within each slit. 
With a typical seeing-convolved profile size of FWHM$\simeq 0\secpoint8- 0\secpoint9$ for
Lyman-break galaxies at 
$z \sim 3$ \citep{law07b,law11}, the average 
resolving power is $R \simeq 800$ (LRIS-B) and $R \simeq 1400$ (LRIS-R) with
the dispersers described above  
(see \citealt{steidel2010}).  

The dichroic split at 5000 \AA\ typically places the location of \lya,
i.e. $1215.67(1+z_{\rm s})$ \AA, near the transition wavelength between
the spectral channels for $z \sim 3$. Longward of \lya\ our primary objective was to resolve 
the width of typical interstellar absorption lines 
(FWHM$\simeq 500-700$ \kms; see, e.g., \citealt{pettini02,shapley06,steidel2010})  
so that the degeneracy between
velocity width and covering fraction could be disentangled. 
The FWHM$\simeq 200$ \kms\ resolution provided by the 600/5000 grating represented a compromise between 
resolution and sensitivity.  
For LRIS-B, high sensitivity (particularly in the wavelength range 3400-4000 \AA) 
was of paramount importance, hence the choice of the 400/3400 grism, which
in combination with the LRIS-B optics produces very high UV/blue throughput
(see \citealt{steidel04}). The spectral resolution of FWHM $\simeq 375$ \kms\ is modest,
but still adequate to resolve typical strong absorption and emission lines.  

\section{Calibrations and Data Reductions} 
\label{sec:cal_and_redux}

\subsection{Calibrations}

We obtained spectroscopic flat-field calibration 
images for LRIS-R and LRIS-B separately. Internal halogen lamp
spectra provided adequate flat-fields for all LRIS-R data, but
were not suitable for LRIS-B, which for our configuration requires very good flats
particularly in the wavelength range 3100-4000 \AA, where there
are substantial spatial variations in quantum efficiency due to non-uniformities 
in the thinning of the silicon during manufacture. 

We found from experience (for LRIS-B) that slitmask spectra of the twilight sky, obtained
at similar elevation and instrument rotator angle to the science observations
through the same mask, are the most effective solution; these were obtained at the beginning and
end of each observing night. The twilight sky spectral flats 
produce adequate signal in the UV, but record
the scattered solar spectrum rather than a featureless continuum.  
To remove the G-star spectrum but preserve the pixel-to-pixel sensitivity
variations, we divide the raw flats by a spatially median-filtered 
1-d spectrum calculated at regular intervals along each slit, producing 
images normalized to an average value of unity but retaining the 
desired pixel-to-pixel sensitivity variations. The issue of scattered light 
associated with flat fielding is addressed in \S\ref{sec:ghosts} below. 

Internal arc lamp 
spectra (Hg, Ne, Zn, and Cd for LRIS-B, Hg, Ne, Ar, Zn for LRIS-R) were used for wavelength calibration
for both LRIS-B and LRIS-R, with 5th order polynomial fits resulting in typical residuals 
of $\simeq 0.10$ \AA\  and $\simeq 0.07$ \AA\ for LRIS-B and LRIS-R, respectively. 
The arc-based wavelength solutions were subsequently shifted by small amounts using measurements
of night sky emission lines recorded in each science exposure. 

Spectrophotometric standard 
stars from the list of \cite{massey88} were observed at the end of each night 
through $1\secpoint0$ slit oriented at the parallactic angle (for all observations, both 
pre- and post-installation of the ADC in August 2007), 
with configuration settings otherwise identical to mask observations.
Absolute flux calibration
uncertainties are estimated to be of order $\simeq 20$\% due to potential 
variations in seeing
conditions (and the associated variation in slit losses) between slitmask 
and standard star observations. However, red-side and blue-side exposures were always obtained simultaneously, for both
science and standard star observations, 
so that with careful reductions of the standards, the relative spectrophotometry between the blue and red channels is
much more precise than the absolute spectrophotometry; the latter is relatively unimportant to our
analysis (see \S\ref{sec:1d_spectra} below). 

\subsection{Data Reduction}
\label{sec:redux}

 The standard LRIS spectroscopic data processing pipeline we have used for 
 previous surveys with LRIS (e.g., \citealt{steidel03,steidel04,steidel2010})
was generally followed in processing KLCS LRIS-R data. However, given the challenge
of measuring very faint flux densities at levels well below the sky background
in the LyC region, we paid particular attention to developing 
procedures for LRIS-B reductions with a goal of minimizing residual systematic
errors wherever possible.  
Given the potential importance of systematics to the final LyC results, we describe the
details of the reduction procedures up to the extraction of 1-D spectra in Appendix~\ref{sec:redux_appendix}.

\label{sec:1d_spectra}
 
The 2-D, background-subtracted, stacked spectrograms for each sequence of LRIS-B or LRIS-R observations with
a given slitmask were reduced (as described in Appendix~\ref{sec:redux_appendix}) so that the centroid of
the trace for a given object on the slit lies at a constant spatial pixel along the dispersion direction, whether
or not the observation was made using the ADC. 
We adopted, conservatively, a 1\secpoint35 boxcar extraction window (10 spatial pixels on the LRIS-B detector) to avoid making assumptions
about variations of the spatial profile with wavelength -- particularly when the trace extends to wavelengths
beyond where there is obvious detected flux. Thus, the extraction aperture for every object is a rectangular region
of size 1\secpoint2 by 1\secpoint35 on the sky.  

For each extracted 1-D spectrum, we used an identical extraction region on the 
2-D ``sky + object'' 2-D spectrograms described above  (\S\ref{sec:noise_model}), 
i.e.,
\begin{equation}
\label{eqn:extraction} 
S[i] = \sum_{j=k-5}^{k+5} s[i,j] 
\end{equation} 
and
\begin{equation}
\sigma[i] =   \left(\sum_{j=k-5}^{k+5} \sigma_{\rm pix}^2[i,j]\right)^{0.5} ~,
\end{equation}
where $k$ is the spatial position (in $j$ pixels) of the object trace, $S[i]$ is the resulting 1-D spectrum at dispersion point $[i]$,  and $\sigma[i]$ is the corresponding one-dimensional error
vector. Section~\ref{sec:2d_bkg} describes in more detail the extent to which the noise model agrees with the data.  

Mask data sets obtained on more than one observing run (see Table~\ref{tab:klcs_fields}) 
were reduced to 1-D separately, and then combined using inverse variance weighting to produce final 1-D spectra.  
The 1-D spectra were wavelength calibrated using the 1-D arc spectra extracted from the same region of the processed 2-D arc frames, 
zero-pointed using night sky emission lines measured in the extracted 1-D object+sky spectra (\S\ref{sec:noise_model}) to correct for small
amounts of flexure and slight differences in illumination between the internal arc lamps and the night sky. The final LRIS-B wavelength scales
are in the vacuum, heliocentric frame with a linear dispersion of 2.14 \AA\ ${\rm pix}^{-1}$ (LRIS-B), covering 3200-5000 \AA.     

Finally, the extracted LRIS-B and LRIS-R spectra were flux calibrated using standard stars from the list of \citet{massey88}, obtained
on the same night as the data comprising each 1-D extracted science spectrum, and corrected for Galactic extinction assuming the
reddening maps of \citet{schlegel98}. The standard star observations on the blue and red sides were
made simultaneously, with the slit oriented at the parallactic angle.  Because the LRIS long slit lies at a fixed position in
the field of view of the instrument, the sensitivity curves in the wavelength transition region of the dichroic beamsplitter 
(where the spectral response of the interference coatings change rapidly with increasing wavelength from
reflection to transmission) are not perfectly matched for slits located far from the focal plane position of the longslit,  
due to small differences in angle of incidence of the incoming light. However, through experimentation we 
found that accurate relative spectrophotometry could be achieved 
by using only LRIS-B spectra
for $\lambda < 5000$ \AA\ and only LRIS-R spectra for $\lambda > 5000$ \AA; i.e., {\it without} averaging over the region of overlap.  

Continuous LRIS-B+R spectra, covering 3200-7200 \AA\, were produced by remapping the individual calibrated spectra onto 
a linear wavelength scale chosen so as to preserve the spectral sampling of the higher resolution red-side data, 1.0 \AA\ ${\rm pix}^{-1}$.
These were used to produce composite spectra using subsets of the KLCS sample (\S\ref{sec:composites}.)
Thus, we made final 1-D flux calibrated spectra for each source. 

\section{Defining the KLCS Sample for Analysis}
\label{sec:final_sample}

 In addition to the procedures described above to ensure that the
observed sample is as free as possible from background subtraction
systematics, we carefully examined all stacked 2-D spectrograms and
extracted 1-D spectra for the sample of 136 observed galaxies
to check for remaining issues that might compromise measurements
of residual LyC flux. 
The following criteria were considered serious enough to warrant removing
galaxies from the analysis sample:
  
1) Instrumental defects were apparent in the two-dimensional
 spectrograms. As discussed above, the effects of non-uniform scattered
light (\S\ref{sec:ghosts}) and/or irreparable slit illumination
 irregularities (\S\ref{sec:slitfunc}) can negatively impact the quality of
the flat-fielding and background subtraction stages. There were
5 cases in which 
such effects were judged to be relevant, 1 in each of the Q0100, Q0256, and B20902
fields, and 2 on mask q0933\_L2 (pre-ADC) in the Q0933 field; all 5 were removed from
the KLCS analysis sample.

2)  The primary target galaxy appeared to have another object near enough 
on the slit that the light from the two objects could not be 
separated with confidence in the 1-D extraction; 
3 such cases were identified (one object from each of Q0100, B20902, and Q0933 [mask q0933\_L2]), 
and removed from the sample. 

\begin{figure}[htbp!]
\centering
\includegraphics[width=8.5cm]{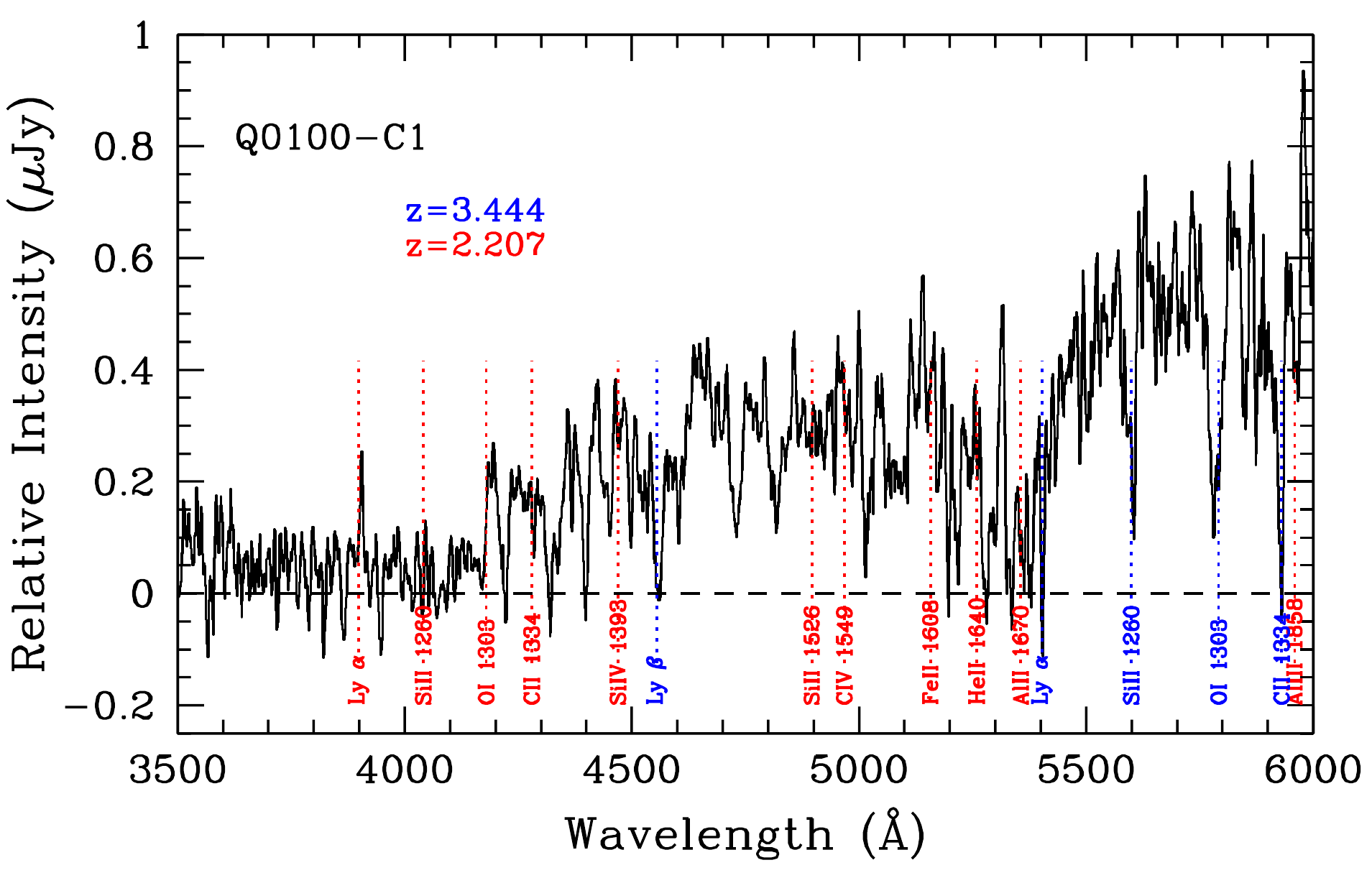}
\includegraphics[width=8.5cm]{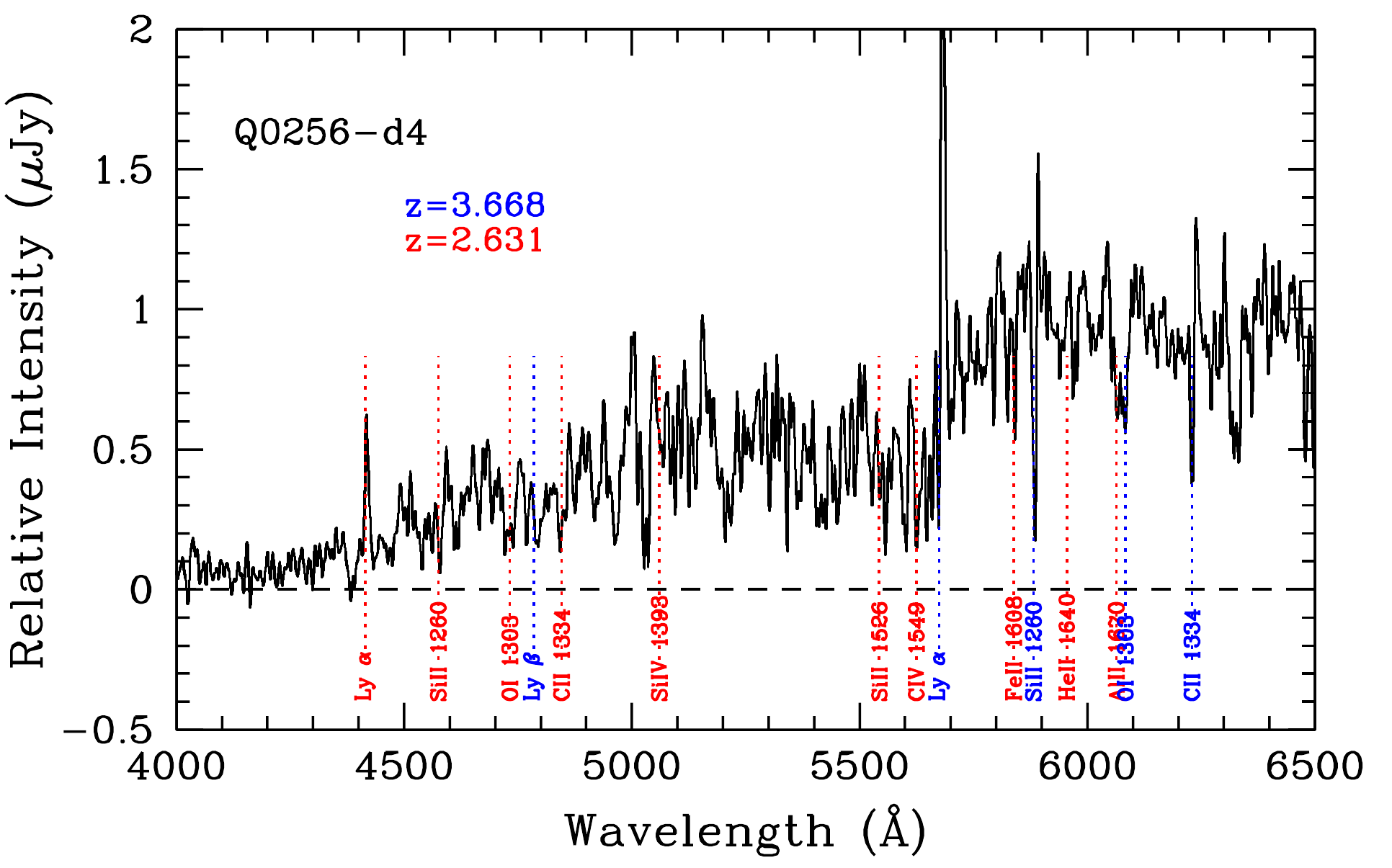}
\includegraphics[width=8.5cm]{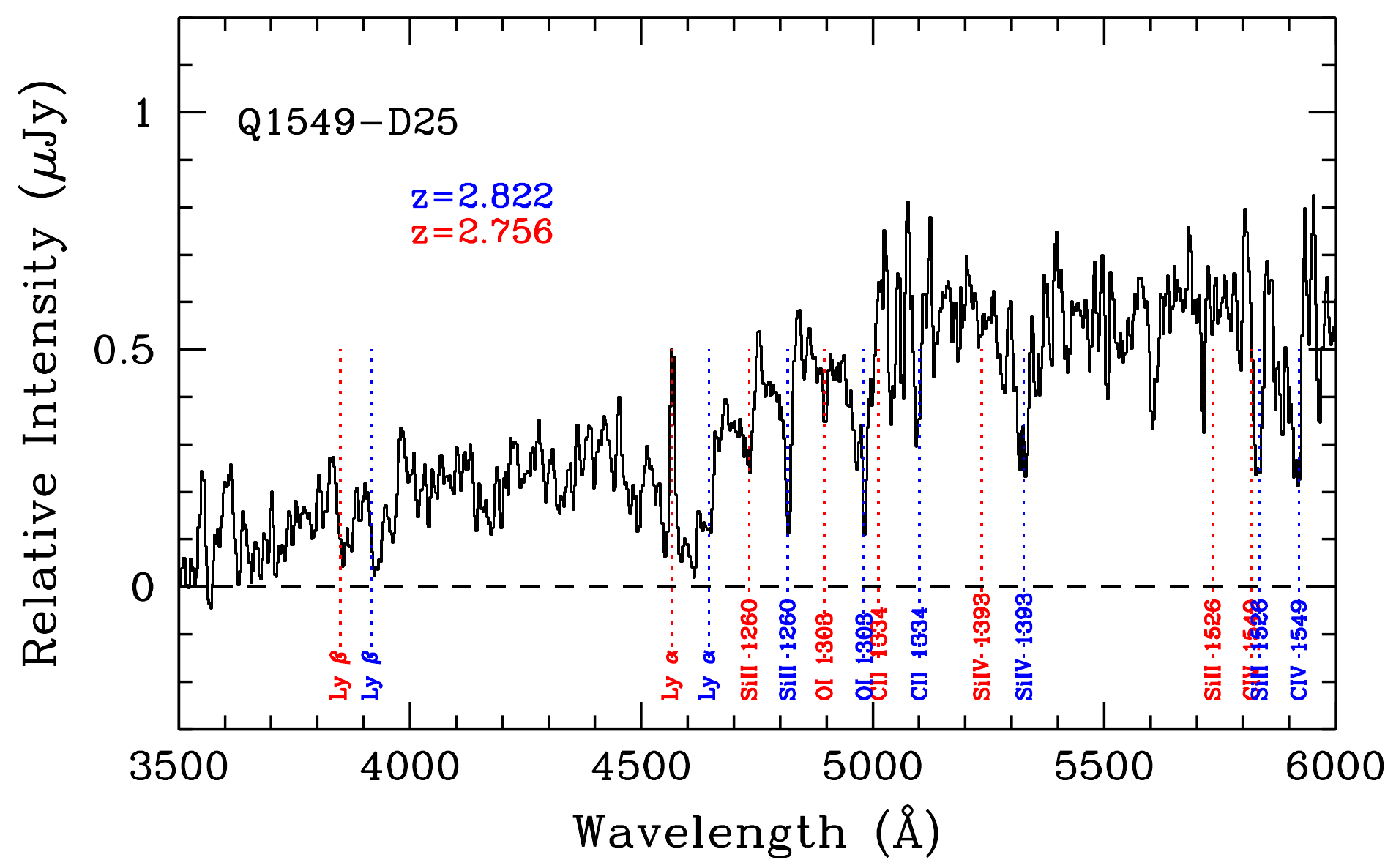} 
\caption {Three examples of targets identified as spectroscopic blends of two redshifts within
the 1-D extraction footprint of the primary target. Each panel has line identifications marked
with colors corresponding to the two redshifts indicated (lower redshift in red, the higher redshift
in blue). Since the lower-redshift features
in the top 2 panels are sufficiently low that non-ionizing flux would contaminate the rest-frame
LyC region of the higher redshift source, they were removed from the analysis sample. The bottom
panel, which shows a 
    spectroscopic blend of galaxies with $z=2.756$ and $z=2.822$, was retained in the sample but 
assigned the lower of the two redshifts. 
    }
\label{fig:d25_blend}
\end{figure}

3) A target was revealed to be a spectroscopic ``blend'' of
 unrelated objects, where the foreground object has 
the potential to cause a spurious (false-positive) detection of LyC flux
if the source redshift is assumed to be the higher of the two redshifts. 
We identified 7 cases of spectroscopic blends, of which 
5 were removed from the sample: Q0100-C1 ($z=3.44/2.21$; see Figure~\ref{fig:d25_blend}), Q0256-d4 ($z=3.67/2.63$; see Figure~\ref{fig:d25_blend}), 
Q0256-m11 ($z=3.090/2.09$), Q0256-md34 ($z=3.130/1.870$),
and Q1009-C41 ($z=3.62/3.22/1.9$). 
The 5 discarded objects, had they been analyzed without the spectroscopic identification 
of the potential contamination, would all have been classified as
LyC detections, with significance ranging from $3\sigma-7\sigma$ and flux
level $f_{900} \simeq 0.050$ $\mu$Jy (m$_{\rm AB}[LyC] \simeq 27.2$).  
In the remaining two cases, multiple redshifts were identified
in the spectrum but the lower of the two redshifts was sufficiently high 
to allow a LyC measurement (i.e., $z \ge 2.75$): these are Q0933-M23 ($z=3.380/3.289$)
and Q1549-D25 ($z=2.822/2.755$; see Figure~\ref{fig:d25_blend}). 
In both cases the target was retained in the sample, 
but the lower of the two redshifts was assigned for subsequent analysis.\footnote{We verified that none of the results of subsequent analysis 
depends significantly on whether or not these two blended sources are included in the final sample.}

Thus, in total, 13 of 136 galaxy targets were removed from the final KLCS sample\footnote{One additional source was identified
on the slit targeting Q0256-m9 ($z=3.28$), $\simeq 2$\arcs\ from the primary and having a nearly identical redshift. Both sources
were subsequently included, called Q0256-m9ap2 and Q0256-m9ap3 (see Table~\ref{tab:klcs}.)}.
The 124 galaxies remaining in the final sample are listed
(along with their coordinates, redshifts, and optical photometry) in Table~\ref{tab:klcs}. 

\begin{deluxetable*}{llllccccrrlrlr}
\tabletypesize{\scriptsize}
\tablewidth{0pc}
\tablecaption{The Keck Lyman Continuum Spectroscopic Sample}
\tablehead{
\colhead{Name} & \colhead{RA} & \colhead{Dec} &
 \colhead{$z_{\rm sys}$} & \colhead{${\cal R}$} & \colhead{$G-{\cal R}$} & \colhead{$U_n-G$} & \colhead{$(G-{\cal R})_{0}$\tablenotemark{a}} & 
\colhead{\wlya\tablenotemark{b}} &  
 \colhead{$f_{900}$} & \colhead{$\sigma_{900}$} &
 \colhead{$f_{1500}$} & \colhead{$\sigma_{1500}$} & \colhead{$(f_{900}/f_{1500})_{\rm obs}$\tablenotemark{c}} \\
 \colhead{} & \colhead{(J2000)} & \colhead{(J2000)} &
 \colhead{} & \colhead{(AB)} & \colhead{(AB)} & \colhead{(AB)} & \colhead{(AB)} & \colhead{(\AA)} & \colhead{($\mu$Jy)} &
 \colhead{($\mu$Jy)} & \colhead{($\mu$Jy)} & \colhead{($\mu$Jy)} 
}
\startdata
     Q1422-c101 &  14:24:42.28  & 22:58:37.86 &  2.8767 &24.17 & 0.85 & 3.79 & 0.73 &-16.5 &  0.003& 0.014& 0.599& 0.012&$ 0.005 \pm0.023$ \\ 
      Q1422-c49 &  14:24:36.11  & 22:52:41.34 &  3.1827 &24.89 & 0.44 & 3.66 & 0.13 &  6.6 & -0.004& 0.013& 0.327& 0.014&$-0.012 \pm0.040$ \\ 
      Q1422-c63 &  14:24:30.18  & 22:53:56.21 &  3.0591 &25.85 & 0.64 & 2.60 & 0.44 & -0.1 & -0.001& 0.014& 0.202& 0.013&$-0.005 \pm0.069 $\\ 
      Q1422-c70 &  14:24:33.63  & 22:54:55.22 &  3.1286 &25.45 & 0.92 & 2.72 & 0.68 &  6.3 &  0.005& 0.014& 0.248& 0.013&$ 0.020 \pm0.056 $\\ 
      Q1422-c84 &  14:24:46.16  & 22:56:51.48 &  2.9754 &24.70 & 0.88 & 3.21 & 0.69 &-13.4 & -0.001& 0.012& 0.383& 0.015&$-0.003 \pm0.031 $\\ 
      Q1422-d42 &  14:24:27.72  & 22:53:50.71 &  3.1369 &25.32 & 0.62 & 2.69 & 0.31 &-10.5 &  0.053& 0.014& 0.373& 0.012&$ 0.142 \pm0.038 $\\ 
      Q1422-d45 &  14:24:32.21  & 22:54:03.02 &  3.0717 &24.11 & 0.31 & 2.49 & 0.11 &  0.3 &  0.000& 0.014& 0.953& 0.014&$ 0.000 \pm0.015 $\\ 
      Q1422-d53 &  14:24:25.53  & 22:55:00.50 &  3.0864 &24.23 & 0.83 & 2.57 & 0.59 & -5.8 &  0.025& 0.014& 0.425& 0.012&$ 0.059 \pm0.033 $\\ 
      Q1422-d57 &  14:24:43.25  & 22:56:06.68 &  2.9461 &25.71 & 0.41 & 2.42 & 0.49 & 52.8 &  0.042& 0.014& 0.143& 0.016&$ 0.294 \pm0.103 $\\ 
      Q1422-d68 &  14:24:32.94  & 22:58:29.13 &  3.2865 &24.72 & 0.39 & 2.56 & 0.55 &153.4 &  0.066& 0.009& 0.346& 0.017&$ 0.191 \pm0.028 $\\ 
      Q1422-d78 &  14:24:40.49  & 22:59:35.30 &  3.1026 &23.77 & 0.95 & 3.40 & 0.74 &  6.3 & -0.008& 0.011& 0.655& 0.012&$-0.012 \pm0.017 $\\ 
      Q1422-d81 &  14:24:31.45  & 22:59:51.57 &  3.1016 &23.41 & 0.51 & 3.53 & 0.53 & 67.1 &  0.025& 0.011& 0.814& 0.015&$ 0.031\pm 0.014 $\\ 
    Q1422-md119 &  14:24:36.18  & 22:55:40.31 &  2.7506 &24.99 & 0.76 & 2.04 & 0.77 & -5.0 &  0.025& 0.025& 0.298& 0.012&$ 0.084\pm 0.084 $\\ 
    Q1422-md145 &  14:24:35.54  & 22:57:19.42 &  2.7998 &24.89 & 0.95 & 2.09 & 0.99 & 13.4 & -0.018& 0.016& 0.220& 0.012&$-0.082\pm 0.073 $\\ 
    Q1422-md152 &  14:24:46.08  & 22:57:52.91 &  2.9407 &24.06 & 1.18 & 2.20 & 1.04 & -8.4 &  0.016& 0.013& 0.397& 0.012&$ 0.040\pm 0.033 $ 
\enddata
\label{tab:klcs}
\tablecomments{Table 2 is published in its entirety in the machine-readable format.
      A portion is shown here for guidance regarding its form and content.}
%\tablenotetext{*}{The complete version of this table will be available in the electronic version
%of the paper. The portion shown here is to provide information on its form and content. }
\tablenotetext{a}{Color after correction for IGM line blanketing and the contribution of \lya\ to the observed $G$ band (see text).}
\tablenotetext{b}{Rest-frame \lya\ equivalent width in \AA, where positive values indicate net emission.}
\tablenotetext{c}{Observed ratio of $f_{900}$ to $f_{1500}$ and its propagated uncertainty.}
\end{deluxetable*}

\section{Measurements of Individual KLCS Spectra}

\label{sec:measure}

\subsection{Residual Interloper Contamination}
\label{sec:contamination}

As discussed in \S\ref{sec:klcs_intro}, foreground contamination leading to false-positive detection
of LyC flux is a major concern for any putative detection of ionizing radiation from high 
redshift galaxies, and recent experience has shown that candidate LyC detections based
on ground-based imaging surveys should be viewed with caution until additional observations -- particularly
multi-band {\it HST} imaging -- can confirm the association of the detected flux with the known $z \sim 3$ galaxy 
or is more likely to be due to an unrelated object at a different (lower) redshift. 
In this section, we estimate the likelihood that our triage of the 1-D KLCS spectra has successfully identified most of the
contaminated sources that would lead to false positive detections of LyC signal. 

Recently, imaging surveys for LyC (\citealt{siana07,inoue08,nestor11,vanzella12,mostardi13}) have used 
Monte Carlo simulations based on 
the surface density of objects in a deep $U$ band images to estimate the probability that a random faint
galaxy with (e.g.) $z<2.75$ would fall close enough to the centroid position of a targeted galaxy
to cause a false-positive LyC detection. 
The probability 
of a chance superposition increases as the sensitivity limit for LyC detections becomes more sensitive; 
any galaxy with UV continuum surface brightness exceeding the typical statistical detection threshold over the
bandpass used for LyC sensing 
is a potential source of contamination. 

For the particular case of UV-color-selected LBGs at $z\sim3$ -- which require the presence of a photometric break between the
observed $U_{\rm n}$ (3550/600) and $G$ (4730/1100) and  a relatively blue color between $G$ and ${\cal R}$ (6830/1000) to have
been selected for observation in the first place --  
the most likely sources of contamination are flat-spectrum (i.e., zero color on the AB magnitude system) 
galaxies with apparent magnitudes bright enough to
yield a photometric or spectroscopic detection at $\lambda_{\rm obs} \simeq 3500$ \AA, but faint enough that
the resulting perturbation of the $U_nG{\cal R}$ photometry does not scatter the object out of the color
selection window. The KLCS spectroscopically-observed sample has apparent $U_n \ge 26.13$\footnote{Only 15\% have $U_n < 27$.}, median $\langle U_n \rangle = 27.7$, 
and typical spectroscopic detection limits in the LyC detection band (rest-wavelength range [880,910] \AA)  of $\sigma_{900} \simeq 0.01 \mu$Jy ($10^{-31}$ ergs s$^{-1}$ cm$^{-2}$ Hz$^{-1}$, or  $m_{\rm AB}[{\rm LyC}] = 28.9$;  
see Figure~\ref{fig:s900} and Table~\ref{tab:klcs}.) 

If we consider objects in the magnitude range $26 < m_{\rm AB}(3500)  < 28.0$ as the most likely potential 
contaminants of the KLCS sample, we can use our deepest available UV images obtained under seeing conditions similar
to those of 
the KLCS spectroscopic observations\footnote{These include 
the NB3420 image in the HS1549 field (\citealt{mostardi13}), NB3640 in SSA22a (\citealt{nestor11}), and the $U_{\rm n}$ images
in the Q1422+23 field (\citealt{steidel03}) and Q0821+31 fields (KBSS; \citealt{rudie12a,reddy12,steidel14}.)} to estimate
the probability of contamination of the KLCS slit apertures.  
The average surface density of detections in the range $26 \le m_{\rm AB}[3500]$\AA\ $\le  28$ is $\Sigma_{\rm avg} = 88.7\pm2.4$ arcmin$^{-2}$, where
the uncertainty represents the scatter in $\Sigma_{\rm avg}$ among the 4 fields.   
Each KLCS spectroscopic extraction aperture subtends 1\secpoint2 by 1\secpoint35, or a solid angle
of 1.6 arcsec$^{2}$ ($4.4 \times 10^{-4}$ arcmin$^{-2}$). The probability that the centroid of a $m_{\rm AB}  = 26-28$ 
object (at any redshift) falls within the 
extraction aperture for any single KLCS target is $P \simeq 88.7 \times 4.4\times10^{-4} = 0.039$; thus, in a sample of
128, we expect $N_c \sim 128 \times 0.039 \simeq 5$ will be affected by such contamination. 

While this type of argument cannot exclude the possibility that there remain unidentified false-positive detections 
in the KLCS sample, the fact that the 
number of slit apertures predicted to be affected by chance superposition of UV-faint foreground galaxies is similar  
to the number of spectra identified as blends 
with foreground galaxies 
suggests that our spectroscopic ``triage'' has likely removed most
of the contaminants that would lead to false positive LyC detections. 

\subsection{Galaxy Systemic Redshifts}
\label{sec:zsys}

Because our primary goal is to measure the residual flux averaged
over a relatively broad window in rest-wavelength, precise systemic redshift measurements are not critical. 
However, since we will be combining individual
spectra into composite stacks (\S\ref{sec:composites}), 
we assigned our best estimate of $z_{\rm s}$ for each galaxy based on the
information in hand. 
Of the 124 sources in the KLCS analysis sample, 55 (44\%) have measurements of nebular
[\ion{O}{3}] emission lines in the rest-frame optical obtained using Keck/MOSFIRE (\citealt{mclean12,steidel14}.) 
In all such cases, the measured $z_{\rm neb}$ is assumed to define the systemic redshift $z_{\rm sys}$,
with an uncertainty of $\simeq 15-20$ \kms.  
 
For objects lacking nebular emission line measurements, we used estimates of $z_{\rm sys}$ based
on the full KBSS-MOSFIRE sample (\citealt{steidel14,strom17}) with $z > 2.0$ and existing high-quality rest-UV spectra.
These were used to calibrate relationships
between $z_{\rm sys}$ and redshifts measured from features in the rest-frame FUV spectra, 
strong interstellar absorption lines ($z_{\rm IS}$) and/or the centroid of Lyman-$\alpha$ emission (\zla).     
As for previous estimates of this kind (e.g., \citealt{adelberger03,steidel2010,rudie12a}), we adopt rules that
depend on the particular combination of features available in each spectrum, where 
spectra fall into one of 3 categories: 
(a) those with measurements of both \zla\ and \zis; (b) those with \zis\ but without \zla\ (generally because the \lya\ feature
appears in absorption); and (c) those with only \zla\ 
available.  

\begin{figure}[htbp!] 
\centerline{\includegraphics[width=8.5cm]{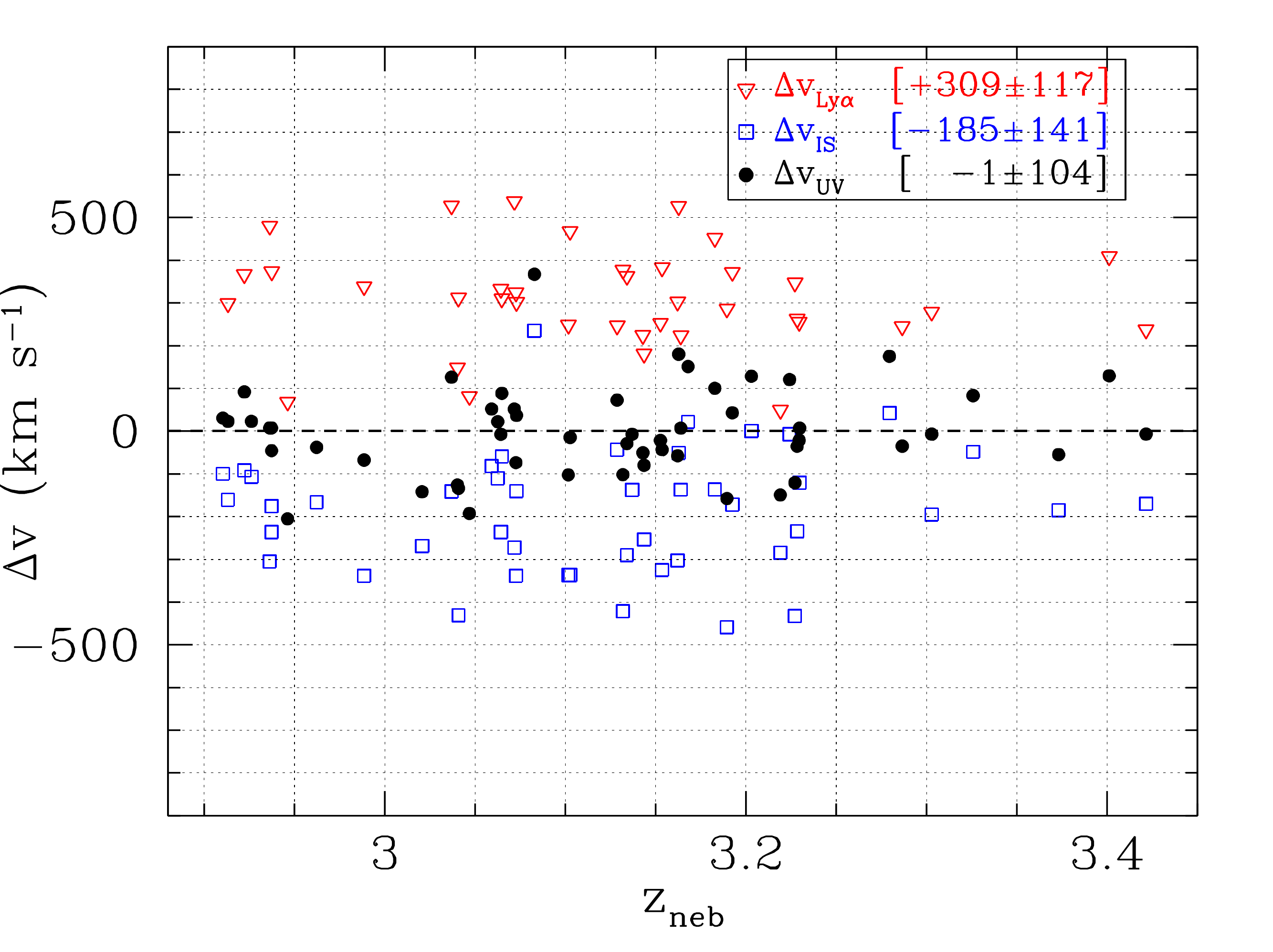}}
\caption{Residuals in the UV-based estimates of $z_{\rm sys}$ (using equations~\ref{eqn:zsys1}-\ref{eqn:zsys3}) (black points) relative to the nebular redshifts for the
55 KLCS galaxies with both UV and nebular redshifts.  The measurements of \zla\ and \zis\ for individual galaxies are represented
by open red triangles and open blue squares, respectively.  The mean residual velocity offset $\langle \Delta v_{\rm uv} \rangle =
-1 \pm 14$ \kms, with rms$= 104$ \kms.  }   
\label{fig:comp_zsys}
\end{figure}

For galaxies in category (a), comprising 
$\simeq 60$\% of the KLCS
sample, 
\begin{eqnarray}
z_{\rm sys} = \zis + 0.40 (\zla-\zis) ~;
\label{eqn:zsys1}
\end{eqnarray} 
for category (b) ($\simeq 30$\% of the KLCS
sample) :
\begin{eqnarray}
z_{\rm sys} = \zis + \frac{\Delta v_{\rm IS}}{c} (1+\zis) ,  
\label{eqn:zsys2}
\end{eqnarray}
with $\dvis\ = 130$ \kms\  ; 
for category (c) ($\simeq 10$\% of the KLCS
sample): 
\begin{eqnarray}
z_{\rm sys} = \zla - \frac{\dvla}{c} (1+\zla)~ ; 
\label{eqn:zsys3}
\end{eqnarray}
with $\dvla = 235$ \kms. 
The residual velocity offset and rms errors after applying these rules to the UV spectra
of the 55 galaxies with measurements of $z_{\rm neb}$ (Figure~\ref{fig:comp_zsys}) 
is 
\begin{equation}
\langle c (z_{\rm sys,uv} - z_{\rm neb})/(1+z_{\rm neb}) \rangle = 
-1 \pm 104~ \kms,   
\end{equation}
where the quantity inside the angle brackets is equivalent to $\Delta v_{\rm UV}$ in Figure~\ref{fig:comp_zsys}. 

The best available value of $z_{\rm sys}$ is given for each galaxy in Table~\ref{tab:klcs}. 

\subsection{LyC Measurements}

\label{sec:measurements}

All quantitative measurements or limits on the flux density
of residual LyC emission 
have been evaluated over a fixed bandpass in the source rest frame, placed
just shortward of the Lyman limit: 
\begin{equation}
\label{eqn:f900_def}
f_{900} \equiv \langle f_{\nu}(\lambda_0)\rangle\quad;\quad 880 \le \lambda_0/{\textrm \AA} \le 910~.
\end{equation}
We also define a rest-frame bandpass to represent the FUV non-ionizing flux density, 
\begin{equation}
\label{eqn:f1500_def}
f_{1500} \equiv \langle f_{\nu}(\lambda_0)\rangle\quad;\quad 1475 \le \lambda_0/{\textrm \AA} \le 1525~.
\end{equation}

The choice of the interval [880,910] for the LyC measurement is a compromise based on two considerations.
First, 
ideally the LyC measurement should be made as close as possible to the rest-frame Lyman limit
of the galaxy, and integrated over the smallest wavelength
range feasible given the noise level of the spectra, in order to minimize the effects of 
\ion{H}{1} in the IGM along the line of sight.  
As discussed in detail in \S\ref{sec:igm_trans}, the opacity of the IGM to LyC photons is 
the largest source of uncertainty in the measurement of the emergent ionizing
photon flux from a galaxy. 
The mean free path of H-ionizing photons emitted at $\langle z_{\rm s} \rangle = 3.05$ corresponds to
a redshift interval of   
$ \Delta z \sim 0.25$ (e.g., \citealt{rudie13}), or 
to $\sim 60$\,\AA\space in the rest frame; i.e., the flux density at $\lambda_0 \simeq 850$ \AA\ is reduced
by an average factor of e ($\sim 2.72$) relative to its emergent value.  
By using the [880,910] interval, to a first approximation the average IGM LyC optical depth   
due to intervening material is $\langle \tau_{\rm igm} \rangle \simlt 0.5$. 

An additional consideration is the wavelength range over which the 
Keck/LRIS-B system sensitivity remains high. Although the UV sensitivity of Keck/LRIS-B is 
the highest among instruments of its kind (\citealt{steidel04}), it decreases with wavelength shortward of 3500\,\AA, particularly
when atmospheric opacity is included. 
The lowest redshift source included in KLCS has 
$z_{\rm s}=2.718$ (Q0100-MD6), placing $\lambda_0 = 880$\,\AA\space at an observed wavelength
of 3272\,\AA\space, where the LRIS-B end-to-end throughput with the d500 beamsplitter
has dropped to $\simeq 20$\% from $\simgt 50$\% near 4000 \AA. 
Including the atmosphere, these numbers reduce to $\sim 11$\% and $\sim 40$\% at a typical
airmass of 1.10 on Mauna Kea. At the mean redshift of the KLCS sample ($\langle z_{\rm s} \rangle = 3.05$), 
[880,910] corresponds to an observed-frame interval [3560,3690] \AA, where the instrumental throughput
averages $\simeq 35$\%. 

Thus, [880,910] is observable over the full range of KLCS, and is narrow enough to minimize the IGM opacity 
against which escaping flux must be detected, but sufficiently broad to allow for improved photon statistics 
while reducing the fluctuations due to discrete \ion{H}{1} absorption lines arising from intervening material,  
superposed on the rest-frame LyC.   

\begin{figure}[htbp]
   \centering
   \includegraphics[width=8.5cm]{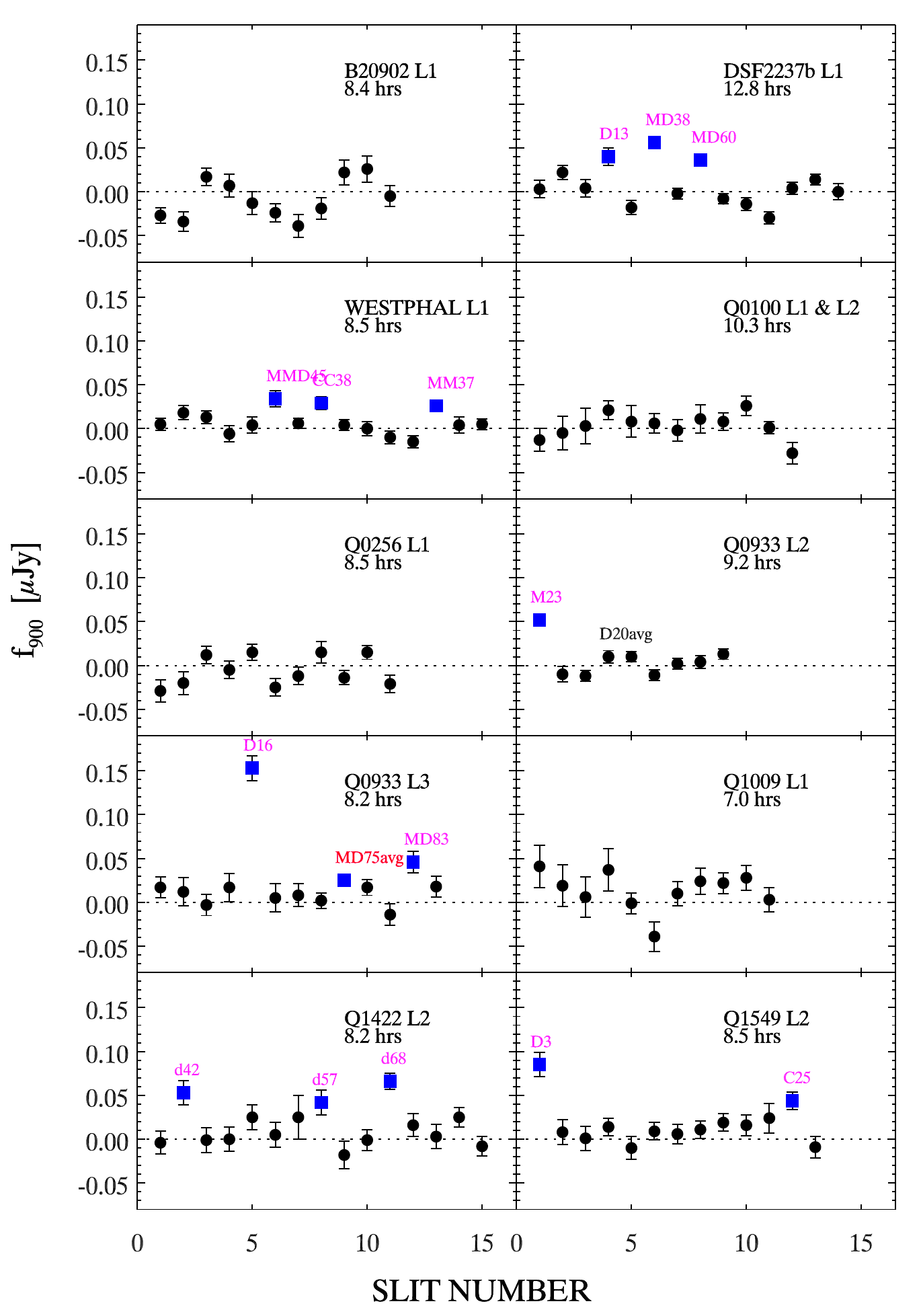}
    \caption{Measurements of the average flux density in the rest-frame
interval [880,910] \AA\ ($f_{900}$) for the 124 LBGs (circles)
      in the KLCS analysis sample.
  Each slit mask used has its own panel, with 
the order of objects following their relative physical placement on the slitmask.
The total exposure time in hours for each mask is 
   indicated under the mask name. 
The errors shown are $\pm1\sigma$ statistical errors derived
accounting for counting statistics and detector noise.
A total of 15 out of 124 galaxies are 
nominal detections, with $f_{900} >  3\sigma_{900}$; these are marked with
square (blue) symbols. 
Note that galaxies Q0933-MD75 and Q0933-D20
were observed twice, on masks q0933\_L2 and q0933\_L3-- points labeled in red represent
the weighted average measurement from the two independent spectra.}
  \label{fig:allmasks}
  \end{figure}

Figure\,\ref{fig:allmasks} presents the measured values of $f_{900}$,
with objects grouped and ordered according to slitmask and the  
relative physical position on the mask (along the slit direction) for each target. Targets having $>3 \sigma$ detections of $f_{900}$ are labeled.   
The values of $f_{900}$ and their
associated statistical error ($\sigma_{900}$) for KLCS galaxies
are listed in Table~\ref{tab:klcs}.  
Also given in Table~\ref{tab:klcs} are measurements of the non-ionizing UV continuum flux density
$f_{1500}$ (equation \ref{eqn:f1500_def}) measured directly from the individual 1-D flux-calibrated spectra.

The measurements and uncertainties for $f_{900}$ and $f_{1500}$ [and their ratio ($f_{900}/f_{1500})_{\rm obs}$] were
obtained directly from the 1-D spectra and associated error arrays based on the noise model
described in \S\ref{sec:noise_model}; we have made no attempt to apply aperture corrections to the spectroscopic
flux density measurements (the photometric measurements in the $U_{\rm n}G{\cal R}$ system are also provided in Table~\ref{tab:klcs})
but we believe that the relative spectrophotometry of the 1D spectra has systematic errors $\simlt 5$\%; uncertainties in the
calibration of LRIS-R relative to LRIS-B spectra may contribute additional $\simlt 10\%$ systematic uncertainty in $(f_{900}/f_{1500})_{\rm obs}$.  

The flux error-bars in Figure~\ref{fig:allmasks} are 
statistical errors based on the noise model presented in \S\ref{sec:noise_model},
with typical values of $\sigma_{900}=0.01$ $\mu$Jy; most objects on each 
mask have $f_{900}$ measurements consistent with zero to within 1-2$\sigma_{900}$. 

Grouping observations by slit-mask allows us to monitor 
the presence of systematic errors from mask to mask.
It is also a useful way to 
inspect the data for correlations with object position on a given
slit-mask. It is apparent from Figure~\ref{fig:allmasks} that 
mask B20902-L1 (and possibly others) has residual systematic errors manifesting
as correlated behavior of $f_{900}$ as a function of position on 
the slitmask. 
Although the maximum deviation from zero in the b20902\_L1 panel is only $\sim 2\sigma_{900}$, 
there appear to be systematic undulations relative to zero with amplitude comparable to the random
errors. The systematics appear to have larger excursions to $f_{900} < 0$, as might
occur from over-subtraction of the background due to contamination of slit regions used
to model it by either scattered light (judged to be the dominant factor for mask b20902\_L1),
or by contributions from unmasked sources falling along the slit.  

Systematic over-subtraction of the background level
was also noted for the sample of 14 sources presented 
by \citet{shapley06} (S06) -- these authors found
that the measured $f_{900}$ for objects without significant detections  was
centered around an unphysical negative value (see their Figure 5).
As discussed in detail in \S\ref{sec:redux} (see also Appendix~\ref{sec:2d_bkg}), considerable effort was invested
in improving the flat fielding (with its tendency to exacerbate problems with non-uniform 
scattered light (\S\ref{sec:ghosts}) and 2-D background subtraction. The tests summarized in appendix~\ref{sec:2d_bkg}
suggest that these were generally successful.   
We show below that the procedures have also reduced the residual systematic errors in extracted 1-D spectra 
to a level significantly smaller than the random errors.  

\begin{deluxetable}{lccl}
\tabletypesize{\scriptsize}
\tablewidth{0pt}
\tablecaption{Objects with Individual LyC Detections}
\tablehead{
 \colhead{Object} &  \colhead{$z_{\rm sys}$} & \colhead{($L/L^{\ast}$)\tablenotemark{a}} & \colhead{$(f_{900}/f_{1500})_{\rm obs}$\tablenotemark{b}}}
\startdata 
       Q0933-D16  & 3.0468  &   0.51  &   $0.54     \pm 0.06$ \\ 
       Q0933-M23  & 3.2890  &   0.92  &   $0.26     \pm 0.03$ \\ 
       Q0933-MD75 & 2.9131  &   0.89  &   $0.10\pm      0.02$ \\ 
       Q0933-MD83 & 2.8790  &   0.60  &   $0.15 \pm     0.04$ \\ 
   Westphal-CC38 &  3.0729  &   1.00  &   $0.06 \pm     0.01$ \\ 
   Westphal-MM37 &  3.4215  &   1.23  &   $0.05 \pm     0.01$ \\ 
  Westphal-MMD45 &  2.9366  &   1.42  &   $0.09 \pm     0.02$ \\ 
       Q1422-d42 &  3.1369  &   0.47  &   $0.14  \pm    0.04$ \\ 
       Q1422-d57 &  2.9461  &   0.30  &   $0.30  \pm    0.10$ \\ 
       Q1422-d68 &  3.2865  &   0.88  &   $0.19  \pm    0.03$ \\ 
       Q1549-C25\tablenotemark{c} &  3.1526  &   0.74  &   $0.08  \pm    0.02$ \\ 
       Q1549-D3  &  2.9373  &   1.16  &   $0.06  \pm    0.01 $ \\
    DSF2237b-D13 &  2.9212  &   0.58  &   $0.08  \pm    0.02$  \\
   DSF2237b-MD38 &  3.3278  &   1.47  &   $0.07   \pm   0.01$  \\
   DSF2237b-MD60 &  3.1413  &   0.67  &   $0.09   \pm   0.02$  
\enddata
\tablenotetext{a}{Based on {$\cal R$} magnitude and assuming 
a characteristic  luminosity $L_{\rm uv}$ corresponds to M$_{\rm AB}^{\ast}(1700 \textrm{\AA}) = -21.0$ (\citealt{reddy09}.)}
\tablenotetext{b}{Observed flux ratio. The error estimate includes
  systematic error.}
\tablenotetext{c}{LyC detected object discussed in detail by \citet{shapley16}.}
\label{tab:fesc_data}
\end{deluxetable}
 
To verify that this is the case, 
we excluded all galaxy measurements for which  
$|f_{900}| > 3\sigma_{900}$ where $\sigma_{900}$ is the statistical error estimate.  
For the remaining sample of 106 measurements, $\langle f_{900}/\sigma_{900} \rangle = 0.31 \pm 1.33$, with median 
$f_{900}/\sigma_{900} = 0.44$, and inter-quartile range [-0.50,+1.41]. 
Excluding the two masks (B20902-L1 and Q1009-L1) with the most obvious systematic issues, the results remain largely unchanged: $\langle f_{900}/\sigma_{900}\rangle
= 0.28 \pm 1.32$, with median $0.44$ and inter-quartile range [-0.66,+1.25]. 
Under the null hypothesis that $f_{900} = 0$ and that systematic errors are small compared 
to normally distributed random errors, one expects $\langle f_{900}/\sigma_{900} \rangle = 0.0 \pm 1.0$.
As we shall see, the true values of $f_{900}$ are likely greater than zero for some fraction of the sample with $|f_{900}/\sigma_{900}| < 3$; since 
the standard deviation of $f_{900}/\sigma_{900}$ is only $\sim 30$\% larger than expected under the null hypothesis, 
we believe that residual systematic errors in $f_{900}$ are sub-dominant compared to statistical errors. In 
the remainder of the analysis below, we continue to assume that our model for random errors in the spectra (\S\ref{sec:noise_model}) 
accurately describes the uncertainties. 

\begin{figure}[htbp!]
    \centering
    \includegraphics[width=8cm]{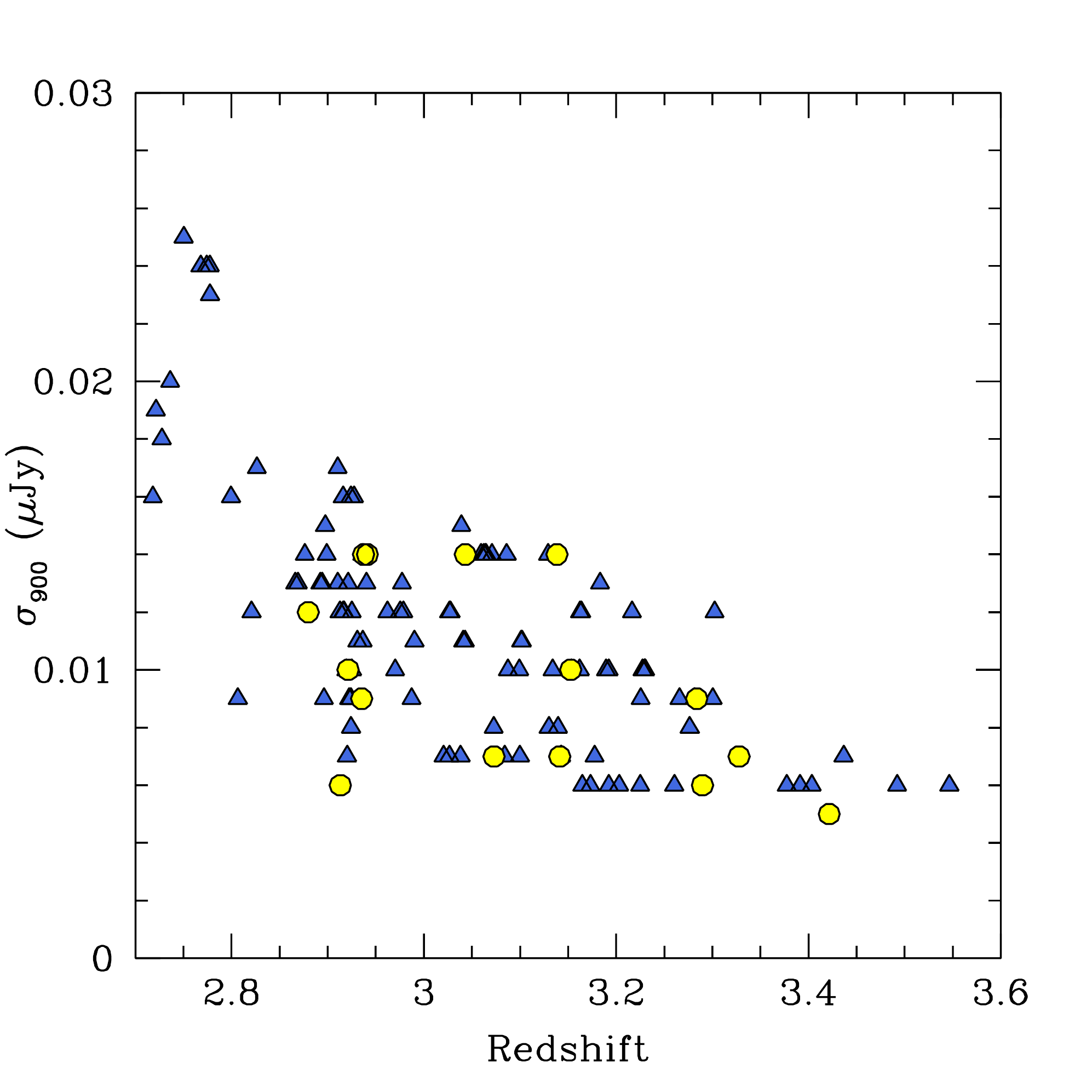}
    \caption{Statistical uncertainties in the measurements of $f_{900}$, 
in units of $\mu$Jy, for all galaxies in the KLCS final sample.  The yellow
circles and blue triangles represent formally detected and formally
undetected objects, respectively.  
    }
  \label{fig:s900}
  \end{figure}
\begin{figure}[htbp!]
\centerline{\includegraphics[width=8cm]{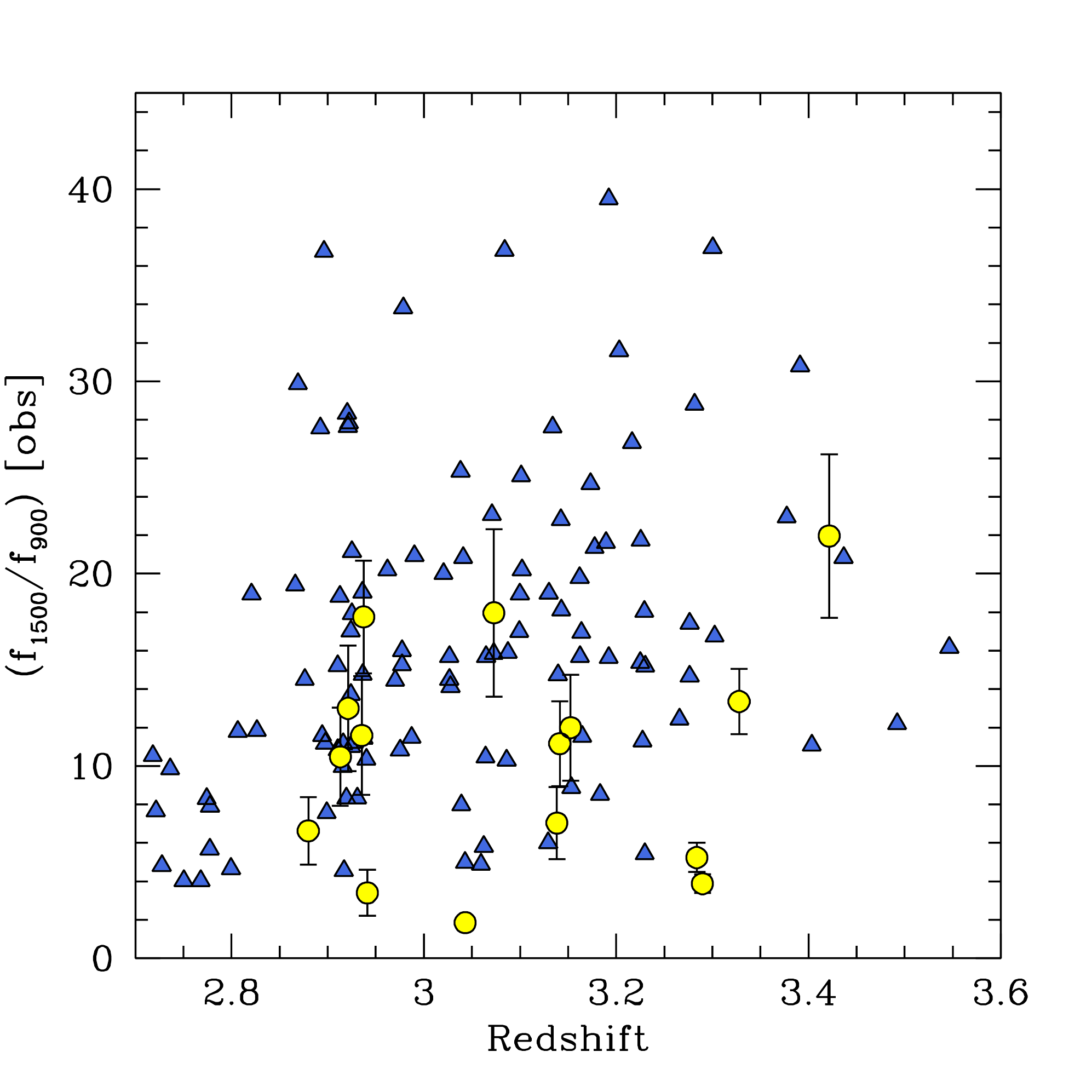}}
\caption{Lower limits ($3\sigma$) on $(f_{1500}/f_{900})_{\rm obs}$ for objects without LyC detections (blue triangles).   
Objects with formal LyC detections are shown as yellow points with $1\sigma$ error bars. 
 } 
\label{fig:f15_f9}
\end{figure}
 
Thus, 15 of 124 galaxies 
have $f_{900} > 3 \sigma_{900}$, which henceforth are referred to
as ``detected''; 
their
properties are listed individually in Table~\ref{tab:fesc_data}. The same objects are marked using blue
squares in Figure~\ref{fig:allmasks}.   
Note that one of the objects in Table~\ref{tab:fesc_data}, Q1549-C25 [$(f_{900}/f_{1500})_{\rm obs} = 0.08$], has been discussed in detail by \citet{shapley16}; in addition
to the LyC flux measurement, it has also been observed with multi-band {\it HST} imaging (see \citealt{mostardi15}) 
which indicates
no evidence for a contaminating foreground source that might have caused a false-positive LyC detection\footnote{There is an approved 
{\it HST} Cycle 25 program (GO-15287, PI: Shapley) to obtain HST imaging for a substantial fraction of the KLCS sample, including all of the objects listed
in Table~\ref{tab:fesc_data}.}.  
The only other confirmed LyC detection of a $z \sim 3$ galaxy is ``Ion-2''  ($z_{\rm s}=3.22$), which has   
$(f_{900}/f_{1500})_{\rm obs} \simeq 0.13$ and rest-frame UV luminosity $\sim L_{\rm uv}^{\ast}$ (\citealt{vanzella15,debarros16}). Most of the objects in Table~\ref{tab:fesc_data} 
have $(f_{900}/f_{1500})_{\rm obs}$ similar to the two confirmed LyC detections\footnote{Most recently (\citealt{vanzella17}), an additional
galaxy with $z_{\rm s} = 3.999$ has been confirmed spectroscopically, with $(f_{900}/f_{1500})_{\rm obs} = 0.052\pm0.011$. After accounting for
the factor of $\sim 1.7$ lower 90th-percentile transmission  at $z_{\rm s} = 4$ compared to $z_{\rm s} = 3.05$ (see Table~\ref{tab:igm_models}), this value is roughly equivalent to
a measurement of $(f_{900}/f_{1500})_{\rm obs} \simeq 0.088 \pm 0.019$ at $\langle z_{\rm s} \rangle = 3.05$, entirely consistent with
the typical KLCS detection listed in Table 2.}.  

Figure~\ref{fig:s900} shows all 124 KLCS targets on a plot of $\sigma_{900}$ versus redshift. Note that
$\sigma_{900}$ shows a trend of increasing toward lower redshift; this is due
to the decreasing instrumental sensitivity at rest-frame wavelengths [880,910] \AA\, which (as we have
seen) changes by a factor of a few over the range $2.75 \simlt z \simlt 3.4$ (where [880,910] falls at observed
wavelengths $3300 \simlt (\lambda/\textrm{\AA}) \simlt 4000$.)  

Figure~\ref{fig:f15_f9} shows a combination of formal detections and -- for the non-detections -- the
adopted 3$\sigma$ lower limits on the ratio $(f_{1500}/f_{900})_{\rm obs}$ using the measurements and
uncertainties given in Table~\ref{tab:klcs}. 
Note that the formally detected objects lie well within the distribution of
$3 \sigma$ upper limits of the non-detections. The implications of these results for the distribution of LyC flux within
the observed sample requires a quantitative assessment of the extent to which \ion{H}{1} gas
outside of galaxies (but along the line of sight) has censored our ability to detect LyC flux 
when it is present; we address this issue in \S\ref{sec:igm_trans}. 

\subsection{LyC Detection vs. Other Galaxy Properties}
\label{sec:uv_color}

In this section we briefly review the properties of galaxies with individual LyC detections versus the majority
that do not have individual detections. We argue below (\S\ref{sec:composites}) that unambiguous interpretation
of LyC detection statistics requires the use of ensembles of galaxies. 

\begin{figure*}[htbp!]
\centerline{\includegraphics[width=5.5cm]{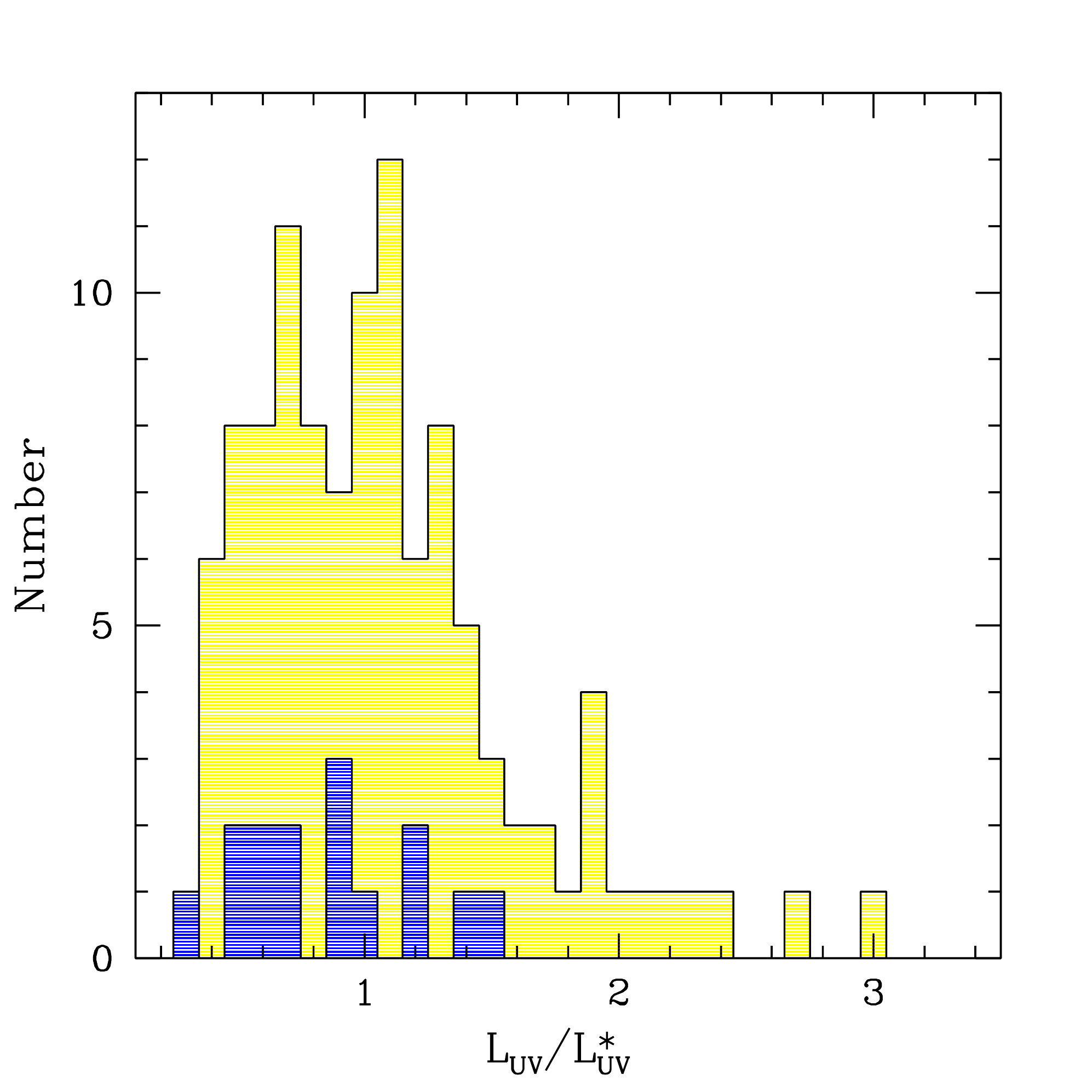}\includegraphics[width=5.5cm]{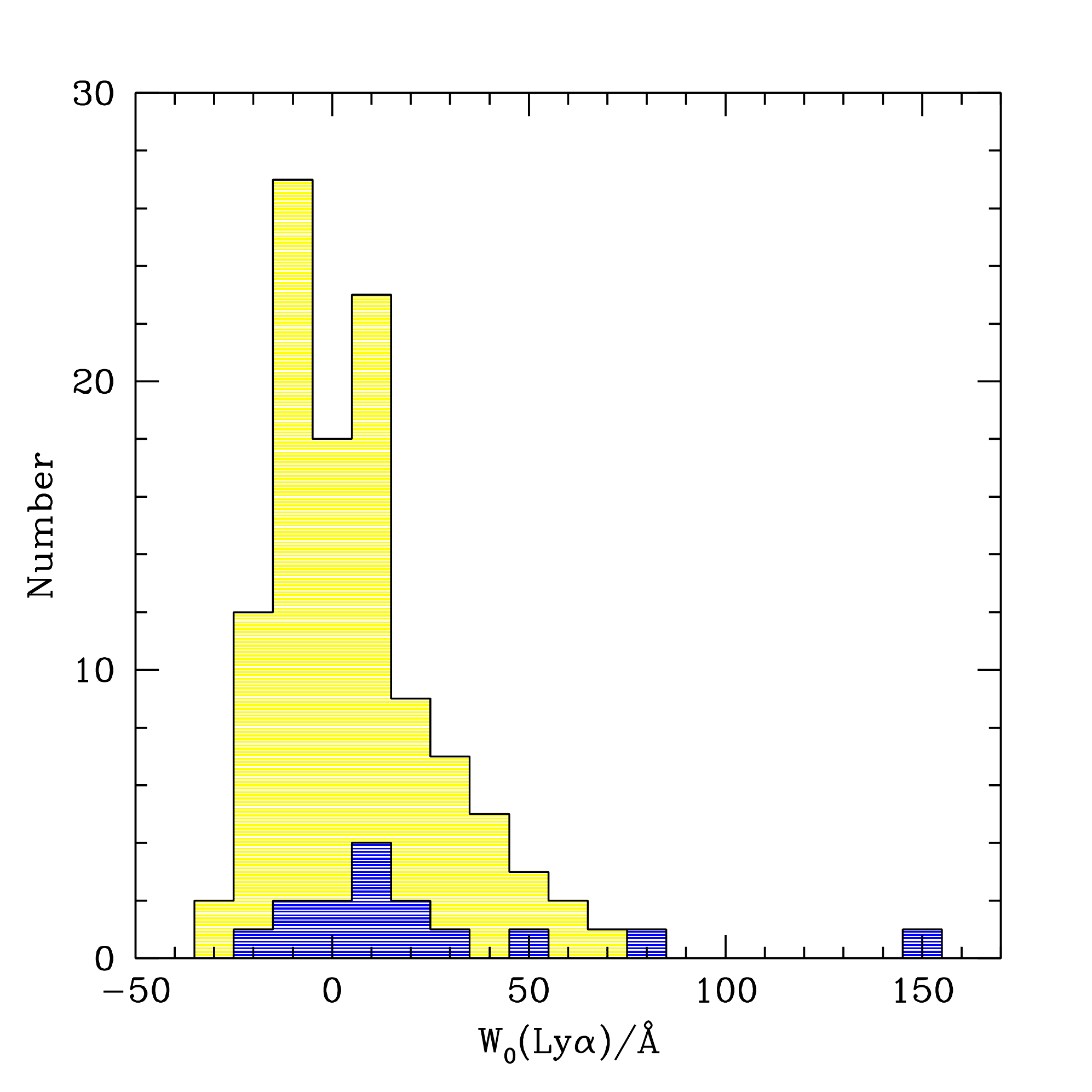}\includegraphics[width=5.5cm]{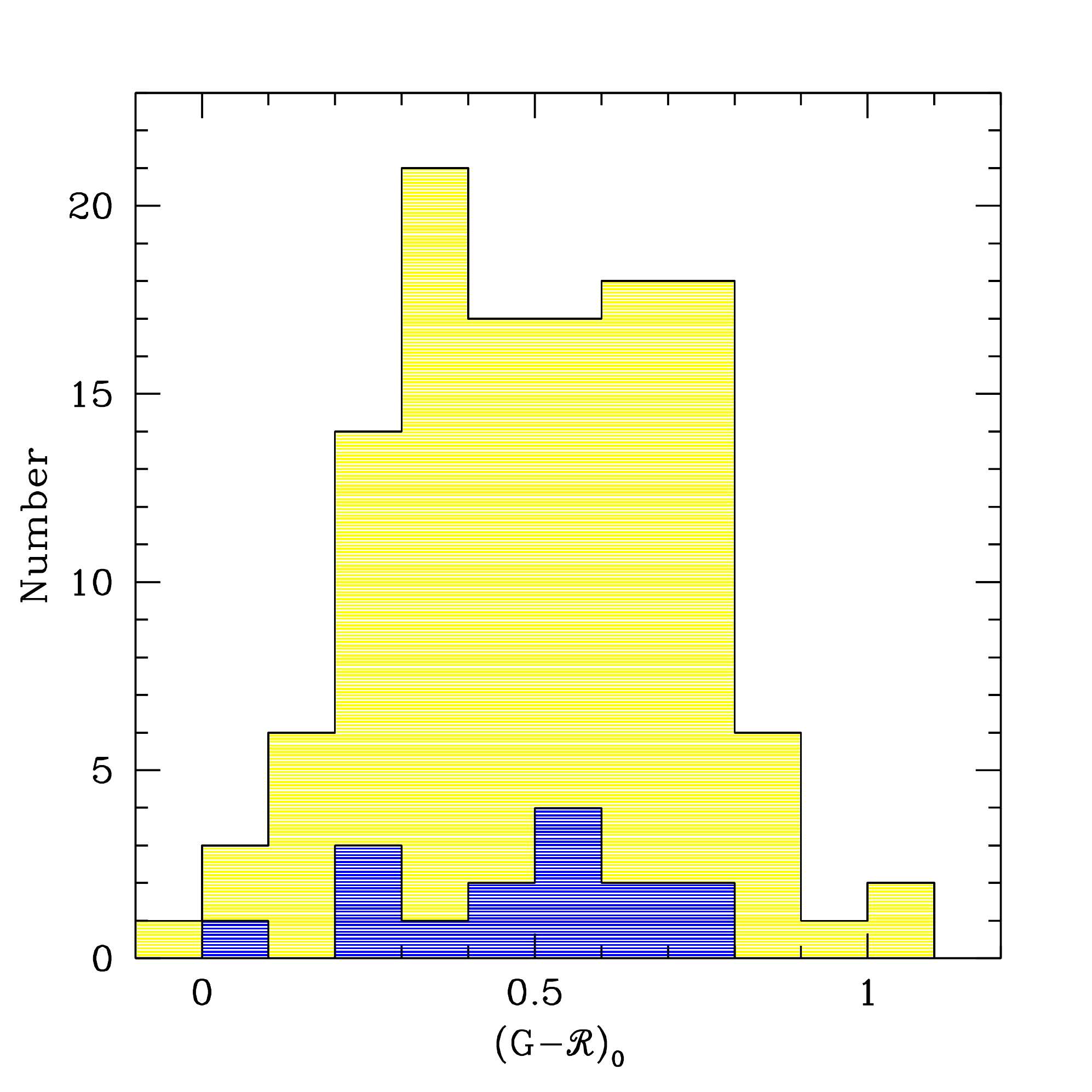}}
\caption{({\it Left:}) Distribution of rest-frame UV luminosity relative to $L^{\ast}_{UV}$ at $z \simeq 3$
for galaxies with individually detected $f_{900}$ (dark/blue) compared to that of the sample with $f_{900} < 3\sigma_{900}$ 
(light/yellow). ({\it Center:}) As for the lefthand panel, comparing the distribution of rest-frame equivalent width of \lya, \wlya.   
({\it Right:}) As for the lefthand panel, comparing the distribution of rest-UV continuum color (see \S\ref{sec:uv_color}). 
2-sample Kolmogorov-Smirnov (KS) tests applied to all 3 cases cannot significantly reject the null hypothesis that the distributions of
formally detected and formally undetected objects are drawn from
the same parent population. }
\label{fig:lum_w0}
\end{figure*}

Figure~\ref{fig:lum_w0} compares the distributions in 3 empirical galaxy properties for the LyC-detected and LyC-undetected 
sub-samples of KLCS (with 15 and 109 objects, respectively.)  The left-most panel of Fig.~\ref{fig:lum_w0} shows
the rest-UV luminosity distribution, relative to $L_{\rm uv}^{\ast}$ in the far-UV (rest-frame 1700 \AA) luminosity
function at $z\sim 3$ (\citealt{reddy09}). As discussed in \S\ref{sec:final_sample} above, most KLCS galaxies
have $L_{\rm uv}$ within a factor of a few of $L_{\rm UV}^{\ast}$.  The sub-sample with formal LyC detections
occupies a slightly narrower range of luminosity, though a two-sample KS test shows that the
two luminosity distributions are statistically consistent with being drawn from the same parent population\footnote{We will
show in \S\ref{sec:composites} below that both UV luminosity and $W_{\lambda}(\lya)$ exhibit clear trends with
$(f_{900}/f_{1500})_{\rm obs}$ based
on more sensitive tests.}.

As one of the most easily-observed and measured spectroscopic characteristics of high redshift star-forming
galaxies, the rest-frame equivalent width of the Lyman-$\alpha$ emission line, $W_{\lambda}(\lya)$, is a useful diagnostic, and has
been shown to correlate strongly with other more subtle spectroscopic features present in the spectra of
LBGs (\citealt{shapley03,kornei10}).  We measured $W_{\lambda}(\lya)$ for the KLCS galaxy sample following the
method described in \citet{kornei10}; the values are listed in Table~\ref{tab:klcs}. 
The center panel of Figure~\ref{fig:lum_w0} (see also Table~\ref{tab:klcs}) shows the distribution of 
$W_{\lambda}(\lya)$ for the full KLCS galaxy sample, divided according to whether or not they are formally detected in the LyC band
[880,910]. 
Although the detected
sub-sample tends  
to have \lya\ in emission, and the fraction of objects with LyC detected appears to be
correlated with \lya\ equivalent width, a two-sample KS test cannot reject the null hypothesis that the sub-samples
are drawn from the same parent population.  

\begin{figure}[htbp!]
\centerline{\includegraphics[width=8.5cm]{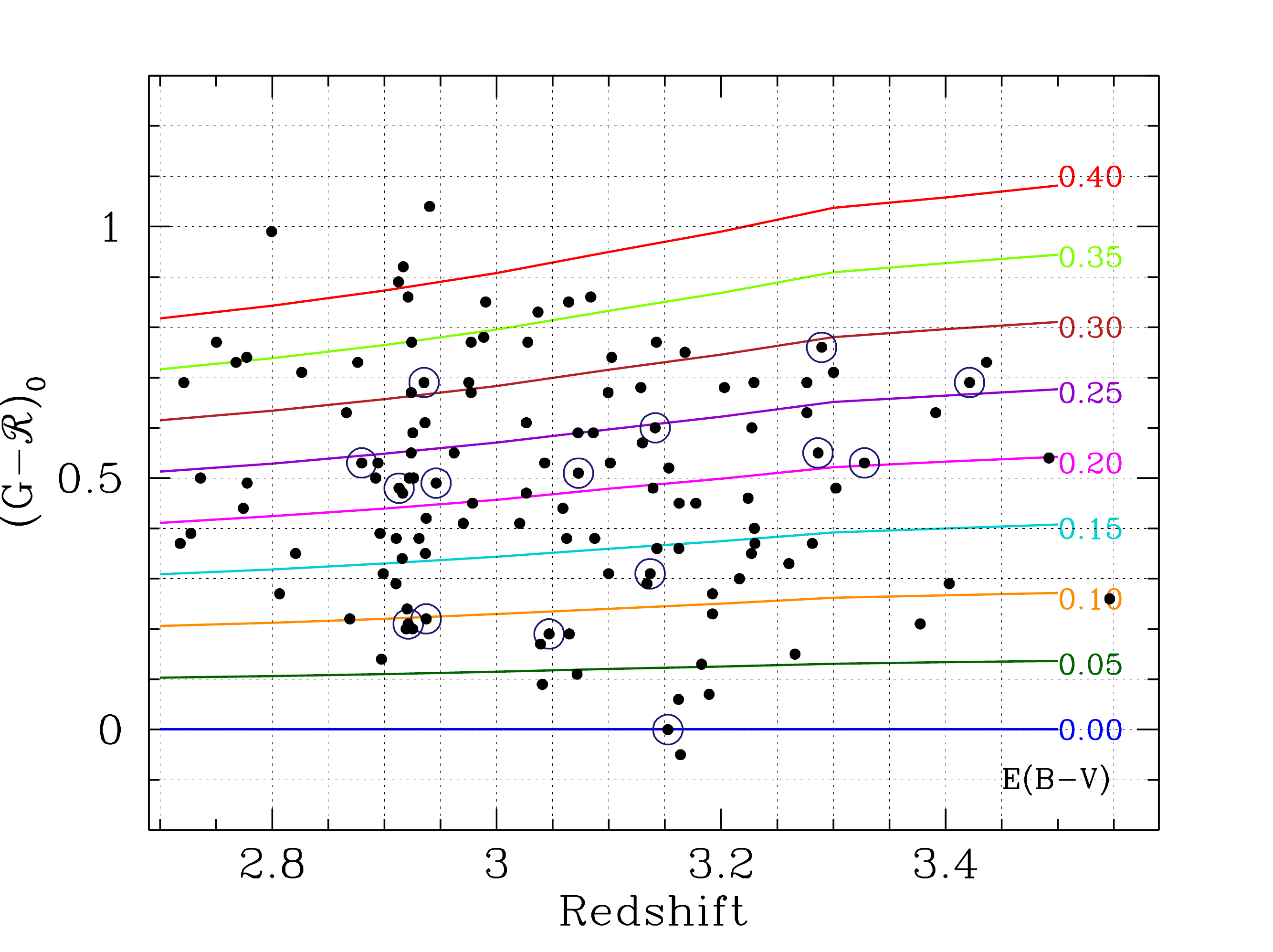}}
\caption{UV color vs. redshift for the KLCS galaxy sample, where the 
$(G-{\cal R})_{0}$ is the observed $G-{\cal R}$ after correcting for the contribution of \lya\ emission or
absorption to the $G$ band, based on each object's spectrum, and accounting for the average line blanketing 
of the $G$ band by the IGM+CGM (see \S\ref{sec:igm_trans}). The color-coded curves are the expected track for
the BPASSv2.1-300bin-Z001 SPS model (\S\ref{sec:sps_models}) after applying reddening according to the \citet{reddy16a} 
attenuation relation with the indicated value of $E(B-V)$. 
The 15 galaxies with $> 3\sigma$ LyC detections are indicated
with larger open circles. 
CGM+IGM . 
}
\label{fig:gmr_vs_z}
\end{figure} 

As described in more detail elsewhere (\citealt{shapley03,steidel03,reddy09}) the $z\sim3$ LBG $U_{\rm n}G{\cal R}$ color selection
imposes small systematic differences in the redshift selection function depending on the intrinsic galaxy properties, in the sense
that intrinsically redder galaxies are less likely to satisfy the selection criteria at the high redshift end of the selection
window. The main reason for this is increased line blanketing
from the \lya\ forest with redshift. Galaxies with very strong \lya\ features
(in either emission or absorption) affect the observed $G-{\cal R}$ color for similar reasons, since \lya\ falls
in the observed $G$ band throughout the KLCS redshift range. 
However, since the full KLCS sample has spectroscopic measurements, we can correct the observed $G - {\cal R}$ color 
of individual galaxies for both effects, thereby yielding estimates of the intrinsic UV continuum color.  Figure~\ref{fig:gmr_vs_z}
shows the measurements for individual KLCS galaxies as a function of redshift in terms of $(G-{\cal R})_0$, the proxy for continuum color after
correction for the mean IGM absorption in the $G$ band and the individual $W_{\lambda}(\lya)$ . 
The KLCS galaxies with individual $>3 \sigma$ LyC detections are circled, and their distribution is compared with the full sample in the 
rightmost panel of Figure~\ref{fig:lum_w0}.  

As for $L_{\rm uv}$ and $W_{\lambda}(\lya)$, the sub-sample having direct individual detections is consistent with being
drawn from the same distribution in $(G-{\cal R})_0$ as the full KLCS sample. 
We discuss the statistical connections between LyC leakage and galaxy properties in more detail in \S\ref{sec:composites}.  

\section{The Opacity of the Intergalactic Medium}

\label{sec:igm_trans}

The opacity of the IGM has been reasonably well-quantified in a statistical
sense (e.g., \citealt{madau95,fauch08,prochaska09,rudie13,inoue14}) from  
high resolution spectroscopic surveys of relatively bright QSOs. 
Modeling the statistics of IGM absorption is 
essential for understanding the implications of any survey seeking to quantify 
the intensity of ionizing photons escaping from high redshift galaxies. 
In order to convert our observations of $(f_{900}/f_{1500})_{\rm obs}$ 
into the more relevant spectrum of emergent ionizing radiation from galaxies, we
used a set of IGM attenuation models using a  
Monte Carlo
technique described by \cite{nestor11} (see also \citealt{bershady99,shapley06,inoue08,inoue14} for
similar models), with \ion{H}{1} distribution function ($f(\nhi,X)$) parameters updated 
based on the KBSS QSO sightlines, including the effects of the CGM (\citealt{rudie13}.) 
The models produce full realizations of individual IGM sightlines toward a source 
with redshift $z_{\rm s}$ by drawing from the empirically-calibrated incidence of intervening \ion{H}{1} 
as a function of \nhi\ and redshift, over the range
$12 \le ({\rm log~(\nhi/cm^{-2}}) \le 22.0$ and $1.6 \le z \le z_{\rm s}$, the details of which are presented in Appendix~\ref{sec:igm_appendix}.
Each simulated spectrum 
includes both line blanketing from Lyman series absorption lines (most relevant for the
low-\nhi\ systems) and LyC opacity (dominated by systems having log~\nhi\ $ \simeq 16 - 18$) 
as a function of observed wavelength over the range $3100 \le \lambda_{\rm obs} \le 1216(1+z_{\rm s})$.   
By creating an ensemble of simulations with the same redshift distribution as the sources in the observed KLCS sample, 
one can make precise statistical statements about the effect of the IGM on the measurements of $f_{\rm 900}$. 

\begin{figure}[htbp]
\centerline{\includegraphics[width=8cm]{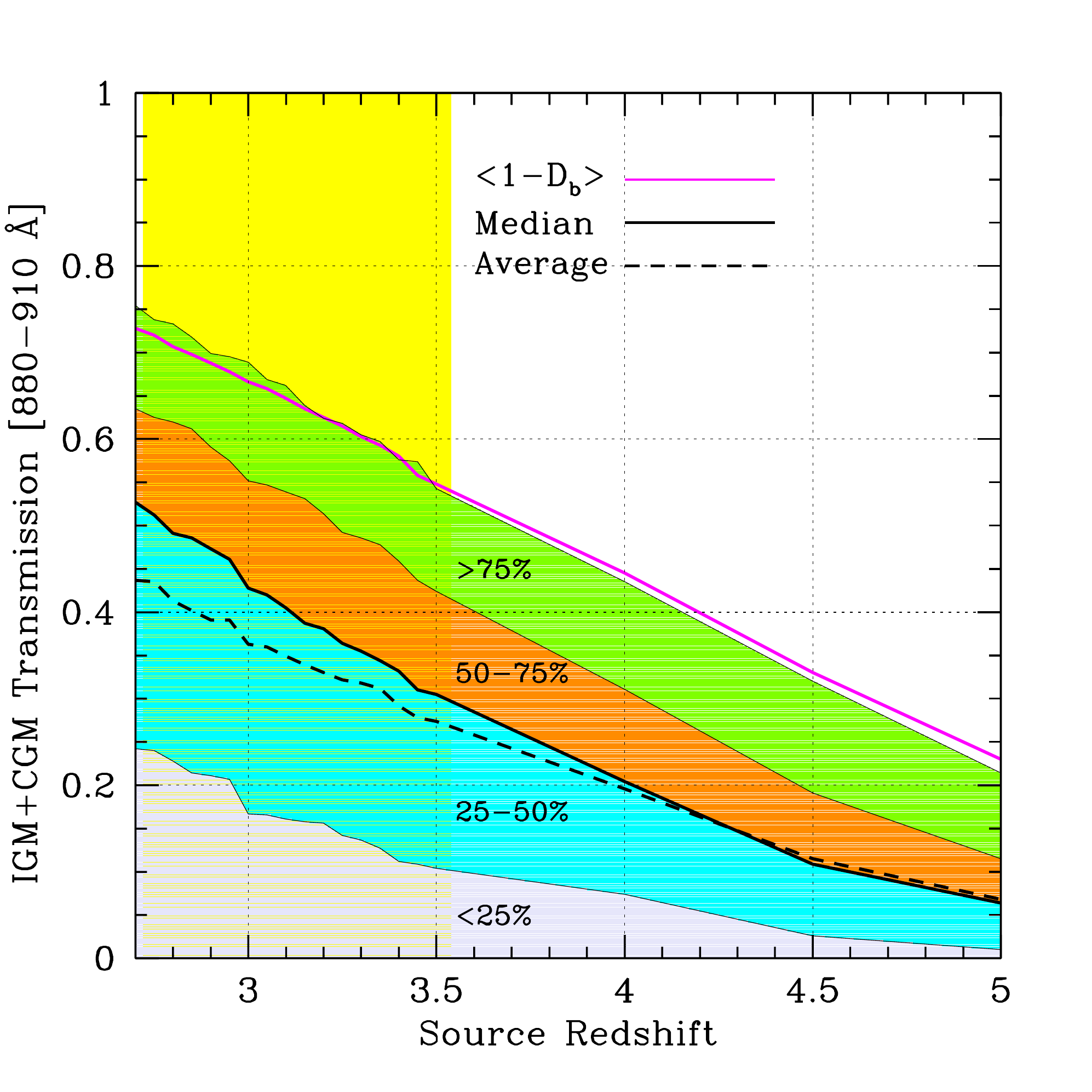}}
\caption{Illustration of the predicted IGM+CGM transmission evaluated at rest-frame [880,910] \AA\ 
as a function of source redshift. The colored bands 
indicate percentile ranges (see Table~\ref{tab:igm_models}), the dashed black curve is the mean transmission, and
the magenta solid curve is $\langle 1 - D_{\rm B} \rangle$, the mean transmission between Ly$\beta$ and the Lyman
limit in the rest frame of the source (see discussion in Appendix\ref{sec:igm_appendix}).  
The KLCS redshift range is (lightly) shaded with a rectangular box. 
 \label{fig:igm_trans_vz}
}
\end{figure}

Monte Carlo simulations were run for two separate models based on the parametrization of
$f(\nhi,X)$ (Figure~\ref{fig:fnx_plot}.) The first, which we call ``IGM-Only'', assumes that every
sightline to a KLCS source is equivalent to randomly selecting \ion{H}{1} absorbers from the distribution function
in a way that depends only on redshift and \nhi, i.e. the ``Average IGM'' 
parametrization in Figure~\ref{fig:fnx_plot}. \citet{rudie12a} showed that regions within 300 kpc (physical) of galaxies at $z\sim 2.4$
give rise to a significantly higher incidence of ${\rm log}\nhi \simgt 14$ absorption. To account for this, 
a second set of realizations, called ``IGM+CGM'', 
includes a model for regions of 
enhanced \ion{H}{1} absorption arising in the circum-galactic medium (CGM) near the source galaxies. As discussed in
Appendix~\ref{sec:igm_appendix}, 
the CGM \ion{H}{1} absorber frequency 
distribution function $f(\nhi,X)[\rm CGM]$ (Figure~\ref{fig:fnx_plot}) is based on the KBSS survey results \citep{rudie12a,rudie13} 
detailing the distribution of \nhi\ along sightlines passing within 50-300 physical kpc (pkpc) and within 700~\kms in redshift (i.e., 
$\left|{\Delta z}\right| \le 0.0023(1+z_{\rm s})$ of    
spectroscopically identified galaxies with $2.0 \simlt z \simlt 2.8$.    
For modeling lines of sight in the 
``IGM+CGM'' simulations we used the CGM parameters for $f(\nhi,X)$ (Table~\ref{tab:fnxtab}) for redshifts $z_{\rm s}-\Delta z \le z \le z_{\rm s}$  
and the ``Average IGM'' formulation 
$z < (z_{\rm s} - \Delta z$), where $\Delta z = 0.0023 (1+z_{\rm s})$ ($c\Delta z = 700$ \kms.) 
For most 
purposes in this paper, the ``IGM+CGM'' is the more relevant of the two; results from both are included in 
Table~\ref{tab:igm_models}. 
The differences between ``IGM+CGM'' and ``IGM-only'' opacity model are discussed below.   

Since we have adopted [880,910] \AA\ in the rest frame of the source as our LyC measurement band, of greatest
interest is the prediction for the statistical reduction by the CGM+IGM of the flux density at observed
wavelengths $880(1+z_{\rm s}) \le \lambda_{\rm obs} \le 910(1+z_{\rm s}$). We define 
\begin{eqnarray}
\langle t_{900} \rangle & \equiv& \langle {\rm exp}(-\tau(\lambda_{\rm obs}) \rangle, \\ 
 880(1+z_{\rm s}) &\le & (\lambda_{\rm obs}/{\textrm \AA}) \le 910(1+z_{\rm s}) .
\label{eqn:t900}
\end{eqnarray}
With this definition of $t_{900}$, 
we can correct observed values of \fobs\ for IGM(+CGM) attenuation 
\begin{equation}
\langle f_{900}/f_{1500}\rangle_{\rm out} = \langle f_{900}/f_{1500}\rangle_{\rm obs}/\langle t_{900}\rangle,  
\label{eqn:f9f15_out}
\end{equation}
where $\langle f_{900}/f_{1500}\rangle_{\rm out}$ is the emergent 
flux density ratio that would be measured by an observer at $z=0$ if there were no opacity contribution
from the CGM and IGM along the line of sight; as discussed in more detail below (\S\ref{sec:implications}, \S\ref{sec:emissivity}), \fout\ is the quantity relevant to
calculations of ionizing emissivity of galaxy populations.  

Figure~\ref{fig:igm_trans_vz} illustrates how the distributions of ``IGM-Only'' and ``IGM+CGM'' transmission $t_{900}$  
depend on $z_{\rm s}$, 
with the range selected for KLCS shaded yellow. 
Table~\ref{tab:igm_models} summarizes the results of the IGM modeling in terms
of the percentiles (10, 25, 50, 75, and 90th) of the IGM or IGM+CGM transmission at 
particular rest wavelength intervals of interest, all as a function of source redshift.
Source redshift values were modeled using small ($\Delta z =0.05$) increments over the KLCS redshift range, 
but we have also included values for sources with $z_s > 3.5$ to provide
some intuition about the rapid decline in IGM transmission as redshift increases\footnote{It is likely that our
Monte Carlo simulations underestimate the LyC opacity for $z > 3.5$, since our assumption about the 
evolutionary parameter $\gamma=1.0$ describes the incidence of LLSs well over the range $2 < z < 3.5$ 
but the slope appears to steepen to $\gamma \simeq 2$ by $z \sim 4$ \citep{prochaska09,songaila2010}.}. 

\begin{figure}[htbp!]
\centerline{\includegraphics[width=9cm]{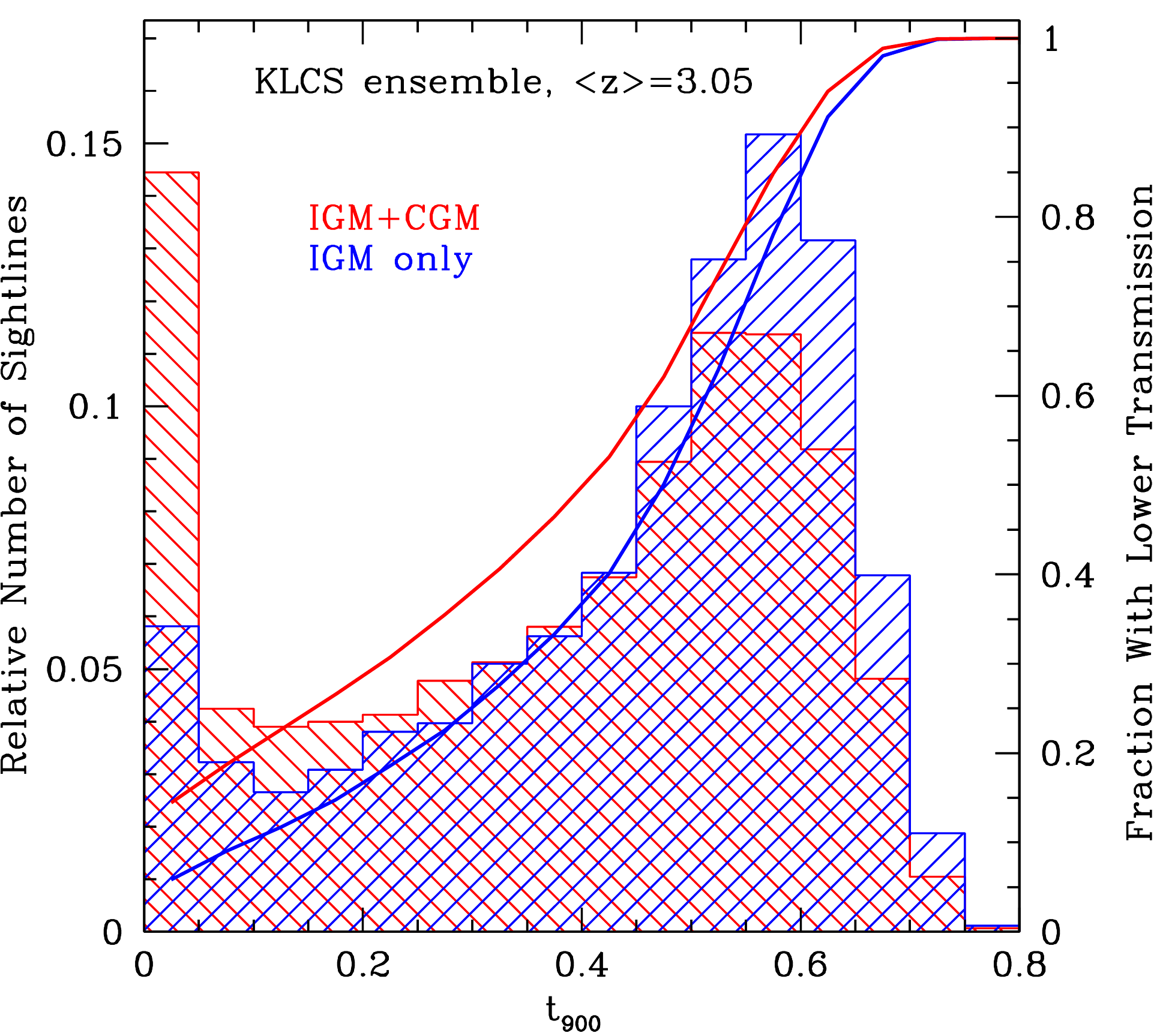}} 
\caption{Histogram of the probability density function of $t_{900}$, the net transmission in the rest wavelength interval [880,910] 
for sources having identical $z_{\rm s}$ distribution as the KLCS ensemble. The histograms show the average of 1000 sets of 124
sightlines: blue-hatched includes only IGM opacity while the red-hatched includes IGM+CGM opacity.  
The continuous curves, color coded in the same way, show the cumulative fraction of sightlines with {\it lower}
$t_{900}$ (both curves refer to the righthand axis). 
Note that the main effect of including the CGM is to flatten 
the distribution for $t_{900} \simgt 0.05$ while increasing the expected number of opaque sightlines ($t_{900} < 0.05$)
by a factor of $\simeq 3$ and decreasing the relative number of sightlines with the highest $t_{900}$. 
\label{fig:igm_trans_histo}
} 
\end{figure} 

Even within the redshift range spanned by the KLCS sample, $\langle t_{900} \rangle$ varies by almost a factor of 2; however, if we treat the
KLCS sample as an ensemble of 124 sightlines with $2.75 \simlt z_{\rm s} \le 3.55$, the IGM+CGM simulations predict that 
the ensemble average $\langle t_{900} \rangle = 0.371$ with 68\% confidence interval $t_{68} = [0.353,0.392]$, i.e. an uncertainty of $\sim 5$\% 
on $\langle t_{900} \rangle$ for the ensemble\footnote{The expected $\langle t_{900} \rangle$ is very close to the average expected if all 
galaxies had $z_{\rm s} = 3.05$, the mean redshift of KLCS.}. 
Fig.~\ref{fig:igm_trans_histo} shows the full distribution of $t_{900}$ expected for ensembles of 124 sightlines
with the same redshift distribution as KLCS, for both the IGM Only and IGM+CGM opacity models.  

One can see from Figure~\ref{fig:igm_trans_histo} that there are two main effects of including the opacity of the CGM: (1) it {\it increases} 
by a factor of $\sim3$ the number
of sightlines with very low transmission ($t_{900} < 0.05$); and  
(2) it significantly {\it decreases} the fraction of sightlines expected to have
transmission near the maximum of $t_{900} \simeq 0.6$. The latter would  
be (all other factors being equal) the most likely to yield detectable LyC signal in the
spectra of individual galaxies. 
This point illustrates the importance of accounting statistically for LyC attenuation 
by gas which is outside of the galaxies, but near enough to be observationally indistinguishable from a case of 
zero LyC photon escape from the galaxy ISM.  

\begin{figure}[htbp]
\centerline{\includegraphics[width=8.5cm]{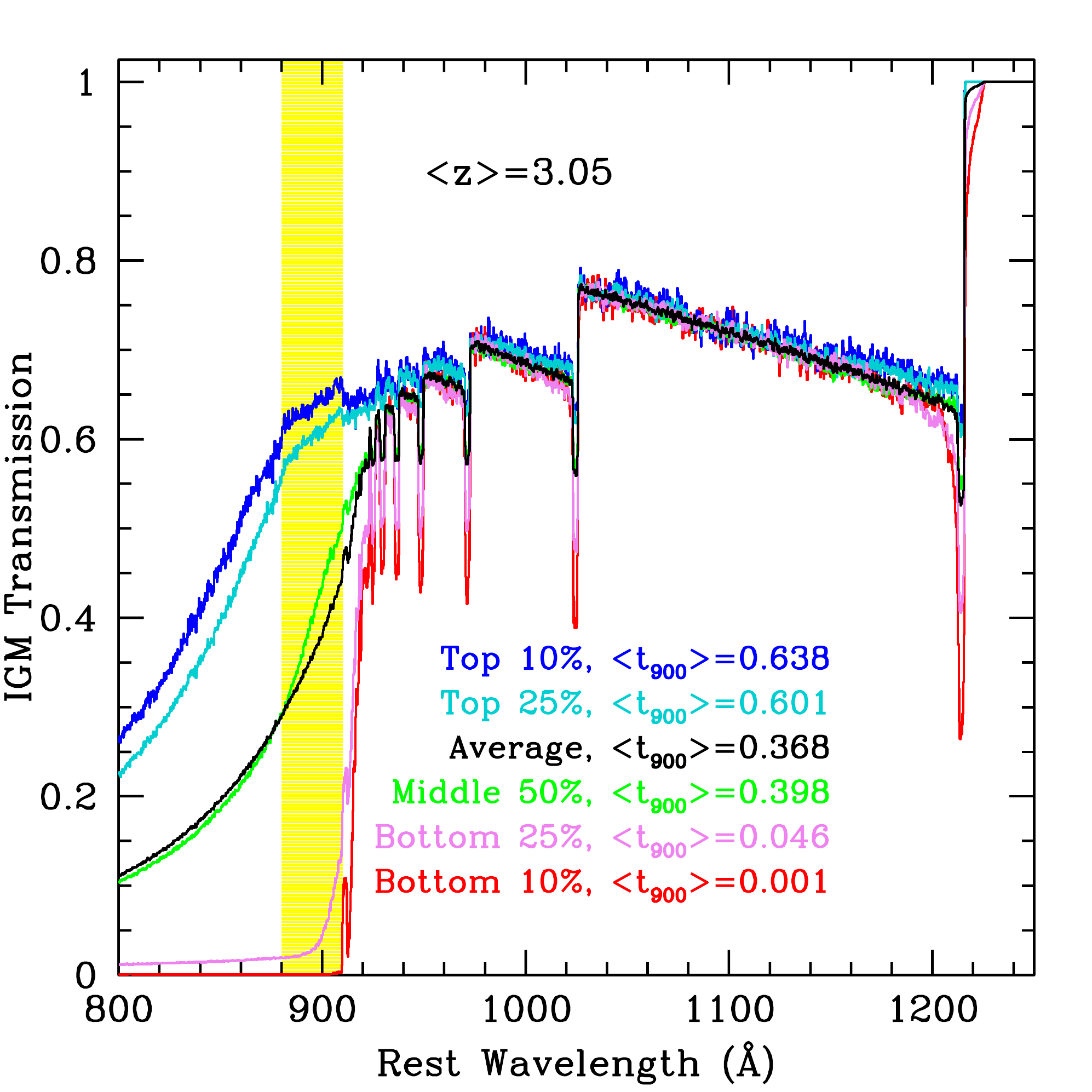}}
\caption{Transmission spectra, in the rest frame of the source, for $\langle z \rangle = 3.05$ for an ensemble of 10000 lines of sight assuming the
CGM+IGM opacity model. 
The various color-coded spectra represent averages within percentile ranges of $t_{900}$. 
Note that the variations in line blanketing from the Ly$\alpha$ forest are small for $\lambda_0 > 940$ \AA\ (except where affected by the CGM, which
manifest here as small regions of enhanced absorption near the rest-frame wavelengths of each Lyman series line).  
The variations are
large in the region of primary interest ([880,910] \AA, shaded yellow) because the effective opacity
is dominated by much rarer absorption systems with high \nhi\ (see Appendix~\ref{sec:igm_appendix}.) 
\label{fig:fakespec_igm}
 } 
 \end{figure}

For some purposes (see \S\ref{sec:composites}), it is useful to form ensemble average transmission spectra covering 
a wider range 
in rest-wavelength than we have used to measure $t_{900}$. 
Figure~\ref{fig:fakespec_igm} 
illustrates the rest-wavelength dependence of ensemble average
transmission spectra 
for $z_{\rm s} = 3.05$.  These spectra were created from an ensemble of 10000 realizations of the full transmission spectra predicted by the Monte Carlo IGM+CGM model, 
sorting them by $t_{900}$, and averaging in percentile bins.

Fig.~\ref{fig:fakespec_igm} shows that 
for $\simgt 10$\% of sightlines at $z_s=3.05$ (``Top 10\%'' spectrum in the figure) there will be little or no 
attenuation over and above that produced by line blanketing. However, given typical detection limits for LyC flux, 
at least the ``Bottom 25\%'' shown in Figure~\ref{fig:fakespec_igm} will {\it appear} 
be be optically thick to their own LyC radiation,  due entirely to attenuation by {\it intervening} \ion{H}{1} at $D > 50$ kpc arising
due to the opacity of the combined CGM+IGM. 

\subsection{IGM Transmission: Sampling Issues}
\label{sec:igm_sampling}

\begin{deluxetable}{lcccccc}
\tabletypesize{\scriptsize}
\tablewidth{0pt}
\tablecaption{Sampling Statistics for LyC Transmission (CGM+IGM)}
\tablehead{
  \colhead{$z_{\rm s}$} & \colhead{$\langle t_{900} \rangle $} & \colhead{$t_{68}$\tablenotemark{a}} & \multicolumn{4}{c}{$\delta t_{900}/\langle t_{900}\rangle$\tablenotemark{b}}   \\
  \colhead{}   &  \colhead{}                           & \colhead{} & \colhead{$N=15$} & \colhead{$N=30$} & \colhead{$N=60$} & \colhead{$N=120$} }
\startdata
 2.75 &  0.453 &  [0.093-0.670] &  0.141 &  0.101 &  0.067 &  0.048\\
 3.00 &  0.391 &  [0.074-0.607] &  0.153 &  0.108 &  0.073 &  0.052\\
 3.25 &  0.325 &  [0.018-0.540] &  0.173 &  0.120 &  0.082 &  0.056\\
 3.50 &  0.271 &  [0.013-0.467] &  0.184 &  0.128 &  0.088 &  0.063
\enddata
\tablenotetext{a}{68\% confidence interval for a single realization of $t_{900}$.}
\tablenotetext{b}{Given an ensemble of $N$ realizations of $t_{900}$ for source redshift $z_{\rm s}$, the ratio of the half-width of the 68\% confidence interval of $\langle t_{900}\rangle$ and the mean value of $\langle t_{900}\rangle$ that would be obtained after a very large number of realizations.}  
\label{tab:sampling}
\end{deluxetable}

Assuming the ``IGM+CGM'' opacity model, at $z \sim 3$, the 68\% confidence 
interval for a single realization of $t_{900}$ at $z_{\rm s} \simeq 3.05$ is $t_{\rm 900}=[0.038, 0.589]$; the corresponding 
68\% confidence intervals are $t_{900} = [0.076,0.665]$ at $z_{\rm s}=2.70$ , and $t_{900}=[0.017,0.471]$ at $z=3.50$.  
The implications of these broad distributions are worth stating explicitly:  
the LyC-detectability of a galaxy with a given intrinsic (i.e., emergent) $(f_{900}/f_{1500})_{\rm out}$ is to a great extent  
controlled by the statistics of the IGM transmission, which is uncertain by factors of between 5 and 10 even over the
limited redshift range $2.70 \simlt z_{\rm s} \simlt 3.50$.  Conversely, a single measurement of $(f_{900}/f_{1500})_{\rm obs}$ cannot
be converted into an intrinsic property of the source, for the same reason (see, e.g., \citealt{vanzella15,shapley16}.) 

However, $\langle t_{900}(z_{\rm s})\rangle$, the {\it mean} transmissivity of the IGM
for a source at $z=z_{\rm s}$, and its uncertainty $\delta t_{900}(z_{\rm s}, N)$ can be quantified for an ensemble of $N$ 
sightlines using the 
Monte Carlo models described in Appendix~\ref{sec:igm_appendix}. For example, assuming $z_{\rm s} = 3.0$, the number of independent IGM sightlines
one must sample in order to reduce $\delta t_{900}(3.0,N)$ to less than 10 (5) percent of $\langle t_{900}(3.0) \rangle$ 
is $N=36$ (150). In other words, if the spectra of 36 sources at $z_{\rm s}=3.0$ are averaged to produce 
a measurement of \fobs,
then $\langle (f_{900}/f_{1500})_{\rm out}\rangle = \langle (f_{900}/f_{1500})_{\rm obs}\rangle/(t_{900} \pm \delta t_{900})$,
and the IGM correction contributes a fractional uncertainty to the inferred emergent flux density ratio of $\simeq 10$\%. 
Some example statistics relevant for the KLCS redshift range and sample size are given in Table~\ref{tab:sampling}. 

This type of analysis is useful when one has observations of a particular class of object (grouped by known property, 
e.g. $L_{\rm uv}$, $z_{\rm s}$, $W_{\lambda}(\lya)$, inferred extinction) 
forming a subset of the full sample: as long as the subset has sufficiently large $N$, the statistical
knowledge of the IGM opacity can be used to derive the average {\it intrinsic} LyC properties of that class. 
The validity of this procedure depends on (1) the assumption that each line of sight in the sample is uncorrelated with any other
line of sight in the same ensemble and (2) that the IGM+CGM opacity model is an accurate statistical description of the true opacity. 

The first assumption -- that lines of sight are independent -- is almost certainly {\it not} valid when a survey is conducted
in a single contiguous field of angular size $\sim 10$\arcm (transverse scale of $\sim 5$ pMpc at $z\sim 3$), as has often been
the case for reasons of practicality (e.g., \citealt{shapley06,nestor11,mostardi13,mostardi15,vanzella10,siana15}). Correlated
sightlines could be especially problematic in fields known to contain significant galaxy over-densities at or just below the
source redshifts.     
As has been discussed by (e.g.) \citet{shapley06}:  if observed sources are located in regions
containing more (or less) gas near \nhi $  = 10^{17}$ cm$^{-2}$ than average, their ``local''
IGM could skew significantly away from expectations if an ``average'' line of sight is assumed.
The effect is likely to be negligible for \lya\ forest blanketing, but could strongly influence 
$f_{900}$, since it relies on the statistics of small numbers through the 
incidence of relatively high column density
\ion{H}{1} over a small redshift path ($\Delta z \simeq 0.14$ for $z_s=3.05$ for our LyC region [880,910] \AA.) 

The full KLCS sample is relatively insensitive to the effects of correlations between IGM sightlines 
by virtue of the fact that it is comprises 9 independent survey regions. Similarly, the IGM opacity model
is based on the statistics of 15 independent QSO sightlines in the KBSS survey \citep{rudie13}, 
and the CGM corrections to the IGM model are based on regions near galaxies selected using essentially 
identical criteria to those used for KLCS (albeit at slightly lower redshift; see \citealt{rudie13}). 

Nevertheless, it is worth pointing out {\it caveats} associated with the CGM+IGM opacity models possibly relevant
even for KLCS. 
First, the adopted IGM+CGM opacity model probably under-estimates the CGM opacity, since it is based on lines of sight to background
QSOs with $50 < D_{\rm tran} \le 300$ pkpc (i.e., projected angular distances 6\arcs$< \theta < 37$\arcs), which
by virtue of the cross-section weighting correspond to an average impact parameter of $\langle D_{\rm tran} \rangle  
\simgt 200$ pkpc. On the other hand, every line of sight to a (source) galaxy includes gas with physical distance
from the source of 50-300 pkpc. If \nhi\ continues to increase with decreasing galactocentric radius 
(as is likely), the transverse sightlines used to estimate the CGM contribution for the opacity model would 
systematically underestimate the total CGM opacity. 
While the CGM+IGM corrections applied to the galaxy spectra in the KLCS sample are likely to be
appropriate for the range of galaxy properties we are sensitive to in this work (due to the similarity in the galaxy
properties between KBSS and KLCS mentioned above), it could be dangerous to apply the corrections to sources
selected using substantially different criteria -- for example,   
if most of the ionizing radiation field is produced by objects with
a different overall environment (e.g., low-density regions harboring fainter sources might be expected to be 
surrounded by less intergalactic
and circumgalactic gas; see \citealt{rakic12,turner14}) then the CGM contribution to the net opacity toward those sources 
might be over-estimated. 

Finally, the absence of a clear distinction between ISM, CGM, and IGM leads to an issue of semantics: in considering the ``escape'' of ionizing photons,
one must also define what constitutes escape, i.e., how far must the ionizing photon travel before
it is counted as having escaped, and what is the probability that it will be observable? 
For the remainder of this paper, we assume that ionizing photons absorbed within a galactocentric 
radius of $r=50$ pkpc  
have by definition {\it not escaped}. We then use the IGM+CGM statistics outlined in Appendix~\ref{sec:igm_appendix} to correct the observations back
to the $r \simeq 50$ pkpc ``surface'' -- our working definition of a galaxy's ``LyC photosphere''.  

\subsection{LyC Detectability: Spectroscopy vs. Imaging}
\label{sec:spec_vs_img}

\begin{figure}[htbp]
\centerline{\includegraphics[width=9cm]{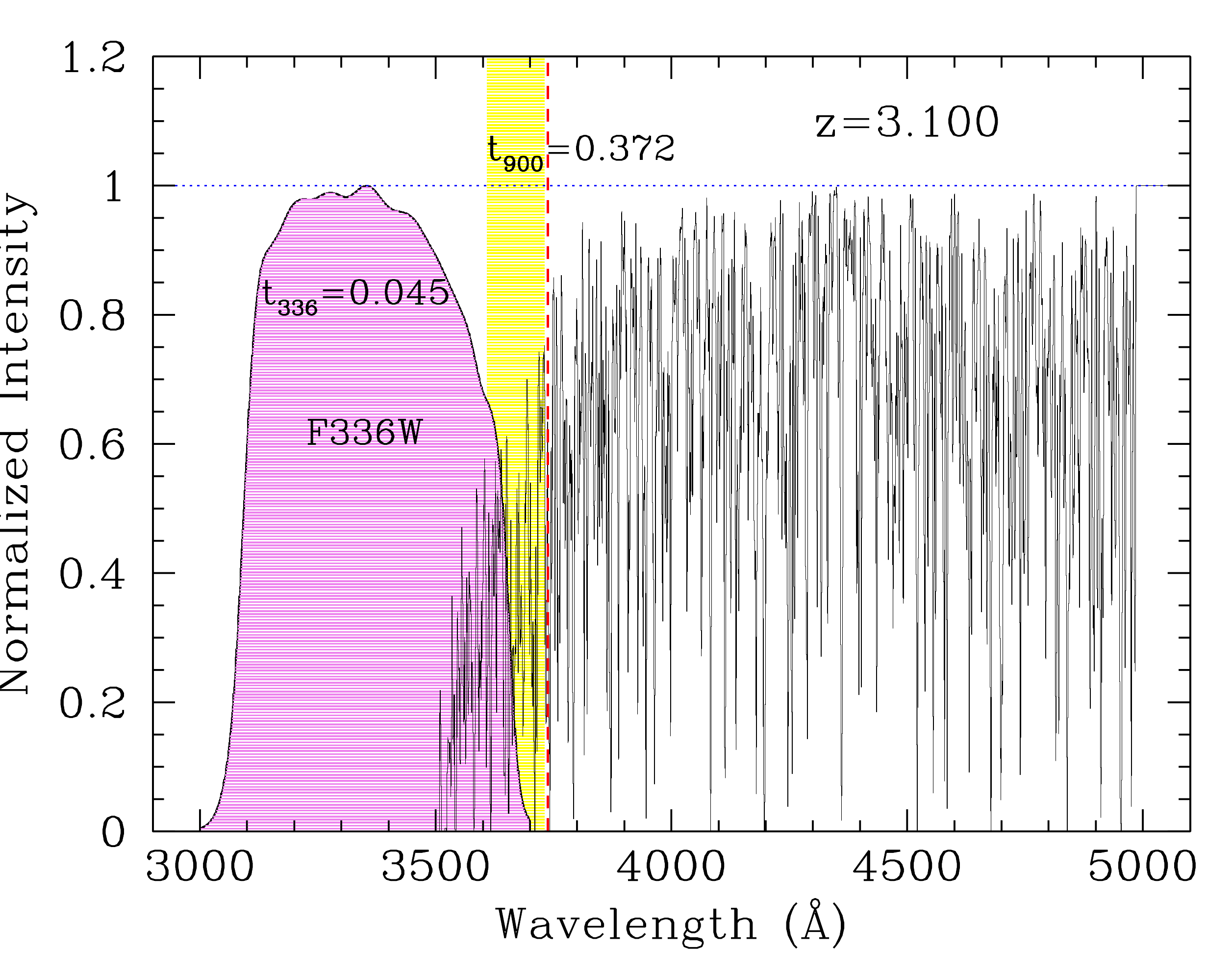}}
\caption{An example of a single IGM sightline, simulated using the Monte Carlo models described in 
the appendix, for a source at $z=3.10$. The IGM transmission vs. observed wavelength is shown, with
the position of the rest-frame Lyman limit indicated with the vertical dashed red line. The light yellow
shaded region illustrates the range of observed wavelength used to calculate $t_{900}$, which in this  
realization has $t_{900} = 0.372$ -- close to the mean value expected for $z \sim 3.1$ (see Table~\ref{tab:igm_models}).
The violet shaded region is the filter response function for the HST WFC3/UVIS F336W filter, used for many
studies of LyC leakage at $z \simgt 3$. The mean transmitted flux integrated over the F336W bandpass has
$t_{F336W} = 0.045$, i.e., a factor of $\simgt 8$ lower than the spectroscopic measure of $t_{900}$. }
\label{fig:compare_methods}
\end{figure}

\begin{figure}[htpb]
\centerline{\includegraphics[width=8.5cm]{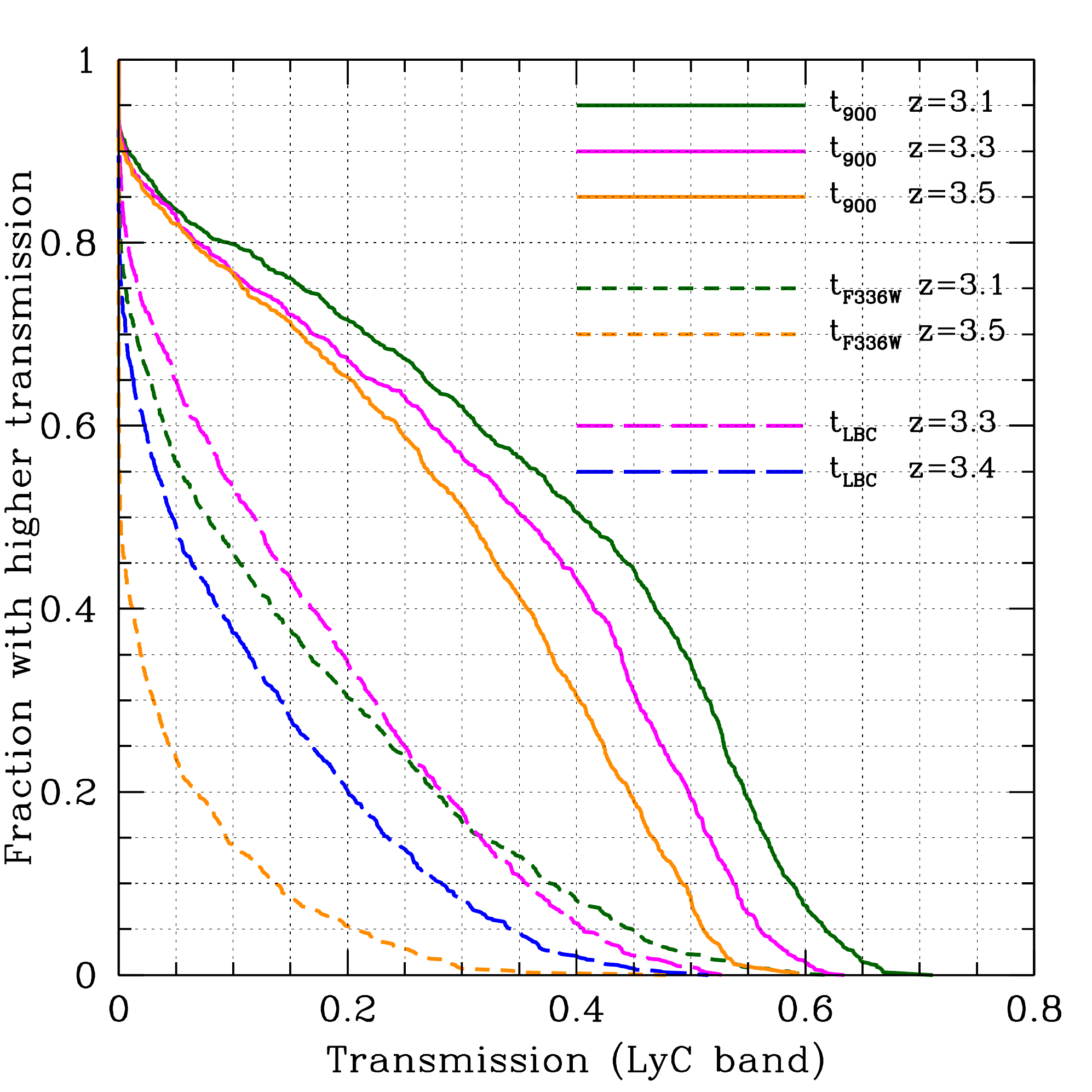}}
\caption{Comparison of the cumulative distribution of IGM+CGM transmission evaluated for $z_{\rm s}=3.1$ (orange), $z_{\rm s}=3.3$ (magenta), and
$z_{\rm s} = 3.5$ (green) using the $t_{900}$ bandpass (solid curves) vs. $t_{\rm F336W}$ or $t_{\rm LBC-U}$ (short-dashed curves), where
F336W is the WFC3-UVIS filter and LBC-U is the U filter described by \citet{grazian16}. Note that the largest differences appear
at relatively high values of transmission.} 
\label{fig:compare_trans}
\end{figure}

There are distinct practical advantages in using a measurement band
that samples a fixed bandwidth in the rest frame of each source,
880-910 \AA, placed just shortward of the intrinsic Lyman limit.  
It is possible to use images taken through a comparably-narrow
bandpass (i.e., $\simeq 100-200\AA$ in the observed frame; e.g., \citealt{inoue08,
nestor11,mostardi13}), but a narrow-band filter will be optimally-placed only
for sources at a fixed redshift, which as discussed above can also be problematic in terms
of sample variance due to gas-phase overdensities and/or correlated sightlines
arising in the intra-protocluster medium.

Similar issues affect observations where a broad-band filter is used for the LyC detection band,
such as surveys conducted using the F336W filter in the WFC3-UVIS camera on board {\it HST} (e.g., \citealt{mostardi15,siana15,vasei16}).
In this case, since the broad bandpass must lie entirely shortward of the
rest-frame Lyman limit of sources to provide unambiguous detections of ionizing photons, it
is confined to source redshifts $z_{\rm s} > 3.07$; however, even for $z_{\rm s} \simeq 3.1$, 
the photon-number-weighted mean flux density measured in the filter bandpass is considerably
reduced relative to the rest-frame [880,910] wavelength interval. An example assuming $z_{\rm s}=3.10$ is shown
in Figure~\ref{fig:compare_methods}, where $t_{900} = 0.372$ (i.e., close to the mean value for random IGM+CGM
sightlines to sources with $z_{\rm s}=3.10$; Table~\ref{tab:igm_models}), but the F336W band-averaged transmission $t_{336} = 0.045$,  $\sim 8.3$ times
smaller (2.3 mag).  

Figure~\ref{fig:compare_trans} shows cumulative distribution functions of $t_{900}$,
$t_{336}$, and $t_{\rm LBC}$ (the broadband $U$ filter used by \citealt{grazian16}) for a large ensemble 
of sightlines to $z_{\rm s} = 3.1$ and $z_{\rm s} = 3.5$.     
At $z=3.1$, the median transmission in the relevant LyC detection band is 5.2 times higher using the [880,910] 
interval than that evaluated through the F336W filter bandpass; the ratio between median transmission values reaches $\sim 200$ by $z=3.5$.  
The 90th percentile transmission $t_{900} = 0.59$ (0.49) for $z_{\rm s}=3.1$ ($z_{\rm s} =3.5$); 
$t_{336W} =0.38$ (0.14) for $z_{\rm s}=3.1$  ($z_{\rm s}=3.5$).  
Similarly, compared to the ground-based $U_{\rm LBC}$ used to measure LyC emission from spectroscopically
identified galaxies at $z\sim 3.3$ by \citet{grazian16,grazian17}, the spectroscopic [880,910] \AA\ bandpass is $\simgt 2$ times
more likely to include sightlines with $t_{\rm LyC} > 0.2$, and $\simeq 9$ times more likely to have sightlines with $t_{\rm LyC} > 0.4$. 
These differences could be quite large if there is limited dynamic range for LyC detection from individual sources --
as has been the case for all surveys to date -- potentially producing large differences in the fraction of sources with
significant LyC detections even for surveys that nominally reach the same flux density limit in the LyC band. 

\begin{deluxetable}{lcccc}
\tabletypesize{\scriptsize}
\tablewidth{0pt}
\tablecaption{Transmission Statistics vs. LyC Detection Band\tablenotemark{a}}
\tablehead{\colhead{LyC Band\tablenotemark{b}} & \colhead{$z_{\rm s}$} & \colhead{$\langle t_{LyC} \rangle$} & \colhead{Median} & \colhead{90th Percentile} }  
\startdata
 [880,910] & 3.10 & 0.352  & 0.393  & 0.589 \\ 
 HST-F336W & 3.10 & 0.139 & 0.078  & 0.377 \\
\hline
 [880,910] & 3.30 & 0.321 & 0.359 & 0.543 \\ 
 LBC-U     & 3.30 & 0.149 & 0.118 & 0.356 \\
 HST-F336W & 3.30 & 0.070 & 0.017 & 0.226 \\ 
\hline
 [880,910] & 3.40 & 0.291 & 0.119 & 0.516 \\ 
 LBC-U     & 3.40 & 0.099 & 0.047 & 0.282 \\ 
 HST-F336W & 3.40 & 0.050 & 0.005 & 0.171 \\
\hline
 [880,910] & 3.50 & 0.264 & 0.302 & 0.492 \\ 
 HST-F336W & 3.50 & 0.039 & 0.002 & 0.137  
\enddata
\label{tab:comp_trans}
\tablenotetext{a}{Mean, median, and 90th percentile transmission in LyC detection passband for
large ensembles of Monte Carlo IGM+CGM transmission spectra with $3.10 \le z_{\rm s} \le 3.50$.}
\tablenotetext{b}{LyC detection passband over which the mean flux density is evaluated: [880,910] is the
spectroscopic band used in this work; the LBC-U filter is close to SDSS $u'$, and is described by \citet{grazian16}; HST-F336W is the {\it HST}/WFC3-UVIS filter F336W. }
\end{deluxetable}

Care must also be exercised when comparing the results of surveys that use different LyC detection bands and/or IGM opacity models: Table~\ref{tab:comp_trans}
compares our determination of the mean IGM+CGM transmission evaluated for 3 different LyC detection passbands. The ensemble of sightlines
used at each $z_{\rm s}$ was identical.  Note that 
we find that the mean IGM+CGM transmission evaluated using the $U_{\rm LBC}$ filter for $z_{\rm s} = 3.30$ ($z_{\rm s}=3.40$) is 
$\langle t_{\rm LBC} \rangle \simeq 0.149$ ($0.099$),
to be compared with the value assumed by \citet{grazian16,grazian17}, $\langle t_{\rm LBC} \rangle = 0.28$. The latter value was based on the 
IGM opacity models of \citet{inoue14}\footnote{At $z = 3.3$, our IGM+CGM (IGM-only) Monte Carlo models have $\langle t_{\rm 880} \rangle = 0.239$ (0.293), compared
to $\langle t_{\rm 880} \rangle = 0.380$ predicted by the analytic models of \citet{inoue14} (their Fig.~10).}. 
Additional ambiguities relevant to the quantitative comparison of LyC results arise
due to differences in the definition of ``escape fraction'', discussed in more detail in \S\ref{sec:fesc}. 

The main point here is that the probability of detecting LyC emission-- or of setting interesting limits on $f_{900}/f_{1500}$-- 
depends very sensitively 
the source redshift $z_{\rm s}$, the bandwidth and relative wavelength/redshift sensitivity of the LyC detection band, and 
the fidelity of the correction for the IGM+CGM opacity. 

\section{Inferences from Composite Spectra of KLCS Galaxies}

\label{sec:composites}

As discussed in \S\ref{sec:igm_trans}, it is potentially misleading
to interpret individual measurements of the quantity $(f_{900}/f_{1500})_{\rm obs}$ because
of the large expected variation in $t_{900}$ from sightline to sightline. 
There are significant advantages associated with considering only ensembles
of galaxies (sharing particular properties) that are large enough that 
the uncertainty in the ensemble average IGM correction is reduced.  Since 
we have relatively high quality spectra of 124 objects remaining
after cleaning the sample of potential contamination or obvious systematic issues,
in this section we combine various subsets of the KLCS spectra to form
high S/N spectroscopic composites. The resulting spectra are then used to obtain sensitive
measurements of $\langle f_{900}/f_{1500}\rangle_{\rm obs}$ for which 
$\langle t_{900} \rangle$ is well-determined. 

\begin{figure*}[htbp!]
\centerline{\includegraphics[width=18cm]{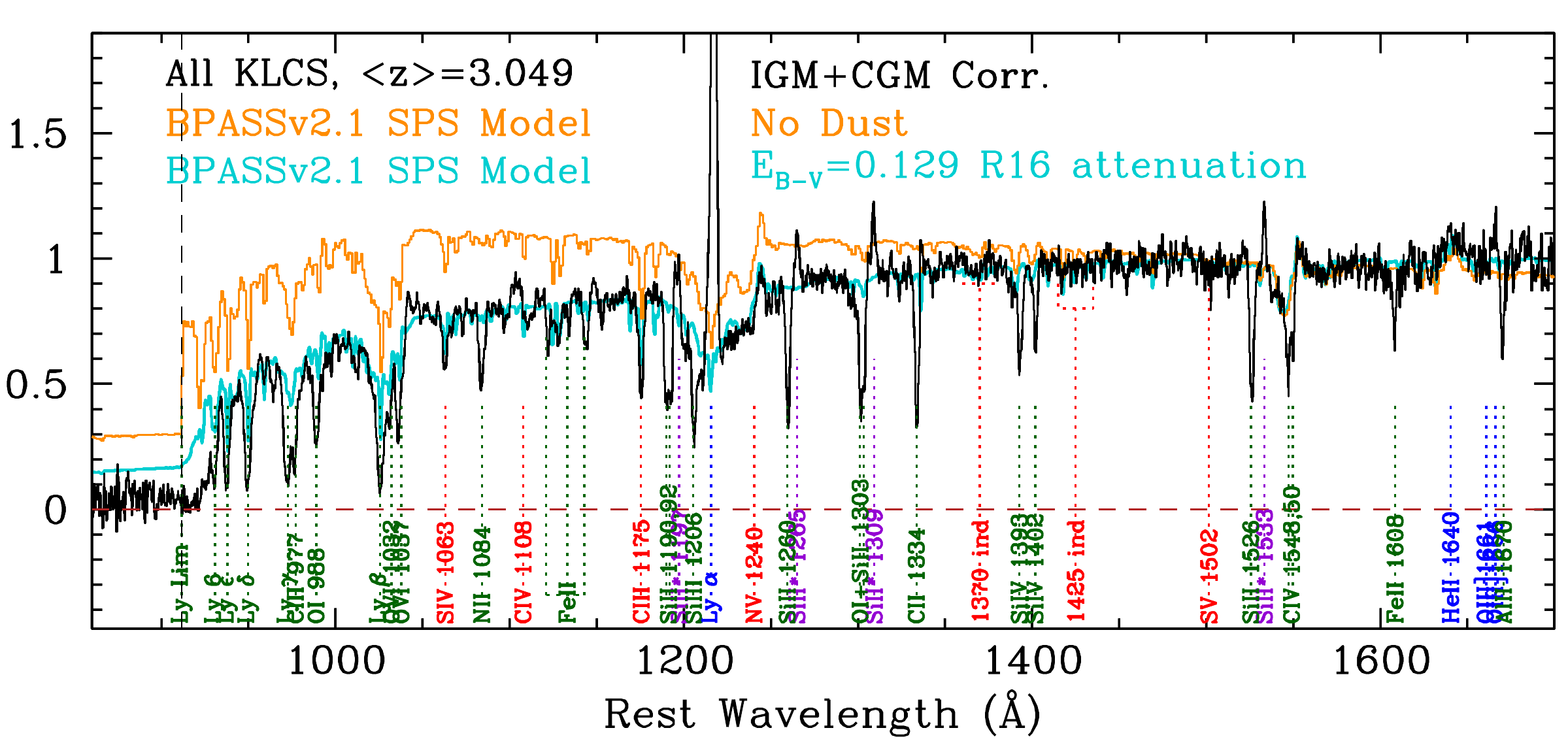}}
\caption{
The composite spectrum of all 124 galaxies included in the KLCS statistical sample (black). The
stacked spectrum has been corrected for the mean IGM+CGM opacity appropriate for
an ensemble with $\langle z_{\rm s} \rangle = 3.05$ as shown in Figure~\ref{fig:fakespec_igm}.
Some of the easily-identified spectral features are indicated, where the labels
have been color-coded according to whether they are primarily interstellar (dark green),
nebular (blue), stellar (red), or excited fine structure emission (purple). 
The orange (turquoise) spectrum is the best-fitting stellar population synthesis model (see \S\ref{sec:sps_models}) 
before (after) 
applying reddening according to \citet{reddy16a} (R16). 
 }
\label{fig:igm_demo}
\end{figure*}
\begin{figure}[!htbp] 
\centerline{\includegraphics[width=9cm]{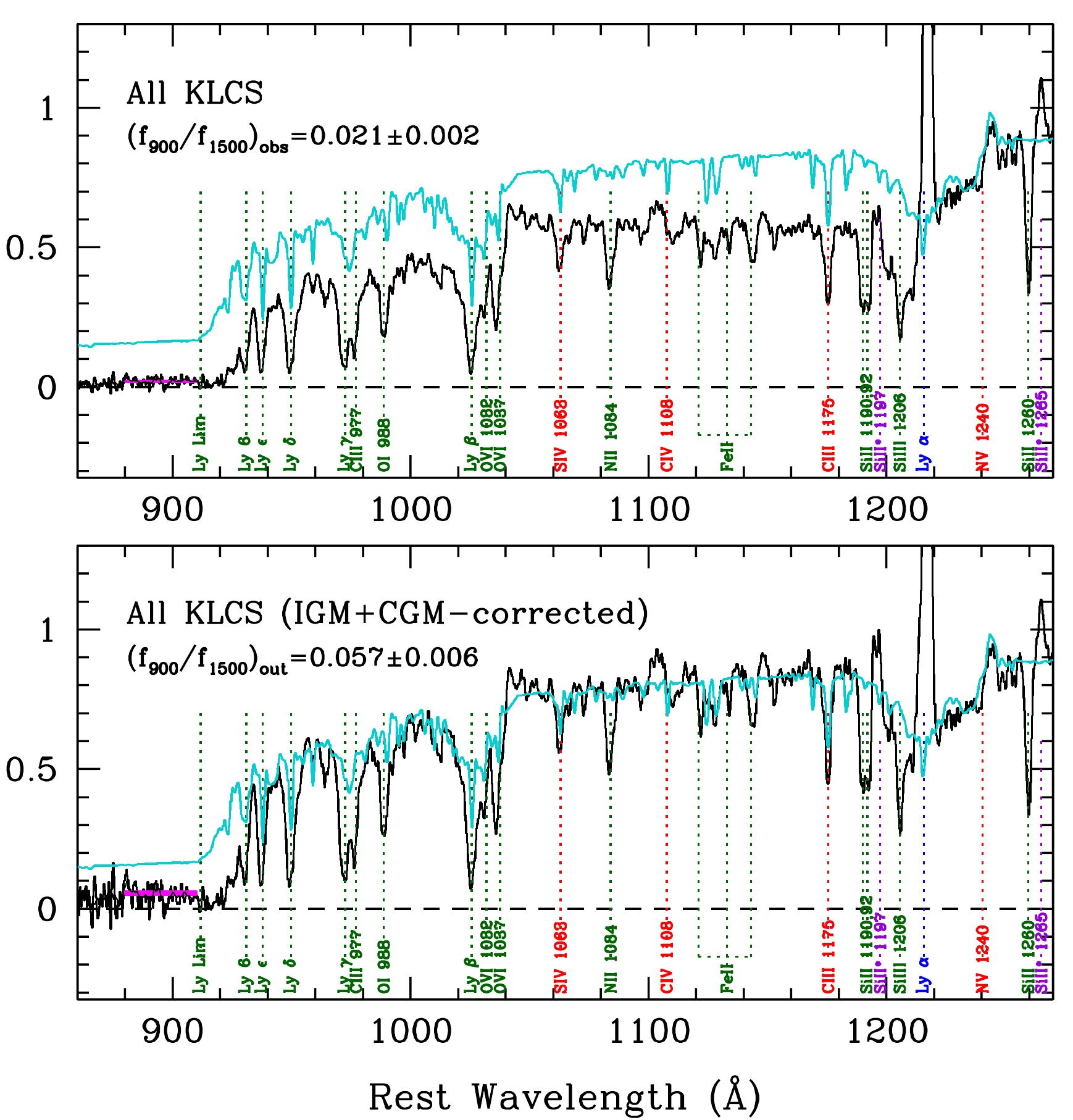}} 
\caption{Zoomed-in version of Figure~\ref{fig:igm_demo}, showing the composite spectrum formed from all 124 galaxies in the
KLCS sample. ({\it Top:}) The composite 
spectrum prior to correction for IGM+CGM opacity, where the measured value of $\langle f_{900}/f_{1500}\rangle_{\rm obs}$ 
is indicated. ({\it Bottom:}) 
Identical to the top panel, except that here the observed spectrum (black histogram) has been corrected  
using the model for the average IGM+CGM transmission spectrum (see Fig.~\ref{fig:fakespec_igm}) appropriate for the KLCS sample distribution of source
redshifts. The value of  
$\langle f_{900}/f_{1500}\rangle_{\rm out}$ given in the bottom panel was calculated using equation~\ref{eqn:f9f15_out}
(see Table~\ref{tab:fesc2_data}.) The values of each, and the wavelength
interval over which they have been evaluated, are indicated by magenta bars in both panels. The model spectrum (cyan histogram) is identical
in both panels, and also the same as that shown in Figure~\ref{fig:igm_demo} using the same color coding. }
\label{fig:igm_demo_zoom}
\end{figure}

The individual KLCS spectra have a range of S/N owing to differences in apparent
magnitude, redshift, and observing conditions. One might be tempted to combine
them so as to maximize the S/N of the resulting composite, but
the necessary weighting involved would invalidate the approach we adopt to
correct for intergalactic opacity, which assumes that every sightline through
the IGM is contributing equally to the net suppression of the intrinsic spectrum. 
Thus, we used the following approach to forming composites.

All extracted one-dimensional spectra (and their associated error spectra) of KLCS galaxies were shifted
to the rest frame using the values of $z_{\rm sys}$ (Table~\ref{tab:klcs}), where each was 
resampled to a common wavelength grid using spline interpolation and scaled
according to its observed continuum flux density $f_{1500}$. 
In forming
a composite spectrum, the mean value at each rest-wavelength dispersion point was calculated
after outliers with values
differing by more than $3 \sigma$ from the median were
rejected. This algorithm is very effective in removing residuals from imperfectly subtracted
night sky lines and other defects present in individual spectra. 
Figure~\ref{fig:igm_demo} shows the spectroscopic stack of all 124 galaxies 
in the KLCS analysis sample. 

Fig.~\ref{fig:igm_demo} also addresses the extent to which the IGM opacity models described in
\S\ref{sec:igm_trans} are consistent with the attenuation 
actually observed in the composite galaxy spectrum. The observed spectrum (black) has been 
corrected using the calculated mean IGM+CGM transmission spectrum appropriate for a 
sample having $\langle z_{\rm s} \rangle \simeq 3.05$ (Figure~\ref{fig:fakespec_igm}), normalized so
that $\langle f_{1500} \rangle = 1$.  
The orange spectrum is the similarly-normalized intrinsic spectrum of
the stellar$+$nebular continuum for the stellar population synthesis (SPS) model found to be most successful
in reproducing the observed composite spectrum (see \S\ref{sec:sps_models}).
The cyan spectrum in Figure~\ref{fig:igm_demo} is the model SPS spectrum 
after reddening using the best-fitting attenuation relation and
color excess E(B-V) (discussed in \S\ref{sec:sps_models} below) normalized so that $\langle f_{1500} \rangle = 1$.
Note that the continua of the model spectrum (cyan) and the IGM+CGM-corrected observed spectrum (black) are
in excellent agreement 
over the rest wavelength range 950-1210 \AA-- the range sensitive to the accuracy
of the IGM \lya\ forest Monte Carlo modeling vs. redshift (\S\ref{sec:igm_trans} and Appendix~\ref{sec:igm_appendix}.) 

Figure~\ref{fig:igm_demo_zoom} shows a zoomed-in version of Figure~\ref{fig:igm_demo}, illustrating the full KLCS sample
before (top panel) and after (bottom panel) applying the IGM+CGM correction to the observed spectrum.  The model spectrum shown in each panel
is identical, and all spectra have been normalized to unity at rest-frame 1500 \AA.  

\begin{deluxetable*}{lccrccrccrr}
\tabletypesize{\scriptsize}
\tablewidth{0pt}
\tablecaption{Statistics Measured from KLCS Composite Spectra}
\tablehead{
  \colhead{Sample} & \colhead {$N$\tablenotemark{a}} & \colhead{$\langle z_{\rm s} \rangle$} & \colhead{$\langle W_{\lambda}(\lya) \rangle$\tablenotemark{b}} &
\colhead{ $\langle L_{\rm uv}/L_{\rm uv}^{\ast} \rangle$} & \colhead{$(G-{\cal R})_0$\tablenotemark{c}} & \colhead{$\langle f_{900}/f_{1500} \rangle_{\rm obs} $\tablenotemark{d}} 
&\colhead{${\rm \langle t_{900} \rangle}$\tablenotemark{e}} 
& \colhead{$\langle f_{900}/f_{1500}\rangle_{\rm out}$} 
& \colhead{${\rm \langle t_{900} \rangle}$\tablenotemark{f}} 
& \colhead{$\langle f_{900}/f_{1500}\rangle_{\rm out}$} \\
\colhead{} & \colhead{} & \colhead{} &  \colhead{[\AA]} & \colhead{} & \colhead{} & \colhead{} & \colhead{[IGM only]} & \colhead{[IGM only]} & \colhead{[IGM+CGM]} & \colhead{[IGM+CGM]}  
  }
\startdata
All  &124 &3.049 &$~+13.1$ &1.04 &0.49 &$~0.021\pm0.002$  &$0.443\pm0.017$ &$0.047\pm0.005$ &$0.368\pm0.020$   &$0.057\pm0.006$  \\
All, detected  &~15 &3.093 &$~+29.2$  &0.86 &0.45 &$~0.134\pm0.007$ &\nodata &\nodata &\nodata &\nodata \\
All, not detected\tablenotemark{g}  &106 &3.044 &$~+10.6$  &1.07 &0.50 &$~0.010\pm0.003$ &\nodata &\nodata &\nodata &\nodata \\
\hline
$z$ (Q1) &~31 &2.843 &$~~+10.6$ &0.91 &0.50 &$~0.022\pm0.007$ &$0.487\pm0.038$ &$~0.045\pm0.015$ &$0.412\pm0.045$  &$~0.053\pm0.018$   \\
$z$ (Q4) &~31 &3.284 &$~+16.7$ &1.18 &0.46 &$~0.019\pm0.003$ &$0.387\pm0.033$ &$~0.047\pm0.009$  &$0.321\pm0.037$ &$~0.056\pm0.011$  \\
\hline
$L_{\rm uv}> L_{\rm uv}^{\ast}$  &~60 &3.079 &$~~+6.6$ &1.44 &0.55 &$~0.002\pm0.003$ &$0.435\pm0.024$ &$~0.005\pm0.007$ &$0.363\pm0.028$  &$~0.006\pm0.008$    \\
$L_{\rm uv}< L_{\rm uv}^{\ast}$  &~64 &3.021 &$~+18.5$ &0.67 &0.43 &$~0.042\pm0.004$ &$0.449\pm0.024$  &$~0.094\pm0.010$ &$0.372\pm0.028$ &$~0.113\pm0.014$   \\
\hline
$L_{\rm uv}$ (Q1)  &~31 &3.064 &$~~+7.4$ &1.73 &0.53 &$~0.002\pm0.003$ &$0.438\pm0.035$ &$0.005\pm0.007$  &$0.364\pm0.040$  &$~0.005\pm0.008$  \\
$L_{\rm uv}$ (Q2)  &~31 &3.086 &$~~+5.7$ &1.12 &0.57 &$~0.000\pm0.004$ &$0.432\pm0.035$ &$~0.000\pm0.007$ &$0.360\pm0.040$  &$~0.000\pm0.011$  \\
$L_{\rm uv}$ (Q3)  &~31 &3.045 &$~+20.0$ &0.81 &0.44 &$~0.042\pm0.005$ &$0.442\pm0.035$ &$~0.095\pm0.014$ &$0.367\pm0.040$ &$~0.114\pm0.018$    \\
$L_{\rm uv}$ (Q4)  &~31 &3.001 &$~+18.1$ &0.51 &0.42 &$~0.052\pm0.007$ &$0.453\pm0.056$ &$~0.115\pm0.020$ &$0.377\pm0.042$ &$~0.138\pm0.024$   \\
\hline
$W_{\lambda}(\lya)$ (Q1) &~31 &3.005 &$~-5.6$ &1.13 &0.54 &$~0.005\pm0.004$ &$0.452\pm0.035$ &$~0.011\pm0.009$ &$0.378\pm0.040$ &$~0.013\pm0.011$   \\
$W_{\lambda}(\lya)$ (Q2) &~31 &3.036 &$~~+2.3$ &1.17 &0.57 &$~0.012\pm0.004$ &$0.441\pm0.035$ &$~0.027\pm0.009$  &$0.368\pm0.040$ &$~0.033\pm0.011$    \\
$W_{\lambda}(\lya)$ (Q3) &~31 &3.055 &$~~+9.3$ &0.96 &0.47 &$~0.017\pm0.005$ &$0.440\pm0.034$ &$~0.039\pm0.012$ &$0.364\pm0.040$  &$~0.047\pm0.015$    \\
$W_{\lambda}(\lya)$ (Q4) &~31 &3.088 &$~+43.2$ &0.92 &0.38 &$~0.058\pm0.006$ &$0.429\pm0.034$ &$~0.138\pm0.018$ &$0.355\pm0.040$  &$~0.166\pm0.025$    \\
\hline
LAEs  &~28 &3.091 &$~+44.1$ &0.87 &0.41 &$~0.063\pm0.006$ &$0.433\pm0.036$ &$~0.145\pm0.018$ &$0.359\pm0.042$ &$~0.175\pm0.026$     \\
non-LAEs  &~96 &3.037 &$~~+3.6$ &1.09 &0.51 &$~0.012\pm0.003$ &$0.445\pm0.020$ &$~0.027\pm0.007$ &$0.370\pm0.023$ &$~0.032\pm0.008$    \\
\hline
$W_{\lambda}(\lya) >0$    &~74 &3.073 &$~+22.4$ &0.99 &0.45 &$~0.031\pm0.003$  &$0.437\pm0.022$ &$~0.071\pm0.008$  &$0.362\pm0.025$ &$~0.086\pm0.010$   \\
$W_{\lambda}(\lya) <0$    &~50 &3.013 &$~-3.8$ &1.19 &0.55 &$~0.007\pm0.003$ &$0.450\pm0.027$ &$~0.016\pm0.007$   &$0.376\pm0.031$  &$~0.019\pm0.008$    \\
\hline
$(G-{\cal R})_0$ (Q1) &~31 &3.095 &$~+19.4$ &0.94 &0.20 &$0.023\pm0.004$ &$0.430\pm0.034$ &$0.053\pm0.010$ &$0.355\pm0.040$  &$0.055\pm0.013$    \\
$(G-{\cal R})_0$ (Q2) &~31 &3.021 &$~+15.2$ &1.02 &0.41 &$0.022\pm0.006$ &$0.447\pm0.035$ &$0.049\pm0.014$ &$0.372\pm0.041$  &$0.059\pm0.017$    \\
$(G-{\cal R})_0$ (Q3) &~31 &3.059 &$~+12.7$ &1.07 &0.57 &$0.030\pm0.005$ &$0.439\pm0.035$ &$0.068\pm0.013$ &$0.373\pm0.040$  &$0.080\pm0.016$    \\
$(G-{\cal R})_0$ (Q4) &~31 &3.023 &$~~+4.4$ &1.15 &0.78 &$0.011\pm0.005$ &$0.447\pm0.035$ &$0.025\pm0.011$ &$0.373\pm0.040$  &$0.029\pm0.016$    
\enddata
\tablenotetext{a}{Number of objects in the sub-sample. }
\tablenotetext{b}{Rest equivalent width of \lya, where positive (negative) values indicate net emission (absorption).}
\tablenotetext{c}{UV color derived from the observed $G-{\cal R}$ color and corrected for the IGM and the strength of the \lya\ feature in the spectrum of each galaxy (see \S\ref{sec:uv_color} and Figure~\ref{fig:gmr_vs_z}).}
\tablenotetext{d}{Observed mean flux density ratio ($f_{\nu}$) between the rest-frame wavelength range 880-910\AA\ and that in the range 1475-1525\AA.}
\tablenotetext{e}{Mean and standard deviation of the ``IGM only'' transmission  
in the rest-wavelength range 880-910\AA\ for ensembles of sources having the same number and redshift distribution 
as the observed sample.}
\tablenotetext{f}{Mean and standard deviation of the ``IGM+CGM'' transmission 
in the rest-wavelength range 880-910\AA\ for ensembles of sources having the same number and redshift distribution
as the observed sample.}
\tablenotetext{g}{Excludes objects with $|f_{900}/\sigma_{900}| \ge 3$ (15 detections and 3 with $>3\sigma$ negative outliers). }
\label{tab:fesc2_data}
\end{deluxetable*}

\subsection{KLCS Subsamples}

\label{sec:subsamples}

We formed a number of subsets of the final KLCS statistical sample based on
empirical criteria that could be measured easily from photometry or spectroscopy of individual
objects; these include $L_{\rm uv}$, $W_{\lambda}(\lya)$, and rest-UV continuum color $(G-{\cal R})_0$ (\S\ref{sec:uv_color}). 
In view of the results of \S\ref{sec:igm_sampling}, a minimum subsample size $N \simgt 30$  was maintained 
so that the uncertainty in the IGM+CGM correction  
is $\simlt 10$\% (see Table~\ref{tab:sampling}).  Thus, for each of the aforementioned parameters
we split the sample of 124 into 4 independent quartiles consisting of 31 galaxies each. 

Additional subsets were formed according to the following criteria: objects with $L_{\rm uv} \ge L_{\rm uv}^{\ast}$ (48\% of the sample, 
very similar to a combination of the $L_{\rm uv}$ (Q1) and $L_{\rm uv}$ (Q2) subsamples) and those with $L_{\rm uv} < L_{\rm uv}^{\ast}$ (52\% of
the sample); according to whether \lya\ 
appears in net emission ($W_{\lambda}(\lya) > 0$; 60\% of the sample) or net absorption ($W_{\lambda}(\lya) \le 0$; 40\% of the sample);
and, finally, grouping together the galaxies exceeding the often-used threshold $W_{\lambda}(\lya) > 20$ \AA\
for ``Lyman Alpha Emitters'' (LAEs). The KLCS LAE sub-sample (28 galaxies, or $\simeq 22.6$\% of the total) 
is based upon the spectroscopically-measured $W_{\lambda}(\lya)$, and happens to be nearly identical to the $W_{\lambda}(\lya)$ (Q4) 
sub-sample of 31 galaxies.     

\begin{figure}[htbp!]
\centerline{\includegraphics[width=9.5cm]{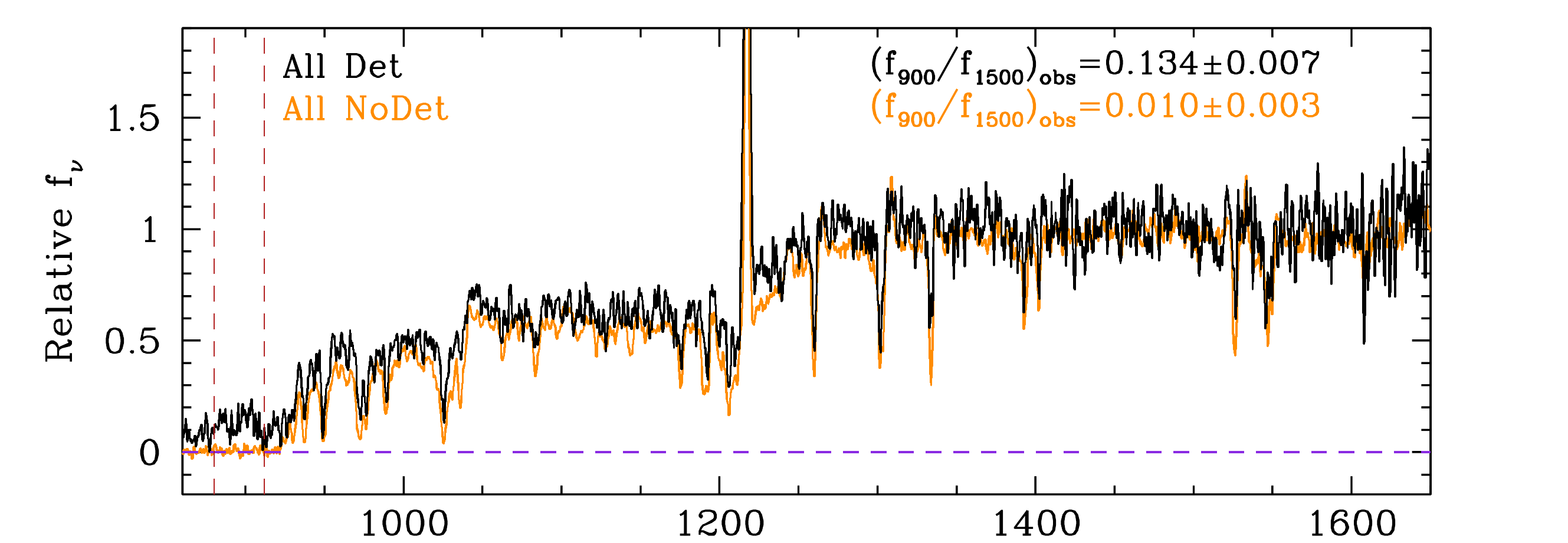}}
\centerline{\includegraphics[width=9.5cm]{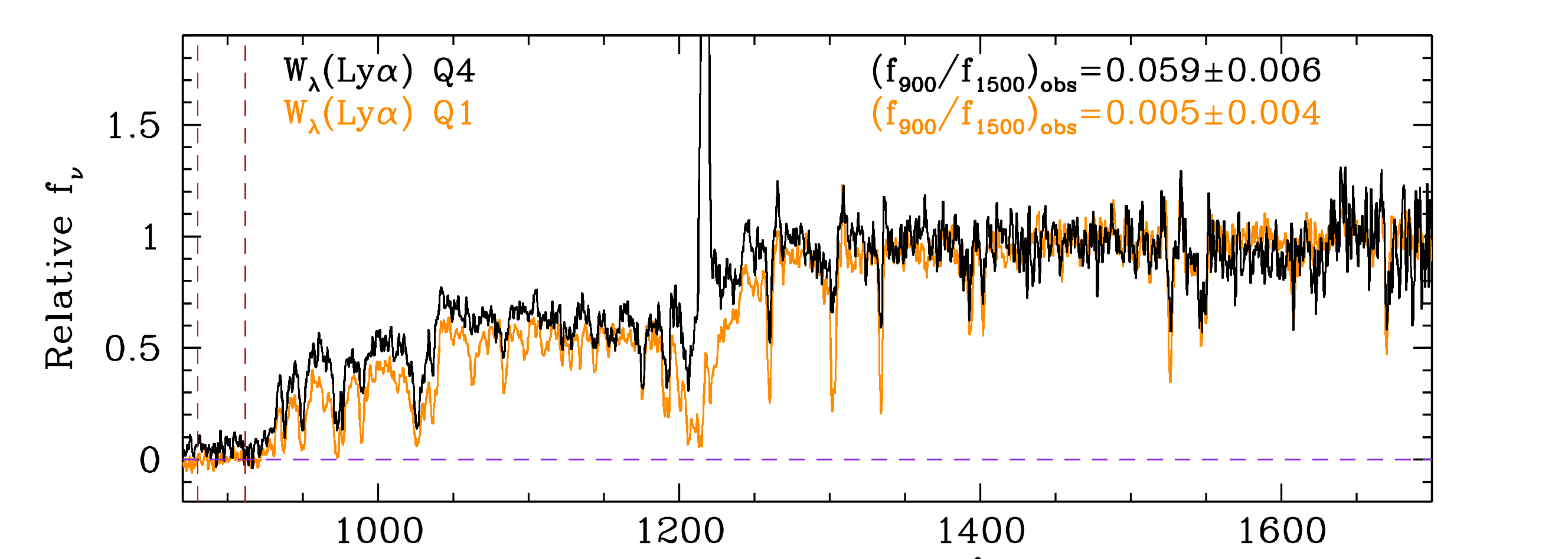}}
\centerline{\includegraphics[width=9.5cm]{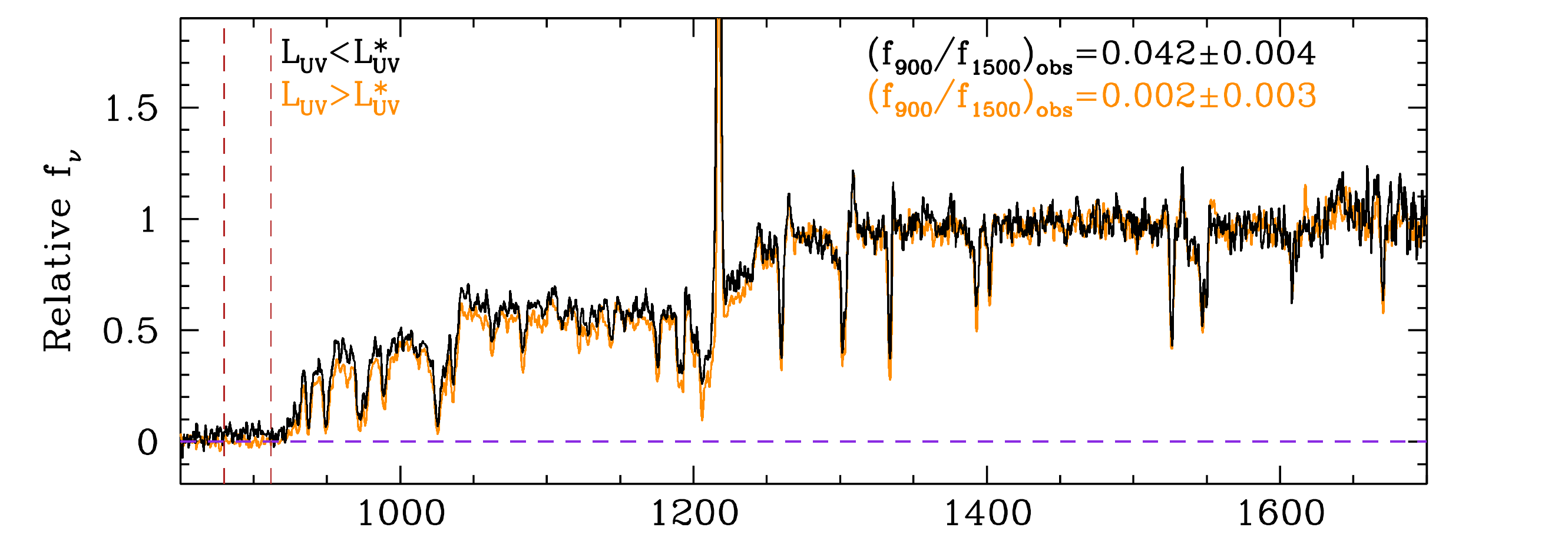}}
\centerline{\includegraphics[width=9.5cm]{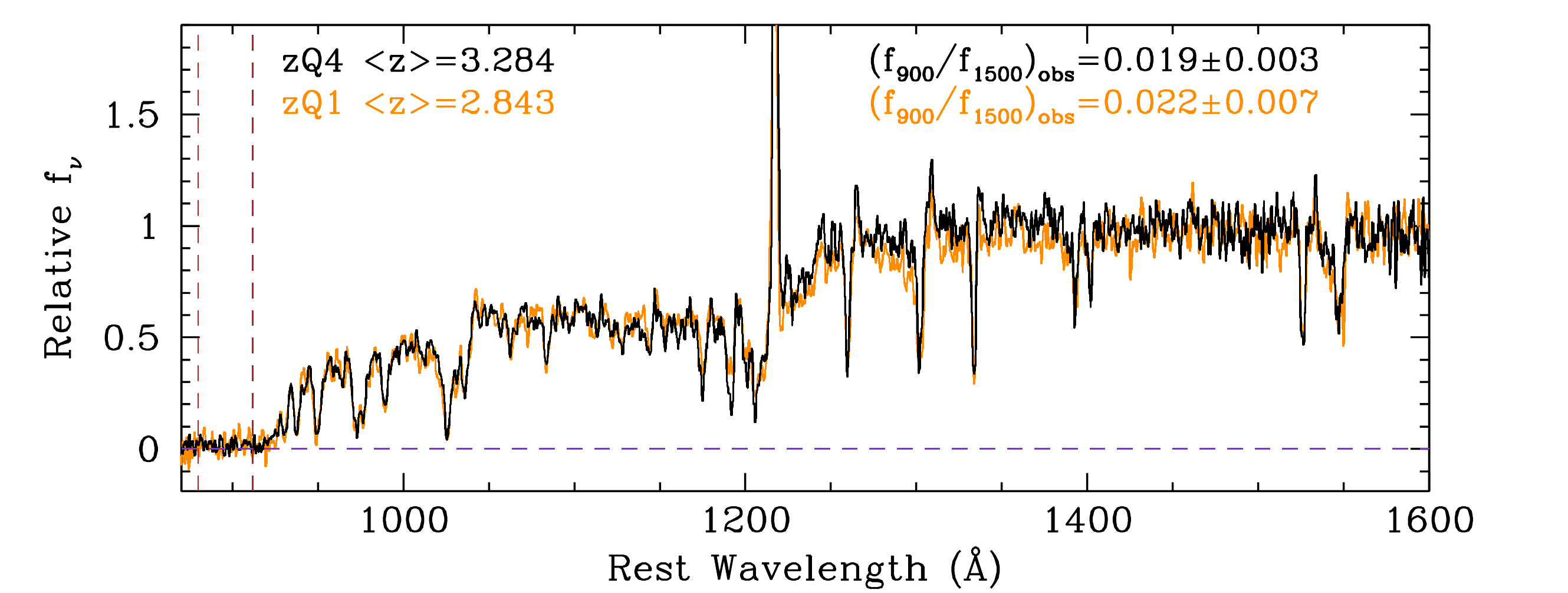}}
\caption{Composite rest-frame spectra of selected KLCS sub-samples, normalized so that $f_{1500} = 1.0$. 
No IGM correction has been applied, i.e., they are the observed flux density as a function of rest
wavelength. 
The subsample names are 
indicated (color-coded) in the top left corner of each panel, with the corresponding measurements of \fobs\ in the upper
right of each panel.  The vertical dashed lines mark the [880,910] rest wavelength interval used for measuring $f_{900}$. 
} 
\label{fig:spec_fig}
\end{figure}

Table~\ref{tab:fesc2_data} includes values of parameters measured directly from the composite spectra or the mean value among the objects comprising
the subsample:  
the number of galaxies in the subsample ($N$), the mean redshift of the objects in the subsample ($\langle z_{\rm s} \rangle$), 
$W_{\lambda}(\lya)$ (measured directly from the composite spectrum), the
mean UV luminosity relative to $L_{\rm uv}^{\ast}$ ($\langle L_{\rm uv}/L_{uv}^{\ast} \rangle$), and the measured flux density ratio \fobs. 
Values of $\langle t_{900} \rangle$ for the transmission in the rest wavelength range [880,910] \AA\ are given for both the ``IGM only'' and ``IGM+CGM'' Monte Carlo
models (\S\ref{sec:igm_trans} and Appendix~\ref{sec:igm_appendix}),  
where each value and its uncertainty were calculated by drawing ensembles of sightlines with the same number and $z_{\rm s}$ distribution as the subsample, averaging
the values of $t_{900}$, and repeating 1000 times. The tabulated uncertainties reflect the 68\% confidence 
interval for the mean transmission of the 1000 ensembles for each subsample.
The composite spectra of several of the sub-samples listed in Table~\ref{tab:fesc2_data} are shown in Figure~\ref{fig:spec_fig}. 

The corrected (emergent) flux density ratio 
\fout\ (equation~\ref{eqn:f9f15_out}) is also listed in Table~\ref{tab:fesc2_data}, 
where the quoted errors  
include uncertainties in both the measurement and in $\langle t_{900} \rangle$. 
Table~\ref{tab:fesc2_data} also includes entries for the full KLCS sample (``All''), and ``All'' divided into subsets according to whether or not
the individual LyC measurement had $|f_{900}/\sigma_{900}| > 3$\footnote{Note that we have not included entries that require a value of 
$\langle t_{900} \rangle$ because the Monte Carlo models assume $t_{900}$ is independent of any property used to select the sample -- clearly invalid in
the case of a known detected or undetected sub-sample.}. 

Finally, Table~\ref{tab:fesc2_data} includes entries for subsamples of KLCS formed according to $z_{\rm s}$, where
$z_{\rm s}$ (Q1) is the lowest-redshift quartile, and $z_{\rm s}$ (Q4) the highest-redshift quartile. These are intended to show that there is no strong 
dependence of the results on source redshift once the redshift-dependent IGM (or IGM+CGM) corrections have been applied; the composite spectra of
these two redshift subsamples are among those shown in 
Figure~\ref{fig:spec_fig}.  

\subsection{Assessment of LyC Results}
\label{sec:lyc_results}

\begin{figure}[htbp]
\centerline{\includegraphics[width=8.0cm]{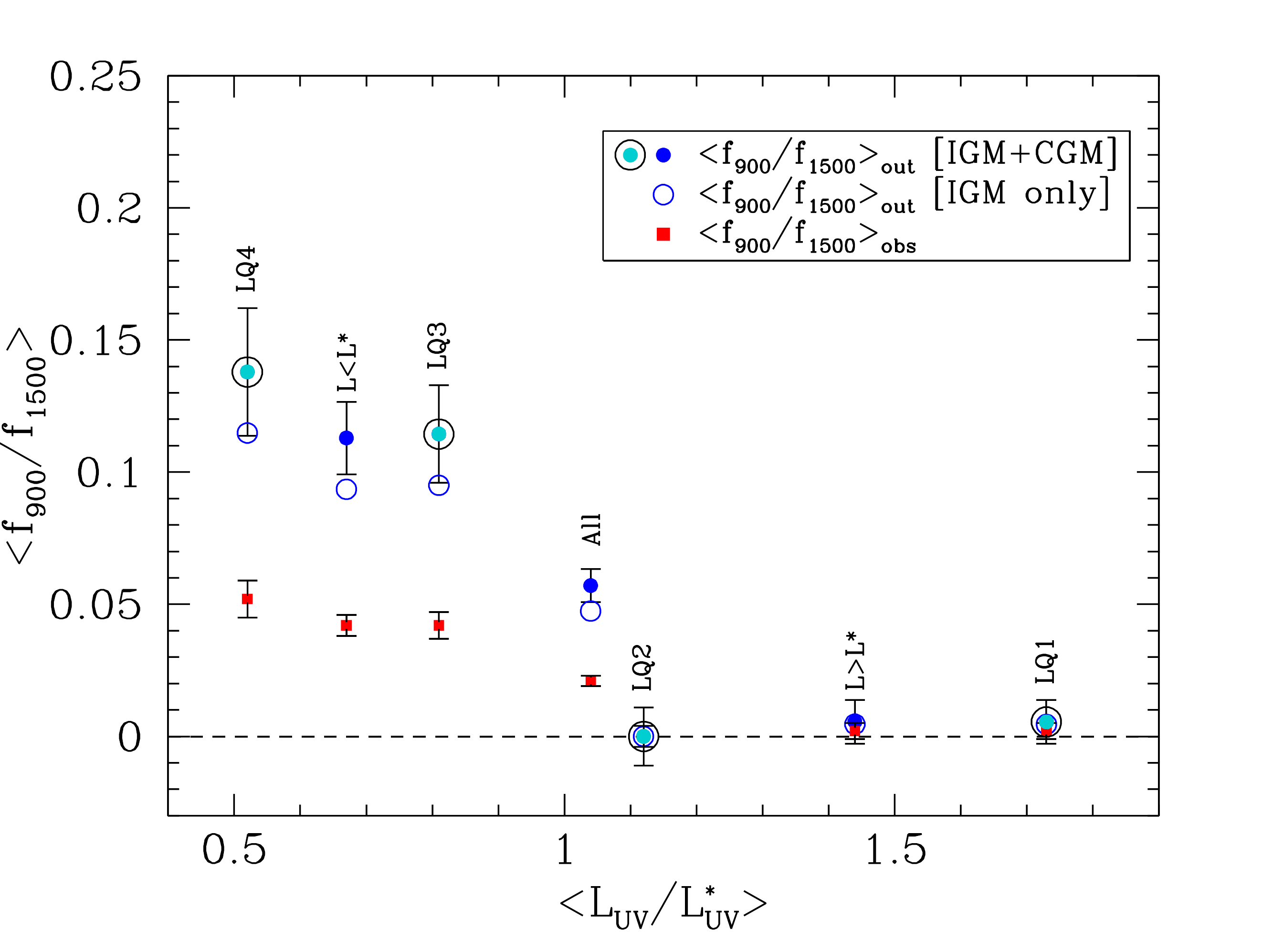}}
\centerline{\includegraphics[width=8.0cm]{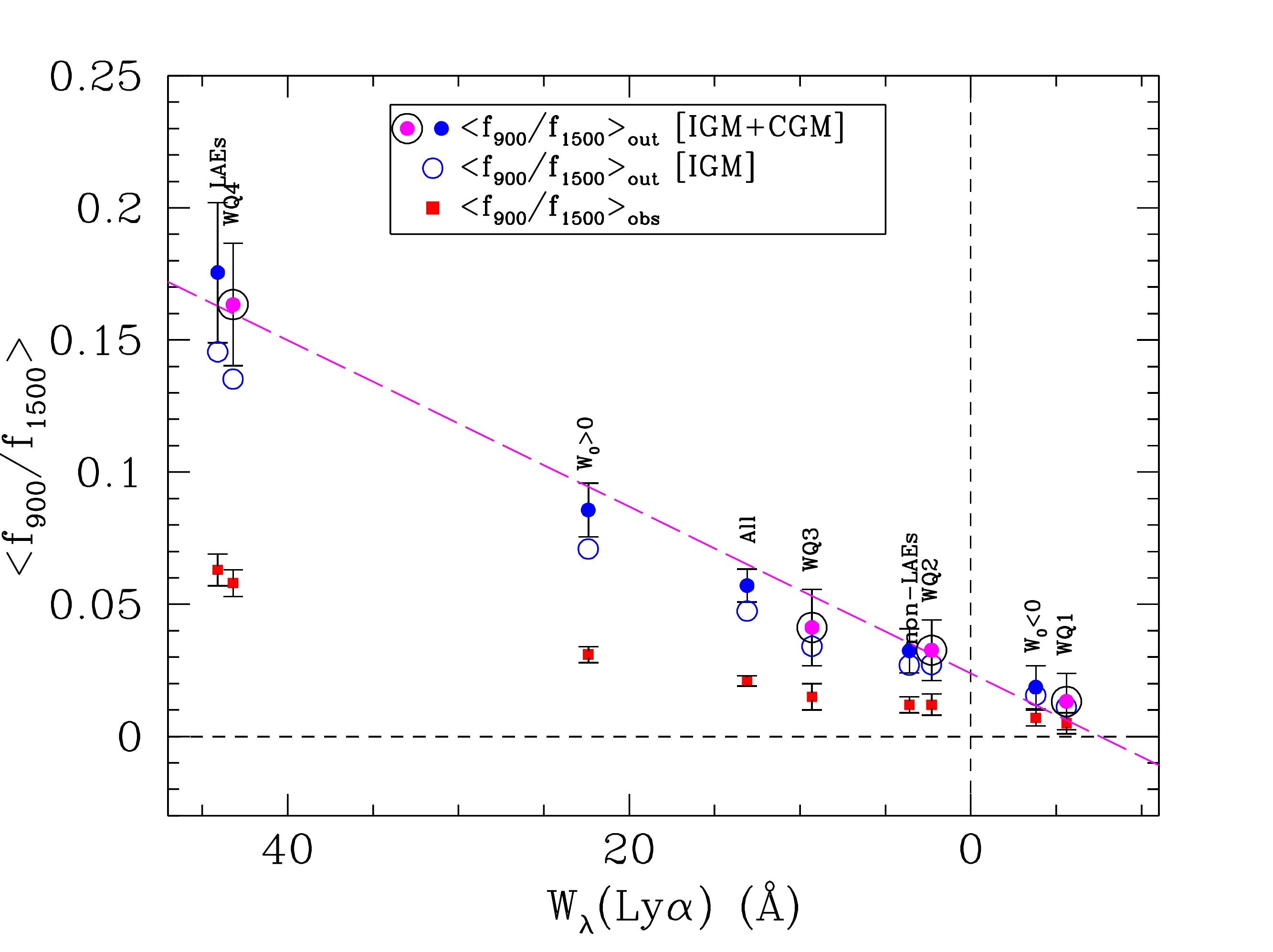}}
\caption{({\it Top:}) Inferred values of $\langle f_{900}/f_{1500} \rangle_{\rm out}$ versus  UV luminosity relative to $L_{\rm uv}^{\ast}$ at
$z \sim 3$.  The four independent quartiles in luminosity are indicated turquoise points surrounded by black circles; other subsamples are
plotted for reference. ({\it Bottom:}) As for the top panel, but showing \fout\ as a function of 
the rest equivalent width of the \lya\ line in the composite spectrum. The independent quartiles in \wlya\ are marked with magenta points 
surrounded by black circles, with other composite sub-samples included for reference, labeled according to
their designation in 
Table~\ref{tab:fesc2_data}. The error bars in both panels 
include both measurement uncertainties and sample variance in the IGM+CGM correction (the latter depends on the size and 
redshift distribution of the sub-sample). The error bars on the \fout\ ``IGM-only'' points (open circles) have been suppressed for
clarity in both panels; the values are listed in Table~\ref{tab:fesc2_data}.   
In the bottom panel, the dashed (magenta) line is a linear relation of the form
\fout$=0.36(W_{\lambda}/110~\textrm{\AA})+0.02$. 
 }
\label{fig:plot_lum_cgm}
\end{figure}

On the basis of the results summarized in Table~\ref{tab:fesc2_data}, there are two easily-measured empirical characteristics that 
correlate most strongly with a propensity to ``leak'' measurable LyC radiation: $L_{\rm uv}$ and
\wlya. Figure~\ref{fig:plot_lum_cgm} illustrates, for selected sub-samples indicated
on the figure, the dependence on $L_{\rm uv}$ and $W_{\lambda}(\lya)$ of the measured
\fobs\ and inferred \fout\ for both IGM transmission models.  

The top panel of Figure~\ref{fig:plot_lum_cgm} suggests an almost bimodal
dependence of $\langle f_{900}/f_{1500} \rangle_{\rm out}$ on $L_{\rm uv}$, where
the 2 lowest-luminosity quartiles (i.e., the lower luminosity half) of the KLCS
sample have \fout$\simeq 0.13$, whereas the UV-brighter half of the sample 
has \fout\ consistent with zero ($3\sigma$ upper limit \fout$\simlt 0.02$.)  The transition
luminosity below which \fout\ appears to increase from zero to $\simeq 13$\% 
is very close to $L_{\rm uv}^{\ast}$, which also happens to lie close to the median $L_{\rm uv}$ of the KLCS sample. 

The trend of \fobs\ and \fout\ with $W_{\lambda}(\lya)$ -- illustrated in the bottom panel of Figure~\ref{fig:plot_lum_cgm} -- is
similar, albeit perhaps exhibiting a more gradual dependence of \fout\ on \wlya\ compared to the relatively abrupt luminosity 
dependence of the subsamples grouped according to $L_{\rm uv}$. 
\begin{figure}[htbp!]
\centerline{\includegraphics[width=8.0cm]{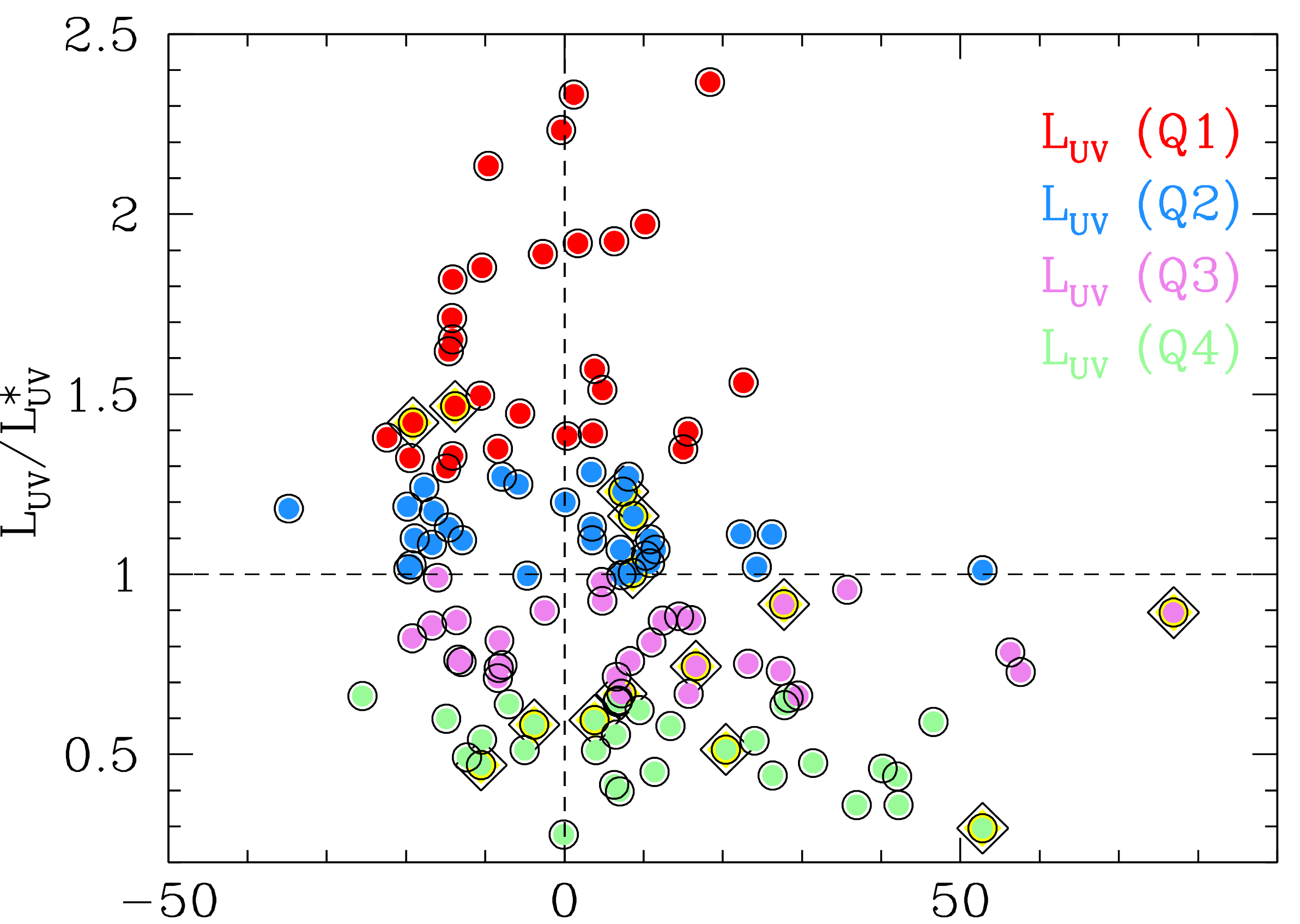}}
\centerline{\includegraphics[width=8.0cm]{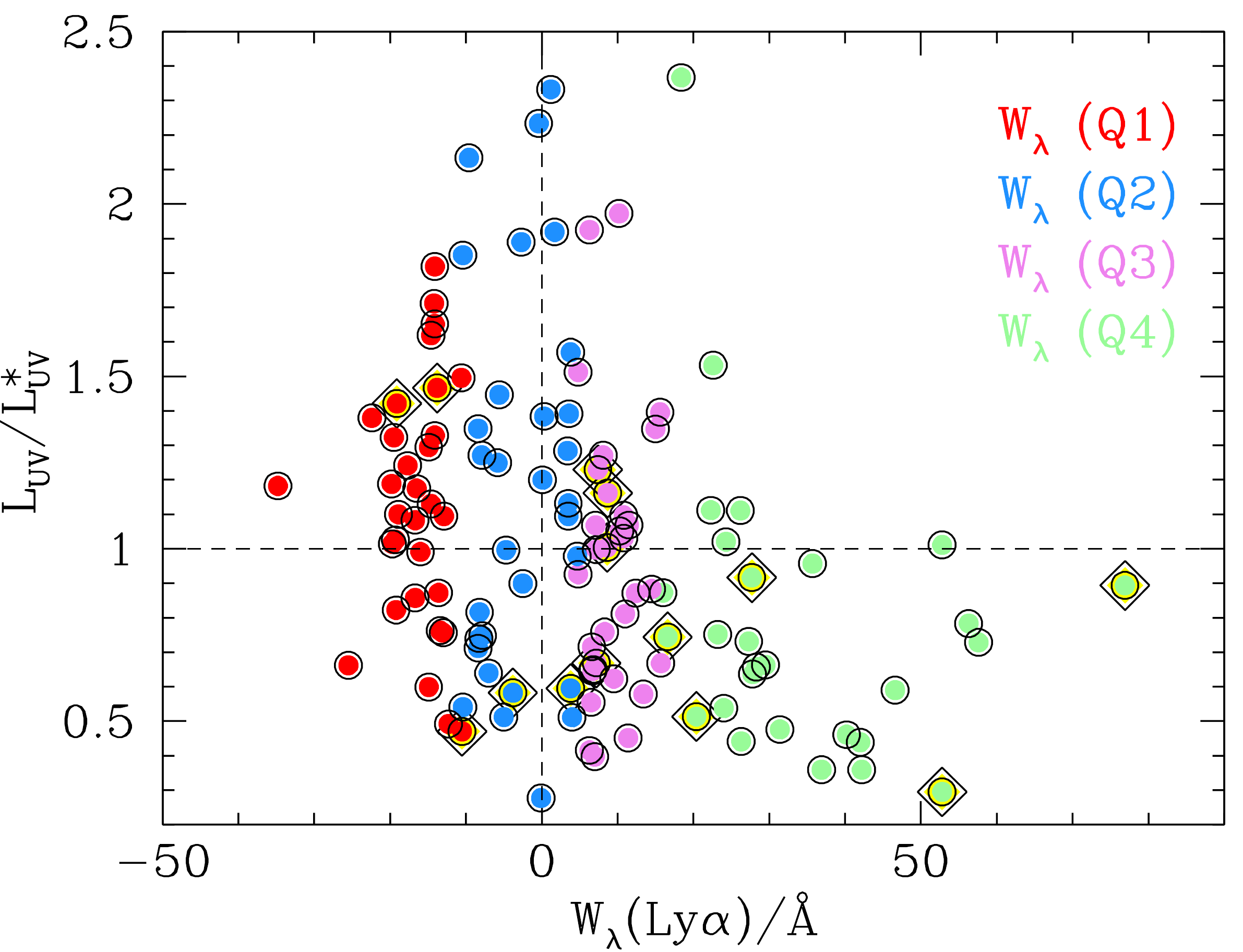}}
\caption{({\it Top:}) $L_{\rm uv}/L_{\rm uv}^{\ast}$ versus the rest-frame \lya\ (\wlya) for the KLCS sample. The points have been
color-coded by quartiles in $L_{\rm uv}$: from lower to higher luminosity, pale green (Q4), violet (Q3), blue (Q2), 
and red (Q1). ({\it Bottom:}) Same as the top panel, except with points color-coded by quartiles in $W_{\lambda}(\lya)$. In both
panels, the galaxies with $3\sigma$ LyC detections are also indicated with diamonds.}
\label{fig:w0_vs_lum}
\end{figure}
Figure~\ref{fig:w0_vs_lum} shows the KLCS sample color-coded according to quartiles in $L_{\rm uv}$ (top) and $W_{\lambda}(\lya)$ (bottom). 
There is clearly an inverse correlation between \wlya\ and $L_{\rm uv}$ in the sense that there is a dearth of galaxies
with bright $L_{\rm uv}$ and large \wlya\ -- e.g., the brighter two quartiles in
$L_{\rm uv}$ (Q1 and Q2) are dominated by galaxies with \wlya$\simlt 10$ \AA\ (median$\simeq 0$ \AA), and most of the galaxies
in the two largest-\wlya\ quartiles (Q3 and Q4) have $L_{\rm uv} < L_{\rm uv}^{\ast}$\footnote{Formally, a Spearman rank correlation test
between $L_{\rm uv}$ and \wlya\ results in a correlation coefficient $\rho = -0.30$, significant at the 3.4$\sigma$ level.}.     
Because of the well-established correlation between $L_{\rm uv}$ and \wlya\ within all LBG samples (including KLCS; see ,e.g., 
\citealt{shapley03,gronwall07,kornei10,stark10}), it is difficult to say with certainty which is the more
reliable indicator of high \fout; however, for reasons discussed in detail in
\S\ref{sec:morphology} and \S\ref{sec:trends} below, we believe that \wlya\ is more directly linked to whether or not a galaxy has 
significant LyC leakage, and that the $L_{\rm uv}$ dependence of \fout\ results from the UV luminosity dependence of the gas-phase covering fraction/column density
along the line of sight. 
 
\section{Galaxy Properties and LyC Escape}
\label{sec:galaxy_properties}

While a determination of the total ionizing emissivity contributed by an ensemble of galaxies can be calculated 
directly from \fout, without knowing anything about the galaxies except for their FUV luminosity function, one
would like to understand the physical causes of the large differences inferred among the sample. More generally,
one would like to place LyC escape in the context of other galaxy properties, including the massive
star populations and the ISM conditions that modulate LyC ``leakage''.  
Understanding which galaxies contribute -- and why -- to the metagalactic ionizing radiation field at $z\sim3$ 
will certainly improve our ability to make quantitative sense of the reionization era, where direct measurements
of \fout\ will not be possible. These questions also bear considerably on our understanding of galaxies, the CGM, and the IGM
at $z \sim 3$ where the impact of LyC leakage is direct and highly relevant.  
In this section, we extend our analysis of the KLCS results to include properties of the galaxies, their stellar populations,
and the radiative transfer of ionizing and non-ionizing FUV light through the ISM. 

We begin by briefly reviewing the commonly-used forms of the LyC ``escape fraction'' 
($f_{\rm esc}$) in the context of the KLCS dataset.

\subsection{Escape Fraction: Definitions}
\label{sec:fesc}

The most common definition of escape fraction, often used in theoretical studies
of re-ionization, is the fraction of ionizing photons produced by stars
in a galaxy that escape into the IGM without being absorbed by \ion{H}{1} within
the galaxy (e.g., \citealt{Wyithe07,Wise09}). This quantity has also been called the ``absolute escape fraction''
($f_{\rm esc, abs}$) to convey the fact that the ionizing photon budget against
which the leaking LyC is compared includes all of the ionizing UV photons whether or not they are evident in observations.  
Thus, in order to determine $f_{\rm esc,abs}$, a measure 
of the intrinsic production rate of ionizing photons is necessary; unfortunately, direct empirical estimates
of the production rate of ionizing photons ($N_{\rm ion}$) are relatively impractical for high redshift galaxies.  Instead, one typically uses SED modeling of  
young stellar populations (including dust extinction, population age, IMF, etc.) 
to estimate the intrinsic ionizing photon production.  
Doing this accurately requires knowledge of the intrinsic SED of the massive stellar populations in
the extreme UV (EUV), especially for photon energies in the range 1-4 Ryd, as well as a detailed understanding
of the distribution and composition of dust grains that attenuate and redden the intrinsic spectrum
of the stars in the galaxy. 

An alternative definition intended to be closer to the most readily-available measurements of high redshift
galaxy populations is the ``relative escape fraction'' ($f_{\rm esc, rel}$; \citealt{steidel01}), 
\begin{equation}
\label{eqn:frel}
f_{\rm esc,rel} = \frac { \langle f_{900} /  f_{1500}\rangle_{\rm out} } {(L_{900}/L_{1500})_{\rm int}}\times10^{\left[A_{\lambda}(900)-A_{\lambda}(1500)\right]/2.5}
\end{equation}
where
\begin{equation}
A_{\lambda} = k_{\lambda}  E(B-V) \quad . 
\end{equation} 
is the attenuation in magnitudes as a function of rest wavelength and $k_{\lambda}$ parametrizes  
the attenuation relation. 
The term $(L_{900}/L_{1500})_{\rm int}$ is  
the intrinsic ratio of the unattenuated stellar population of the galaxy, prior to transfer through 
the ISM, the CGM, and the IGM. Note that there are inconsistencies in the definition of
$f_{\rm esc, rel}$ in the literature having to do with whether or not dust attenuation is included; some authors 
have implicitly assumed that $A_{\lambda}(900) - A_{\lambda}(1500) = 0$, i.e. that the attenuation by dust affects ionizing
and non-ionizing UV equally, or that dust affects only the non-ionizing UV [i.e., $A_{\lambda}(900~\textrm{\AA}) = 0$].  
The definition of $f_{\rm esc,rel}$ used by \citet{grazian17}, expressed using the notation we have adopted here, 
is $f_{\rm esc,rel} = (f_{900}/f_{1500})_{\rm out}/(L_{900}/L_{1500})_{\rm int}$.  In order to avoid ambiguity, 
in what follows below we have attempted to be clear about any assumptions made in mapping \fout\ to more
model-dependent quantities whenever relevant;  
\S\ref{sec:screen} and \ref{sec:holes}
consider ISM models both with and without dust attenuation of emergent LyC flux. 

\subsection{Stellar Population Synthesis Models}
\label{sec:sps_models}

The intrinsic energy distribution of the stellar sources, parametrized by $(L_{900}/L_{1500})_{\rm int}$,
is a crucial ingredient to understanding the relationship between ionizing photon
production and escape into the IGM. The intrinsic break 
at the Lyman limit in the integrated spectrum of stars depends on the age, metallicity, initial mass
function (both slope and upper mass limit), and the effects of binary evolution of massive stars. 
Depending on the assumptions, SPS models predict spectra 
falling in the range $0.15 \simlt (L_{900}/L_{1500})_{\rm int} \simlt 0.75$ (see, e.g., \citealt{leitherer99,eldridge17}). 
In the absence of external
constraints, the factor of $\sim 5$ uncertainty on $(L_{900}/L_{1500})_{\rm int}$ translates directly into uncertainties on estimates
of escape fraction. 

Fortunately, high quality far-UV composite spectra such as those of the KLCS sub-samples significantly constrain
the likely range of stellar population parameters consistent with the observations.
Using similar far-UV composites constructed
from the spectra of $\langle z \rangle \simeq 2.4$ galaxies that were also observed
in the near-IR as part of KBSS-MOSFIRE, \citet{steidel16} (S16) showed that, among the SPS
models considered, the only set which could simultaneously match the details of the
far-UV stellar spectrum --  {\it and} correctly match the observed ratios of strong
nebular emission lines in the spectra of the same galaxies --  were those with low stellar metallicity,
${\rm (Fe/H) \simeq 0.1~ (Fe/H)_{\odot}}$, and which include the evolution of
massive stars in binary systems (BPASSv2.1; \citealt{eldridge17}).  

Prior to fitting models to the observed spectra, the mean IGM transmission spectrum for
the CGM+IGM Monte Carlo models described in \S\ref{sec:igm_trans} and Appendix~\ref{sec:igm_appendix}
(similar to that shown in Figure~\ref{fig:fakespec_igm}) were calculated separately for each subsample by drawing 1000 sightline ensembles 
with the same number and $z_{\rm s}$ distribution. We divided the observed subsample spectrum by the mean transmission spectrum and
propagated the 68\% confidence interval around the mean at each rest-wavelength point as the contribution
of the IGM correction to the error vector for $\lambda_0 < 1216$ \AA. 
 
Following the procedure detailed in S16, we fit each of the composite KLCS spectra
in Table~\ref{tab:fesc2_data} with a range of SPS models similar to that described by S16, with the following 
additions:  we tested 3 different attenuation relations for reddening the SPS model spectra, in addition
to varying the metallicity of the stars and the upper mass limit and slope of the IMF. We found that the newest 
publicly released 
BPASSv2.1 \citep{eldridge17} models with stellar metallicity $Z_{\ast}=0.001$ ($\simeq 0.07~ Z_{\odot}$, assuming the solar abundance
scale of \citealt{asplund09}), IMF slope $\alpha=-2.35$, and upper stellar mass limit of 300 M$_{\odot}$ (hereafter referred
to as BPASSv2.1-300bin-z001), provided the lowest $\chi^2$ (i.e., the best fit) for
every KLCS composite listed in Table~\ref{tab:fesc2_data}. 
For all of the fits, we assumed a continuous star formation history, constant star formation rate, and an age of
$t=10^8$ years, close to the median age inferred from SED fits to photometry from the UV to mid-IR (e.g., \citealt{reddy12,du18}.) The
effects of the star-forming age on the EUV-FUV spectra in the context of the BPASS-v2.1-300bin-z001 models are discussed in \S\ref{sec:implications}.
The continuum reddening was allowed to vary over the range $0 \le E(B-V) < 1.0$ for each subsample; 
however, the attenuation relation accompanying the best-fitting model 
varied among the sub-samples; Table~\ref{tab:models} summarizes the 
parameters of the best fit for each sub-sample listed in Table~\ref{tab:fesc2_data}.

\begin{deluxetable*}{llccccrrrr}
\tabletypesize{\scriptsize}
\tablewidth{0pt}
\tablecaption{Statistics Derived from SPS Model Fits to Spectra}
\tablehead{
\colhead{Sample} & \colhead{Att.\tablenotemark{a}} & \colhead{E(B-V)\tablenotemark{b}} & \colhead{$(L_{900}/L_{1500})_{\rm mod}$\tablenotemark{c}} 
& \colhead{C1500\tablenotemark{d}} & \colhead{$\langle L_{\rm bol}\rangle$\tablenotemark{e}} &  
\colhead{$\langle f_{\rm esc,rel}\rangle$\tablenotemark{f}} & 
\colhead{$\langle f_{\rm esc,abs}\rangle$\tablenotemark{g}} & \colhead{$\langle f_{\rm esc,rel}\rangle$\tablenotemark{f}} & \colhead{$\langle f_{\rm esc,abs}\rangle $\tablenotemark{g}} \\
\colhead{} & \colhead{} & \colhead{} & \colhead{} & \colhead{} & \colhead{($10^{11}L_{\rm \odot}$)} & \multicolumn{2}{c}{[IGM only]} & \multicolumn{2}{c}{[IGM+CGM]}} 
\startdata
All &                               R16 & 0.129 & 0.167 & 2.88 & 2.98 & $0.29\pm0.03$   &  $0.09\pm 0.01$ & $~0.36\pm0.04$ & $~0.12\pm0.01$  \\
All, detected\tablenotemark{h}  &   SMC & 0.045 & 0.183 & 1.76 & 1.50 &  $1.04\pm 0.10$  & $ 0.60 \pm  0.06$ &  $~1.21\pm0.11$ & $~0.70\pm0.07$  \\
All, not detected  &                R16 & 0.135 & 0.163 & 3.03 & 3.17 &  $0.13\pm  0.04$  &  $0.04\pm0.01$ &  $~0.17\pm0.05$ & $0.06\pm0.02$  \\
\hline	
z(Q1) &                             SMC & 0.060 & 0.160 & 2.12 & 1.92 & $0.27\pm 0.08$ & $0.12\pm 0.04$ &  $~0.34\pm0.11$ & $~0.16\pm0.03$  \\
z(Q4) &                             R16 & 0.080 & 0.202 & 1.93 & 2.26 & $0.25\pm 0.04$ & $0.12\pm 0.02$ &  $~0.31\pm0.05$ & $~0.16\pm0.03$  \\
\hline	
$L_{\rm uv} > L_{\rm uv}^{\ast}$ &  R16 & 0.141 & 0.160 & 3.18 & 4.55 &  $0.03\pm 0.04$ &  $0.01\pm 0.01$ & $~0.05\pm0.05$ & $~0.02\pm0.02$  \\
$L_{\rm uv} < L_{\rm uv}^{\ast}$ &  SMC & 0.047 & 0.180 & 1.80 & 1.20 & $0.52\pm 0.04$ &  $0.27\pm 0.03$ & $~0.66\pm0.09$ & $~0.36\pm0.04$  \\
\hline	
$L_{\rm uv}$ (Q1) &                 R16 & 0.133 & 0.165 & 2.98 & 5.11 &  $0.01\pm  0.04$   &  $0.02\pm 0.01$ & $0.03\pm0.05$ & $~0.01\pm0.02$   \\
$L_{\rm uv}$ (Q2) &                 R16 & 0.154 & 0.152 & 3.54 & 3.94 & $-0.01\pm  0.06$ &  $0.00\pm  0.02$ &  $-0.01\pm0.07$& $ -0.00\pm0.02$    \\
$L_{\rm uv}$ (Q3) &                 SMC & 0.047 & 0.180 & 1.80 & 1.44 &  $0.54\pm  0.07$   &  $0.29\pm  0.04$ &  $0.67\pm0.10$ & $~0.38\pm0.05$    \\
$L_{\rm uv}$ (Q4) &                 SMC & 0.044 & 0.185 & 1.73 & 0.88 &  $0.62\pm  0.11$   &  $0.34\pm  0.07$ & $0.79\pm0.13$ & $~0.45\pm0.07$   \\
\hline	
$\wlya$ (Q1)  &               R16 & 0.158 & 0.150 & 3.66 & 4.10 &	 $ 0.07\pm  0.06$   &  $0.02\pm  0.02$ & $~0.08\pm0.07$ & $~0.02\pm0.02$    \\
$\wlya$ (Q2) &                R16 & 0.176 & 0.140 & 4.24 & 4.93 &	$  0.21\pm  0.06$   &  $0.05\pm  0.02$ &  $~0.26\pm0.08$ & $~0.06\pm0.02$   \\
$\wlya$ (Q3) &                SMC & 0.049 & 0.177 & 1.85 & 1.76 &	$  0.21\pm  0.07$   &  $0.11\pm  0.04$ &  $~0.26\pm0.08$ & $~0.14\pm0.05$    \\
$\wlya$ (Q4) &                SMC & 0.027 & 0.215 & 1.40 & 1.29 &	$  0.61\pm  0.08$   &  $0.43\pm  0.06$ &  $~0.76\pm0.11$ & $~0.55\pm0.08$    \\
\hline	
LAEs &                              SMC & 0.027 & 0.215 & 1.40 & 1.21  &  $0.66\pm  0.08$    &   $0.46\pm  0.06$ &  $~0.85\pm0.12$ & $~0.62\pm0.09$   \\
non-LAEs &                          R16 & 0.146 & 0.157 & 3.31 & 3.58  & $0.19\pm 0.04$  &   $0.05\pm  0.01$ & $~0.23\pm0.05$ & $~0.07\pm0.02$   \\
\hline	
$\wlya >0$ &                 SMC & 0.042 & 0.188 & 1.69 & 1.66 &  $ 0.37\pm  0.04$  &  $0.21\pm  0.03$ & $~0.48\pm0.05$ & $~0.28\pm0.03$   \\
$ \wlya <0$ &                  R16 & 0.165 & 0.146 & 3.87 & 4.57 &  $ 0.11\pm  0.05$  &  $0.03\pm  0.01$ &  $~0.13\pm0.05$ & $~0.03\pm0.01$    \\
\hline	
$(G-{\cal R})_0$ (Q1) &            SMC & 0.015 & 0.240 & 1.21 & 1.12 &  $0.21\pm  0.04$    &    $0.17\pm  0.03$ & $~0.28\pm0.05$ & $~0.23\pm0.04$    \\
$(G-{\cal R})_0$ (Q2) &            R16 & 0.113 & 0.178 & 2.53 & 2.56 &  $0.26\pm  0.08$    &    $0.10\pm  0.03$ & $~0.33\pm0.10$ & $~0.13\pm0.04$    \\
$(G-{\cal R})_0$ (Q3) &            R16 & 0.159 & 0.149 & 3.69 & 3.91 &  $0.48\pm  0.09$    &    $0.12 \pm 0.02$ & $~0.60\pm0.11$ & $~0.16\pm0.03$    \\
$(G-{\cal R})_0$ (Q4) &            R16 & 0.221 & 0.117 & 6.13 & 7.00 &  $0.19\pm  0.12$    &    $0.03 \pm 0.02$ & $~0.25\pm0.14$ & $~0.04\pm0.02$    
\enddata
\tablenotetext{a}{Attenuation relation for the best-fit SPS model: SMC: \citep{gordon03,reddy16a}; R16: \citep{reddy15,reddy16a}.}
\tablenotetext{b}{Formal uncertainties for a given attenuation and SPS model are small, $\sigma[{\rm E(B-V)}]\simeq 0.001-0.002$}
\tablenotetext{c}{Flux density ratio for stellar population synthesis model, after reddening
according to the specified extinction relation. The intrinsic value is $(L_{900}/L_{1500})_{\rm int} \simeq 0.28$. }
\tablenotetext{d}{Inferred extinction correction at $\lambda_0=1500$ \AA\, for the best-fit
SPS model. }
\tablenotetext{e}{Inferred bolometric luminosity from the attenuation-corrected $L_{1500}$, in solar luminosities.}
\tablenotetext{f}{Ratio between $(f_{\rm 900}/f_{1500})_{\rm out}$ and $(L_{\rm 900}/L_{1500})_{\rm mod}$, for
the best-fitting SPS model. }
\tablenotetext{g}{Inferred absolute escape fraction, where $f_{\rm esc,abs} \equiv f_{\rm esc,rel}/C1500$ (see Fig.~\ref{fig:absfesc}). }
\tablenotetext{h}{Values of $f_{\rm esc, rel}$ and $f_{\rm esc, abs}$ for the subsample with individual detections assuming that
the IGM corrections are drawn from the top 12\% of the $t_{900}$ distribution function at $\langle z \rangle = 3.093$ (see Table~\ref{tab:fesc2_data}).}
\label{tab:models}
\end{deluxetable*}

Although three different attenuation relations were tested in the fitting, 
none of the composite spectra was best-fit by 
the \citet{calzetti00} attenuation relation. 
As indicated in Table~\ref{tab:models}, 13 of the 23 composite spectra were best fit by the attenuation relation constructed by combining
the results of \citet{reddy15} with the far-UV extension of \citet{reddy16a} (R16), while the remaining 10 spectra favored the steeper,
SMC attenuation curve which combines the empirical line-of-sight SMC extinction curve from \citet{gordon03} (assuming $R_{\rm V} = 2.74$) with 
an extension to the FUV derived as in \citet{reddy16a}.  

The values of $(L_{900}/L_{1500})_{\rm mod}$ listed in Table~\ref{tab:models} assume (for the time being) that the same attenuation curve and $E(B-V)$ value
affects both the ionizing and non-ionizing UV
light, and that the far-UV attenuation curves can be extrapolated
shortward of $\lambda_0 \le 950$~\AA, the shortest wavelength over which they are empirically constrained\footnote{The extrapolation need only 
extend to $\lambda_0=880$ \AA, the shortest wavelength used for any measurements.} (see \citealt{reddy16a}).  
Here we remind the reader that the values of $\langle f_{\rm esc,rel} \rangle$ listed in Table~\ref{tab:models} are as defined in equation~\ref{eqn:frel}. 
Although the fits to the stellar continuum exclude all wavelength pixels shortward of 1070~\AA,
the fitted spectra appear to be excellent representations of the observed spectra all the way down to $\sim 930$ \AA, where
the confluence of the stellar and interstellar Lyman series lines in the wavelength interval $910 -930$ \AA\ depresses the galaxy
spectrum relative to the model; an example is shown in Figures~\ref{fig:igm_demo} (see also Figure~\ref{fig:igm_demo_zoom}.)  

The BPASSv2.1-300bin-z001 continuous star formation SPS model, including the contribution of the nebular continuum as described by S16, 
has $(L_{900}/L_{1500})_{\rm int} \simeq 0.28\pm 0.03$\footnote{The quantity $(L_{900}/L_{1500})_{\rm int}$ is closely related to
the quantity $\xi_{\rm ion}$ (see, e.g., \citealt{robertson15}), the number of H-ionizing photons produced per unit
non-ionizing UV specific luminosity at rest-frame 1500 \AA. The BPASSv2.1-300bin-z001 model used in the present case
has $\xi_{\rm ion} \simeq 25.5\pm0.1$ for continuous star-formation ages of $7.5 \simlt {\rm log}({\rm t/ yr}) \simlt 8.7$.}, prior to reddening according to the
values in Table~\ref{tab:models}; our assumption that the entire spectrum is reddened by the same attenuation relation
further reduces the intrinsic $L_{900}/L_{1500}$ by an amount that depends on $E(B-V)$ and the attenuation curve. Thus, the
values of $(L_{900}/L_{1500})_{\rm mod}$ in Table~\ref{tab:models} correspond to the predicted emergent spectrum that would be observed if
there were reddening by dust, but no photoelectric absorption by \ion{H}{1} in the ISM along the line of sight. 

\begin{figure}[htbp!]
\centerline{\includegraphics[width=8.5cm]{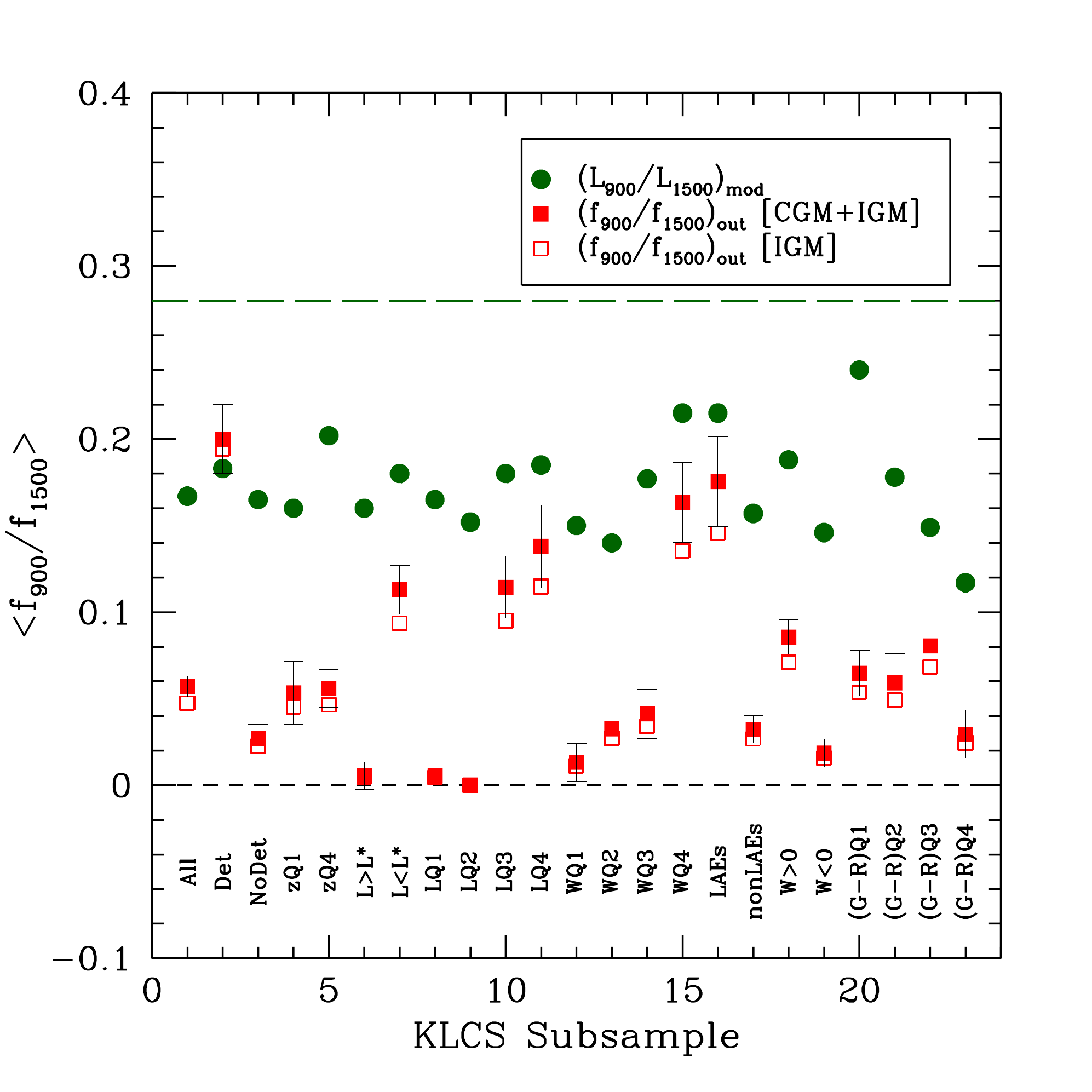}}
\caption{Comparison of IGM+CGM (filled red squares with error bars) and IGM-only (open red squares) of \fout\ (see Table~\ref{tab:fesc2_data}),  
with the values of $(L_{900}/L_{1500})_{\rm mod}$ based on the best-fit continuum-reddened stellar population synthesis model. The
ratio between the values of the red (squares) and filled green dots is defined as $f_{\rm esc,rel}$ (Table~\ref{tab:models}). The error bars on the
IGM-only points (open squares) have been suppressed for clarity (see Table~\ref{tab:fesc2_data} for values). The dashed horizontal line shows
the value of $(L_{900}/L_{1500})_{\rm int}$ for the SPS model. }
\label{fig:f9f15_models}
\end{figure}

\begin{figure}[htbp!]
\centerline{\includegraphics[width=8.5cm]{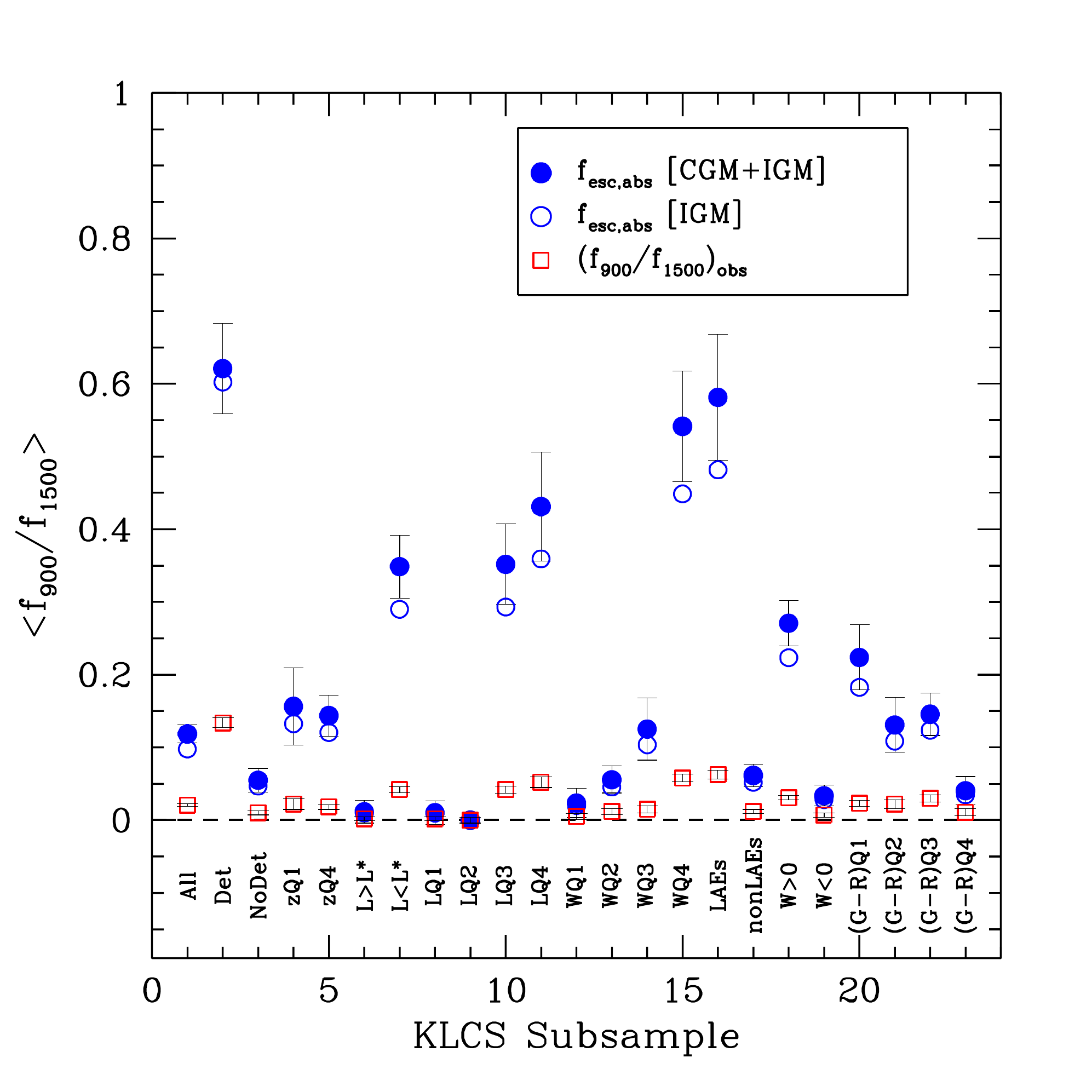}}
\caption{Comparison of the inferred value of $f_{\rm esc,abs}\equiv f_{\rm esc,rel}/{\rm C}1500$ (solid and open blue circles) 
with the {\it observed} ratio (open red squares) $(f_{900}/f_{1500})_{\rm obs}$ (see Table~\ref{tab:fesc2_data}).  
As for Figure~\ref{fig:f9f15_models}, the error bars have been suppressed for $f_{\rm esc,abs}$ values for the ``IGM-only'' opacity
model;  
the values of the points are collected in Table~\ref{tab:models}. 
 }
\label{fig:absfesc}
\end{figure}

Table~\ref{tab:models} lists the inferred values of $f_{\rm esc, rel}$, equal to the ratio between
\fout\ (Table~\ref{tab:fesc2_data}) and $(L_{900}/L_{1500})_{\rm mod}$ (see equation~\ref{eqn:frel}), the flux
density ratio for the best-fit (reddened) SPS spectrum.
Figure~\ref{fig:f9f15_models} illustrates the associated values for each of the subsample composite spectra. 

At this point, the value of $f_{\rm esc, rel}$ has accounted for the relative attenuation between $900$ and $1500$ \AA, but
not for the absolute attenuation affecting $f_{1500}$.  In the context of the modeled UV SEDs summarized in Table~\ref{tab:models},
this factor is simply 
\begin{equation}
\label{eqn:att_factor}
{\rm C1500} \equiv 10^{E(B-V)\times k_{\lambda}(1500)/2.5}
\end{equation}
where $k_{\lambda}(1500) = 8.91, 13.05$ for the R16 and SMC attenuation relations, respectively.  

We then define the absolute escape fraction as
\begin{equation}
\label{eqn:fesc_abs}
\langle f_{\rm esc, abs} \rangle = \langle f_{\rm esc, rel}\rangle /{\rm C1500}~;
\end{equation}
both ${\rm C1500}$ and $f_{\rm esc, abs}$ are included in Table~\ref{tab:models}. Figure~\ref{fig:absfesc} 
compares the values of this inferred quantity with the {\it observed} $\langle f_{900}/f_{1500}\rangle_{\rm obs}$ 
for each KLCS subset composite spectrum. 
Note that all of the KLCS subsets with $f_{\rm esc,abs} \simgt 0.20$ -- whether based on UV color, UV luminosity, or $W_{\lambda}(\lya)$ -- 
 are best-fit
using the ``line-of-sight'' SMC extinction curve. This can probably be attributed to the relatively blue UV color of
these composites, as the bluest/youngest star forming galaxies are known to be most consistent with 
a steep UV attenuation curve (\citealt{reddy10,reddy17}), rather than grayer extinction curves such as
those of \citet{calzetti00} or \citet{reddy16a}, perhaps due to the prevalence of small dust grains (or the absence of large ones). 

We note that fitting reddened SPS models to the observed spectra to estimate
$f_{\rm esc,abs}$ (Table~\ref{tab:models} and Figure~\ref{fig:absfesc}) rather
than the more empirical \fout\ reveals that, in addition to luminosity and
\wlya\ dependence noted in \S\ref{sec:composites} for \fout\ (each of which becomes steeper relative to 
$f_{\rm esc, abs}$), accounting for continuum reddening 
corrections suggests that $f_{\rm esc,abs}$ also
depends strongly on UV color ($(G-{\cal R})_0$), in qualitative agreement with the analysis presented by \citet{reddy16b}.   

We return to a discussion of the implications of the various measures of escape fraction in \S\ref{sec:discussion}. 

\subsection{Far-UV Spectral Properties and LyC Escape}
\label{sec:morphology}

As described in \S\ref{sec:sps_models} above, the relative escape fraction is 
$f_{\rm esc,rel} \simeq 0.5$ for the most-leaking KLCS subsets,
with corresponding $f_{\rm esc,abs} \simeq 0.3-0.4$; both values are model-dependent in the sense that they
are evaluated relative to assumed SPS models (whereas $(f_{900}/f_{1500})_{\rm out}$ is SPS model-independent).  
We have shown that there is a strong correlation between \wlya\ and $f_{\rm esc}$, and it has been
known for some time (\citealt{shapley03}) that $W_{\lambda}(\lya)$ is anti-correlated with the strength of low-ionization (e.g., 
 \ion{C}{2}, \ion{Si}{2}, \ion{O}{1})
interstellar absorption features observed in the same spectra. The presence of strong lines of neutral and singly-ionized species
in gas seen in absorption against the UV continuum from massive stars suggests significant neutral H column densities, $N_{\rm HI} > 10^{17}$ cm$^{-2}$,
and the relatively constant $W_{\lambda}/\lambda$ for lines of the same species but varying $\lambda f$, where $\lambda$ is the rest wavelength
of the transition and $f$ is the oscillator strength (e.g., \ion{Si}{2} $\lambda 1190$, 1193, 1260, 1304, 1526)   
indicates the presence of saturation\footnote{For example, assuming a linear curve of growth, one expects $W_{\lambda}(1260)/W_{\lambda}(1526) \simeq 6$,
whereas the observed ratio is typically $\simeq 1$; the ratio reaches a maximum of $\simeq 1.8$ 
for the ``LyC Detected'' sub-sample, which has the weakest metal lines overall.}.

The observation that strong IS lines often do not become black at maximum optical depth then suggests partial covering of the continuum
source (hereinafter, $f_{\rm c} < 1$) by the low-ion-bearing gas. If the same partial covering were to apply to the \ion{H}{1}, 
there would follow an obvious relationship between gas covering fraction $f_{\rm c}$ and the probability that significant
LyC could ``escape'' via the uncovered regions.  Conversely, 
if a galaxy spectrum's low-ionization IS lines are black over a considerable range in 
velocity, one would expect no significant LyC photon leakage in our direction (e.g., \citealt{steidel01,pettini02,shapley03,shapley06,quider09,steidel2010,heckman11,jones12}). 
However, as recently shown (\citealt{henry15,vasei16,reddy16b}), 
$f_c < 1$ for low-ionization metal lines is a necessary, but not sufficient, predictor of significant LyC leakage from a
galaxy.  The metal lines would be relatively 
insensitive to low-metallicity \ion{H}{1}, and possibly also to metal-enriched gas with LyC optical depths $1 \simlt \tau_{\rm LyC} \simlt 10$ 
where neutral and singly-ionized metallic species might not be the dominant ionization stages in the gas (see also \citealt{rivera-thorsen15}.) 

\begin{figure*}[htbp!]
\centerline{\includegraphics[height=5.0cm]{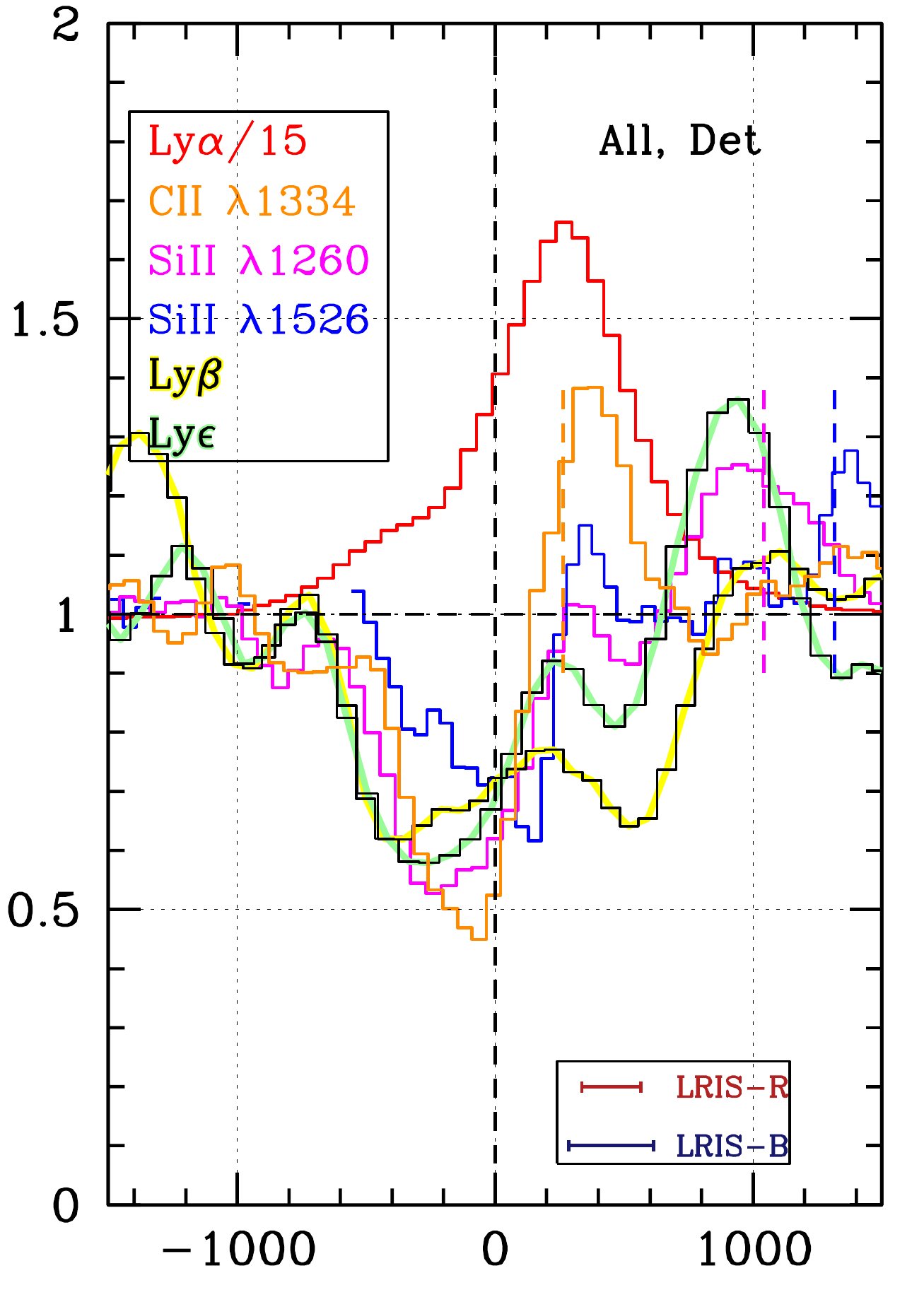}\includegraphics[height=5.0cm]{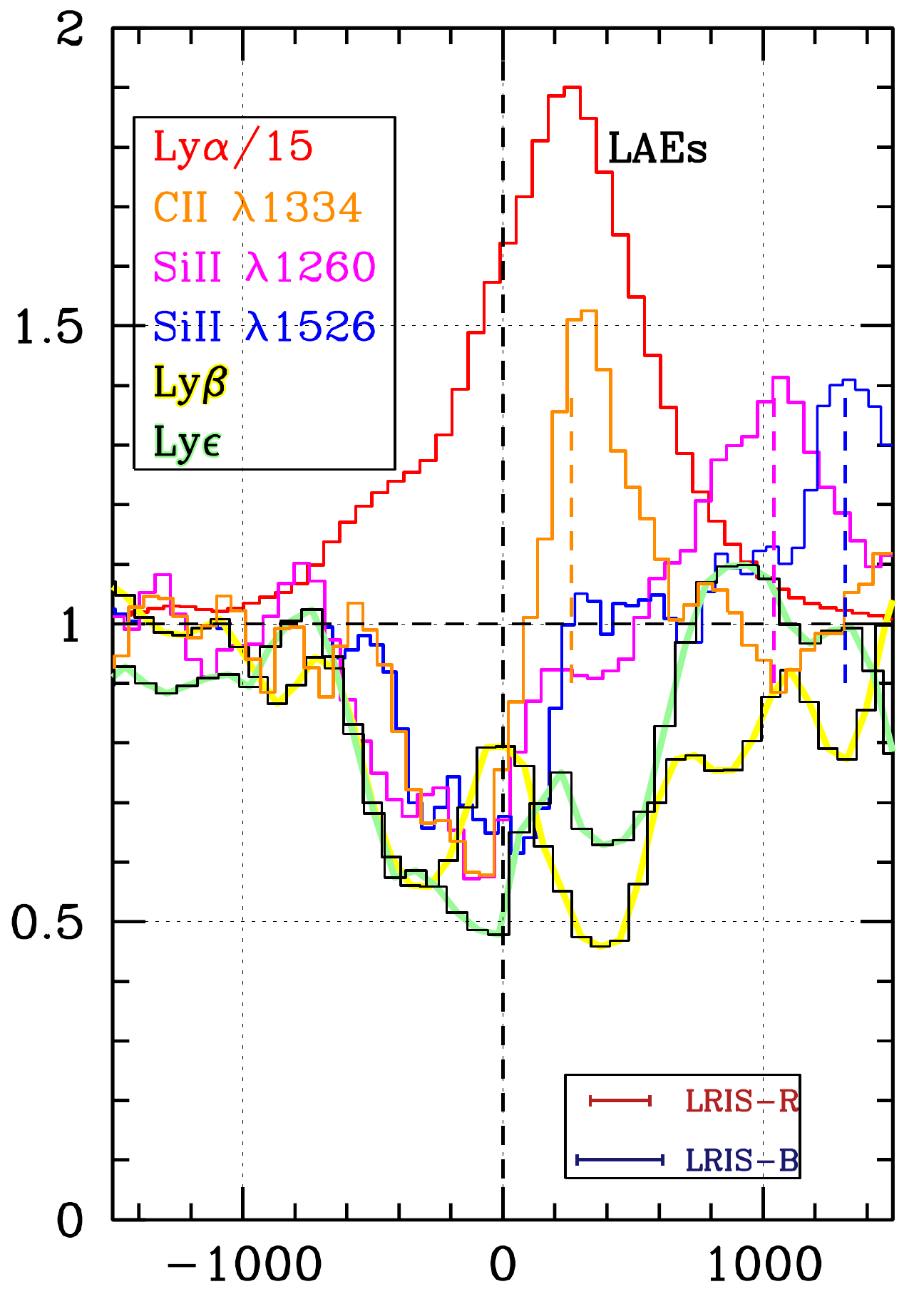}\includegraphics[height=5.0cm]{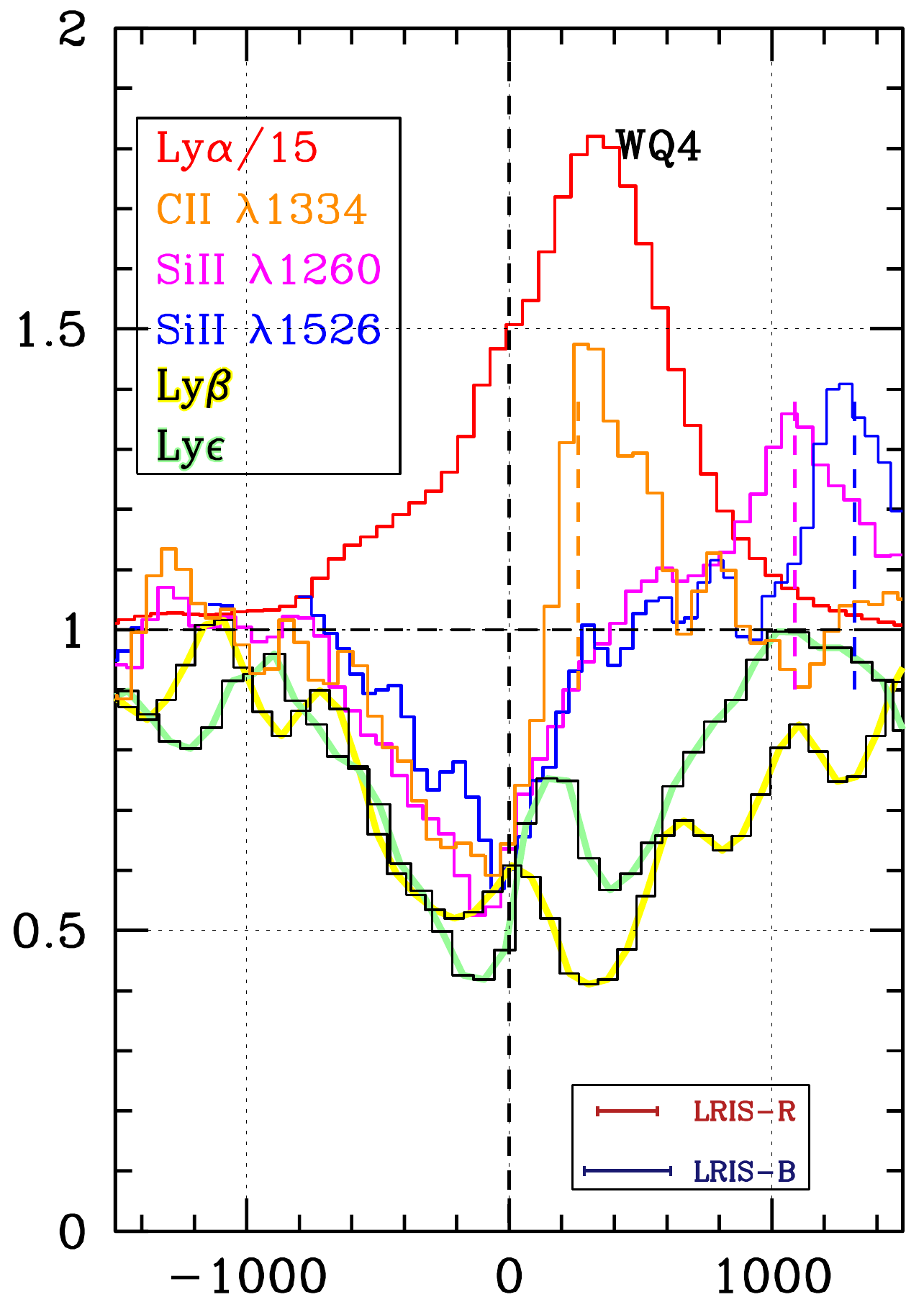}\includegraphics[height=5.0cm]{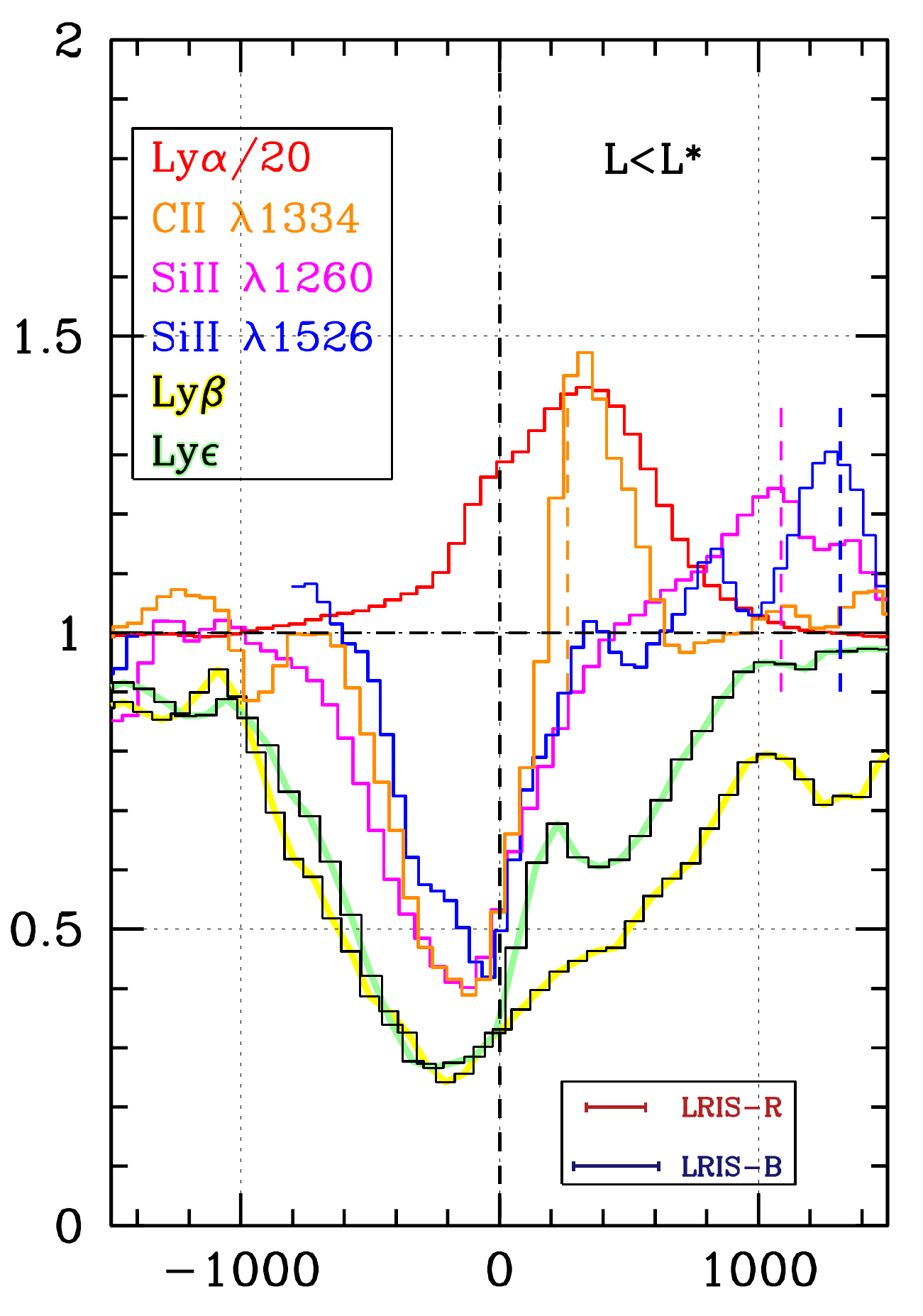}} 
\centerline{\includegraphics[height=5.0cm]{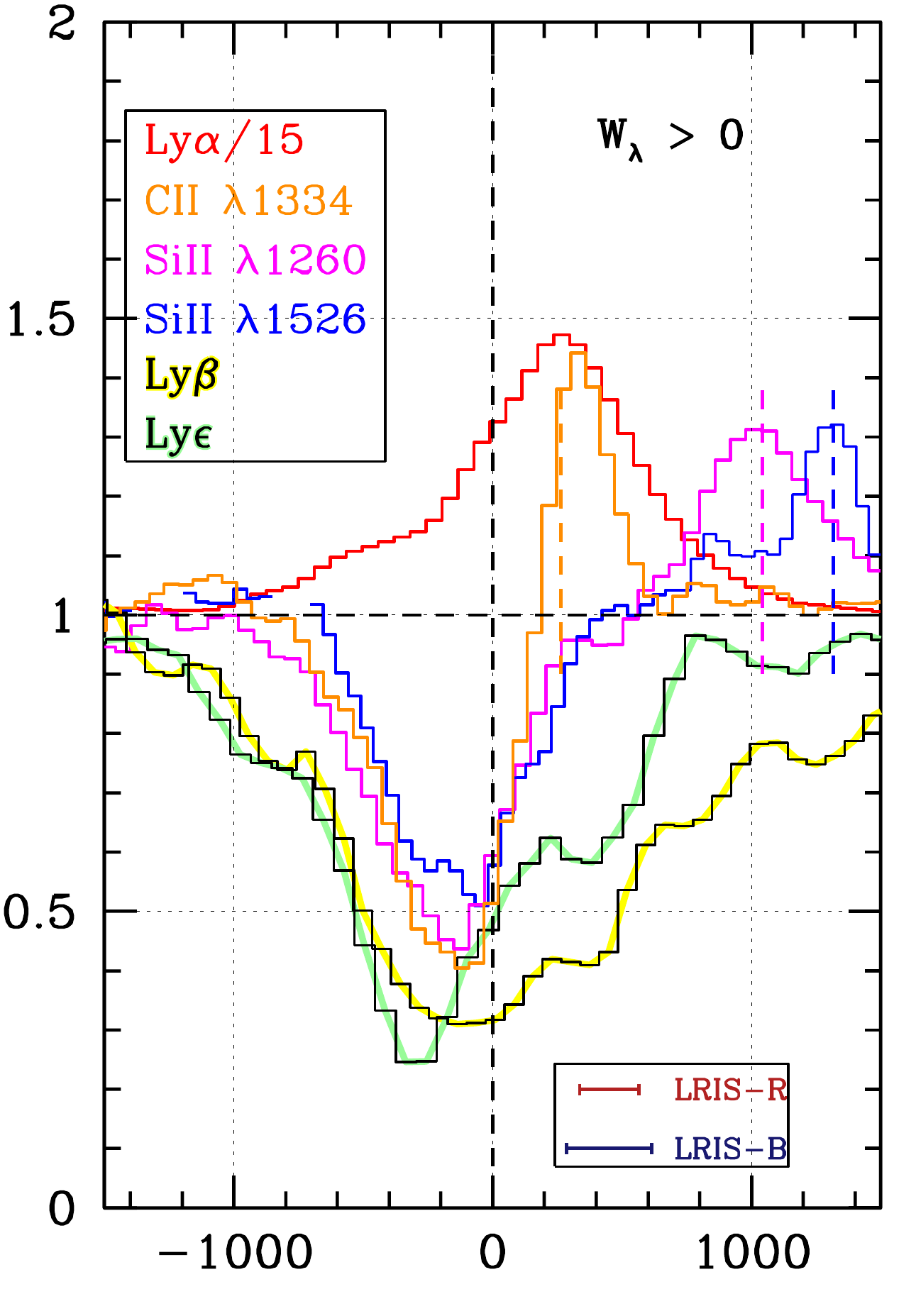}\includegraphics[height=5.0cm]{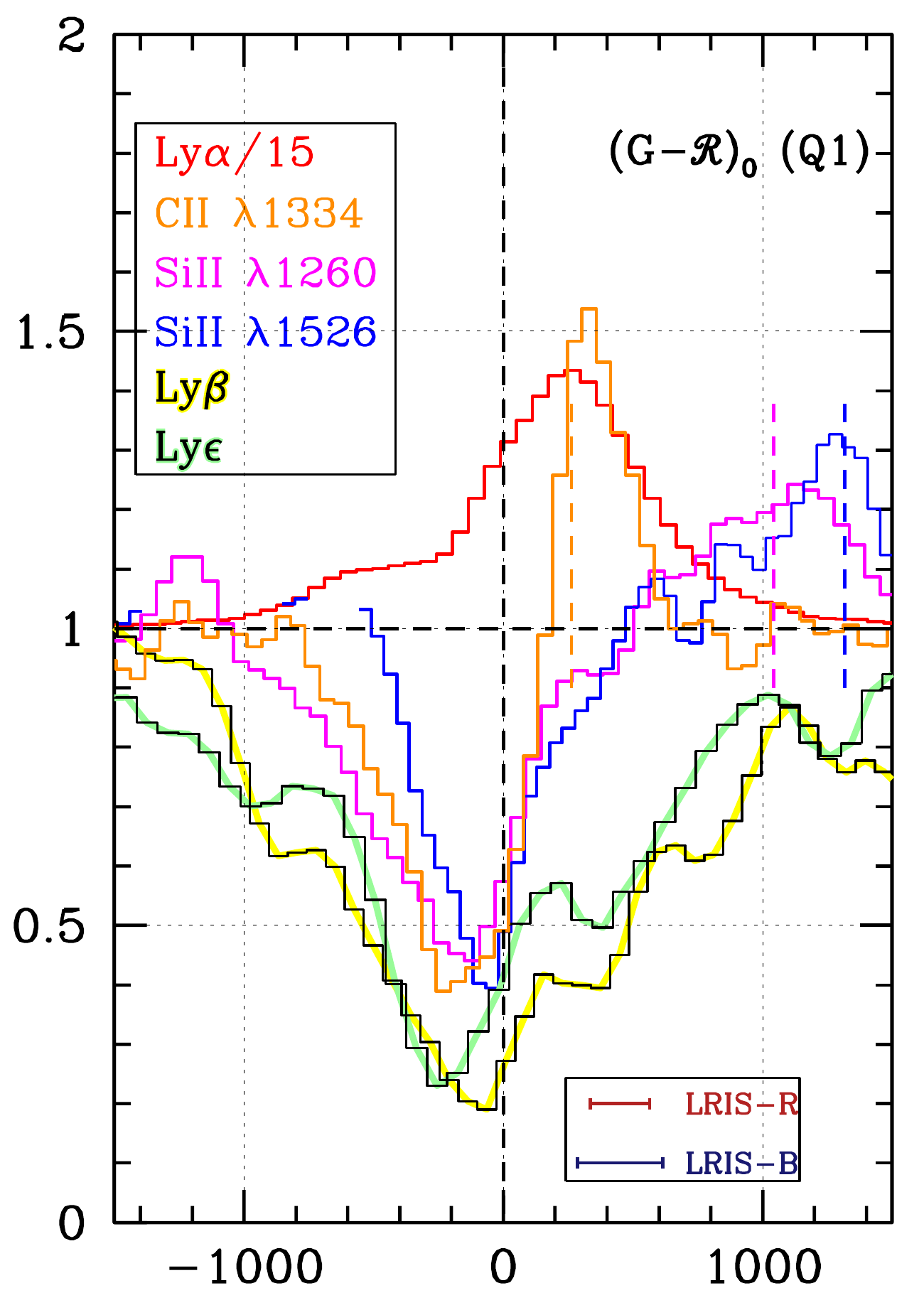}\includegraphics[height=5.0cm]{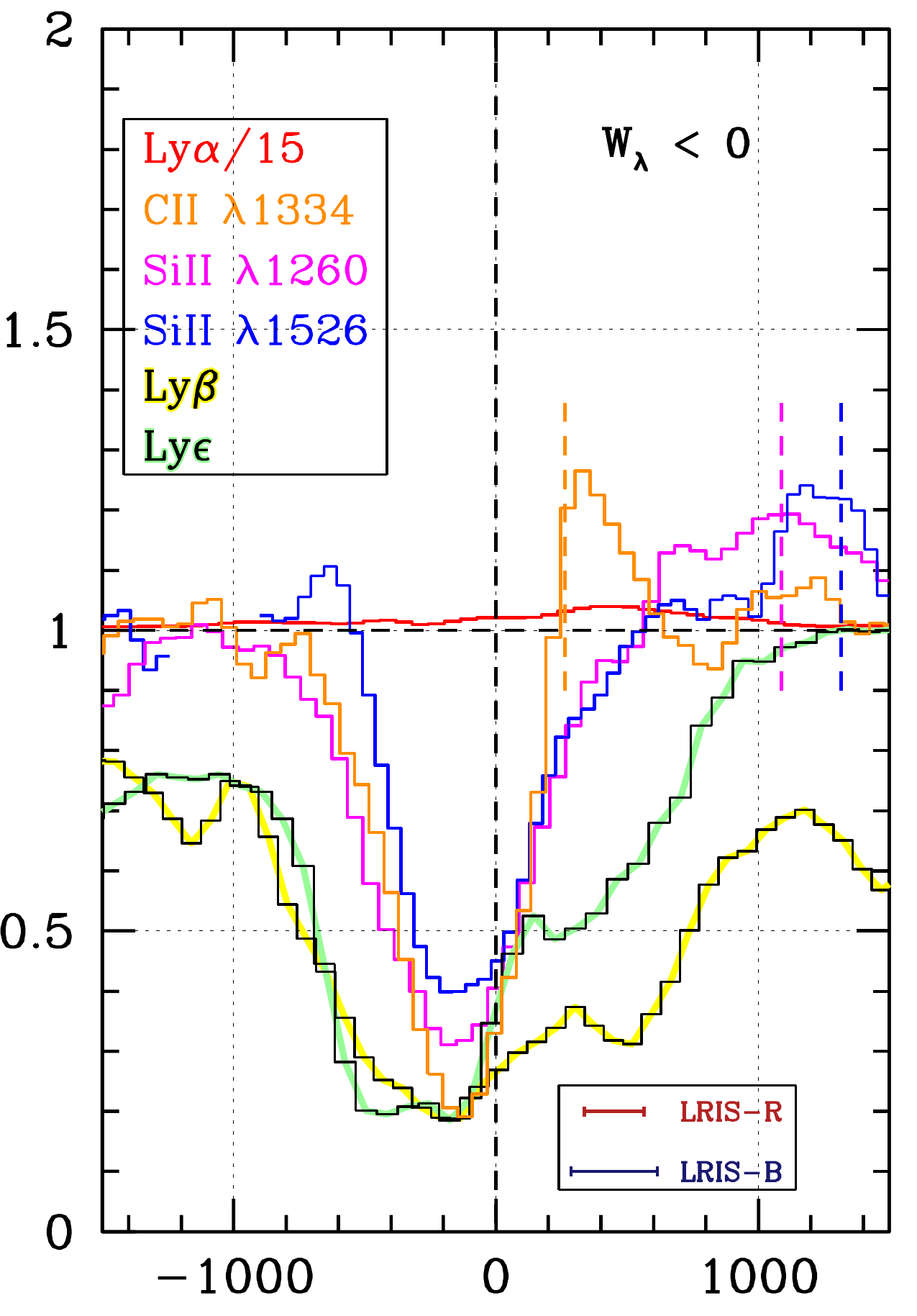}\includegraphics[height=5.0cm]{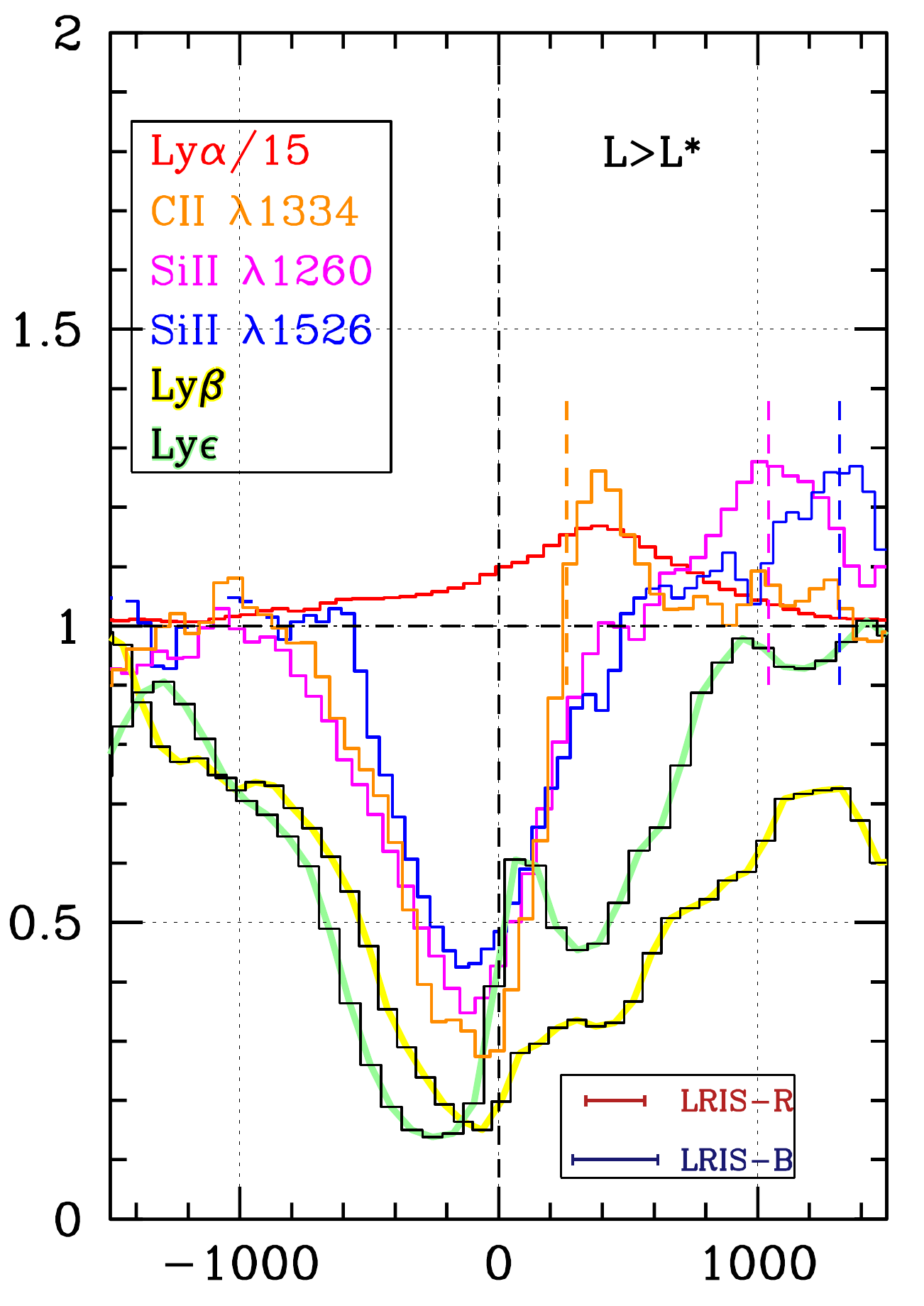}} 
\caption{Profiles of normalized intensity vs. systemic velocity for selected IS absorption lines for selected KLCS sub-samples after normalizing by the best-fit SPS model as described
in \S\ref{sec:sps_models}: 
\ion{C}{2}~$\lambda 1334.5$ (orange), \ion{Si}{2}~$\lambda 1260.4$ (magenta), \ion{Si}{2}~$\lambda 1526.7$ (blue), \lyb$~\lambda 1025.7$ (yellow), 
Ly$\epsilon~\lambda 937.8$ (pale green), and \lya\ emission (red). 
The panels are arranged (left to right and top to bottom) in order of decreasing $f_{\rm esc,abs}$ (last column of Table~\ref{tab:models}.) 
\lya\ emission profiles are based on the SPS continuum-subtracted spectrum; each has been scaled down by a factor of 15 and shifted by +1.0  
for display purposes. The three additional emission profiles at velocities $>0$  \kms\ are non-resonant emission lines \ion{C}{2}$^{\ast}~\lambda 1335.7$ (orange),
\ion{Si}{2}$^{\ast}~\lambda 1264.8$ (magenta), and \ion{Si}{2}$^{\ast}~\lambda 1533.4$ (blue). Short vertical dashed lines, color coded in the same way,
mark the position of $v_{\rm sys} = 0$. The legend in the lower right of each panel shows the approximate spectral resolution for LRIS-R (relevant
for the metal lines) and LRIS-B (relevant for the Lyman series lines).  
\label{fig:vel_lyc}
}
\end{figure*}

To examine trends in the residual depth and the velocity extent of interstellar absorption features, 
we divided each of the KLCS composites by its best-fit SPS model,
and formed continuum-normalized spectra with the strongest stellar features removed.  
Figure~\ref{fig:vel_lyc} shows the line profiles 
of \ion{Si}{2} $\lambda 1260$ and $\lambda 1526$, \ion{C}{2} $\lambda 1334$, and 
the \lyb~$\lambda 1025.7$ and Ly$\epsilon$~$\lambda 937.8$ for several of the KLCS subsets. 
Also shown in each panel of Figure~\ref{fig:vel_lyc} is the profile of \lya\ emission, after
subtracting the stellar continuum, dividing by a factor of 15, and adding 1 (both of the latter for display purposes). The spectral
resolution (FWHM) for the metal lines (labeled ``LRIS-R'') and for the Lyman lines (labeled ``LRIS-B'') are 
shown in the bottom right of each panel. 
Figure~\ref{fig:vel_lyc} is arranged according to the measured
$f_{\rm esc,abs}$, from largest (top left) to smallest (bottom right).  

In addition to the trends already noted between \wlya\ and \fout,
there is a similar trend between the apparent depth of the Lyman series and low-ionization metallic absorption lines and $f_{\rm esc}$. 
In this context, it is noteworthy 
that the metallic lines shown in Figure~\ref{fig:vel_lyc} were recorded using LRIS-R, which have (as discussed in \S\ref{sec:klcs_specobs}) 
spectral resolving power $R \sim 1400$ ($\sigma_{\rm inst} \simeq 90$ \kms), whereas the Ly series absorption lines were
measured using LRIS-B, with $R \sim 800$ ($\sigma_{\rm inst} \simeq 160$ \kms). In spite of this, the Lyman series absorption lines in all cases
have equal or larger apparent optical
depth than the metal lines.  The similarity of the apparent depth of Lyman lines (at least, for $v_{\rm sys} \simlt 0$)  
in all of the composites shown suggests ${\rm log(\nhi/cm^{2})} \simgt 18$ -- in which case all of the observed Ly lines would be strongly saturated -- 
coupled with partial covering of the far-UV continuum. In general, the maximum depths of both \ion{H}{1} and metallic absorption lines
occur at $-300 \simlt v \simlt -100$ \kms, consistent with the hypothesis that the bulk of the cool, metal-enriched gas along the sightline 
``down the barrel'' is out-flowing. The Ly line profiles exhibit more varied behavior for velocities $v > 0$, where
absorption is less-deep overall, and where Ly$\beta$ absorption is significantly stronger than Ly$\epsilon$, 
suggesting that both $f_{\rm c}$ and
${\rm log(\nhi)}$ are lower for gas with $v > 0$ physically located on the near-side of the stars comprising the UV continuum.   

The shapes and depths of the metal lines at $v > 0$ are difficult to interpret in the same way, since all of the metallic species shown
in Fig.~\ref{fig:vel_lyc} are resonance lines that may manifest ``emission filling'' from scattered line photons re-emitted in our direction
(see, e.g., \citealt{prochaska11,erb12,martin12,scarlata15}). However, all 3 of the lines illustrated
have nearby non-resonant transitions (i.e., transitions from the same excited state to excited fine structure levels just above the ground state) which can act as ``relief valves'' for the resonance lines, so that a substantial fraction of absorbed resonance
photons may be re-emitted in the non-resonance emission lines -- in this case, \ion{Si}{2}$^{\ast}\lambda 1264.9$, 
\ion{C}{2}$^{\ast} \lambda 1335.7$, \ion{S}{2}$^{\ast} \lambda 1533.4$. Each panel of Figure~\ref{fig:vel_lyc} marks the position of
the galaxy systemic redshift relative to the profiles of the non-resonant emission lines -- with the same color coding used for the absorption profiles of the
nearby resonance line. The implications of the non-resonant emission lines for LyC escape are discussed in \S\ref{sec:efs}. 

In any case, the Ly series absorption lines between the Lyman limit (911.75 \AA) and Ly-$\beta$ 
provide the most direct
constraints on \nhi\ and/or the continuum covering fraction ($f_{\rm c}$) of \ion{H}{1} near the galaxy systemic redshift along
the line of sight. They are particularly useful for determining \nhi\ when ${\rm log~(\nhi/cm^{-2}) < 18}$ -- where the LyC would be optically thin regardless of
$f_{\rm c}$) --  or when ${\rm log~(N_{HI}/cm^2) > 20}$, where prominent damping wings of \lya\ can be recognized even in spectra
of modest spectral resolution. In between these limits (i.e., for ${\rm 18 \simlt log(\nhi/cm^{-2}) \simlt 20}$), \nhi\ is not
well-constrained by the spectra, but the fact that the observable Lyman series lines are strongly saturated provides redundant constraints
on the covering fraction of optically-thick \ion{H}{1}. If 
these saturated Lyman series lines are resolved in velocity space, 
the residual intensity at maximum line depth relative to the continuum should be constant and equal to 
$1-f_{\rm c}$, where $f_{\rm c}$ is
the fraction of far-UV continuum covered by optically thick material. 
At the resolution of the KLCS galaxy spectra  shortward of rest-frame
\lya\ ($\sigma_{\rm inst} \simeq 160$ \kms) 
the cores of the higher Lyman series lines with $\sigma_{\rm v} \simlt 100$ \kms\ would not be resolved, in which case
any apparent residual flux in the cores of the Lyman series lines would correspond to a lower limit on the fraction
$f_{\rm c}$ of the stellar FUV continuum covered by optically thick \ion{H}{1}. 

\begin{deluxetable}{lll}
\tabletypesize{\scriptsize}
\tablewidth{0pt}
\tablecaption{Wavelength Regions Used for Spectral Fits\tablenotemark{a}}
\tablehead{
\colhead{$\lambda_{1}$}  &  \colhead{$\lambda_{2}$} & \colhead{Comments} \\
\colhead{(\AA)}  & \colhead{(\AA)} & \colhead{} } 
\startdata
\cutinhead{Regions Masked for Stellar Continuum Fits}
1080 & 1087 & \ion{N}{2} IS abs\\
1099 & 1107 & \ion{Fe}{2}$^{\ast}$ em.\\
1119 & 1123 & \ion{Fe}{2} IS abs complex \\
1131 & 1155 & \ion{Fe}{2} IS abs complex \\
1172 & 1178 & \ion{C}{3} \\
1186 & 1231 & \ion{Si}{2}, \ion{Si}{3} IS abs. \\
1254 & 1270 & \ion{Si}{2} IS abs, \ion{Si}{2}$^{\ast}$ em \\
1291 & 1312 & \ion{Si}{2}, \ion{O}{1} abs, \ion{Si}{2}$^{\ast}$ em \\
1328 & 1340 & \ion{C}{2} IS abs, \ion{C}{2}$^{\ast}$ em\\
1363 & 1373 & \ion{O}{5} stellar wind\\
1389 & 1396 & \ion{Si}{4} IS abs \\
1398 & 1404 & \ion{Si}{4} IS abs \\
1521 & 1528 & \ion{Si}{2} IS abs \\
1531 & 1536 & \ion{Si}{2}$^{\ast}$ em\\
1541 & 1552 & \ion{C}{4} IS abs \\
1606 & 1610 & \ion{Fe}{2} IS abs \\
1657 & 1675 & [\ion{O}{3}] neb em, \ion{Al}{2} IS abs \\
1708 & 1711 & \ion{Ni}{2} IS abs \\
\cutinhead{Regions Used for ${\rm N_{HI}}$, $f_{\rm c}$ Fits}
880   & 910 & LyC region\\
929.0  & 931.0 & Ly6 core\\
936.6   & 937.8 & Ly$\epsilon$ core\\
948.0   & 950.8 & Ly$\delta$ core \\
970.8   & 973.6 & Ly$\gamma$ core \\
1013.3  & 1020.0 & Ly$\beta$ blue wing\\
1023.5  & 1026.0 & Ly$\beta$ core \\
1178.8  & 1183.3 & Ly$\alpha$ blue damping wing\\
1223.7  & 1248.5 & Ly$\alpha$ red damping wing 
\enddata
\label{tab:spec_regions}
\tablenotetext{a}{Spectral fits to the stellar continuum were performed over the wavelength range 1070-1740 \AA\ (except for masked
regions) for all composites but z(Q4), for which the range
1070-1640 \AA\ was used.}
\end{deluxetable}

\begin{deluxetable}{llccccr}
\tabletypesize{\scriptsize}
\tablewidth{0pt}
\tablecaption{ISM Fit Results: Screen Model\tablenotemark{a}}
\tablehead{
\colhead{Sample}  &  \colhead{Att.} & \colhead{E(B-V)} &  \colhead{log($N_{\rm HI})$} &  \colhead{$f_{\rm c}$\tablenotemark{b}}  & \colhead{$f_{\rm esc,abs}$\tablenotemark{c}} \\
\colhead{} & \colhead{} & \colhead{} & \colhead{(${\rm cm}^{-2}$)}} 
\startdata
All & 	R16 &  0.129 &	20.61 & 0.70 & $0.12\pm0.02$  \\
All, detected\tablenotemark{d} & SMC & 0.045 & (17.41) & 0.47 & $0.69\pm0.04$ \\
All, not detected & R16 & 0.135	& 20.65 & 0.75 & $0.05\pm0.01$ \\ 
\hline
z(Q1) &	SMC & 0.060 & 20.61 & 0.71 & $0.15\pm0.05$ \\ 
z(Q4) & R16 & 0.080 & 20.39 & 0.69 & $0.15\pm0.03$ \\
\hline
$L_{\rm uv} > L_{\rm uv}^{\ast}$ & R16 & 0.141	& 20.75	& 0.79 & $<0.02$ \\
$L_{\rm uv} < L_{\rm uv}^{\ast}$ & SMC & 0.047	& 20.43	& 0.62 & $0.36\pm0.04$ \\
\hline
$L_{\rm uv}$ (Q1) & R16 & 0.133	& 20.68	& 0.82 & $<0.03$ \\
$L_{\rm uv}$ (Q2) & R16 & 0.154	& 20.74	& 0.75 & $< 0.05$ \\
$L_{\rm uv}$ (Q3) & SMC & 0.047 & 20.43	& 0.63 & $0.37\pm0.05$  \\
$L_{\rm uv}$ (Q4) & SMC & 0.044 & 20.48	& 0.57 & $0.44\pm0.06$ \\
\hline
$W_{\lambda}(\lya)$ (Q1) & R16 & 0.158 & 21.10 & 0.81 & $<0.04$ \\
$W_{\lambda}(\lya)$ (Q2) & R16 & 0.175 & 20.79 & 0.76 & $0.06\pm0.02$ \\
$W_{\lambda}(\lya)$ (Q3) & SMC & 0.049 & 20.00 & 0.80 & $0.13\pm0.05$ \\
$W_{\lambda}(\lya)$ (Q4) & SMC & 0.027 & 20.07 & 0.45 & $0.54\pm0.06$  \\
\hline
LAEs              & SMC & 0.027	& 20.02 & 0.40 & $0.57\pm0.06$ \\ 
non-LAEs          & R16 & 0.145	& 20.71 & 0.77 & $0.07\pm0.02$ \\
\hline
$W_{\lambda}(\lya) >0$  & SMC & 0.041	& 20.12 & 0.66 & $0.28\pm0.03$ \\ 
$W_{\lambda}(\lya) <0$  & R16 & 0.165	& 21.02 & 0.79 & $<0.03$ \\ 
\hline
$(G-{\cal R})_0$ (Q1) & SMC & 0.015 & 20.29 & 0.74 & $0.21\pm0.04$  \\
$(G-{\cal R})_0$ (Q2) & R16 & 0.112 & 20.57 & 0.73 & $0.13\pm0.04$ \\
$(G-{\cal R})_0$ (Q3) & R16 & 0.162 & 20.62 & 0.73 & $0.15\pm0.03$ \\
$(G-{\cal R})_0$ (Q4) & R16 & 0.222 & 21.00 & 0.59 & $0.04\pm0.02$ 
\enddata
\label{tab:fc}
\tablenotetext{a}{Models assume that dust affects both the ionizing and non-ionizing emission from galaxies.}
\tablenotetext{b}{Fraction of the EUV and FUV stellar continuum covered by optically-thick (in the Lyman continuum) \ion{H}{1}. }
\tablenotetext{c}{Assuming CGM+IGM correction, with measurements with $<2\sigma$ significance 
replaced by $2\sigma$ upper limits. }
\tablenotetext{d}{Assuming 90th percentile $t_{900}$, for CGM+IGM correction, as in Table~\ref{tab:models}. }
\end{deluxetable}

\subsection{Geometric Models of the ISM}

In \S\ref{sec:sps_models} above, we described fitting SPS spectra to the far-UV spectra of the
observed KLCS composites in order to estimate $\langle f_{\rm esc,rel}\rangle$ and $\langle f_{\rm esc,abs}\rangle$ in a self-consistent
manner. These spectral fits assumed that both ionizing and non-ionizing components of the integrated UV light from stars are reddened 
by the same dust screen with attenuation that is a smooth function of rest-wavelength governed by
one of three model attenuation relations (Calzetti, R16, and revised SMC). However, introducing the parameter $f_{\rm c}$ to
describe the fraction of the FUV stellar continuum covered by optically thick \ion{H}{1} requires a decision about how
to handle dust reddening -- i.e., whether the ``covered'' and ``uncovered'' portions of the stellar continuum have been
reddened by the same dust opacity.  It is easy to imagine that the dust column density could be lower along lines
of sight that do not intersect any ISM gas with ${\rm log(\nhi/cm^{-2}) \simgt 18}$; if LyC escape requires geometric ``holes''
that have been cleared of {\it all} gas (neutral and ionized) between the stars and the IGM, then it may be that
any residual LyC flux has not been attenuated or reddened by dust at all (a counter-example in a LyC-emitting
galaxy at low redshift has been discussed by \citet{borthakur14}.) In sections~\ref{sec:screen}~and~\ref{sec:holes}, we consider
two simple geometric models of the ISM intended to bracket the range of possibilities in extinction/reddening. 
In the first, which we call the ``screen model'', all light is attenuated and reddened by the same foreground
screen- the standard assumption implicit in estimating dust extinction from far-UV photometry or spectroscopy, and the
one used in \S\ref{sec:sps_models} for calculations summarized in Table~\ref{tab:models}. The second alternative,
which we call the ``holes model'', assumes that the covered portion of the FUV continuum is reddened by dust, but
that the uncovered portion passes through the ISM without significant attenuation from dust or 
photoelectric absorption.  We discuss the two cases separately below.  

\subsubsection{The ``Screen'' Model}
\label{sec:screen}

 The emergent spectrum after passing through the ISM and CGM in the screen model context
can be expressed as
\begin{equation}
\label{eqn:screen}
S_{\nu,{\rm obs}} = 10^{-A_{\lambda}/2.5}S_{\nu,{\rm int}} \left[(1-f_{\rm c}) + f_{\rm c}\langle e^{-\tau(\lambda)}\rangle\right]
\end{equation} 
where $A_{\lambda} = k_{\lambda}E(B-V)$, $S_{\nu,{\rm int}}$ is the intrinsic stellar spectrum prior to attenuation and absorption in the ISM, 
and ${\rm e}^{-\tau(\lambda)}$ is the transmission spectrum through the ISM due to line and continuum absorption which depends on \nhi\ and (to a lesser extent) the kinematics of
the absorbing gas. 
\begin{figure*}[htbp!]
\centerline{\includegraphics[width=15cm]{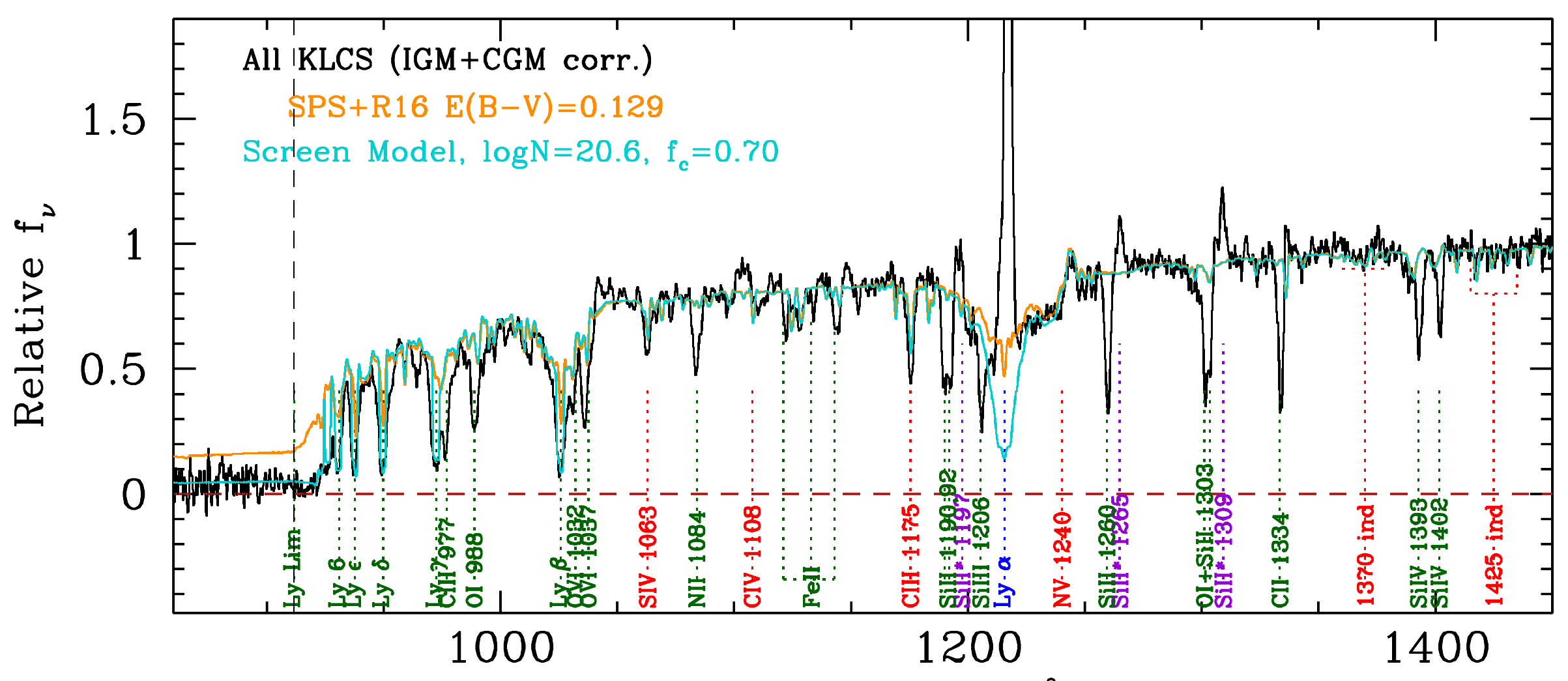}}
\centerline{\includegraphics[width=15cm]{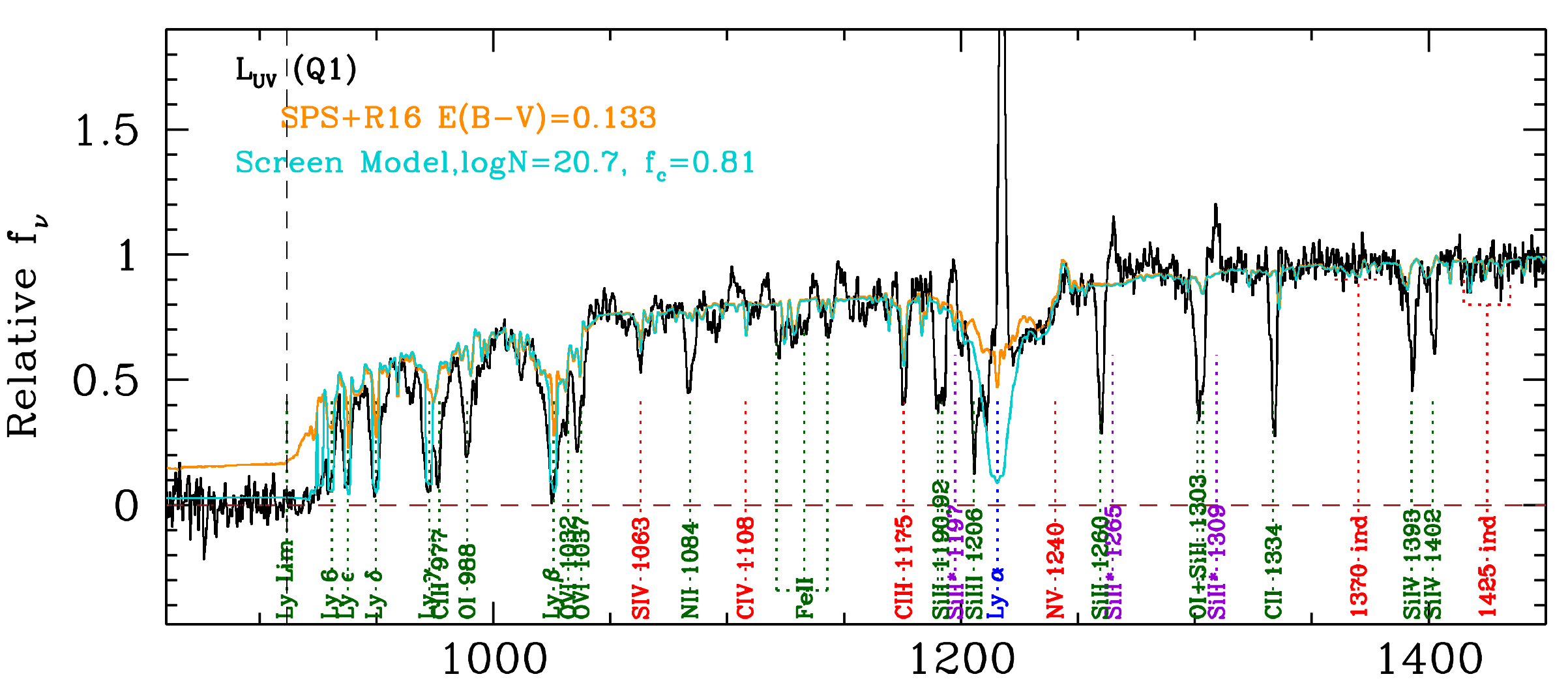}}
\centerline{\includegraphics[width=15cm]{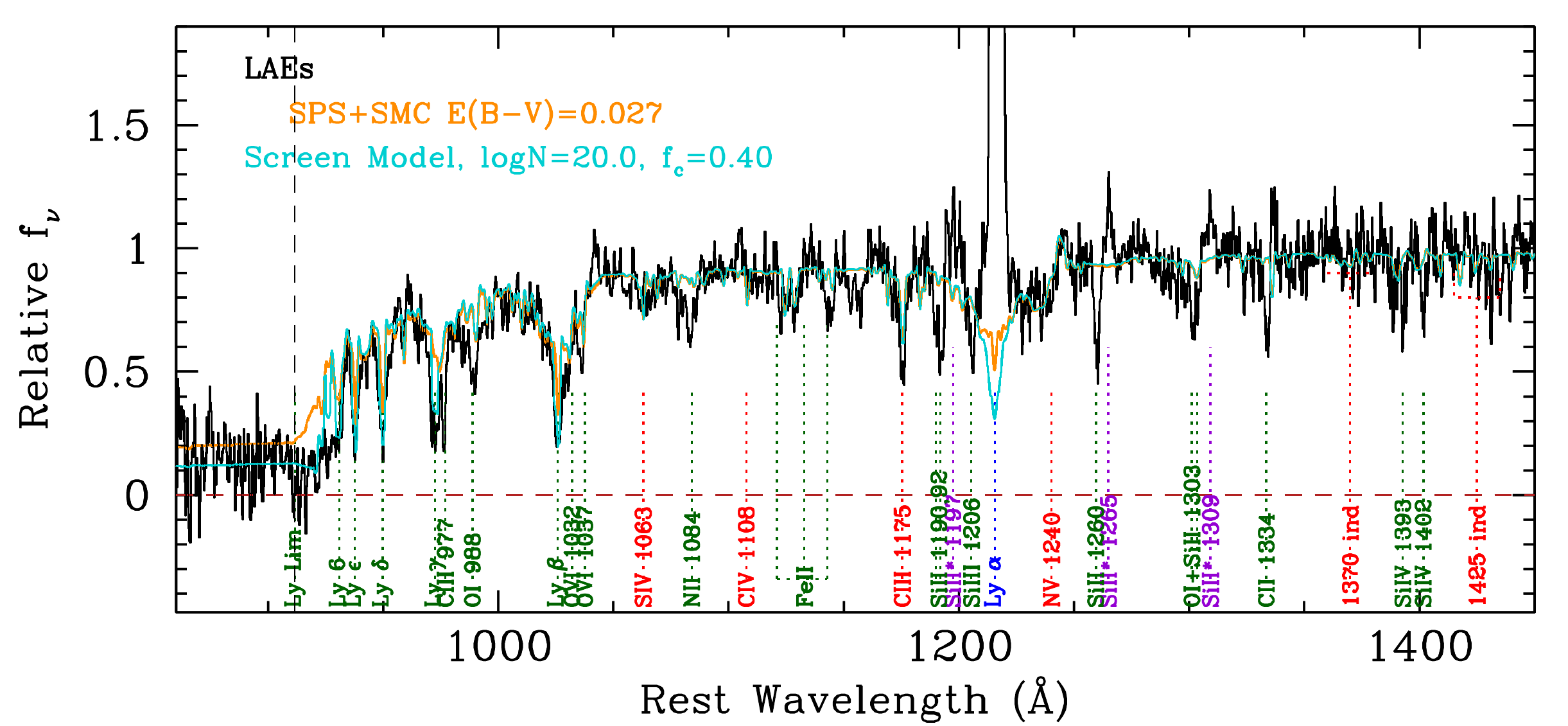}}
\caption{Example fits using the ``screen'' model for the galaxy ISM, which includes \ion{H}{1} absorption with column density ${\rm log (\nhi/cm^{-2})}$ 
covering
a fraction $f_{\rm c}$ of the stellar UV continuum. The values of ${\rm log(\nhi/cm^{-2})}$ and $f_{\rm c}$ (see Table~\ref{tab:fc}) are constrained
by the damping wings of \lya\ and \lyb\, the depth of the higher
order Lyman series lines, and the residual LyC flux in the range $\lambda_0 = 880-910$ \AA; the model spectrum is otherwise identical to that used for the fits
listed in Table~\ref{tab:models}. Each of the observed spectra (black histograms) has been corrected using the CGM+IGM model appropriate for the source
redshift distribution of the galaxies in the sub-sample (thus, adjusted so that $f_{900}/f_{1500} =$\fout\ (see Table~\ref{tab:models}.)  
}
\label{fig:fc}
\end{figure*}

Using an approach similar to that described by \citet{reddy16b}, 
we modeled the ISM\footnote{ISM is defined as any \ion{H}{1} not accounted for by the IGM+CGM correction, and thus
assumed to be due to interstellar material within a galactocentric distance of 50 pkpc.} for each of the KLCS composite
spectra by beginning with the best-fit SPS model (Table~\ref{tab:models}) and fitting for additional \ion{H}{1}
line and continuum absorption shortward of
\lya. For modeling the Lyman series absorption lines, we assumed a single component with Doppler width $b =125$ \kms, 
with velocity relative to systemic $v_{\rm sys} = -100$ \kms\ (the typical velocity corresponding to the minimum residual flux density of absorption profiles
of \ion{H}{1} and low-ionization metal lines), and smoothed the artificial line profiles to the resolution of the observed spectra. 
As described below (see Table~\ref{tab:spec_regions}), only  
the cores of the first 6 Lyman series lines, the damping wings of \lya\ and \lyb, and the Lyman continuum region ($f_{900}$) were used in evaluating the fits, 
thus they are insensitive to the precise line profile shapes for saturated lines without damping wings.
     
Aside from the ISM treatment, the fits are identical to those described in \S\ref{sec:sps_models}, where the SPS model is reddened using the attenuation relation that
provides the best fit to the CGM+IGM-corrected spectrum in the rest wavelength range 1070-1740 \AA, 
after masking the regions listed in the top portion
of Table~\ref{tab:spec_regions}. Simultaneously, we fit for the values of $f_{\rm c}$ and ${\rm log(\nhi)} $ by comparing the
predicted model spectrum with the observed spectrum, where the additional parameters are constrained in the wavelength intervals listed
in the bottom part of Table~\ref{tab:spec_regions}. The regions near \lya\ (and, to a lesser extent, near \lyb) 
are sensitive to \ion{H}{1} damping wings for ${\rm log(N_{HI}/cm^{2}) \simgt 20.0}$, while the depths of Lyman series lines
and the residual LyC emission constrain $f_{\rm c}$. Assuming the same SPS model and attenuation relations used in Table~\ref{tab:models}, 
we recorded the values of ${\rm E(B-V)}$, $f_{\rm c}$, and ${\rm log(N_{HI}/cm^{2})}$ that minimized the residuals with 
respect to the data; these are listed in Table~\ref{tab:fc}. We estimated the uncertainties on each parameter by
repeatedly perturbing every pixel of the observed spectrum by an amount consistent with the error spectrum (which 
includes contributions from both sample variance and shot noise) and refitting. The typical uncertainties on
${\rm E(B-V)}$, $f_{\rm c}$ and ${\rm log(N_{HI}/cm^{2})}$ were found to be $\pm 0.002$, $\pm 0.03$, and $\pm 0.10$, respectively. 
The best-fit value of $f_{\rm esc,abs}$ computed directly from the models, and the RMS scatter of the best-fit values
obtained from repeated fits of the perturbed observed spectra, are listed in the last column of
Table~\ref{tab:fc}. 
Example fits for the full KLCS galaxy sample (``All''), the highest $L_{\rm uv}$ quartile (LQ1), and the subset with
$W_{\lambda}(\lya) > 20$ \AA\ (LAEs), are shown in 
Figure~\ref{fig:fc}.  

As noted by \citet{reddy16b}, $f_{\rm c}$ and \nhi\ generally have little covariance because one ($f_{\rm c}$)
is determined by the minima at the cores of Ly series lines and the LyC region, while ${\rm log} N_{\rm HI}$ is constrained by the broad
damping wings of \lya. However, as discussed above, column densities in the range ${\rm 18 \simlt log (N_{HI}/cm^{-2}) \simlt 20}$ are relatively poorly constrained because
there are no discernible \lya\ damping wings, yet the resolved Lyman series lines have saturated cores. This affects the fit for the ``All, detected''
sample, where $f_{\rm c}$ and ${\rm N_{HI}}$ for a fixed \fout\ are less well-determined\footnote{Formally, the uncertainties 
are $\pm 10$\% and $\pm 0.75$ dex, respectively.}. 

The tabulated $f_{\rm esc,abs}$ values are generally very similar to those listed in Table~\ref{tab:models}; the difference
in the present case is that the fits include constraints provided by the 
the depth of the Ly series lines and the \lya\ damping wings {\it in addition} to the measurement of $(f_{900}/f_{1500})_{\rm out}$.  
Thus, the values listed in Table~\ref{tab:fc} constitute more stringent checks on the internal consistency of the geometrical
model of the ISM: rather than calculating $f_{\rm esc,abs}$ from
the SPS fit to the stellar continuum and $(f_{900}/f_{1500})_{\rm obs}$ alone, we now require consistency between the diminution of LyC
flux relative to the SPS model and the other observational constraints on the column density and geometric covering fraction of
the gas-phase ISM.   
In this sense, the fits summarized
in Table~\ref{tab:fc} provide additional support for our simplified model for the galaxy ensembles.   We return to a discussion
of the implications in \S\ref{sec:implications} below. 

\subsubsection{The ``Holes'' Model}
\label{sec:holes}

\begin{deluxetable}{llcccclr}
\tabletypesize{\scriptsize}
\tablewidth{0pt}
\tablecaption{ISM Fit Results: Holes Model\tablenotemark{a}}
\tablehead{
\colhead{Sample}  &  \colhead{Att} & \colhead{E(B-V)$_{\rm cov}$} &  \colhead{log($N_{\rm HI})$} &  \colhead{$f_{\rm c}$\tablenotemark{b}}  & 
\colhead{$f_{\rm esc,abs}$\tablenotemark{c}} \\
\colhead{} & \colhead{} & \colhead{} & \colhead{(${\rm cm}^{-2}$)} & \colhead{} & \colhead{} } 
\startdata
All      &    SMC &  0.068 & 20.57 &  0.91 & $0.09\pm0.01$ \\ 
All, detected\tablenotemark{d}  &    SMC &  0.085 & (17.9) & 0.80 & $0.31\pm0.03$ \\ 
All, not detected  &    R16 &  0.163 & 20.60 &  0.95 & $0.05\pm0.01$ \\ 
\hline
z(Q1) &                   SMC &  0.076 & 20.56 &  0.92 & $0.08\pm0.01$ \\ 
z(Q4) &                   R16 &  0.114 & 20.34 &  0.88 & $0.12\pm0.02$ \\ 
\hline
$L_{\rm uv} > L_{\rm uv}^{\ast}$ &   R16 &  0.165 & 20.71 &  0.96 & $<0.04$  \\ 
$L_{\rm uv} < L_{\rm uv}^{\ast}$ &   SMC &  0.068 & 20.39 &  0.87 & $0.13\pm0.03$ \\ 
\hline
$L_{\rm uv}$ (Q1)     &    R16 &  0.153 & 20.64 &  0.96 & $<0.04$ \\ 
$L_{\rm uv}$ (Q2)     &    R16 &  0.195 & 20.63 &  0.96 & $<0.04$ \\ 
$L_{\rm uv}$ (Q3)     &    SMC &  0.065 & 20.39 &  0.87 & $0.13\pm0.03$ \\ 
$L_{\rm uv}$ (Q4)     &    SMC &  0.069 & 20.44 &  0.84 & $0.16 \pm 0.03$ \\ 
\hline
$W_{\lambda}(\lya)$ (Q1)    &    R16 &  0.182 & 21.05 &  0.97 & $<0.03$ \\ 
$W_{\lambda}(\lya)$ (Q2)    &    R16 &  0.208 & 20.74 &  0.97 & $<0.04$ \\ 
$W_{\lambda}(\lya)$ (Q3)    &    SMC &  0.059 & 19.93 &  0.93 & $0.07\pm0.02$ \\ 
$W_{\lambda}(\lya)$ (Q4)    &    SMC &  0.054 & 20.05 &  0.73 & $0.27\pm0.02$ \\ 
\hline
LAEs         &    SMC &  0.058 & 20.05 &  0.71 & $0.29\pm0.03$ \\ 
non-LAEs  &    R16 &  0.174 & 20.66 &  0.96 & $0.04\pm0.02$ \\ 
\hline
$W_{\lambda}(\lya) > 0$     &    SMC &  0.059 & 20.11 &  0.86 & $0.14\pm0.02$ \\ 
$W_{\lambda}(\lya) < 0$     &    R16 &  0.193 & 20.97 &  0.97 & $<0.03$ \\ 
\hline
$(G-{\cal R})_0$ (Q1) &    SMC &  0.019 & 20.28 &  0.85 & $0.15\pm0.02$ \\ 
$(G-{\cal R})_0$ (Q2) &    R16 &  0.142 & 20.52 &  0.94 & $0.06\pm0.02$ \\ 
$(G-{\cal R})_0$ (Q3) &    R16 &  0.195 & 20.56 &  0.95 & $<0.08$  \\ 
$(G-{\cal R})_0$ (Q4) &    R16 &  0.296 & 20.87 &  0.97 & $<0.06$   
\enddata
\label{tab:fc2}
\tablenotetext{a}{Models assume that dust and photoelectric absorption affects only the covered fraction $f_{\rm c}$ of the UV continuum. }
\tablenotetext{b}{Fraction of the EUV and FUV stellar continuum covered by the assumed $N_{\rm HI}$ and dust. }
\tablenotetext{c}{Inferred absolute escape fraction of LyC photons.} 
\tablenotetext{d}{Assuming a CGM+IGM correction as in Table~\ref{tab:models}.}
\end{deluxetable}

\begin{figure*}[htbp!]
\centerline{\includegraphics[width=15cm]{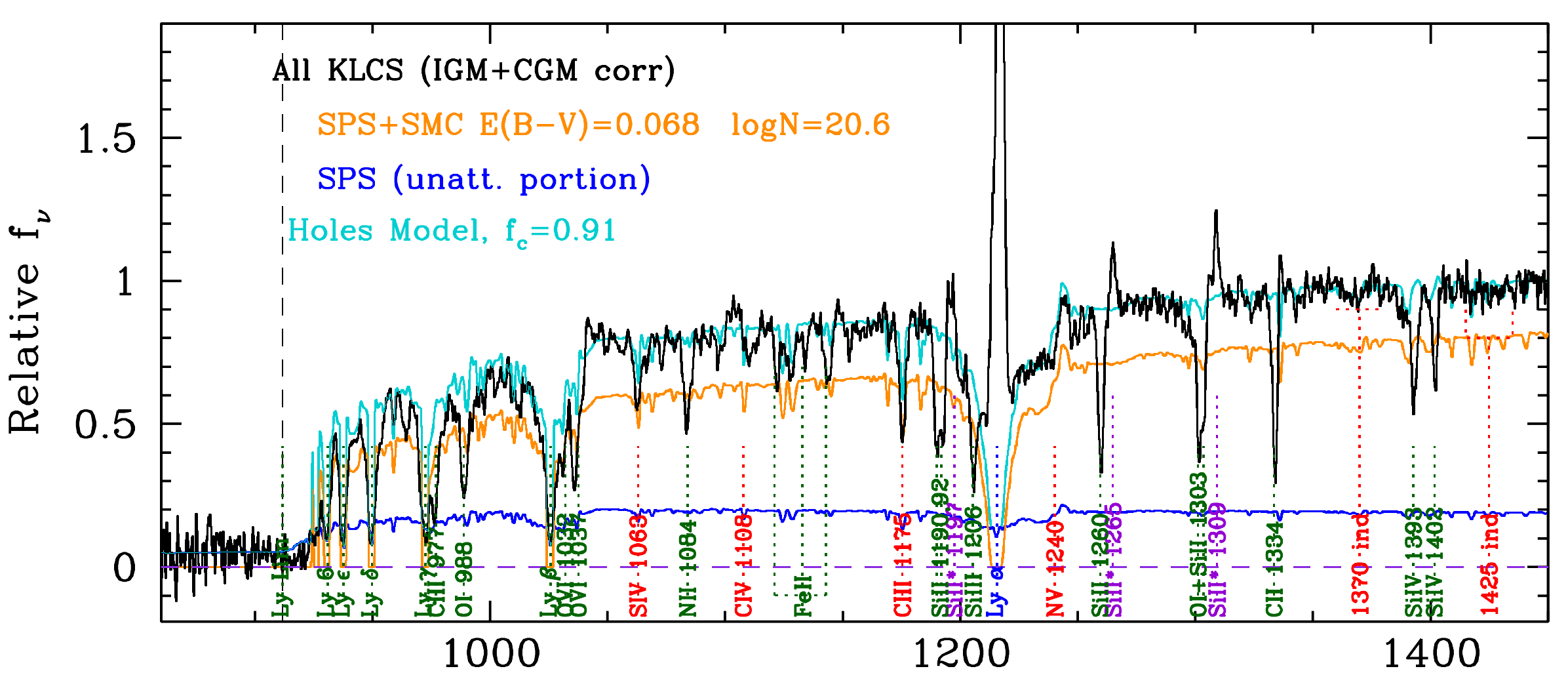}}
\centerline{\includegraphics[width=15cm]{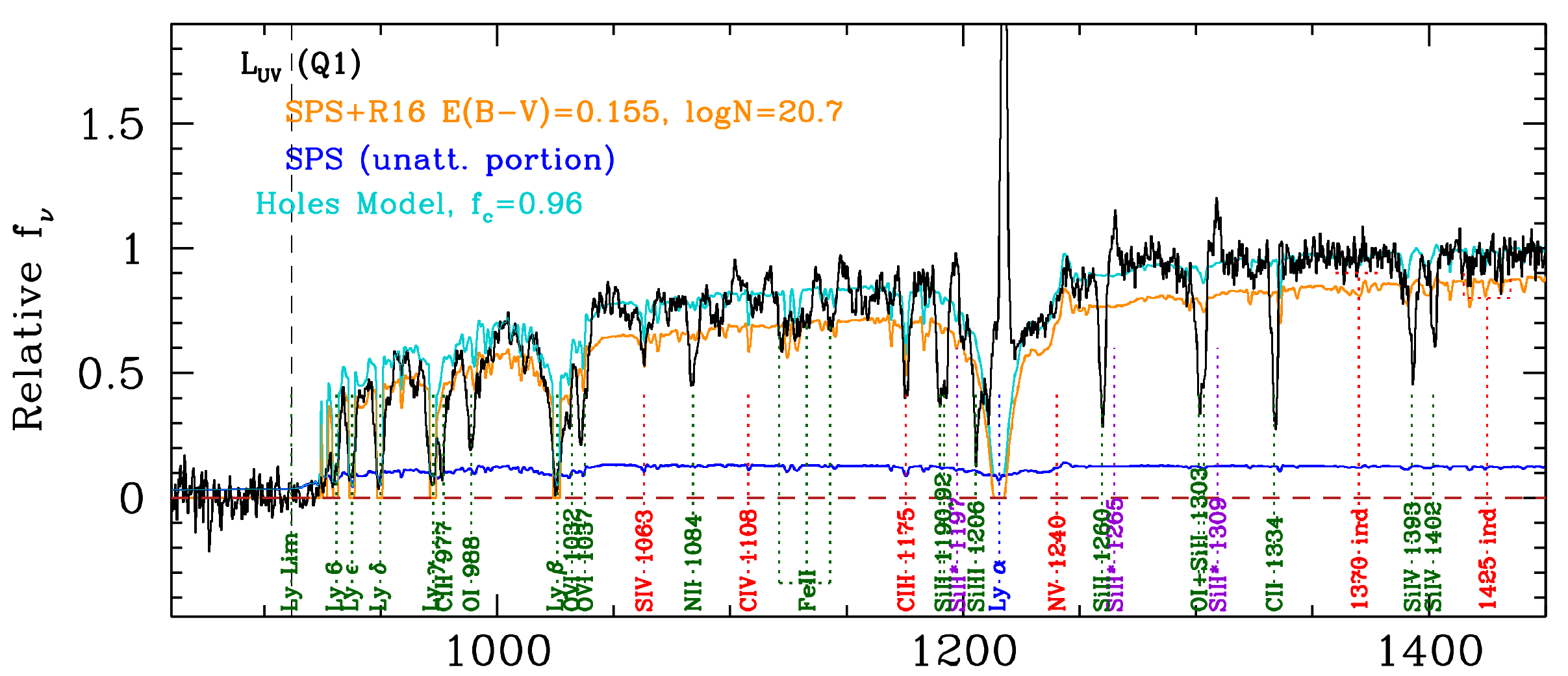}}
\centerline{\includegraphics[width=15cm]{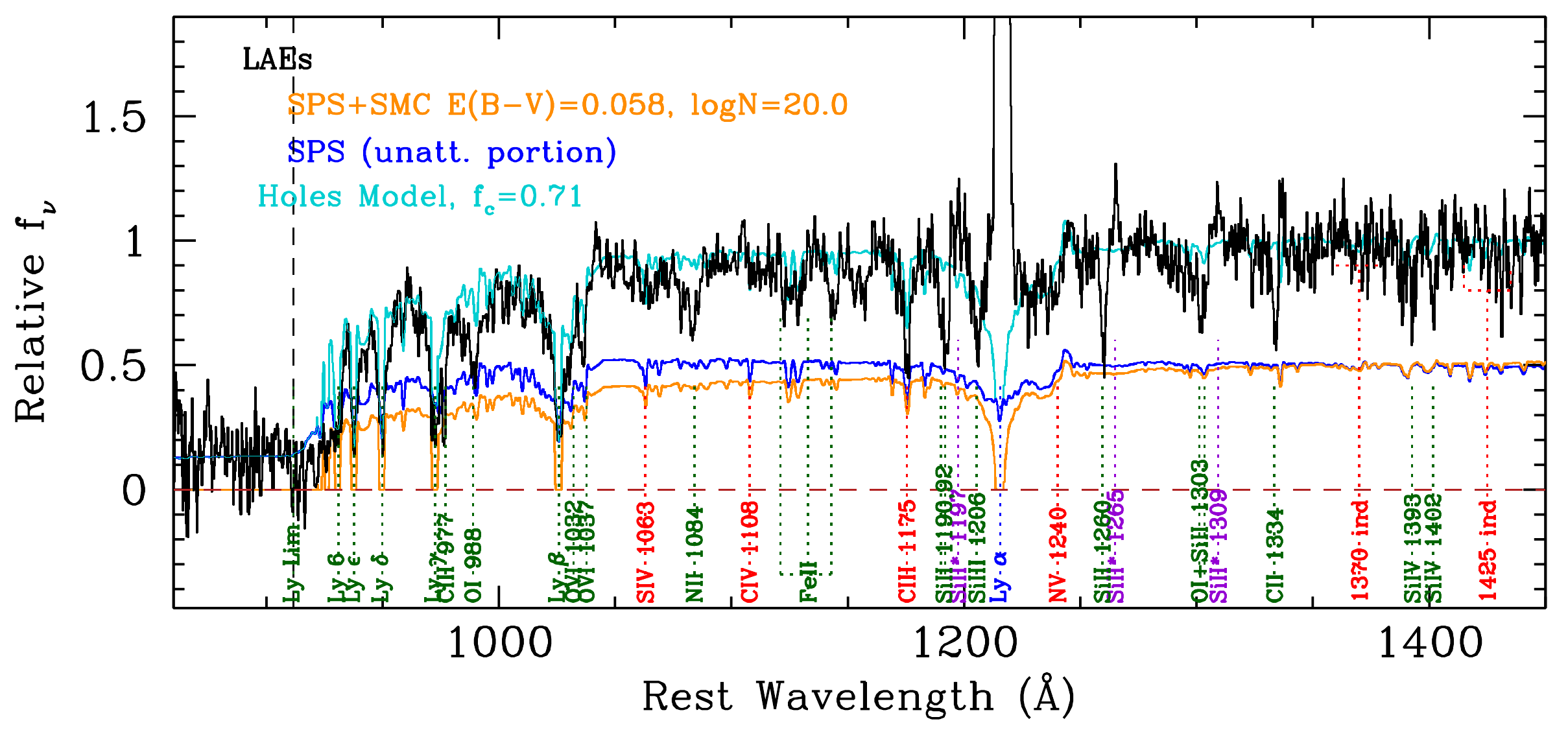}}
\caption{Fits to the same observed spectra as in Figure~\ref{fig:fc}, but using ``holes'' models for the galaxy ISM. 
The model spectra in this case are linear combinations of
a component which escapes the galaxy without attenuation (with relative amplitude $1-f_{\rm c}$; blue) and a component (orange) that is both reddened
by dust and subject to line and continuum opacity due to \ion{H}{1}; the attenuated component has an initial (unabsorbed) relative amplitude $f_{\rm c}$. 
For this type of model (see Table~\ref{tab:fc2}), since the reddened
portion of the spectrum must pass through gas with large $N_{\rm HI}$ (with $\tau_{\rm LyC} \simgt 1000$ for typical values of ${\rm log~\nhi}$), $f_{\rm esc,abs} \simeq (1-f_{\rm c})$. 
}
\label{fig:fc_holes}
\end{figure*}

An alternative model for the effect of the galaxy ISM on the far-UV spectrum is one in which leaking LyC photons escape via
completely transparent ``holes'' in the ISM, as discussed by \citet{reddy16b} (see also \citealt{zackrisson13}). This type
of model is qualitatively supported by the fact that the equivalent widths of interstellar lines of low-ionization species
are correlated with inferred reddening by dust, but the same is not true for higher ionization species like \ion{C}{4} (e.g., \citealt{shapley03}.)  
In the ``Holes'' model 
the spectrum
is decomposed into two distinct components, one of which is reddened and absorbed by foreground dust and gas, and the other unattenuated and
unabsorbed -- and is therefore a scaled version of the intrinsic SPS spectrum. The parameter $f_{\rm c}$ is the fraction of the intrinsic
stellar spectrum covered by optically-thick absorbing material along our line of sight, and the net observed spectrum is given by 
the linear combination 
\begin{equation}
S_{\nu,{\rm obs}} = S_{\nu,{\rm int}}\left[(1-f_{\rm c})+ f_{\rm c} ~{\rm e^{-\tau(\lambda)}} 10^{-A_{\lambda}/2.5} \right]  
\label{eqn:holes_model}
\end{equation}
where $A_{\lambda} = k_{\lambda}E(B-V)_{\rm cov}$ and $E(B-V)_{\rm cov}$ is the reddening applied to the covered portion of the galaxy
The second term on the righthand
side of equation~\ref{eqn:holes_model} also accounts for \ion{H}{1} line and continuum opacity as described in the previous section.
For each KLCS composite spectrum, we fitted for the model values of 
$f_{\rm c}$, ${\rm log(\nhi/cm^{-2})}$, and $E(B-V)_{\rm cov}$ that minimized $\chi^2$; the best-fit model parameters are given in Table~\ref{tab:fc2}. 
Figure~\ref{fig:fc_holes} shows the 
decomposition of the best fitting holes models for the same composites shown in Figure~\ref{fig:fc}.  

The main difference between the screen and holes models is the interpretation of $f_{\rm c}$: for the holes
case, the differential attenuation between the covered and uncovered components requires accounting for extinction as part of the calculation 
of $f_{\rm c}$; this is easily seen in the top panel of Figure~\ref{fig:fc_holes}, which shows that the 
relative contributions of the attenuated and unattenuated components 
to the observed (non-ionizing) spectrum are in the ratio $\sim 3:1$, while geometrically the ratio is $f_{\rm c}/(1-f_{\rm c}) \sim 10$. 
The concept $f_{\rm esc, rel}$ is probably less useful in the context of the holes models, 
since \fout\ now has a more complex relationship to the intrinsic SED of the SPS spectrum than in the screen case (because of the assumption
of two independent components, one affected by dust and the other unaffected.)    

However, in the holes models, $f_{\rm esc,abs}$ is very straightforward to estimate provided $\tau(\rm {LyC}) \gg 1$ for the covered component,
in which case
\begin{equation}
\label{eqn:fc}
f_{\rm esc, abs} \simeq 1 - f_{\rm c}~ .
\end{equation} 
Equation~\ref{eqn:fc} applies to all of the models listed in Table~\ref{tab:fc2} except for the composite of individually-detected galaxies (``All, detected''). 
As discussed
previously, the appropriate CGM+IGM correction for this sub-sample is much more uncertain than for any of the other composites; 
with the assumptions adopted earlier -- that the detected subsample has IGM+CGM transmission drawn from the highest 12\%  -- 
the best-fit model spectrum has $1-f_{\rm c} \approx 0.20$ but $f_{\rm esc,abs}$ has an additional contribution  
from ionizing photons
that have passed through the \ion{H}{1} layer without being completely attenuated (because the LyC optical depth is of order unity). 

\subsection{Model Assessment}

Models fits for both the screen and the holes models can simultaneously reproduce the overall spectral shape, the depth of the Lyman series
absorption lines, the wings of the interstellar \lya\ absorption feature, and the residual LyC flux using the same underlying SPS model.  
There is a slight tendency for both sets of ISM models to over-estimate 
the residual LyC flux for composites having the lowest values of \fout~: e.g., the $L_{\rm uv} > L_{\rm uv}^{\ast}$ sub-sample
has \fout$ = 0.006 \pm 0.008$ (Table~\ref{tab:fesc2_data}) but the best-fit model gives \fout$ \sim 0.03$ (Tables~\ref{tab:fc} and
\ref{tab:fc2}). Such discrepancies would be expected if the covering fraction of the optically thick ISM layer is slightly underestimated
by the cores of the Lyman series lines, or if there is \ion{H}{1} with $17.5 \simlt {\rm log(N_{HI}/cm^2)} \simlt 20$ that
does not contribute significantly to the \lya\ damping wings and whose Lyman series lines are not fully resolved in the spectra (and thus their depth
is underestimated). Since the parameter $f_{\rm c}$ is constrained in our modeling by simultaneously matching the LyC residuals (which should not
depend on spectral resolution) and the Lyman series
line depths (which may depend on resolution), larger reduced $\chi^2$ evaluated over those spectral regions would flag inconsistencies.

Another possibility of course is that \fout\ is underestimated due to residual systematics in the background subtraction, in this case
over-subtraction; this possibility cannot be ruled out entirely, but seems less likely than the presence of unresolved intermediate column density \ion{H}{1}. 

Broadly speaking, the success of the simple models presented above demonstrates that there are real trends 
within the KLCS sample among the depth of the Lyman series lines, the residual LyC flux, and the column density of the dominant component of ISM $N_{\rm HI}$ 
absorption; the trends -- increased LyC leakage in galaxies with lower \nhi, lower inferred $f_{\rm c}$, 
and larger $W_{\lambda}(\lya)$
 -- are in the direction one might have anticipated. We consider a simple physical interpretation of the results in \S\ref{sec:trends}, where
we conclude that the ``Holes'' model does surprisingly well in explaining the observed interconnections between the observable quantities.    

\section{Implications of Escape Fraction Results}
\label{sec:implications}

The term ``escape fraction'' has come to represent several different quantities,
depending on the context and the underlying assumptions. We have gone through the
exercise in the previous sections of calculating a number of different quantities
related to LyC escape, justifying underlying assumptions by demonstrating that
they do not lead to internal inconsistencies. Here we briefly digress to review
the different quantities and the extent to which they are model- or assumption-dependent. 
As above, we begin  
with the most model-independent quantities, and proceed in order of increasing model dependence:

\subsection{$\langle f_{900}/f_{1500}\rangle_{\rm obs}$}  
\label{sec:fesc_obs}

Assuming that the observations are free of systematics
that would invalidate the results, this quantity is entirely model-independent; however,
it is of only limited use for individual objects because it is so strongly modulated by
the variations of CGM+IGM opacity among different lines of sight (\S\ref{sec:igm_trans}). With ensembles in which
more than $\simeq 30$ objects with similar redshifts and common observed properties are
considered, the mean
value of $(f_{900}/f_{1500})_{\rm obs}$ can be corrected using empirically-calibrated IGM+CGM models to obtain \fout.
For the KLCS sample, $0 \le \langle f_{900}/f_{1500} \rangle_{\rm obs} \le 0.14$ among the galaxy subsamples (Table~\ref{tab:fesc2_data}),
with a sample mean of  $\langle f_{900}/f_{1500}\rangle_{\rm obs} = 0.021\pm 0.002$. 

\subsection{$\langle f_{900}/f_{1500}\rangle_{\rm out}$}  
\label{sec:fesc_out}

This parameter is the most important one, in practical terms, for estimating
the contribution of a galaxy population of known far-UV (e.g., at rest-frame 1500 \AA) luminosity function  
to the metagalactic ionizing radiation field (see \S\ref{sec:emissivity}.) 
By our definition, this ratio characterizes the flux density of ionizing photons averaged over
the 880-910 \AA\ window, relative to the non-ionizing UV flux density near 1500 \AA, statistically 
corrected for IGM+CGM opacity back to the ``galaxy photosphere'' at $r_{\rm gal} = 50$ kpc. 
As discussed in \S\ref{sec:igm_trans}, the statistical uncertainty on the IGM correction is drastically
reduced for an ensemble of galaxies provided the lines of sight are independent and the sample selection
does not depend significantly on the IGM properties at fixed redshift. Remaining systematic uncertainties 
in the values of $\langle t_{900} \rangle$ for an ensemble will stem primarily from the treatment of the CGM and possible 
systematic differences in the large-scale IGM environment.  

Table~\ref{tab:fesc2_data} includes values of $\langle t_{900}\rangle$ 
under two different assumptions for the \ion{H}{1} column density distribution for distances $ r > 50$ pkpc from
a galaxy; $\langle t_{900} \rangle$ values with and without an enhancement in \nhi\ from the CGM typically differ by 
$\sim 20$\%, which likely over-estimates the true level of systematic uncertainty, but would be a conservative
upper limit.  As Table~\ref{tab:fesc2_data} indicates, such systematics, if present, would affect the inferred
values of $f_{\rm esc,rel}$ and $f_{\rm esc,abs}$ by a similar factor. 

\begin{figure}[htbp!]
\centerline{\includegraphics[width=8.5cm]{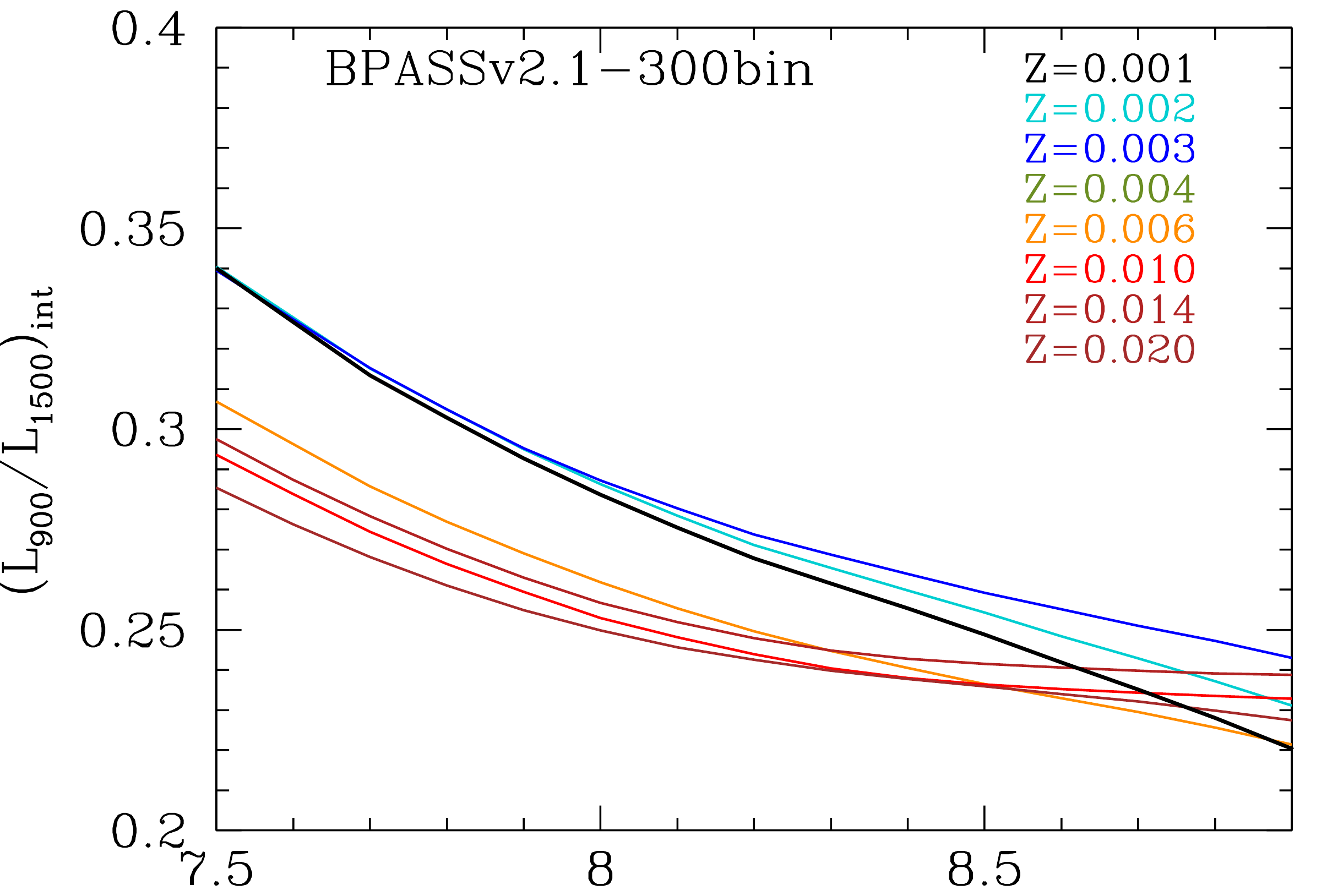}}
\centerline{\includegraphics[width=8.5cm]{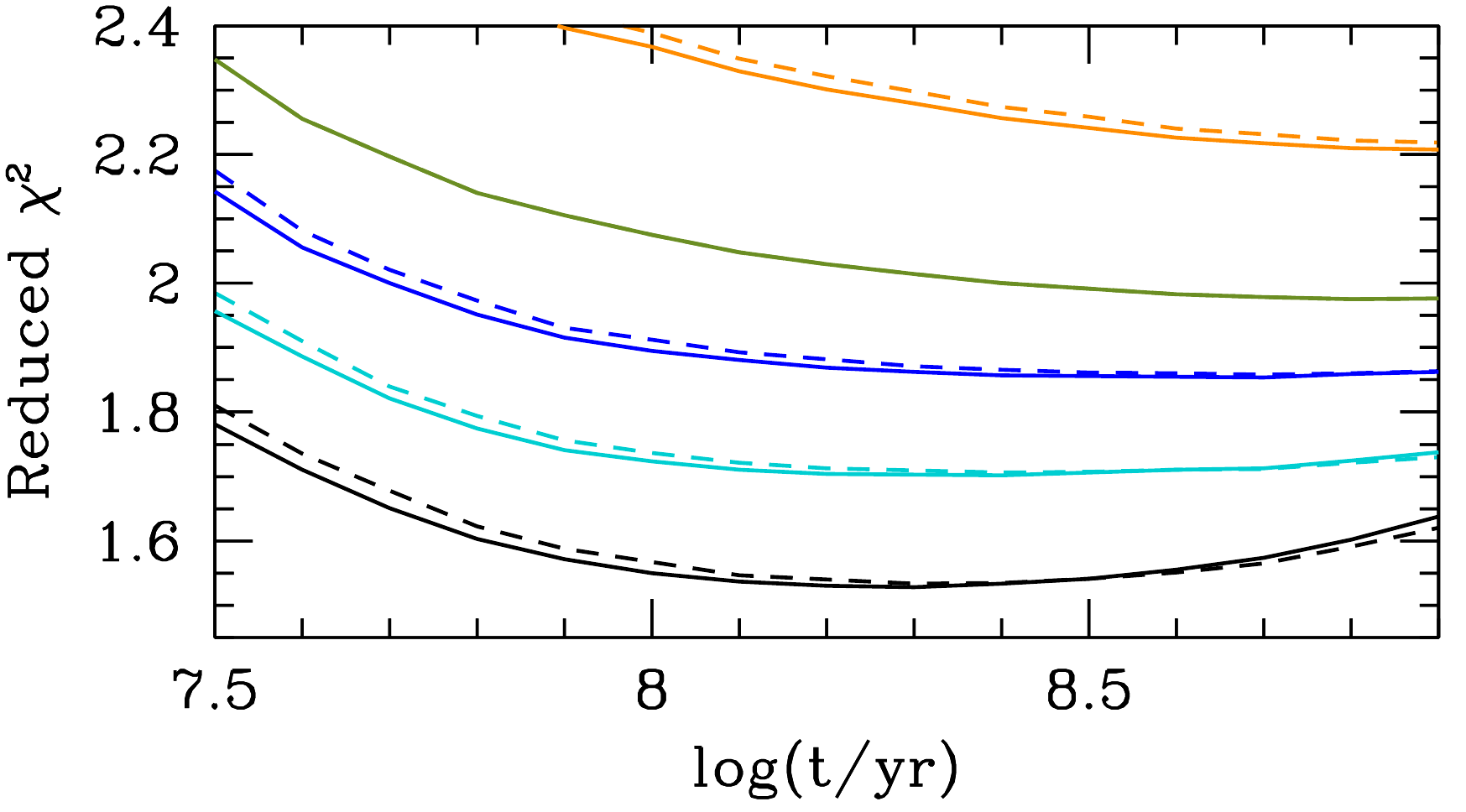}}
\caption{({\it Top:}) Model predictions for the intrinsic flux density ratio $(L_{900}/L_{1500})_{\rm int}$ for 
BPASSv2.1-300bin-IMF135 continuous star formation models (\citealt{eldridge17}) evaluated as a function of assumed age and stellar metallicity;
the latter 
varies from 
$Z_{\ast}=0.001$ ($\sim 0.07 Z_{\odot}$) to $Z_{\ast}=0.020$ ($\sim 1.4 Z_{\odot}$), or $-1.15 \le {\rm [Fe/H]} \le +0.15$. 
({\it Bottom:}) Relative values of $\chi^2/\nu$ for models versus age, color-coded according to stellar metallicity as in the top panel,
for the best-fitting model relative to the full KLCS composite spectrum (``All''). 
The solid curves correspond to the ``screen'' models, and the dashed curves to the ``holes'' models. The $Z=0.001$ models provide by far the
best fits 
at all ages; the best-fitting $Z=0.001$ models have ${\rm log(t/yr) \simeq 8.2-8.3}$ ($\simeq 160-200$ Myr).  
 }
\label{fig:f9_f15_models}
\end{figure}

\subsection{$\langle f_{\rm esc,rel} \rangle$} 
\label{fesc_rel}

As discussed in \S\ref{sec:fesc} above, $f_{\rm esc,rel}$ is a useful concept 
if the intrinsic $(L_{900}/L_{1500})_{\rm int}$ of a galaxy's integrated stellar population is known, since the latter
ratio is the absolute maximum possible $(f_{900}/f_{1500})_{\rm out}$; i.e., $f_{\rm esc,rel}$ approaches unity only
when there is no \ion{H}{1} with appreciable LyC optical depth between the observed stars and $r_{\rm gal} \sim 50$ pkpc.   

If $f_{\rm esc, rel} \le 1$, $(L_{900}/L_{1500})_{\rm int} \simeq 0.28$ (\S\ref{sec:sps_models}; Figure~\ref{fig:f9_f15_models}), 
and $\langle t_{900} \rangle \simeq 0.37$
(Table~\ref{tab:fesc2_data}), then the maximum value of $\langle f_{900}/f_{1500}\rangle_{\rm obs}$ expected for an ensemble of galaxies with 
$z_{\rm s} \simeq 3.05$ is $\langle f_{900}/f_{1500}\rangle_{\rm obs,max} \simeq 0.11$. For an {\it individual} galaxy with $z_{\rm s}=3.05$, $t_{900} < 0.67$
(the $\sim 99$th percentile IGM+CGM transmission; see Table~\ref{tab:igm_models}), 
$(f_{900}/f_{1500})_{\rm obs, max} \simeq 0.30 \times 0.67 \simeq 0.20$. Accounting for dust\footnote{For the attenuation relation that
provides the best fit to the bluest galaxies in the KLCS sample (SMC), 
$k_{\lambda}(900) - k_{\lambda}(1500) = 10.14$; e.g., for $E(B-V)=0.05$, $A(900)-A(1500) \simeq 0.51$ mag, reducing 
$(f_{900}/f_{1500})_{\rm obs,max}$ to $\sim 0.13$ and $f_{\rm esc,rel} \simlt 0.19$.} or assuming a more typical IGM sightline
would immediately reduce the expected upper limits on observed values.  These small numbers would be reduced by large factors
if the LyC regions were not sampled optimally (\S\ref{sec:spec_vs_img}), or the source redshift were significantly higher (\S\ref{sec:igm_trans}).  
A corollary to these considerations is that one should be suspicious of individual measurements of LyC leakage if $(f_{900}/f_{1500})_{\rm obs} \simgt 0.2$, 
and of ensemble measurements with $\langle f_{900}/f_{1500}\rangle_{\rm obs} \simgt 0.11$.

Clearly, assuming a different SPS model, with a different $(L_{900}/L_{1500})_{\rm int}$,  
would also change the inferred $f_{\rm esc, rel}$ given a measurement of, or limit on, $(f_{900}/f_{1500})_{\rm out}$. Figure~\ref{fig:f9_f15_models} 
shows the metallicity and age dependence of $(L_{900}/L_{1500})_{\rm int}$ (including self-consistently the contribution of nebular continuum emission to
the non-ionizing UV light) for constant SFR BPASSv2.1-300bin models. In the context of such models, typical $z \sim 3$ galaxies have SED-inferred ages
of $\sim 50-500$ Myr (${\rm 7.7 \simlt log(t/yr) \simlt 8.7}$); the $Z=0.001$ model that best reproduces the details of the
observed spectra has $0.24 \simlt (L_{900}/L_{1500})_{\rm int} \simlt 0.33$ over the same age range. 
Assuming younger ages for the SPS models would, naively, lower the required value of $f_{\rm esc, abs}$ to match an observed $(f_{900}/f_{1500})_{\rm out}$;
however, younger SPS models are somewhat bluer in the non-ionizing FUV, so that matching an observed galaxy SED requires more reddening by dust -- 
and all reasonable dust attenuation relations have $k_{\lambda}(900) > k_{\lambda}(1500)$, meaning that additional reddening reduces 
$f_{900}$ relative to $f_{1500}$. 
For the screen models, fits of BPASSv2.1 models to the full KLCS galaxy composite with the same IMF and stellar metallicity ($Z_{\ast}=0.001$)
but varying the age over the range ${\rm log (t/yr) = [7.5\rightarrow 9.0]}$ results in monotonic variation in best-fit parameter values: $E(B-V)$ [$0.145\rightarrow0.085$], C1500 [$3.29 \rightarrow 2.01$], \nhi\ [$20.80 \rightarrow 20.05$], 
and $f_{\rm esc,abs}$ [$0.097 \rightarrow 0.191$], but only small changes in $f_{\rm esc,rel}$ [$0.32\rightarrow 0.38$] and $f_{\rm c}$ [$0.73 \rightarrow 0.68$]. 
However, the best fits to the spectra-- both in terms of reproducing the details of the stellar continuum and matching the observed LyC residuals, 
Lyman series line depths, and \lya\ damping wings --  are obtained  
for ${\rm 7.9 \simlt log(t/yr) \simlt 8.5}$, where the parameters are consistent with those listed in Table~\ref{tab:fc} for the fiducial 
${\rm log(t/yr) = 8.0}$ model (see the bottom panel of Figure~\ref{fig:f9_f15_models}.)   

\subsection{$f_{\rm esc,abs}$} 
\label{fesc_abs}

In this section we have tried to emphasize that the parameter often seen as the ``holy grail'' of LyC studies  --  
the absolute escape fraction $f_{\rm esc,abs}$ -- is several steps removed from an observational measurement, even at $z \sim 3$ where
direct measurements of residual LyC emission are feasible.  
Each of the steps is model-dependent, beginning with uncertainty in the shape and absolute intensity of 
the unattenuated EUV and FUV light from stars, followed by the net effect of dust on both the ionizing and non-ionizing
UV, and, finally, the rather poorly-defined notion of what constitutes ``escape'' of an ionizing photon from the galaxy in which it was produced.  

However, we have also attempted to leverage recent progress in understanding the ionizing sources through joint analysis
of the stellar and nebular spectra (e.g., \citealt{steidel16,strom17}) and to follow a self-consistent thread capable 
of reconciling the detectability of LyC
radiation leakage with all other available empirical constraints. 
With these caveats, we have measured  
$f_{\rm esc, abs}$ for a sample of $\simeq L_{\rm uv}^{\ast}$ galaxies at $z\sim3$. Focusing on the ``holes'' models, for reasons discussed 
in more detail in the next section, the bottom line is that
some galaxies appear to leak a substantial fraction of the ionizing radiation they produce-- up to
30\% for those in the highest quartile of $W_{\lambda}(\lya)$, while others - notably the brightest galaxies, and those which 
have \lya\ strongly in absorption -- leak at most imperceptibly ($< 3$\% at 2$\sigma$.) 
Averaged over the KLCS sample, all of which has $L_{\rm uv}$ within 
a factor of a few
of $L_{\rm uv}^{\ast}$ at $z \sim 3$, $\langle f_{\rm esc,abs} \rangle \sim 9$\% given our assumed SPS and reddening models (Table~\ref{tab:fc2}).

It is important to keep in mind the degree to which $f_{\rm esc,abs}$ depends on the latter assumptions -- if, for example, we had 
assumed that the attenuation relation is that of \citet{calzetti00} and that the appropriate SPS model is a solar-metallicity Starburst99 model
(\citealt{leitherer14}; see \citealt{steidel16} for comparisons with the BPASSv2.0 models), then 
the best-fitting model parameters give $f_{\rm esc,abs} = 0.25$\footnote{The same best-fit model has $f_{\rm esc,rel} = 0.99$, log\nhi$=20.9$, and 
$f_{\rm c} = 0.71$, and is a very poor fit to the observed spectrum, with reduced $\chi^2$ larger by more than a factor of 3 compared
to our fiducial model.}, a factor of more
than two times {\it higher} than our fiducial model (Table~\ref{tab:models}), but still under-predict the observed LyC flux by a factor of $\simgt 3$. Changing the attenuation relation to R16 but keeping the SPS
model fixed, $f_{\rm esc,abs} = 0.33$, with similarly poor $\chi^2$ relative to the observed spectrum (Figure~\ref{fig:f9_f15_models}.) 
{\it With the wrong choice of SPS model, 
it is not possible to simultaneously match the shape of the FUV spectrum, the Lyman series line depth, and the residual LyC emission} (see also \citealt{reddy16b}.)       

\subsection{Comparison to Other Recent Results}
\label{sec:compare_results}
\subsubsection{High Redshift}

Given the many quantities commonly used to describe LyC measurements, 
it is useful to compare our results
to others from the recent literature. 
The most directly comparable, in that the results involve composite spectral stacks, have been discussed
by \citealt{marchi17,marchi18}, using spectra obtained as part of the VUDS Survey (\citealt{lefevre15}). 
Due to the minimum wavelength of $\simeq 3800$ \AA\
the samples include only galaxies having $3.5 \le z_{\rm s} \le 4.3$ (median $z_{\rm s} = 3.8$.) 

\citet{marchi17} use a sample of 33 galaxies which were conservatively culled from an original sample of 46 using
multi-band HST imaging to eliminate any significant possibility of foreground contamination. Their spectral stack, 
using a weighted average, has $\langle f_{900}/f_{1500} \rangle_{\rm obs} = 0.008\pm0.004$ at median redshift $\langle z \rangle = 3.80$. If
we apply our own determination of the mean IGM+CGM transmission at $z \simeq 3.8$, $\langle t_{900} \rangle \simeq 0.25$ ($\simeq 10$\%
lower than the transmission assumed by \citealt{marchi17}) to facilitate a direct direct comparison with LLS, one obtains $\fout \simeq 0.032\pm0.016$, 
somewhat smaller than that of the full KLCS sample ($\fout = 0057\pm0.006$; Table~\ref{tab:fesc2_data}.). However, the \citet{marchi17} sample
is on average $\simeq 1$ magnitude brighter (in terms of $M_{\rm uv}$) than KLCS; in addition, only
12 of 33 galaxies (36\%) have \lya\ appearing in emission, compared with 74/124 (60\%) in the KLCS sample, in spite of the significantly
higher mean redshift of the former. As discussed above, the KLCS results would predict that a brighter sample, dominated by galaxies with 
\lya\ observed in absorption, would have relatively low $\fout$ (see Table~\ref{tab:fesc2_data}.)  

\citet{marchi18} use a different subset of VUDS to investigate the dependence of $\fout$ on properties believed to 
be correlated with LyC leakage based on existing detections at low and high redshift. They conclude that galaxies 
with large \wlya\ and compact physical sizes (in the UV continuum and/or in \lya\ emission) 
show evidence for significantly higher $\fout$ than the remainder of the sample. 
Again adopting the measurements of $\langle f_{900}/f_{1500} \rangle_{\rm obs}$ and applying our estimated mean IGM+CGM transmission 
at $z \simeq 3.8$ for consistency, we find that the subsample with $\wlya > 50$ \AA\ has $\fout \simeq 0.3$, a factor of $\simeq 2$ higher
than the KLCS subsamples with the largest \fout\ measured at $z=3.05$. However, all subsets of the $\langle z_{\rm s} \rangle = 3.8$ 
spectra imply $\fout \simgt 0.17$, and the authors
are careful not to claim to have measured uncontaminated values of $\fout$. In fact, if one adopts
lowest-measured \fout\ as the zero point, the differential between their highest and lowest \fout\ subsamples would yield
$\fout \simeq 0.16$, entirely consistent with the maximum $\fout$ among the KLCS sub-samples (Table~\ref{tab:fesc2_data}.). 
Thus, we conclude that the results of both studies are consistent with those of KLCS. 

\subsubsection{Low Redshift} 

As mentioned in \S\ref{sec:klcs_intro}, several recent programs targeting analogs of high redshift
galaxies have successfully detected LyC emission from low-redshift ($z \simeq 0.3$) galaxies using HST/COS
The highest success rate, and the largest values of \fabs, have
been measured for galaxies selected on the basis of unusually high values of the strong emission line index
[\ion{O}{3}]$\lambda\lambda 4960$,5008/[\ion{O}{2}]$\lambda\lambda 3727$,3729$\equiv$O32 (\citealt{izotov16a,izotov16b}.) The rationale behind 
the selection is that high O32 may indicate the presence of density-bounded \ion{H}{2} regions signifying local escape of significant LyC emission. 
(\citealt{jaskot13,nakajima14}). Of five objects with ${\rm O32} > 4$, all 5 are LyC detections
, with $\fabs = 0.06-0.13$, similar to the mean value for the KLCS sample. Subsequently, a more extreme low-redshift galaxy with ${\rm O32 = 11.5}$
and $\fabs = 0.46$, has been identified by \citet{izotov18}, 
comparable to \fabs\ estimated for the directly detected sub-sample of KLCS (Table~\ref{tab:fc2}.) The \lya\ properties of the O32-selected
low-z galaxies (\citealt{verhamme17,izotov18}), also found to be unusual, are discussed in \S\ref{sec:lowz} below.

\section{A Model for Lyman Continuum Escape}
\label{sec:lyc_escape}

\begin{figure*}[thbp!]
\centerline{\includegraphics[width=8.5cm]{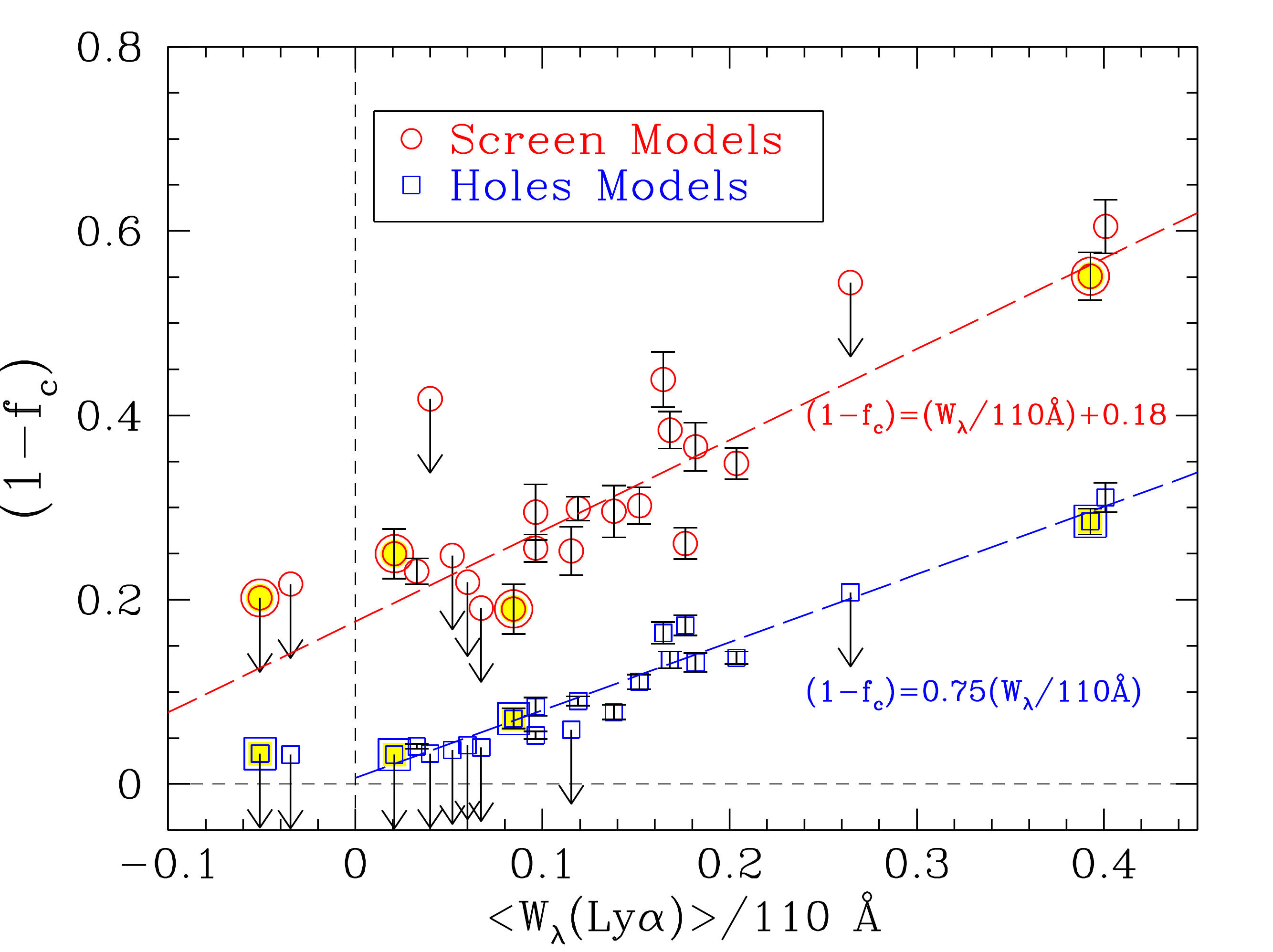}\includegraphics[width=8.5cm]{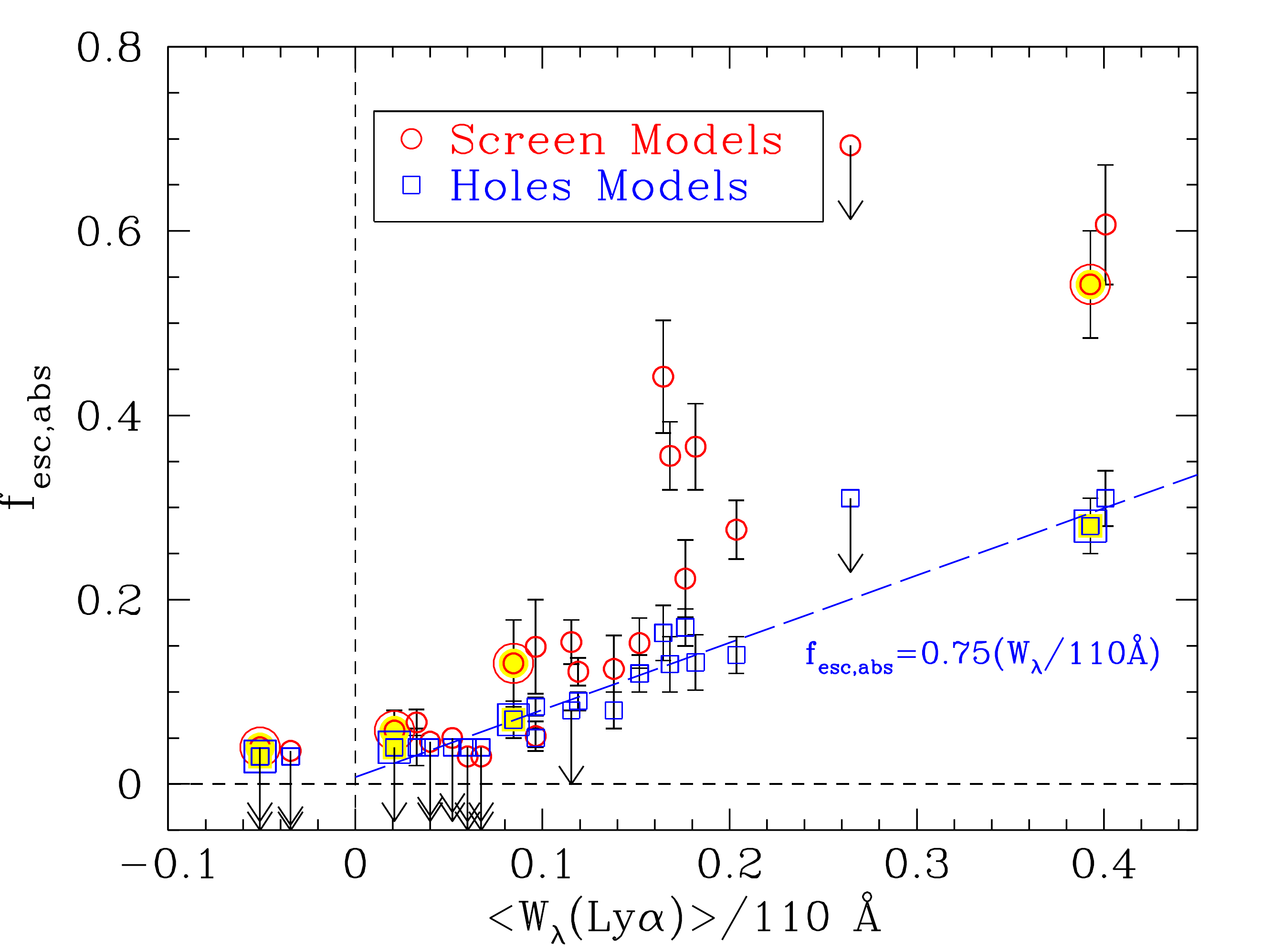}}
\caption{({\it Left:}) The best-fit value (see Tables~\ref{tab:fc} and \ref{tab:fc2}) (1-$f_{\rm c}$) for each of the composite spectra, where $f_{\rm c}$ 
is the inferred fraction of the stellar continuum 
covered by optically thick \ion{H}{1} based on the joint constraints provided by the depth of the Lyman series absorption and
the residual LyC emission (\fout). ({\it Right:}) As for the lefthand panel, where (1-$f_{\rm c}$) has been replaced by
$f_{\rm esc,abs}$ under the Screen and Holes models. Note that, because of the way $f_c$ is defined for Holes models, (1-$f_{\rm c}$) is essentially identical to 
$f_{\rm esc,abs}$ except for the "All, Detected" subsample.  The points representing the 4 independent quartiles in \wlya\ are highlighted with
lightly-shaded larger symbols. 
}
\label{fig:wlya_vs_fesc}
\end{figure*}

\subsection{Trends with $W_{\lambda}(\lya)$}
\label{sec:trends}

%It now seems firmly established that the fraction of galaxies with strong \lya\ emission ($W_{\lambda}(\lya) > 20$ \AA) is $\sim 0.20-0.25$ at $z\simeq 3$ 
%(\citealt{steidel00,shapley03}; the fraction in the KLCS sample is 26/124$= 22.6$\%). The fraction of LBGs with strong
%\lya\ emission also appears to rise monotonically with increasing redshift up to $z\sim 6$ (e.g.,
%\citealt{stark10,stark11}). This trend  -- the increasing fraction with $W_{\lambda}(\lya) > 20$ \AA\ -- may be attributed, at least
%in part, to an increase in the rate of production of H-ionizing photons relative to the non-ionizing FUV (i.e., an increase in $\xi_{\rm ion}$;  
%see, e.g., \citealt{trainor16}), or to systematic changes in the ISM geometry/opacity
%and/or the structure of the CGM surrounding typical star-forming galaxies; both would cause a systematic increase in 
%\fout\ with increasing $z_{\rm s}$, all else being equal.   

In \S\ref{sec:composites}, we observed a strong dependence of the residual LyC emissivity \fout\ on
spectroscopically-measured $W_{\lambda}(\lya)$, and noted (Figure~\ref{fig:plot_lum_cgm}) that the relation
could be captured approximately by a linear relation
\begin{equation}
{\rm \fout \simeq 0.36 \left[ \frac{\langle W_{\lambda}(\lya)\rangle}{110~\textrm{\AA}}\right]+0.02}
\label{eqn:fout_vs_wlya}
\end{equation}
The parametrization of equation \ref{eqn:fout_vs_wlya} was chosen in part because the 
fiducial BPASSv2.1-300bin-t8.0 continuous star formation models used to fit the sub-sample composite spectra 
in \S\ref{sec:screen} and \S\ref{sec:holes} predict that $W_{\lambda}(\lya) \simeq 110$ \AA, assuming ``Case B'' (ionization bounded) 
recombination and no selective attenuation of \lya\ photons
relative to the FUV continuum near \lya\footnote{The predicted $W_{\lambda}(\lya)$ varies by only $\sim \pm10$\% over the age range 
$7.7 \simlt {\rm log(t/yr)} \simlt 8.7$ for the same models.}. 
The term inside square brackets is therefore the ratio of observed \wlya\ to the maximum value
expected under the most favorable conditions for \lya\ production, i.e.,  where every LyC photon has been absorbed and
converted to \lya\ or nebular continuum in accordance with Case B expectations. An observation of the maximum \wlya\ would also require 
that, within the spectroscopic aperture, the intensity of the 
\lya\ emission line relative to the continuum flux
density has not
been significantly altered by resonant scattering of \lya\ (i.e, spatial redistribution and/or 
selective absorption of \lya\ photons by dust grains; see, e.g., \citealt{steidel11}).  

The lefthand panel of Figure~\ref{fig:wlya_vs_fesc} illustrates the nearly-linear relationship between $1-f_{\rm c}$ (Tables~\ref{tab:fc}~and~\ref{tab:fc2} for
the Screen and Holes models, respectively)
and $\langle \wlya\rangle$ among the composite spectra. This behavior strongly supports 
the hypothesis of a direct causal link between \wlya\ in emission and the depth/strength of interstellar 
absorption features observed in the same galaxy spectra (e.g., \citealt{shapley03,steidel2010,steidel11,jones12}); it also lends credence
to the use of simple geometric models of the galaxy ISM to understand the interplay between residual LyC emission, 
the depth of Lyman series absorption lines, and \lya\ escape. 
The sense of the relationship
is that the fraction of the Case B \wlya\ (110 \AA\ in our model) measured within a spectroscopic aperture is directly
related to the portion of the stellar UV continuum that is {\it not} covered 
by optically-thick \ion{H}{1} ($1-f_{\rm c}$) along the same line of sight.    

Of course, one does not expect the production rate of \lya\ photons to be compatible with Case B assumptions if there is
significant detection of residual LyC flux, since any detected 
ionizing photons will not have
produced the corresponding \lya\ photons expected for an ionization-bounded geometry. 
For a typical value
$1-f_{\rm c} \simeq 0.1$, the reduction in the \lya\ source function would be small; however, in the limit
of very high $1-f_{\rm c} \sim 1$, an observer would see \lya\ emission along the same line of sight only if the photons 
have been scattered in our direction after (initially) being emitted in another. In all likelihood this suggests 
that the correlation between $(1-f_{\rm c})$ or $f_{\rm esc, abs}$ and \wlya\ would turn over $f_{\rm esc,abs} \simgt 0.5$. 
Conversely, galaxies with $W_{\lambda}(\lya) \sim W_{\lambda}(\lya,{\rm CaseB}) \simeq 110$\AA\
are {\it a priori} unlikely to exhibit substantial LyC leakage. 

In other words, equation~\ref{eqn:fout_vs_wlya} should not be extrapolated beyond
the range covered by the KLCS composites, which all have $\langle \wlya\ \rangle < 50$ \AA; only 6 individual KLCS sources  
have spectroscopically measured $W_{\lambda}(\lya) > 50$ \AA\,
and only one has $W_{\lambda}(\lya) > 100$ \AA\ (see Table~\ref{tab:klcs}). 

The relationships shown in Figure~\ref{fig:wlya_vs_fesc} are surprisingly tight, given the complexity
of \lya\ radiative transfer compared to that of ionizing photons; for example, \lya\
can be scattered by \ion{H}{1} with \nhi$\ll 10^{17}$ cm$^{-2}$, and \lya\ resonant scattering may increase 
the probability of absorption by dust or of scattering into a diffuse \lya\ halo, in which case an aperture 
tailored to the apparent angular scale of the FUV continuum light (as for KLCS spectra) might miss a large
fraction of the emergent \lya\ flux (see, e.g., \citealt{steidel11,hayes14,wisotzki16}). 
However, a ``down the barrel'' spectrum which includes a LyC detection will also capture the fraction of
\lya\ photons that pass through the same low optical depth holes in the ISM. The tightness of the relationships shown in Figure~\ref{fig:wlya_vs_fesc} 
suggests that most \lya\ photons detected within the spectroscopic aperture may have passed through the same holes in
the ISM.  
This scenario is easier to explain in the context of the ``holes'' models, as shown by the small scatter in both
panels of Figure~\ref{fig:wlya_vs_fesc} -- evidently, applying dust attenuation to the entire EUV/FUV continuum makes it more difficult
to account for the tightness of the correlation between \fout\ and \wlya. 
We also expect that using a larger aperture to measure the same galaxies would increase the scatter between \wlya\ and \fout\ as an increasing
fraction of the observed \lya\ photons will have been scattered to large galactocentric radii via less direct channels through the ISM than the
LyC photons. 

In the discussion that follows, we continue to adopt the ``holes'' model as the simplest picture that remains consistent with the 
current observations.  

\begin{figure}[htbp!]
\centerline{\includegraphics[width=7.5cm]{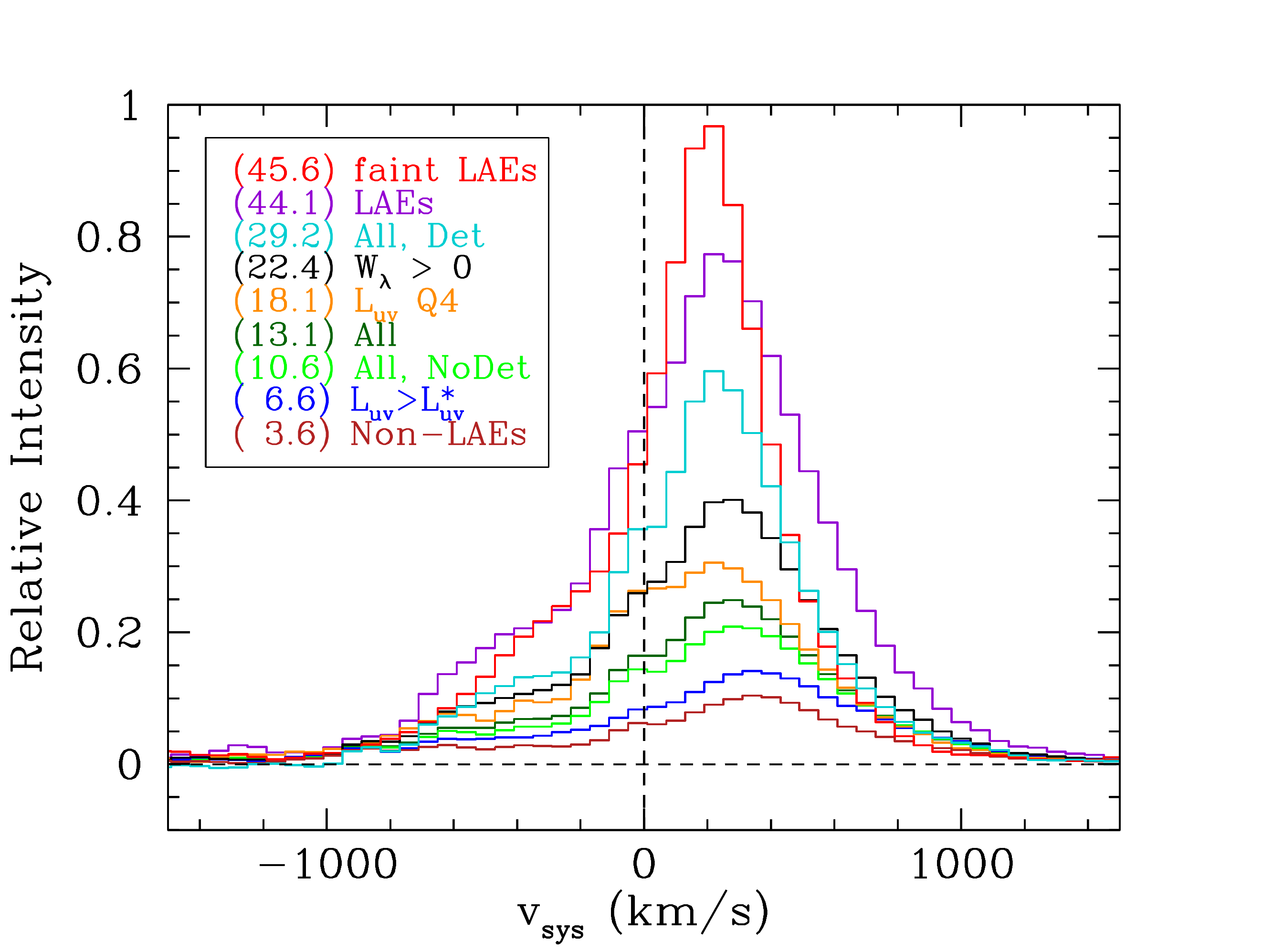}}
\centerline{\includegraphics[width=7.5cm]{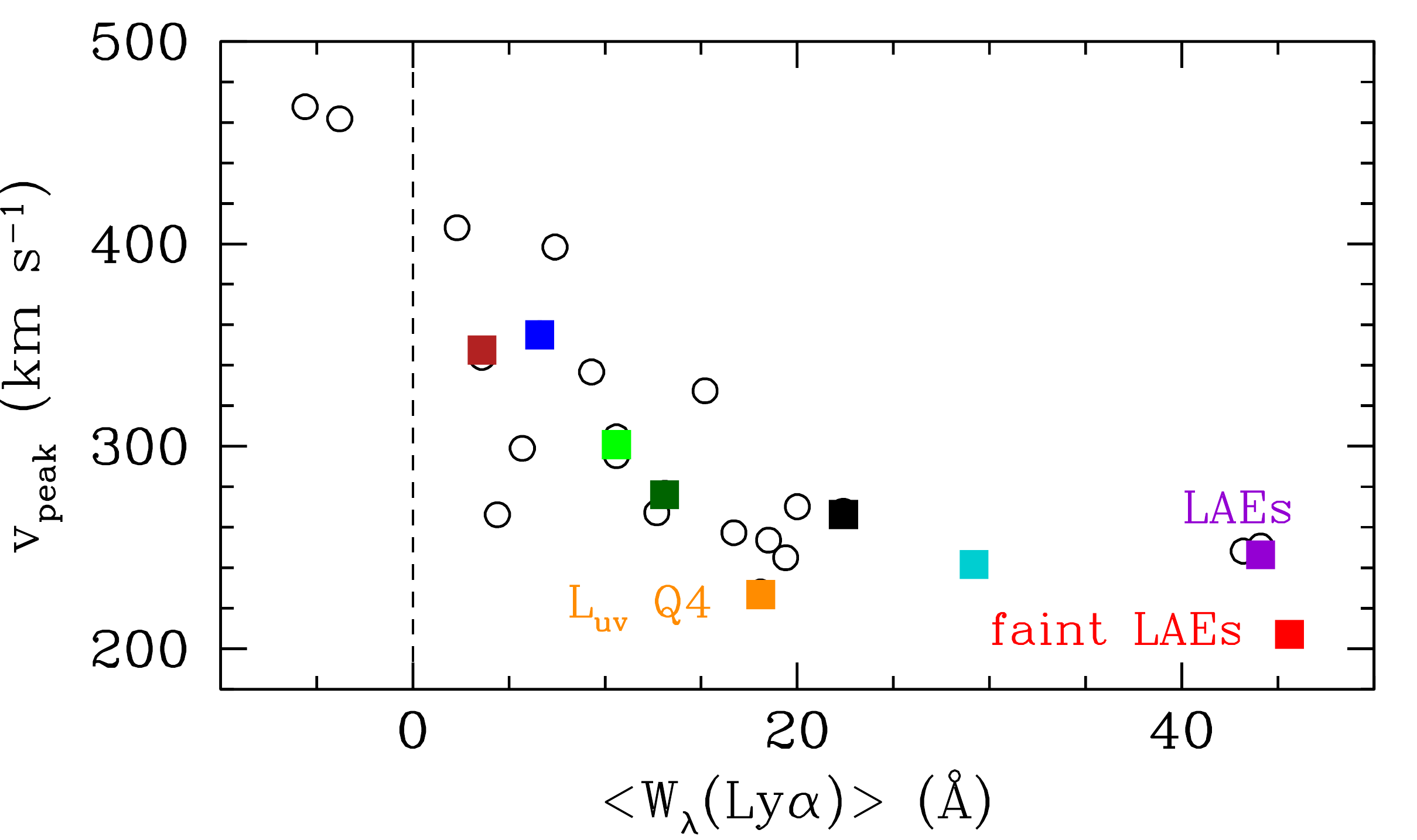}}
\centerline{\includegraphics[width=7.5cm]{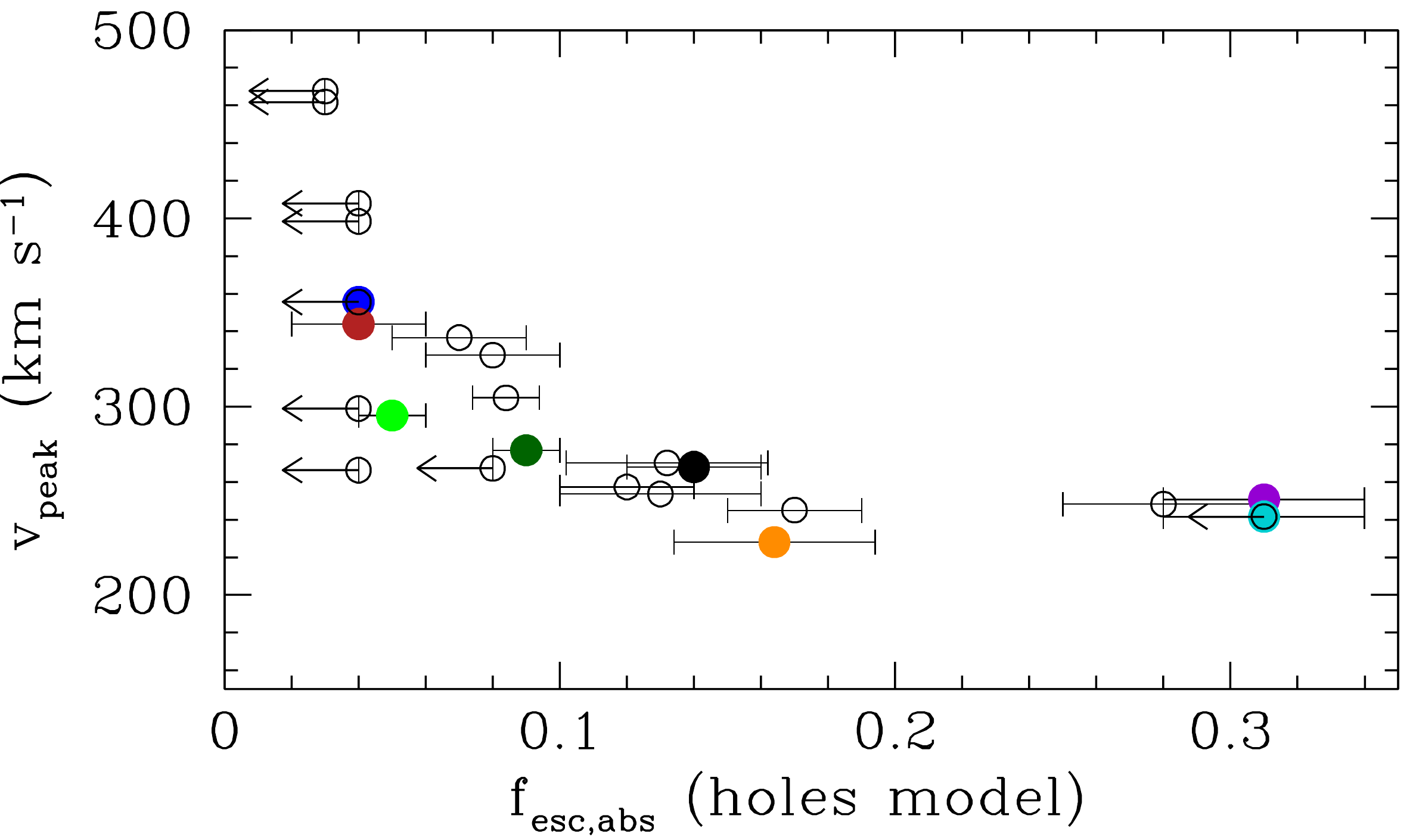}}
\caption{({\it Top:}) Stellar continuum subtracted \lya\ emission line intensity profiles (i.e., the area under the profile is
proportional to \wlya) for selected KLCS sub-samples. 
The profiles are color-coded according to the legend, which also includes  
the value of \wlya/\AA\ as given in Table~\ref{tab:fesc2_data}. 
The spectrum labeled ``faint LAEs'' is the stacked profile of 300 continuum-faint \lya-selected LAEs at $z \sim 2.7$ from \citet{trainor15}
with $M_{\rm uv}\sim -18$ (i.e., $\sim 10$ times fainter than the LAEs in the KLCS sample). 
({\it Center:}) The systemic velocity of the red peak of \lya\ emission, $v_{\rm peak}$, color-coded as 
in the legend for the top panel. Although \wlya\ varies by a factor of $\sim 15$ and $v_{\rm peak}$ is clearly correlated with $\langle \wlya \rangle$, 
the fraction of the total \lya\ emission with $v < 0$ is remarkably constant among all of the composite profiles:  
$f(\lya,{\rm blue}) = 0.31 \pm 0.02$. ({\it Bottom}:) As in the center panel, but with the measured values
of $\langle \wlya\rangle$ replaced by the inferred \fabs. 
}
\label{fig:lya_profiles}
\end{figure}

\subsection{\lya\ Emission Line Kinematics}

Theoretical predictions concerning the relationship between the kinematics and escape of \lya\ emission
and the leakage of significant LyC emission have varied depending on the physical model assumed
for the structure and kinematics of the ISM (e.g., \citealt{yajima14,verhamme15,dijkstra16,verhamme17}). In general, 
most models predict that if the ISM is porous enough to allow the escape of ionizing photons along
a particular line of sight, then one should also observe \lya\ emission with a significant component
near $v_{\rm sys} \simeq 0$ and/or a double-peaked profile with small velocity separation, which
might indicate relatively low \nhi\ along the line of sight. 
However, the picture that seems
indicated by the KLCS results is that, while there is an overall correlation between the inferred
value of \nhi\ associated with the covered portion of the stellar continuum and   
$f_{\rm esc,abs}$ (see, e.g., Table~\ref{tab:fc2}), it is the covering fraction ($f_{\rm c}$) that appears
most closely linked with LyC leakage (as well as with \wlya). Since the measured values of \nhi\ are far too
high to have allowed {\it any} detectable LyC leakage if the entire stellar UV continuum were covered, 
the apparent connection between \nhi\ and \fabs\ may simply indicate that the incidence of clear channels
through the ISM is higher when the characteristic \nhi\ (with covering fraction $f_{\rm c}$) is lower. 
Similar results have recently been obtained for a sample of LyC-detected low-redshift galaxies with \nhi\ measurements (\citealt{gazagnes18,chisolm18}). 

Interestingly, among the KLCS subsample composites, none exhibits a \lya\ peak near $v\sim0$ -- including the composite
formed entirely of individual LyC detections -- as shown in Figure~\ref{fig:lya_profiles}. 
Indeed, the overall shapes of the \lya\ profiles -- including the subsample comprised of individual LyC detections -- 
are remarkably similar, with a fraction $f(\lya,{\rm blue}) \simeq 0.31 \pm 0.02$ of the total \lya\ emission emerging with $v_{\rm sys} < 0$. 
The center panel of Figure~\ref{fig:lya_profiles} shows that the velocity relative to systemic of the \lya\ red peak ($v_{\rm peak}$) 
decreases with increasing $\langle \wlya \rangle$, as found by \citet{erb14,trainor15} for similar-sized samples of LBGs and LAEs at $z \sim 2-3$. 
Nevertheless, the bottom panel of Figure~\ref{fig:lya_profiles}) shows
that the KLCS subsamples with $\fabs \simgt 0.1$ tend to have a nearly constant 
$v_{\rm peak} \sim 250$ \kms, whereas those with $\fabs \simlt 0.1$ exhibit a wide range in $v_{\rm peak}$.  
The faint LAE sample of \citet{trainor15} (whose LyC properties are unknown, but which have Lyman $\beta$ line depths similar
to the KLCS LAE sample) continues a trend of slightly smaller \lya\ velocity shifts for intrinsically fainter galaxies at fixed
\wlya\ (center panel of Figure~\ref{fig:lya_profiles}); this, along with the accompanying smaller maximum blueshifted velocity
of the low-ionization metals, is attributed to lower outflow velocities. It is therefore quite possible that the characteristic $v_{\rm peak}$ 
at fixed $\fabs$ would
also be smaller for galaxies with $M_{\rm uv} > -19.5$.  

Clearly, \lya\ peaking at $v = 0$ is not a necessary condition for significant LyC leakage among
galaxies with $z \simeq 3$. The non-zero vales
of $v_{\rm peak}$ among LyC-detected subsamples suggests that most \lya\ photons escaping along the same line of
sight as LyC photons have been scattered at least once from gas moving with $v_{\rm sys} > 0$. This suggests that
the holes through the ISM allowing LyC photons to escape are not generally optically thin to \lya\ photons without the
benefit of a velocity kick to move them off resonance of gas near $v_{\rm sys} = 0$. 

\subsubsection{Comparison to Low-Redshift Results}

\label{sec:lowz}

\citet{verhamme17} analyze the statistics of \lya\ emission among
the sample of known LyC emitters at low redshift, and compare them to larger samples of potential analogs to high redshift galaxies, including 
LBAs (\citealt{heckman11}), Green Peas (GPs; \citealt{jaskot14,henry15}), 
and galaxies drawn from the \lya\ Reference Sample (LARS; \citealt{ostlin14,hayes14,rivera-thorsen15}.)  With the recent addition of the
more extreme object  
J1154+2443 (\citealt{izotov18}), there are 6 high-O32-selected low-z galaxies with LyC detections. All have reported $\wlya \ge 75$ \AA, and, perhaps more
significantly,  
\lya\ profiles with no discernible absorption component and relatively small velocity differences between the blueshifted and redshifted peaks of \lya\ 
($\Delta v_{\lya} \sim 200-400$ \kms). \citet{verhamme17} 
argue that a small $\Delta v_{\lya}$ is a model-independent signature of low foreground \nhi, 
and therefore a reliable predictor of significant LyC emission. \citet{izotov18} further suggests that there is a dependent relationship 
between \fabs\ and
both O32 and $\Delta v_{\lya}$. We are currently unable to test the O32 dependence in the KLCS sample, though we note that
${\rm O32} > 4$ is satisfied by a substantial fraction ($\simeq 40$\%) of $z \sim 3$ galaxies in the KBSS-MOSFIRE sample, which were selected using
continuum criteria identical to those of KLCS.   

Caution must be used in comparing values of \wlya\ between samples, since $\wlya$ 
is known to depend on aperture size for both low-z (\citealt{hayes14}) and high-z (\citealt{steidel11,wisotzki16}) galaxies. With this caveat in mind,  
the O32-selected low-z galaxies have larger \wlya\ than any of the KLCS composite sub-samples; two of the individual KLCS galaxies have spectroscopically-measured  $\wlya > 75$ \AA, both of which are among the 15 individually-detected LyC sources; see Table~\ref{tab:fesc_data}. 
We also note that the composite spectrum of the individually-detected galaxies is the only subsample for which the best-fit value of 
\nhi$\ll 10^{20}$ cm$^{-1}$ (see Tables~\ref{tab:fc} and \ref{tab:fc2}), thus resembling the \lya\ profiles of the low-z sample.

The values of $\Delta v_{\lya}$ reported for the low-z LyC leakers is the velocity separation between red and blue \lya\ peaks; while these peaks are
clearly present in the \lya\ profiles of the KLCS sub-samples shown in the top panel of Figure~\ref{fig:lya_profiles}, only the red peak can be measured
reliably. The 6 O32-selected low-z LyC emitters all have \lya\ red peaks with $v_{\rm peak} \le 150$ \kms, smaller than the characteristic
$v_{\rm peak} \simeq 250$ \kms\ of the KLCS subsamples with \fabs$\simgt 0.1$ (bottom panel of Figure~\ref{fig:lya_profiles}). Thus, it is not yet clear whether identical criteria can be used to select LyC-emitting galaxies at $z \sim 0.3$ and $z \sim 3$; observations of low-z galaxies
with less stringent pre-selection criteria, and O32 measurements for the KLCS sample, would help to improve the comparison. At present, we can
say that high O32 ratios are very common among $z \sim 3$ galaxies, perhaps for reasons not directly related to LyC escape (see, e.g., \citealt{trainor16,steidel16,strom17}). Our results at $z \sim 3$ also suggest that very small \lya\ velocity offsets are not necessary for significant LyC escape, though
\fabs\ is low when $v_{\rm peak} \simgt 300$ \kms.

\subsection{Resonantly-Scattered Metal Lines}
\label{sec:efs}

As briefly mentioned in \S\ref{sec:morphology}, complementary insight into the geometric/kinematic distribution of optically thick material can
be obtained from observations of the non-resonant (sometimes referred to as ``fluorescent'') emission lines that share the
same excited state as the resonance line, but involve de-excitation to fine structure levels just  
above the ground state.  Since the interstellar
\ion{Si}{2}$\lambda 1260$ and \ion{Si}{2}$\lambda 1526$ absorption lines are resonance transitions, modulo the selective absorption of line photons
by dust grains, there should be no net
emission or absorption provided that the measurement aperture includes the entire scattering medium (e.g., \citealt{shapley03,erb12,henry15}). 
For this reason, it is potentially interesting to measure the 
equivalent widths of the resonance absorption and the associated non-resonance emission 
within the (fixed) spectroscopic apertures used for KLCS. 

\begin{figure*}[htbp!]
\centerline{\includegraphics[height=8cm]{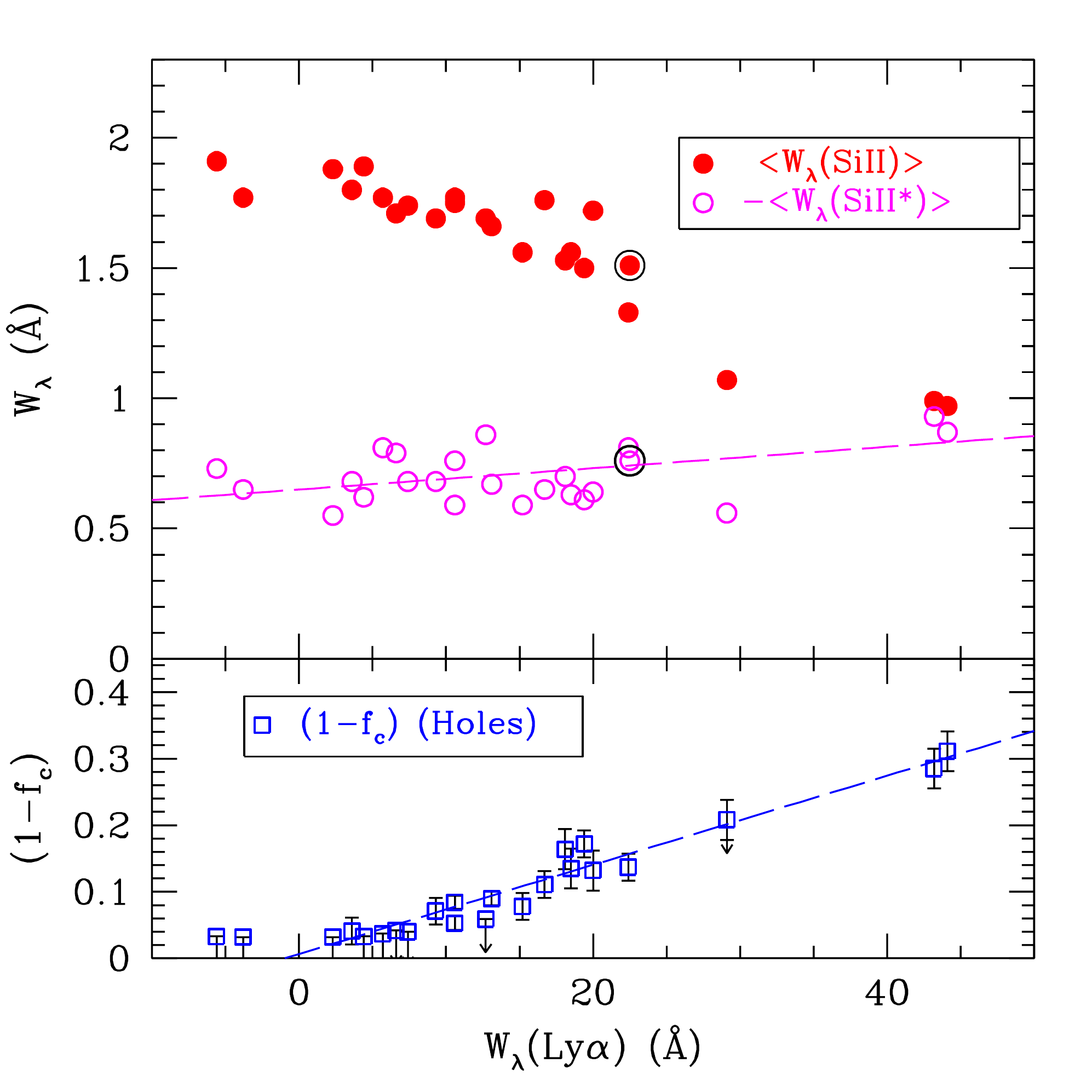}\hspace{0.5cm}\includegraphics[height=8.0cm]{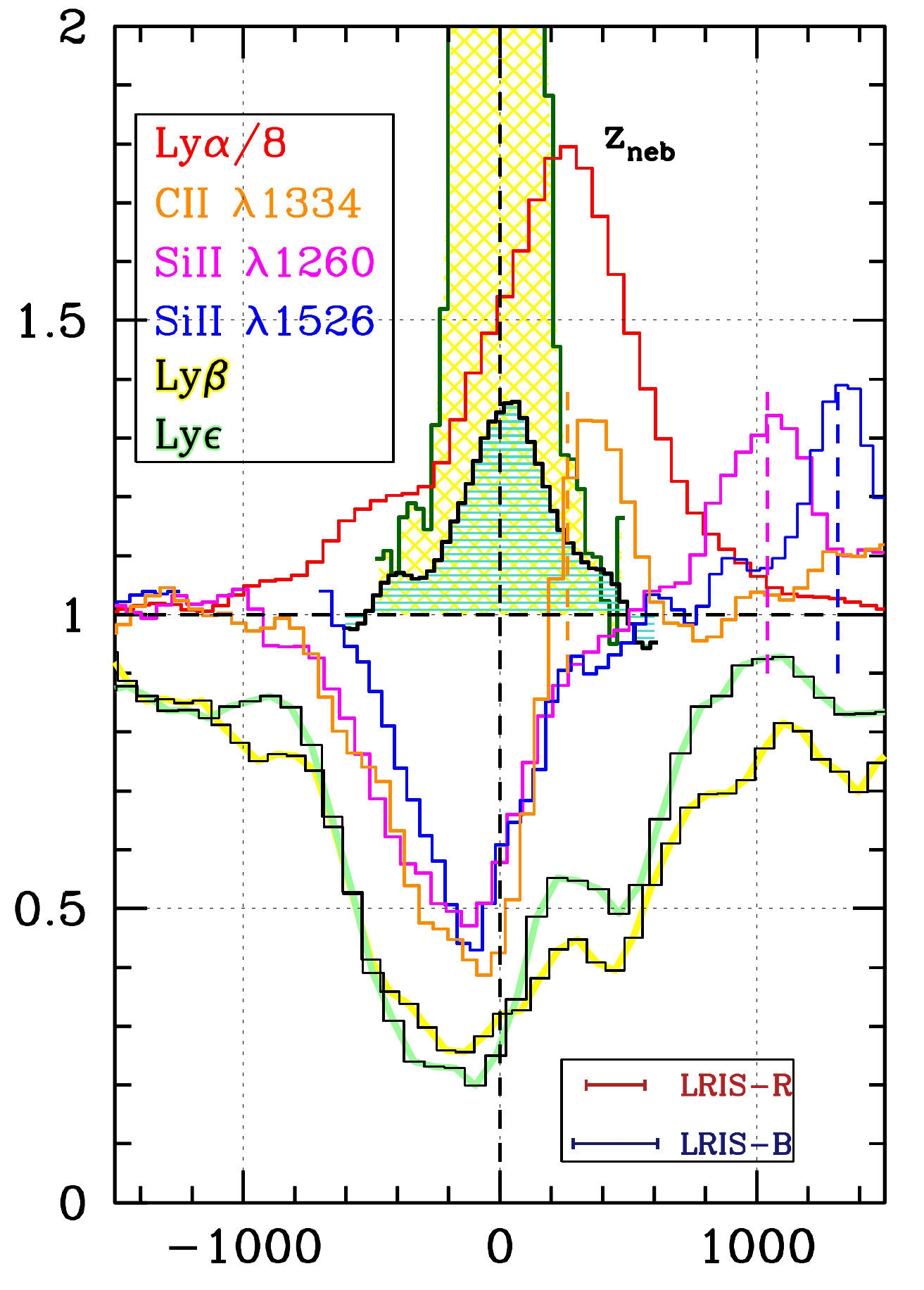}}
\caption{({\it Left:}) The relationship between the strength of interstellar resonance absorption lines (the average of \ion{Si}{2} $\lambda 1260$ and $\lambda 1526$
(solid red points))
and that of the associated excited fine structure 
emission lines (\ion{Si}{2}$^{\ast} \lambda 1264$ and $\lambda 1533$; open magenta points) versus $\langle \wlya \rangle$ (see
Figure~\ref{fig:vel_lyc}.) There is a strong trend of non-resonance emission strength increasing relative to resonance absorption line strength as
$\langle \wlya \rangle$ increases. 
The blue open squares in the bottom panel are the values of $1-f_{\rm c}$ determined using the ``holes'' model (\S\ref{sec:holes}),
and are the same as those shown in the top panel of Figure~\ref{fig:wlya_vs_fesc}. 
({\it Right:}) As in Figure~\ref{fig:vel_lyc}, for the subset
of KLCS with nebular redshifts to show the non-resonant emission line kinematics with respect to \lya\ emission and the strong IS resonance absorption features. 
The green shaded emission feature is the average of the \ion{Si}{2}$^{\ast}\lambda 1265$ and \ion{Si}{2}$^{\ast}\lambda 1533$ 
profiles in the same composite; the equivalent width measurements for this particular composite are indicated with black open symbols in the lefthand panel.
The yellow-shaded emission line is the mean [\ion{O}{3}]$\lambda 5008$ emission line of the same subsample.}
\label{fig:si2star}
\end{figure*}

Figure~\ref{fig:si2star} compares the mean equivalent widths of the resonance lines \ion{Si}{2}$\lambda 1260$ and \ion{Si}{2}$\lambda 1526$ 
with the average equivalent width of the non-resonant emission lines \ion{Si}{2}$^{\ast} \lambda 1265$ and \ion{Si}{2}$^{\ast} 1533$ as a function
of \wlya\ for the KLCS sub-samples. The lefthand panel of Figure~\ref{fig:si2star} shows the 
well-established anti-correlation between \wlya\ and $W_{\lambda}$(LIS) (e.g., \citealt{shapley03,du18}) 
where LIS refers to low-ionization resonance absorption lines (red filled points). In addition, the ratio of \ion{Si}{2}$^{\ast}$ emission 
(skeletal points) to \ion{Si}{2} absorption increases with $\langle \wlya \rangle$ -- there is a weak trend of increasing strength of
$W_{\lambda}$(\ion{Si}{2}$^{\ast}$ with $\langle \wlya \rangle$ (magenta dashed line in Figure~\ref{fig:si2star}) but a strongly decreasing
trend of $W_{\lambda}$(\ion{Si}{2}) with increasing $\langle \wlya \rangle$ causes the ratio $W_{\lambda}$(\ion{Si}{2}$^{\ast})/W_{\lambda}$(\ion{Si}{2}) to vary  
from $\sim 0.3$ to $\sim 0.9$ across the sample.   

Although a detailed discussion of the implications of the non-resonant emission is deferred to future work, one interpretation of the trends illustrated in
Figure~\ref{fig:si2star} is that the physical extent of the scattering medium  
producing resonance absorption lines in ``down the barrel'' spectra
is more compact for galaxies with higher LyC escape fractions. In such a scenario, the KLCS spectroscopic aperture  
(projected physical size $9.2 \times 10.3$ kpc at $\langle z_{\rm s} \rangle$) contains a significantly larger fraction of the gas with log~$\nhi \simgt 10^{17}$ cm$^{-2}$ 
in the CGM of galaxies having the highest LyC escape fraction. In other words -- not surprisingly -- galaxies are more likely to leak along
an observed line of sight when they exhibit more spatially compact distributions of optically-thick gas in the interface between the ISM and CGM.  

An interesting question is whether the more compact distribution of optically thick material is the ``cause'' or the ``effect'' of higher LyC escape. 
It is
quite possible that the low-ionization CGM is reduced by a more intense local
ionizing radiation field in the inner CGM. That is, the reduced spatial extent of optically-thick low-ionization material
surrounding LyC leaking galaxies might be caused at least in part by self-ionization. For example, we consider the \wlya\ (Q4) subsample, which
has $\nu L_{\nu}(1500\textrm{\AA}) \simeq 1.6\times10^{11} L_{\odot}$ and $\fout \simeq 0.17$ (Table~\ref{tab:fesc2_data}), thus producing an average 
escaping ionizing luminosity $\langle \nu L_{\nu}(\textrm{LyC}) \rangle \simeq 10^{44}$ ergs s$^{-1}$. The average specific intensity at the Lyman limit
as a function of 
galactocentric distance $r$ 
would be $\langle I_{\nu}(r) \rangle \simeq 10^{-19}(r/50~{\rm kpc})^{-2}$ ergs s$^{-1}$ cm$^{-2}$ Hz$^{-1}$: at $r=50$ kpc, 
the local ionizing radiation field intensity would exceed that of the metagalactic background by a factor of $>10$, 
and could dominate the ionization state of the CGM to $r \simgt 150$ kpc around individual galaxies.     

\subsection{Ionizing Emissivity of $z \sim 3$ LBGs}
\label{sec:emissivity}

Quantitative measurement of LyC flux from the KLCS sample of galaxies at $z \sim 3$ is the first
step in addressing the issue of the connection between directly-observed galaxy properties
and the propensity to leak significant LyC radiation.  We have shown that even this step has required
combining $N \sim 30$ independent galaxy spectra according to various easily-measured empirical criteria in order
to accurately correct the ensemble for the effects of intergalactic and circumgalactic gas along our line of sight. 

If the KLCS galaxy sample were a representative subset of the population of star-forming galaxies that contributes
to the metagalactic ionizing radiation field
at $z \sim 3$, then the values of \fout\ in Table~\ref{tab:fesc2_data} are all that is needed to convert a measurement of
the far-UV (non-ionizing) emissivity 
$\epsilon_{\rm uv} = \int L_{\rm uv} \Phi(L_{\rm uv}) dL_{\rm uv}$,
where $\Phi(L_{\rm uv})$ is the galaxy luminosity function evaluated in the rest-frame FUV (typically 1500-1700 \AA), into 
the ionizing emissivity contributed by the same population of galaxies, 
\begin{equation} 
\epsilon_{\rm LyC} \simeq  \int_{L_{\rm uv,min}}^{L_{\rm uv,max}} \langle f_{900}/f_{1500}\rangle_{\rm out} \times L_{\rm uv} \Phi(L_{\rm uv}) dL_{\rm uv} 
\label{eqn:emissivity}
\end{equation} 
as discussed by many authors (see, e.g., \citealt{steidel01,shapley06,nestor11,nestor13,mostardi13,grazian17}.)
This method is especially useful because $\Phi(L_{\rm uv})$ has been measured over a substantial
range in $M_{\rm uv}$ at redshifts $z\simeq 2-9$ (e.g., \citealt{reddy09,oesch10,bouwens15,finkelstein15,parsa16}.)

Note that \fout\ appears inside the integral in equation~\ref{eqn:emissivity}; as discussed in the previous
section, although   
KLCS samples only a limited range in $M_{\rm uv}$ ($-19.5 \simgt M_{\rm uv} \simgt -22.1$;  
$0.25 \simlt L_{\rm uv}/L_{\rm uv}^{\ast} \simlt 3$), evidently only the UV-fainter half of the sample 
has significant \fout.  We argued in \S\ref{sec:trends} that a surprisingly tight relationship between 
the covering fraction $f_{\rm c}$ and the measured \wlya\ in the spectroscopic aperture suggests that 
the most direct signature of leaking LyC emission appears to be \wlya, and that the apparent
dependence on $L_{\rm uv}$ is secondary. However, it is straightforward to obtain approximate numbers for the contribution of KLCS-like
galaxies to $\epsilon_{\rm LyC}$ using the empirical dependence of \fout\ on either $L_{\rm uv}$ or \wlya. 

\subsubsection{Weighting by $L_{\rm uv}$}
\label{sec:emissivity_lum}

If we adopt the rest-frame FUV galaxy luminosity function (UVLF) determined by \citet{reddy09} for star-forming galaxies at $2.7 \simlt z \simlt 3.4$,
\begin{equation}
\Phi(L_{\rm uv}) = \Phi^{\ast} \left(L_{\rm uv}/L_{\rm uv}^{\ast}\right)^{\alpha} {\rm exp}\left(-L_{\rm uv}/L_{\rm uv}^{\ast}\right) , 
\label{eqn:lumfun}
\end{equation}
with $\alpha = -1.73$, $\Phi^{\ast} = 1.71\times10^{-3}$ Mpc$^{-3}$, and $M_{\rm uv}^{\ast} = -21.0$,
then the total FUV emissivity of the population of galaxies with $M_{1700} < -17.4$ (i.e., $L_{\rm uv} > 0.036 L_{\rm uv}^{\ast}$)
is  $\epsilon_{\rm uv} = 3.28\pm0.24 \times 10^{26}$ ergs s$^{-1}$ Hz$^{-1}$ Mpc$^{-3}$.  
The KLCS dataset includes only galaxies with $M_{\rm uv} \le -19.5$, sampling a range in $L_{\rm uv}$ that accounts for 
$\simeq 0.52\times \epsilon_{\rm uv} \simeq 1.71\times10^{26}$ ergs s$^{-1}$ Hz$^{-1}$ Mpc$^{-3}$. 

Note that the UV emissivity of the $M_{\rm uv} < -19.5$ population is relatively 
insensitive to the choice of UVLF: for example,
if instead of \citet{reddy09} we were to use
the parameters advocated by \citet{parsa16}, i.e., $M_{\rm uv}^{\ast} \simeq -20.5$, $\alpha \simeq -1.37$, and $\epsilon_{\rm uv} \simeq 3.16\times10^{26}$
ergs s$^{-1}$ Hz$^{-1}$ Mpc$^{-3}$ integrated down to $M_{\rm uv} \simeq -15.5$, 
then galaxies with $M_{\rm uv} < -19.5$ produce $\epsilon_{\rm uv} \simeq 0.48\times3.16\times10^{26} \simeq 1.52\times10^{26}$ ergs s$^{-1}$ Hz$^{-1}$ Mpc$^{-3}$ 
at $\langle z_{\rm s} \rangle = 3.05$, a difference of only $\simeq 11$\%. 
Similarly, the fractional contribution of galaxies with $M_{\rm uv} < -19.5$ to the total emissivity integrated
down to a common $M_{\rm uv} \simeq -15.5$ are 43\% (\citealt{reddy09}) and 48\% (\citealt{parsa16}.) 

If we assume for the moment that \fout$= 0$ for galaxies with $M_{\rm uv} \le -21.0$ and \fout$=0.11$ for $-21 < M_{\rm uv} \le 19.5$ 
(Table~\ref{tab:fesc2_data} and Figure~\ref{fig:plot_lum_cgm}), then the galaxies that contribute significantly to $\epsilon_{\rm LyC}$
at $z \sim 3$ have $\epsilon_{\rm uv} \simeq 1.21\times10^{26}$ ergs s$^{-1}$ Hz$^{-1}$ Mpc$^{-3}$ and 
thus contribute
$\epsilon_{\rm LyC} \simeq 0.11 \times 1.21\times10^{26} \simeq 13.0 \times 10^{24}$ ergs s$^{-1}$ Hz$^{-1}$ Mpc$^{-3}$.
The contribution would be slightly lower, $\epsilon_{\rm LyC} \simeq 11 \times 10^{24}$ ergs s$^{-1}$ Hz$^{-1}$ Mpc$^{-3}$,
assuming \fout$=0.094$ as for the ``IGM Only'' transmission model. 

\subsubsection{Weighting by \wlya}
\label{sec:emissivity_wlya}

Alternatively, we can use our derived empirical relation between \wlya\ and \fout\ to estimate the contribution of KLCS-like galaxies 
to $\epsilon_{\rm LyC}$. In this case, we 
take \fout\ outside the integral in equation~\ref{eqn:emissivity} and
use the distribution function $n[\wlya]$ of UV continuum-selected $z \sim 3$ galaxies to estimate contributions to $\epsilon_{\rm LyC}$ 
relative to the integral over the UVLF, $\epsilon_{\rm uv}$.

We begin with equation~\ref{eqn:fout_vs_wlya}, the best linear relation between \fout\ and \wlya; where again 
the statistic \fout\ is adopted as the least model-dependent link between the FUV luminosity of galaxies and their
contribution to the ionizing emissivity, since it does not depend on assumptions regarding SPS models, ISM radiative
transfer, or attenuation by dust. 
However, for the present purpose, we assume that $W_{\lambda}(\lya,{\rm CaseB}) \simeq 110$ \AA\
as discussed in \S\ref{sec:trends} above, and a model for mapping between \fout\ and \wlya\ given by
equation~\ref{eqn:fout_vs_wlya}. For 
$\wlya > 0.5W_{\lambda}(\lya,{\rm CaseB}) = 55$ \AA, which is unconstrained by the KLCS composite spectra, we assume
that \fout\ reaches a maximum when $(\wlya/110~\textrm{\AA}) \sim 0.5$, since at some point the increasing escape
of LyC photons would result in a decreasing \lya\ source function compared
to Case B expectations. A naive expectation is that the source function of \lya\ emission along a particular line of sight 
would be reduced by a factor (1-$f_{\rm c}$) relative to Case B, followed by a further reduction proportional
to $f_{\rm c}$ due to scattering of \lya\ photons by the covered portion of the UV continuum. 
In such a simplified
picture, one might expect $W' \equiv \wlya/W_{\lambda}(\lya,{\rm CaseB}) \approx 2f_{\rm c}(1-f_{\rm c})$, which has a maximum
$W'\approx 0.5$ when $f_{\rm c} \simeq 0.5$  and goes to zero as $f_{\rm c} \rightarrow 0$ and as $f_{\rm c} \rightarrow 1$. Solving
for $f_{\rm c}$ in terms of $W'$ gives
\begin{equation}  
(1-f_{\rm c}) = 0.5 - \left[|0.5W'-0.25|\right]^{1/2}.
\end{equation}
Using equation~\ref{eqn:fout_vs_wlya} and the holes model fit $(1-f_{\rm c}) \approx 0.75 W'$ (Figure~\ref{fig:wlya_vs_fesc}) to 
connect \fout\ to $1-f_{\rm c}$ over the range constrained by the data results in a model which maps
the expected \fout\ to $W'$ over the full range of each:
\begin{equation}
\label{eqn:fout_vs_w}
\fout \simeq 0.28-\left[|0.146W'-0.073|\right]^{1/2} .
\end{equation}

\begin{figure}[htbp!]
\centerline{\includegraphics[width=8.5cm]{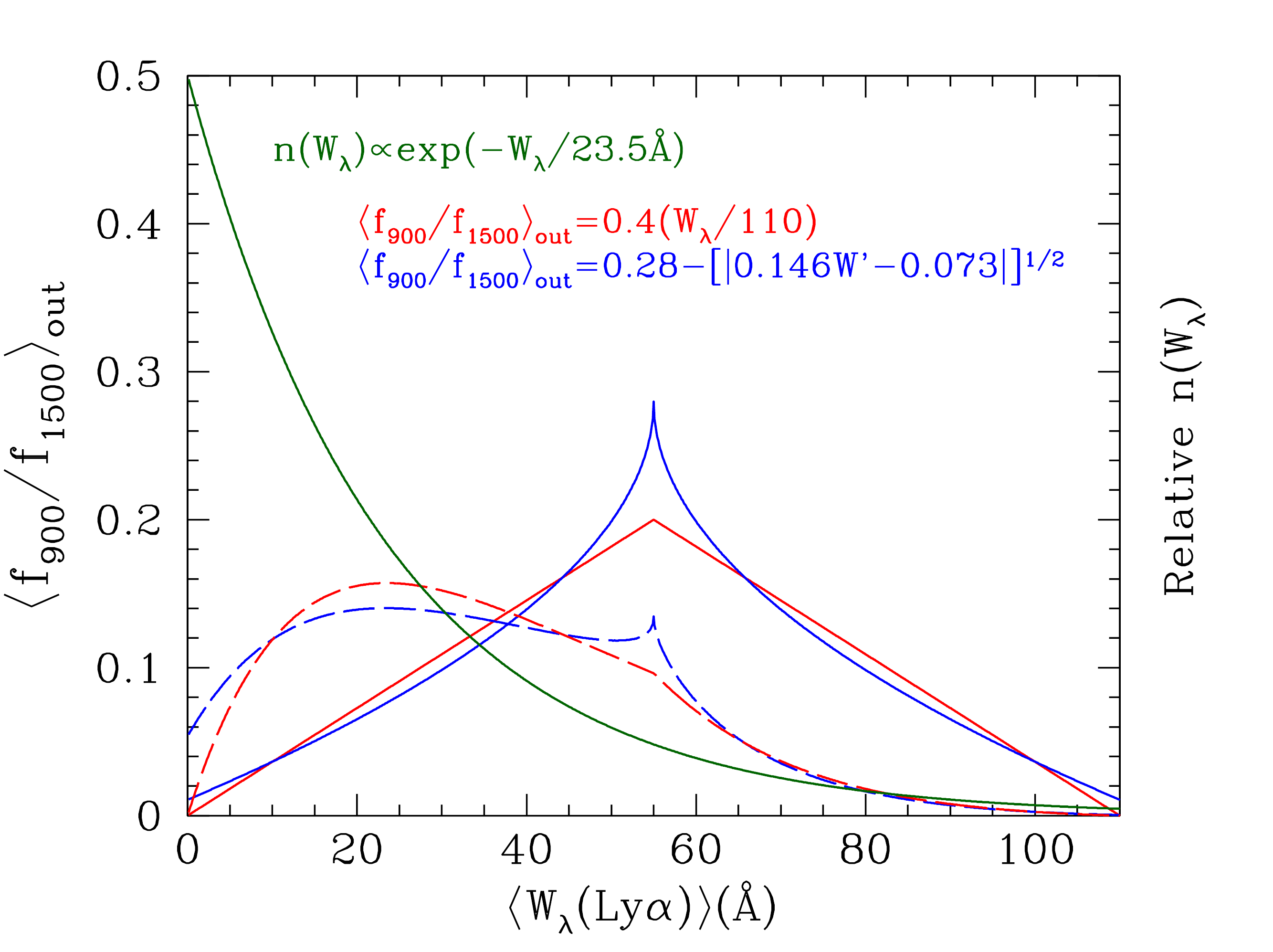}}
\centerline{\includegraphics[width=9.0cm]{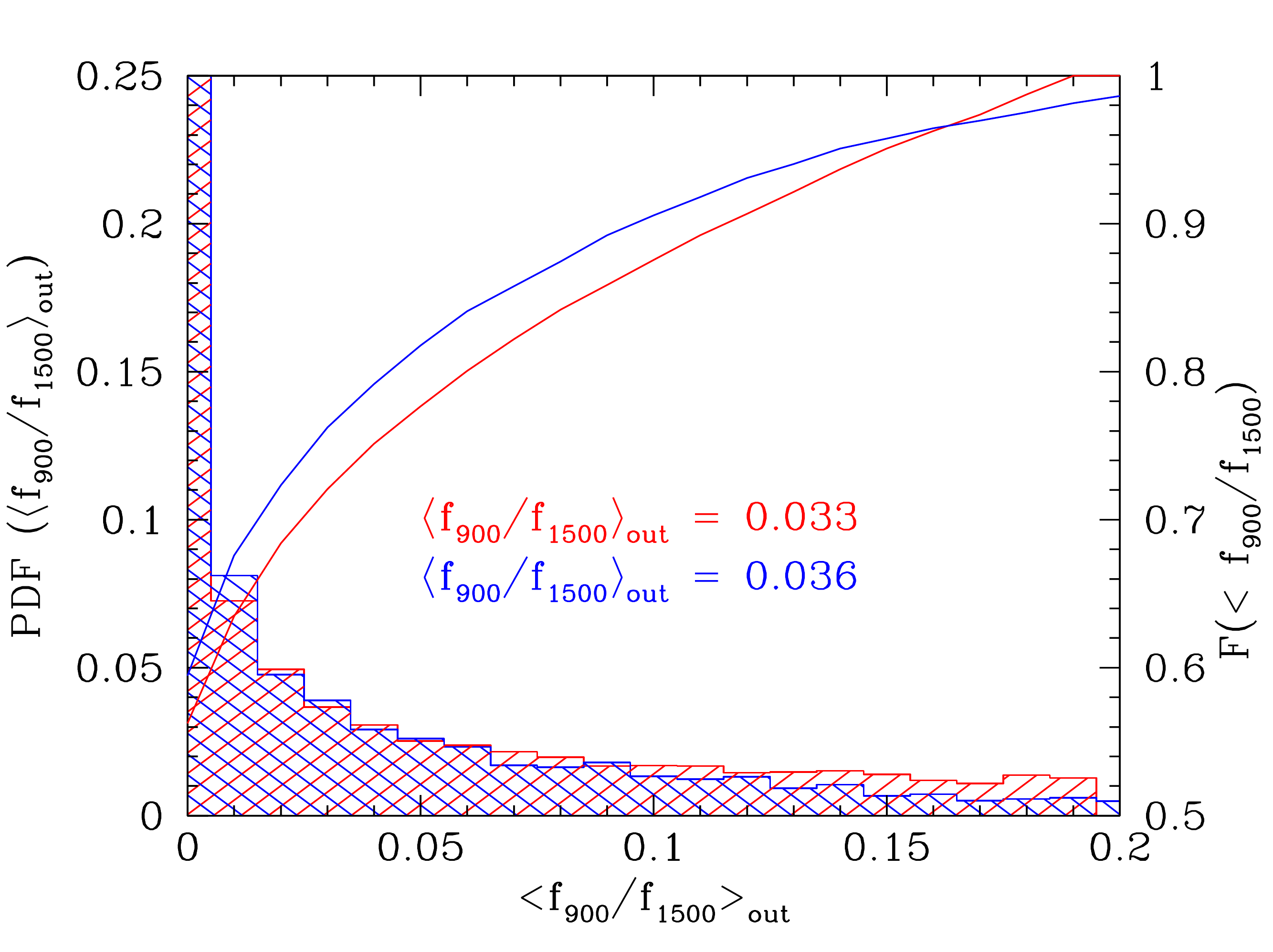}}
\caption{({\it Top:}) Estimated \fout\ as a function of \wlya. The red and blue solid curves correspond to
the distribution functions of the same color coding in the legend. The dark green curve 
shows the relative incidence of \wlya\ for a continuum-selected spectroscopic sample, which weights the incidence
of the values of \fout\ expected in an ensemble. The dashed curves show the expected distribution of \wlya\ weighted
by the relative contribution to the total LyC emissivity, i.e., proportional to $n[W_{\lambda}] \times \fout$.  
({\it Bottom:}) The probability distribution function for \fout\ for an ensemble of LBGs, accounting for the
distribution functions in the top panel for $\langle W_{\lambda}(\lya) \rangle > 0$ and assuming that 40\% of the ensemble has
spectroscopic $\langle W_{\lambda}(\lya) \rangle < 0$ (i.e., \lya\ in net absorption.) The solid curves refer to the righthand
axis, showing the cumulative fraction with lower \fout. The ensemble averages for the two 
distributions are indicated with the same color coding.
}
\label{fig:lyc_num}
\end{figure} 

Two models mapping \fout\ to \wlya\ are illustrated in the top panel of Figure~\ref{fig:lyc_num}; one is the best
linear relation from equation~\ref{eqn:fout_vs_wlya}, reflected about $W'  = 0.5$, the assumed maximum. The second
is based on equation~\ref{eqn:fout_vs_w}.
Also shown in the top panel of Figure~\ref{fig:lyc_num}  
is the relative incidence
of \wlya\ parametrized as an exponential of the form $n(W_{\lambda}) \propto {\rm exp}(-W_{\lambda}/W_0)$, with $W_0 =23.5$ \AA, very close to the
{\it observed} distribution function in the KLCS sample, as well as that obtained from previous spectroscopic samples at $z \sim 3$ (\citealt{shapley03,kornei10}.) 
\citet{reddy08a} showed
that the observed distribution of \wlya\ at $z \sim 3$ must be close to the intrinsic distribution for the galaxies contributing to the 
FUV continuum luminosity function.  
We further assume that $\simeq 40$\% of continuum-selected galaxies have spectroscopic $\wlya \le 0$ (and therefore have \fout$\simeq 0$). 
It is then straightforward to compute the probability distribution of \fout\ expected for an ensemble of continuum-selected
galaxies with $M_{\rm uv} < -19.5$ ($L_{\rm uv} > 0.25 L_{\rm uv}^{\ast}$) as in the KLCS sample. The resulting PDFs
of the quantity \fout\ for an ensemble of galaxies with the assumed $n[W_{\lambda}(\lya)]$ are shown in the bottom panel of Figure~\ref{fig:lyc_num}.  
Note that just over 50\% of the PDF falls in 
lowest bin of \fout\, but the ensemble average is \fout$\simeq 0.035$.  

Returning to the estimate of the contribution to $\epsilon_{\rm LyC}$,
the \wlya-based $\fout = 0.035$ implies that galaxies with $M_{\rm uv} < -19.5$ contribute 
$\epsilon_{\rm LyC} \simeq 0.035\times \times 1.71\times10^{26} \simeq 5.9\times10^{24}$ ergs s$^{-1}$ Hz$^{-1}$ Mpc$^{-3}$. 
Because of the strong empirical connection between \fout\ and \wlya, and because previous work has shown that spectroscopic
samples are unbiased with respect to \wlya, we are more confident of the \wlya-based estimate of $\epsilon_{\rm LyC}$ 
than of the luminosity-weighted estimate in \S\ref{sec:emissivity_lum}. 

For comparison, recent indirect estimates of the total ionizing emissivity at $z \sim 3$ inferred from modeling the transfer of ionizing radiation
through the IGM coupled with measurements of the mean transmissivity of the Lyman $\alpha$ forest 
\citep{becker13} find $\epsilon_{\rm LyC} \simeq 10\times10^{24}$ ergs s$^{-1}$ Hz$^{-1}$ Mpc$^{-3}$ with a systematic uncertainty of $\sim 0.4$ 
dex (factor of 2.5). \footnote{The systematic uncertainties are caused primarily by degeneracies among the 
spectral index of the ionizing sources and the  
effective mean free path of ionizing photons, both of which affect the conversion between a measurement of $\epsilon_{\rm LyC}$ and the corresponding 
photoionization rate $\Gamma$; see \citet{becker13} for detailed discussion.}.  
Thus, without extrapolating our results to lower luminosities that are not directly constrained by the KLCS sample, the ionizing emissivity 
$\epsilon_{\rm LyC}[L_{\rm uv} > 0.25L_{\rm uv}^{\ast}] \simeq 5.9\times10^{24}$  
ergs s$^{-1}$ Hz$^{-1}$ Mpc$^{-3}$ comprises $\sim 60$\% of the estimated total. In addition, it 
exceeds the total ionizing emissivity of 
QSOs at $z \sim 3$, estimated to lie in the range $\simeq 1.6-5 \times 10^{24}$ ergs s$^{-1}$ Hz$^{-1}$ Mpc$^{-3}$ (\citealt{hopkins07,cowie09}.)

A more detailed discussion of the implications of our result for the ionizing emissivity $\epsilon_{\rm LyC}$ and photoionization rate $\Gamma$ 
at $z \sim 3$, as well as a comprehensive comparison to other observational constraints, is  
deferred to a separate paper. 

\section{Summary and Discussion}
\label{sec:discussion}

We have presented the results of a deep spectroscopic survey of $\langle z \rangle = 3.05\pm0.18$ Lyman-break-selected sources 
obtained using LRIS-B+R on the Keck 1 telescope.
The spectra cover the observed wavelength range $3200 - 7200$ \AA\, with spectral resolving power of $R=800-1400$, 
enabling direct measurement of the rest-frame LyC region
$880 \le \lambda_0/\textrm{\AA} \le 910$ for sources with redshifts $z > 2.75$, as well as the far-UV spectrum over the range
$930 \simlt \lambda_0/\textrm{\AA} \simlt 1800$.  Very careful attention was paid to minimizing systematic
errors associated with flat-fielding, background subtraction, and spectral extraction. In addition, the initial sample of 136 spectra 
was carefully examined to remove those contaminated by foreground (lower-redshift) galaxies or affected by other potential 
systematic errors. The final KLCS sample of 124 galaxies spans a UV luminosity
range $-19.5 \simgt M_{\rm uv} \simgt -22.0$ ($0.25 \simlt (L_{\rm uv}/L^{\ast}_{\rm uv}) \simlt 3$), and is representative of the larger spectroscopic parent samples from
which it was drawn. 

Of the 124 individual galaxies
in the final KLCS sample, 15 (12\%) have formal spectroscopic detections ($f_{900} \ge 3\sigma_{900}$) of
residual LyC flux;
the remainder have typical 2$\sigma$ upper limits $f_{900} \simlt 0.014~\mu$Jy ($m_{\rm AB} \simgt 28.5$) 
averaged over the narrow rest-frame bandpass $880 \le \lambda_0 \le 910$ \AA.  
%A composite rest-frame spectrum of the full KLCS sample has \fobs$=0.021+/-0.002$, which corresponds to 
%to an average residual LyC flux density of $m_{\rm AB}[880,910] \simeq 28.8 \simeq 100$ nJy. 

Measurements of the emergent LyC emissivity of galaxies are limited both by the dynamic range of individual spectra 
and -- more importantly -- by large and 
stochastic variations in IGM transmission for individual lines of sight averaged over the LyC measurement band [880,910] \AA. 
Because of the importance of an accurate statistical model of the opacity of gas lying outside of the galaxy ISM, but
between the source and the observer, to measurement of the emergent ratio of ionizing to non-ionizing 
flux density ratio \fout, we used the statistics of \ion{H}{1} measured in both the general IGM
and in the CGM of galaxies from the Keck Baryonic Structure Survey (KBSS; \citealt{rudie13}) to produce new Monte Carlo
simulations of the IGM transmission as a function of source redshift.  These simulations were used to model the transmission
for individual sightlines at a given source redshift, and to predict the average transmission for 
ensembles of sources with the same redshift distribution as the real KLCS sample.

We focused our analysis on composite spectra formed from subsets sharing common attributes that 
are easily measured from individual spectra. The ensemble spectra have much higher S/N than individual measurements, thus
substantially increasing the dynamic range for LyC detection; in addition, the sub-sample sizes have been chosen so that
the marginalized IGM+CGM transmission expected for the ensemble is known accurately. The mean transmission appropriate to
each ensemble was used to correct the observed flux density ratio 
$\langle f_{900}/f_{1500}\rangle_{\rm obs}$ to the more relevant emergent flux density ratio \fout\ without adding significant
additional uncertainty. 

Our main results may be summarized as follows:   \\

$\bullet$ Averaged over the full KLCS sample, we find $\langle f_{900}/f_{1500}\rangle_{\rm out} = 0.057\pm0.006$. 
This value is independent of the nature of the ionizing sources within the galaxies, and is model-independent
except for small residual uncertainties in the IGM+CGM transmission; the same value, within statistical
uncertainties, is found for galaxies in the lowest and highest quartile in redshift.  
However, we find two other galaxy characteristics that correlate strongly with
the propensity to leak measurable LyC radiation: rest-frame UV luminosity ($L_{\rm uv}$) and rest-frame
Lyman-$\alpha$ equivalent width [$W_{\lambda}(\lya)$]. In particular, galaxies belonging to the lowest
quartile ($\langle L_{\rm uv}/L^{\ast}_{\rm uv} \rangle \simeq 0.5$) have $\langle f_{900}/f_{1500}\rangle_{\rm out} = 0.138\pm0.024$,
while the highest quartile ($\langle  L_{\rm uv}/L^{\ast}_{\rm uv} \rangle \simeq 1.7$) have $\langle f_{900}/f_{1500}\rangle_{\rm out} = 0.000
\pm 0.003$). Similarly, galaxies in the highest quartile of \lya\ equivalent width ($\langle W_{\lambda}(\lya) \rangle
=42$ \AA) have $\langle f_{900}/f_{1500}\rangle_{\rm out} = 0.166\pm0.025$, while the lowest quartile ($\langle W_{\lambda}(\lya) \rangle
= -18$ \AA) have $\langle f_{900}/f_{1500}\rangle_{\rm out} = 0.013\pm0.011$. \\

$\bullet$ We show that trends in LyC emission are strongly linked to the apparent depth, relative to the continuum, of
low-ionization metal lines and, especially, the depth of Lyman series absorption lines (other than \lya), which always indicate
a higher covering fraction than the metal lines (\S\ref{sec:morphology}.) In most cases,
there is evidence for strong saturation among these ISM absorption features, suggesting that the apparent depth
of the features is directly related to the continuum covering fraction of neutral gas with appreciable \nhi. The \lya\
emission line equivalent width $W_{\lambda}(\lya)$, long known to be inversely correlated (and probably causally connected) 
to the low-ion absorption line strength (e.g., \citealt{shapley03,
steidel2010,steidel11}) is a manifestation of the effect of the same gas on the emergent EUV/FUV spectrum.  \\

$\bullet$ We use the results above to explore more model-dependent characterizations of the LyC emission from 
the galaxies, including the relative and absolute escape fraction of ionizing photons from the parent galaxies, 
$f_{\rm esc, rel}$, $f_{\rm esc, abs}$ respectively.  We show (\S\ref{sec:sps_models}) that the non-ionizing (i.e., 920-1750 \AA) 
composite FUV spectra can be very well matched by SPS models
with low stellar abundance ([Fe/H]$ \sim -1.15$; $Z_{\ast}/Z_{\odot} \simeq 0.07$) which include binary evolution of massive stars reddened by
plausible far-UV attenuation relations. These models have intrinsic $L_{900}/L_{1500} \simeq 0.28\pm0.03$ over
the plausible range of star formation age $7.5 \simlt {\rm log(t/yr)} \simlt 8.7$, corresponding to $\xi_{\rm ion} \simeq 25.5\pm0.1$. \\

$\bullet$ We show (\S\ref{sec:screen} and \S\ref{sec:holes}) that the same SPS and attenuation models can be used with simple geometrical models of the
ISM to simultaneously account for the depth of the Lyman series absorption lines, the residual LyC emission, 
the damping wings of the \lya\ absorption profile (which constrains \nhi\ in the dominant absorbing gas) 
as well as the stellar FUV spectrum.  The best-fit models provide constraints on \nhi, the continuum
covering fraction $f_{\rm c}$, and the intrinsic and escaping fraction of both ionizing and non-ionizing stellar
photons.\\ 

$\bullet$ We examine in more detail (\S\ref{sec:trends}) the tight relationship between $\wlya$ and  
the inferred uncovered portion $(1-f_{\rm c})$ measured from model fits to the composite spectra. 
For our favored model geometry, in which both LyC and a fraction of \lya\ photons escape along the line
of sight through the same ``holes'' in the ISM, 
the average LyC escape fraction for KLCS galaxies at  
$z \sim 3$ is $\langle f_{\rm esc,abs}\rangle = 0.09\pm0.01$. Within the sample, 
$\langle f_{\rm esc,abs} \rangle \simeq \langle 1-f_{\rm c}\rangle$ varies linearly
with the observed $\langle \wlya \rangle$ (\S\ref{sec:trends}), over the range $0 \simlt \langle f_{\rm esc,abs} \rangle \simlt 0.29$
and $0 \le \wlya/{\textrm \AA} \simlt 50$.  We argue that the {\it kinematics} of \lya\ emission are not uniquely connected with the
propensity to leak LyC radiation into the IGM; rather, it is the presence of holes (parametrized by $\langle 1-f_{\rm c} \rangle$) 
in the scattering/absorbing  medium along our line
of sight which apparently controls both $\wlya$ and \fout.  \\

$\bullet$ Finally, we use the observed interrelation of $L_{\rm uv}$, $f_{\rm c}$, \wlya, and \fout\ to estimate
the contribution of galaxies with $M_{\rm uv} < -19.5$ to the total ionizing emissivity $\epsilon_{\rm LyC}$ at $z\sim 3$ (\S\ref{sec:emissivity}). Because
the correlation between the spectroscopically-measured \wlya\ and inferred \fabs\ is both strong and linear, our most confident
estimate comes from integrating the non-ionizing UVLF weighted by the product $n[\wlya]\times \fout$, where
$n[\wlya]$ is the relative incidence of \wlya\ and \fout\ is the empirically-measured ionizing to non-ionizing
continuum ratio evaluated as a function of \wlya. We find a mean effective \fout$\sim 0.035$ for galaxies with $-19.5 \ge M_{\rm uv} \ge -22.1$, 
accounting for $\simgt 50$\% of the total $\epsilon_{\rm LyC}$ estimated using indirect methods, and
exceeding by a factor of $\sim 1.2-3.7$ the total ionizing emissivity of QSOs at $z \sim 3$. \\

Throughout this paper, we have deliberately concentrated on the least model-dependent, most easily-measurable parameters -- 
UV luminosity and $W_{\lambda}(\lya)$ -- in seeking significant trends between galaxy properties
and the propensity to leak significant LyC flux. There has recently been a great deal of interest in identifying 
observational signatures that strongly correlate with LyC escape but are more amenable to measurement at very high redshift
where direct measurements of LyC flux become difficult
(e.g., \citealt{steidel11,heckman11,zackrisson13,nakajima14,erb14,trainor15,trainor16,verhamme17}). Among the most promising
indirect indicators discussed are anomalous nebular line ratios, possibly due to density-bounded \ion{H}{2} regions, where
nebular emission from lower-ionization species (such as [OII] or [SII]) may be suppressed. One proposed manifestation is
a large value of the ratio ${\rm O32 \equiv I([OIII] 4960+5008)/I([OII] 3727+3729)}$, which has the advantage of being
easily-observable by JWST in the redshift range $z \simeq 6-9$.  The subset of the KLCS sample with 
$z > 2.92$, comprising 94 of 124 galaxies in the sample (76\%), allows measurement of [\ion{O}{3}] $\lambda\lambda 4960$, 5008 
and [\ion{O}{2}] $\lambda\lambda 3727$,3729 in the K and H atmospheric windows from the ground\footnote{Redshifts $z > 3.0$ (55\% of
the KLCS sample) also permit 
measurement of H$\beta$.}.
We are currently completing Keck/MOSFIRE observations of the relevant KLCS sub-sample 
to investigate correlations between these excitation/ionization-sensitive line ratios and direct LyC observations. The results
will be presented in future work. 

Deep ground-based near-IR photometry, Spitzer/IRAC coverage, and selected coverage with deep HST WFC3-IR and ACS imaging
are also well underway - these observations, when completed, will provide stellar population parameters for all of KLCS. The
HST observations will serve as a final verification that the sample has been cleaned of all contaminating sources. 
Additionally, 
in view of the apparently-strong anti-correlation with UV luminosity and UV reddening (\S\ref{sec:composites}; \citealt{reddy16b})
and positive correlation with $W_{\lambda}(\lya)$
of the LyC escape fraction,  we are in the process of extending KLCS to fainter UV luminosities, bluer UV colors, and
stronger $W_{\lambda}(\lya)$ using Keck/LRIS-B+R. 

Continued development of state-of-the-art stellar population synthesis models (e.g., \citealt{eldridge17,choi17}), particularly those
focusing on the rest-frame UV, will clearly be important to modeling the sources of LyC emission in a fully-consistent treatment 
including the stars, gas, and dust in forming galaxies. ``Rules of thumb'' established at $z \simeq 3$ will be crucial to
interpreting sources at much higher redshift where only indirect measurements are possible. 

\acknowledgements

This work has been supported in part by the US National Science Foundation through grants
AST-0606912 and AST-0908805, and AST-1313472 (MB, CCS, GCR, RFT, ALS). 
CCS acknowledges additional support
from the John D. and Catherine T. MacArthur Foundation, the David and Lucile Packard Foundation, and
the JPL/Caltech President's and Director's program.
We thank J. J. Eldridge and Elizabeth Stanway for their continued work developing the BPASS SPS models,
and for many illuminating conversations. 
We are grateful to 
the dedicated staff of the W.M. Keck Observatory who keep the instruments
and telescopes running effectively.
We wish to extend thanks to those of Hawaiian ancestry on whose sacred mountain we are privileged
to be guests.

\appendix

\section{KLCS Data Reduction Details}
\label{sec:redux_appendix}

\subsection{Scattered Light and Flat-Fielding}
\label{sec:ghosts}

As discussed in \S\ref{sec:cal_and_redux}, adequate illumination for spectroscopic flat-fields necessitates  
the use of twilight sky, observations. However, because of the rapidly increasing system sensitivity
as wavelength increases from 3100 \AA\ to 4000 \AA, it is difficult in practice (particularly given
the very short periods of evening and morning twilight on Mauna Kea) to obtain good exposures 
in the deep UV while controlling the amount of stray light reaching the LRIS-B detector. 
 
Extensive tests performed during several LRIS observing runs
established that stray light at a very low level
was adversely affecting the LRIS-B observations, particularly at the shortest wavelengths
where the sky background is faintest. We found that most of the problem could be attributed to stray light 
present in the twilight sky flat fields, which were invariably obtained just after sunset (or just before
sunrise). Evidently scattering of sky light inside the telescope dome 
allowed low-level, undispersed stray light to reach the LRIS-B detector, which would then
imprint diffuse undulations in the background level in flat-fielded science exposures. These
subtle fluctuations had a significant effect on the precision of subsequent background subtraction.  

Once the problem was identified, we adopted an improved strategy for obtaining
twilight sky
flat-fields: for each slitmask we obtained two sets of exposures, one of which
had the LRIS-B $U_{\rm n} = u'$ filter (bandpass 3500/600 \AA) inserted in the
beam.  
The $U_n$ filter cuts off 
longward of  $\sim3900$\,\AA\space and efficiently blocks the
vast majority of stray light, since the LRIS-B filters
are inserted just in front of the spectrograph camera. 
The presence of the filter made it possible to safely obtain
flat field exposures with higher signal-to-noise ratio (S/N) 
shortward of $3900$\,\AA. 
The $U_{\rm n}$ twilight spectroscopic images were then 
combined with shorter twilight sky exposures taken 
without the filter in place. Since most of the stray light problem was confined to
regions with $\lambda < 3900$ \AA, the two sets of flats were ``spliced'' together 
to create composite
twilight sky flats with high S/N throughout
the $3200$\,\AA$-5000$\,\AA\space wavelength range, with
greatly reduced contamination by scattered light. 

\subsection{Image Rectification}
\label{sec:rectify}

 After bias subtraction and CCD gain correction,
 the relevant ``footprint'' on the spectroscopic frames 
are identified by tracing the slit edges across the dispersion
direction, fitting the edge positions with a 3rd order polynomial 
that is then used to extract each slit and transform it into
a rectilinear image by applying nearest-integer pixel shifts in
the spatial direction only. This produces a rectified slit image
that requires no re-sampling of pixel intensities.    

In the absence of differential atmospheric refraction, the previous step
would make object traces run parallel to the CCD columns and thus simple
linear traces could be used to extract one-dimensional (1D) spectra. 
For the data obtained prior to the ADC installation in July 2007, the wavelength
dependence of objects within slits was accounted for, as described in \S\ref{sec:tracing}. 

\subsection{Slit Illumination Correction}

\label{sec:slitfunc}

Imperfect milling or cleaning of LRIS slitmasks prior to insertion in the instrument 
can cause spatially-abrupt 
changes in slit illumination along the spatial direction, since  
(if everything is set up correctly) the slit plane is where both the telescope and
the spectrograph camera are focused. 
This slit illumination function will also incorporate any intrinsic illumination gradients 
due to field-position-dependent vignetting within the telescope or instrument optics. 
Since our observations are attempting to detect
signal at the level of $\simlt 1\%$ of the night sky background, the typical
1-2\% divots in the slit function need to be corrected before the sky subtraction stage. 
Left uncorrected, small systematic deviation from zero mean might 
remain in the background subtracted images; 
in short exposures this problem can easily go unnoticed,
as other sources of random error are dominant.
However, its systematic influence becomes
clear in the average of many exposures. 

Custom procedures were developed to correct for variations in slit function 
for both science and calibration exposures, accounting for small amounts of instrument
flexure between each and the images used for illumination correction (i.e.,
shifts in the positions of the slit edges, and therefore of any slit defects, on the
detector.) 
The slit function for each slit on a mask was obtained from rectified
two-dimensional spectrograms (\S\ref{sec:rectify}) of the twilight sky by collapsing
in the dispersion direction over the wavelength range deemed unaffected by 
residual scattered light,  
and averaging with
iterative $\sigma$-rejection. 
The slit illumination correction was then applied to 
all rectified science and calibration exposures.

\subsection{Atmospheric Dispersion Correction and \\
 Object Tracing}
\label{sec:tracing}

Since we are interested both in detecting LyC flux when present, and obtaining
sensitive upper limits when it is not, it is advantageous to be confident about
the location of an object's trace whether or not flux is evident. An often-used
technique for tracing object positions in 2-D spectrograms is to determine spatial
centroids regularly along the whole dispersion axis, which of course relies on
the presence of signal
over the same range.  There is a tendency for traces extrapolated
beyond the point where there is detectable flux to ``wander'' in search of upward
fluctuations in the noise level. This is clearly undesirable in our particular application. 

 Prior to the deployment of the ADC on Keck 1,  the trace of an object spectrum
 would differ from the trace of its slit edge by an amount 
 determined by differential atmospheric refraction (DAR).
Slitmask alignment for KLCS was always performed using direct imaging mode through the mask with a
G-band filter 
in the LRIS-B beam, and guiding was maintained using an offset guide field CCD camera 
with peak sensitivity in V band.
 The exact value of the apparent spatial shift with time  
 of source positions within slits depends on wavelength, airmass, slit position
angle with respect to the elevation direction (the parallactic angle) and
 atmospheric conditions (temperature, pressure, water vapor pressure) 
 at the time of observation. 
For reasons of practicality, KLCS observations were obtained (both before
and after installation of the ADC) using a single mask and sky PA over
a range of hour angle, albeit as close to the meridian as possible. 

 As a consequence, individual 
 exposures of the same source could not be spatially
 registered at all wavelengths simultaneously. The difference in the
 apparent spatial position of the object over the 3200-5000 \AA\ range of LRIS-B
 can be as large as $\sim 1$\arcsec ($\sim7-8$ LRIS-B pixels) if the slits have been oriented 
near the parallactic angle to minimize light losses due to DAR.  

 We developed an approach that deals with this problem 
 in the earliest KLCS masks taken prior to August 2007.
 Using the atmospheric dispersion model of
 \citet{filippenko82} we predict the value of the
 apparent spatial offset of the object position as a function
of wavelength, and implement the shifts during the slit rectification
step so that the object position in the rectified image is parallel
to the slit edges, at a constant pixel location.  
 Tests performed on observations of bright sources
 demonstrated that using the average atmospheric 
 conditions for a given night on Mauna Kea provided sufficient
 precision in predicting the wavelength-dependent image shift from atmospheric dispersion.
 After August 2007, this procedure became unnecessary, as  
the ADC removes all but $\simlt 0.1$\arcs\ of differential
shift over the full LRIS-B wavelength range. 

\subsection{Background Subtraction}
\label{sec:lrisb_bkg}

Once careful slit illumination and flat-field corrections have been applied
(\S\ref{sec:slitfunc} and \S\ref{sec:ghosts}), 
 a low-order polynomial becomes an adequate description  
 of the background intensity along the spatial direction.
 The accuracy of the background estimation depends to a large
 extent on the number of spatial pixels at each dispersion point
that sample only ``background''
and not the outer parts of either the principal target or any part of
nearby objects that happen to fall on the slit. 
Our experience has been that, at the level of sensitivity reached in the KLCS
 ($m_{AB}\sim28.0$, or $\sim 2 \times 10^{-31}$ ergs s$^{-1}$ cm$^{-2}$ Hz$^{-1}$,)
 the fraction of ``clean'' pixels remaining after source masking can be in the minority. 
We found it helpful, in terms of the number of spectra ultimately useful for LyC measurements,
to enforce relatively long individual 
slits centered on high priority targets (slit lengths $> 15$ arcsec), sometimes at the expense of 
the number of targets that could
be accommodated on a single slitmask. 

We also deliberately lowered the priority for slit assignment
of objects lying in obviously crowded regions on the deep images. 
In spite of this pre-screening of potential targets,   
the majority of the slits included
serendipitous sources detected in the vicinity of the primary target.
In fact, after combining $\ge8$ hours of LRIS spectroscopic observations
 the sensitivity is high enough to clearly detect the spectra of all sources 
 seen in our deepest ground-based broad-band $G$ or ${\cal R}$ images, which typically
reach $1\sigma$ limits of $G\sim29.5$ and ${\cal R}\sim28.8$ mag arcsec$^{-2}$; 
 \citep{steidel03,steidel04}.
 Using faint object positions predicted from the deep images we were in a position to 
 exclude most or all pixels along the slit that might be contributing
 something other than sky background.  

 Background subtraction was performed in two stages:
 on the first pass we used object masks created from objects with significant
detections in single 1800s LRIS-B exposures; 
in the second stage we compared the stack of the 2D background-subtracted 
spectrograms with broad band images and masked additional
 slit regions containing any source visible in either.
 An iterative fit with a polynomial function of order 2 to 4 was 
 used to determine the final background model, which was subtracted separately for
each 1800s exposure. The sky-subtracted spectrograms were then combined in 2-D, after
shifting into spatial and spectral registration, by averaging, with bad pixel and
outlier rejection. 

The effectiveness of the above background subtraction procedures is discussed in more detail in 
\S\ref{sec:2d_bkg}. 

\subsubsection{Noise Model} 
\label{sec:noise_model}

Throughout the reduction process, care was taken to preserve the original pixel sampling
and number of photoelectrons per pixel in order to track formal uncertainties through
all steps.  A stack of all science frames made just prior to the 2-D background subtraction
stage was made as a record of the number of sky+object photoelectrons recorded in
every pixel.  These images are used to estimate the variance  
each pixel $[i,j]$ in the 2d spectrogram, 
\begin{equation}
\label{eqn:noise_model}
\sigma_{\rm pix}^2[i,j] = N_{\rm exp}[i,j]\times (C[i,j] + \rho^2)~,
\end{equation}
where $N_{\rm exp}$ is the number of 1800s exposures averaged to produce the value of pixel $[i,j]$, 
$C[i,j]$ is the number of photoelectrons recorded in the pixel, and $\rho$ is the detector read noise,
assumed to be $\rho = 3.8$e$^{-}{\rm pix}^{-1}$ for LRIS-B.

\subsection{Testing for Systematic Errors in 2D Spectrograms}
\label{sec:2d_bkg}

\begin{figure*}[htbp]
   \centering
   \includegraphics[width=17cm]{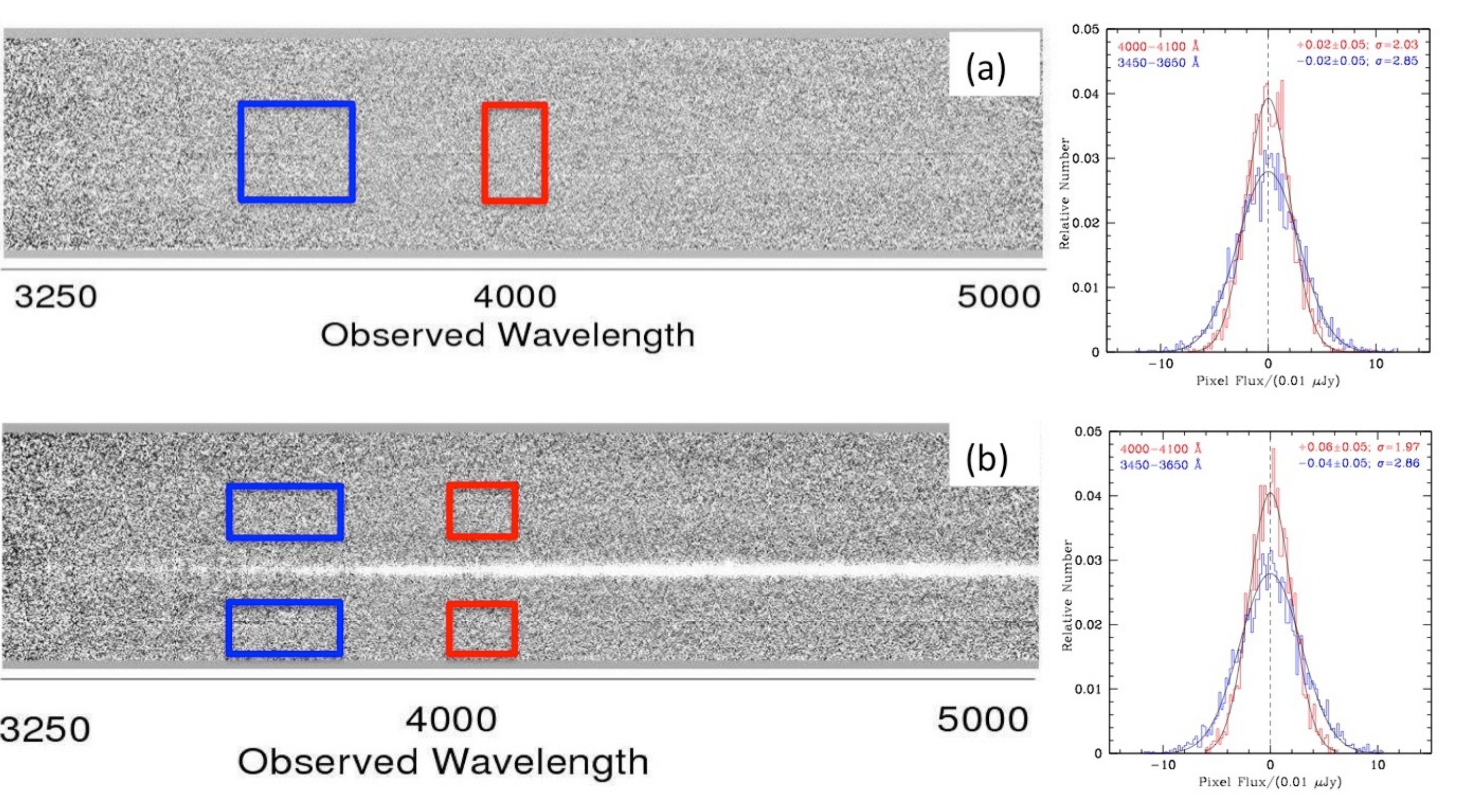}
\caption{ (a) Two-dimensional composite spectrogram, combined using the object masks created 
during the background subtraction second pass, described in \S\ref{sec:lrisb_bkg}, for the 10 slits
containing nominally detected LyC sources. Histograms of pixel fluxes, in units of 
0.01 $\mu$Jy ($10^{-31}$ ergs s$^{-1}$ cm$^{-2}$ Hz$^{-1}$), are shown to the right for the regions indicated
with colored boxes on the image: blue corresponds to the observed wavelength range 3450-3650 \AA\ and
red to the range $4000-4100$ \AA, and both windows are 80 spatial pixels ($\simeq 10.8$ arcsec) wide. The
values of the mean and standard error in the mean are shown for each histogram, along with the fitted value
of $\sigma$ assuming zero mean and a normal distribution, as expected for
regions on the 2-D images without background subtraction systematics and with no residual flux from 
unmasked objects. (b) Similar to (a), except that in this case, the primary
target galaxy was left unmasked, and the 2-D spectrograms were registered spatially to place the target
at a constant Y-pixel location. The pixel statistics within the colored boxes (shown at right) 
should be sensitive to possible systematic sky subtraction errors related to the position of the primary
target along the slit. As the histograms show [based on the same total number of pixels in each sample as in
(a)], the residuals are consistent with Gaussian noise with zero mean, and $\sigma_{\rm pix}$
values in the crucial 3450-3650 \AA\ range are identical to within 0.3\% to those in the same wavelength interval in (a).  
}
 \label{fig:2d_bkg}
 \end{figure*}

In this section, we outline tests performed to assess the 
quality of data reduced as in \S\ref{sec:redux}. 
Most importantly, we tested for the presence of residual systematic 
errors in the 2-D background subtraction by creating
an average two-dimensional spectrum for an ensemble of final 2-D background-subtracted
spectrograms for a set of KLCS targets which each led to $>3\sigma$ detections of
residual flux in the rest-frame [880-910] \AA.  
Fully processed, background-subtracted two-dimensional rectified 
 spectrograms  
 were wavelength calibrated and then put onto a common observed wavelength grid; 
 We used the same object masks that had been created in the second
pass of the sky subtraction phase for each slit (\S\ref{sec:lrisb_bkg}), so that when the spectrograms
were averaged, the masked pixels in each were excluded. 
 The effective exposure time for the resulting stack [Figure~\ref{fig:2d_bkg}] is $\sim 90$
 hours, so that the random noise (due to photon statistics and detector read noise) is expected to be 
lower by a factor of $\simeq 3$ compared to any 
single observation in KLCS, thus more sensitive to possible systematic errors. 

 The rms noise values within the boxes shown in Figure~\ref{fig:2d_bkg} are higher in the $3450-3650$ \AA\ range 
due to lower system sensitivity (since the pixel intensity values are calibrated onto a flux density scale.) 
If the background subtraction is accurate, with systematics that are small compared to random errors, 
the distribution of pixel values is expected to well-described by a normal distribution with zero mean.  
The histograms to the right of the image show that this is the case for both wavelength regions. 
In addition, the
measured rms noise level $\sigma$ is entirely consistent with the noise model described in \S\ref{sec:noise_model} (see below). 
  
As a second test for background subtraction systematics, we re-made the stacked composite spectrum using
the same slits, but 
without masking the primary target galaxy (all other object masks remain the same). In this case, the
spectrograms were spatially shifted before averaging  
so as to align the slit position of the primary target in the final image.
The resulting two-dimensional stacked composite is 
presented in Figure \ref{fig:2d_bkg}(b); in this case,
the stack would be sensitive to possible systematic errors
that might be correlated with the position of the target object (generally the brightest object on the slit.) 
Unlike the previous case, which tested our ability to mask 
all significant sources in performing 2-D background subtraction, systematic errors
in sky subtraction should become 
more prominent as more high-sensitivity spectrograms are averaged.
To the right of Figure~\ref{fig:2d_bkg} are histograms of intensities drawn from 
regions on either side of the main target, beginning $\pm$4\secpoint0 from the target
centroid, evaluated in the same rest-wavelength intervals as in Figure~\ref{fig:2d_bkg}(a) (with the same
total number of pixels, divided into two separate spatial regions). 
The results are very similar --  with zero mean and essentially identical $\sigma$ within each wavelength
region --  to those obtained in Figure~\ref{fig:2d_bkg}(a). 

\begin{figure*}[htbp!]
\centerline{\includegraphics[width=9.0cm]{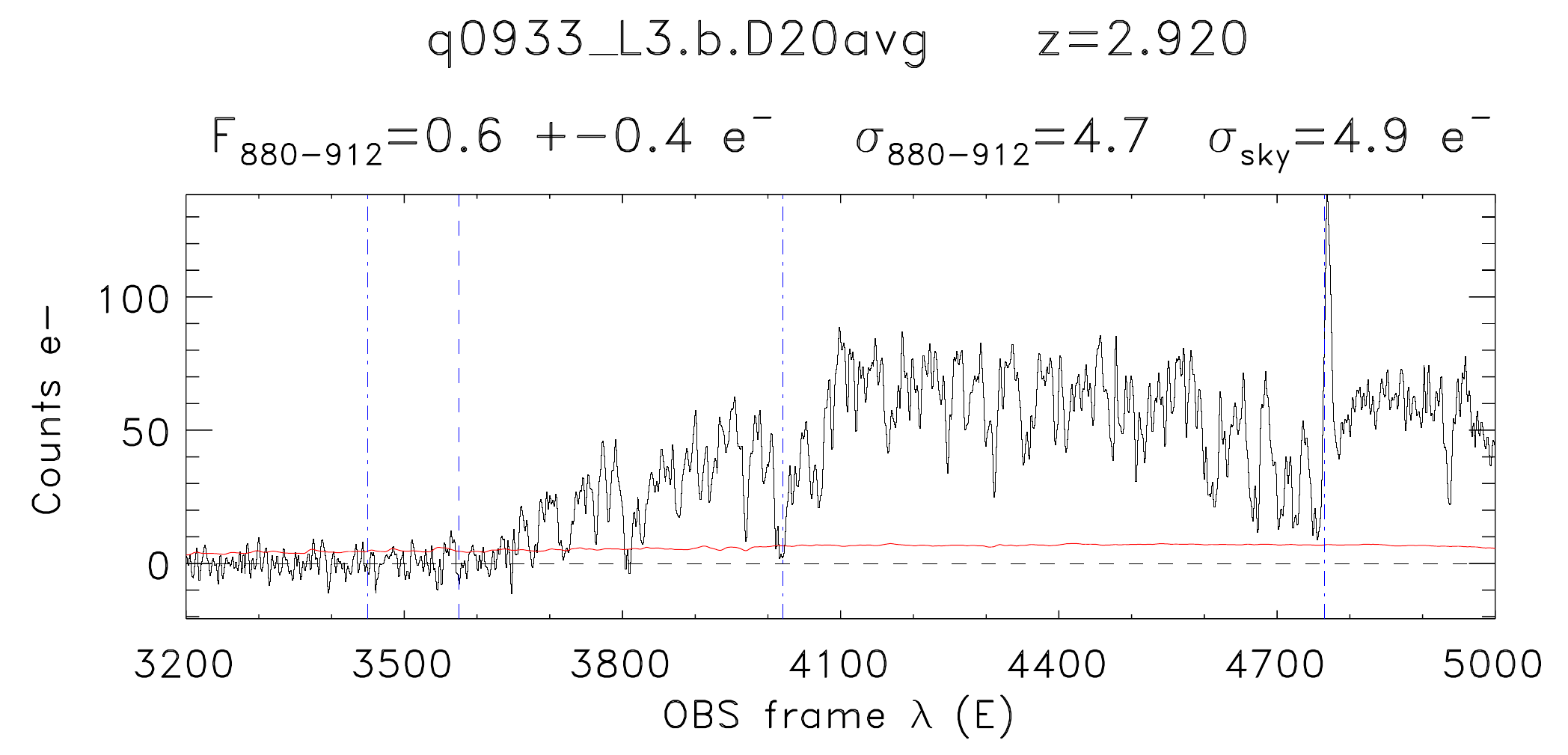}\includegraphics[width=9.0cm]{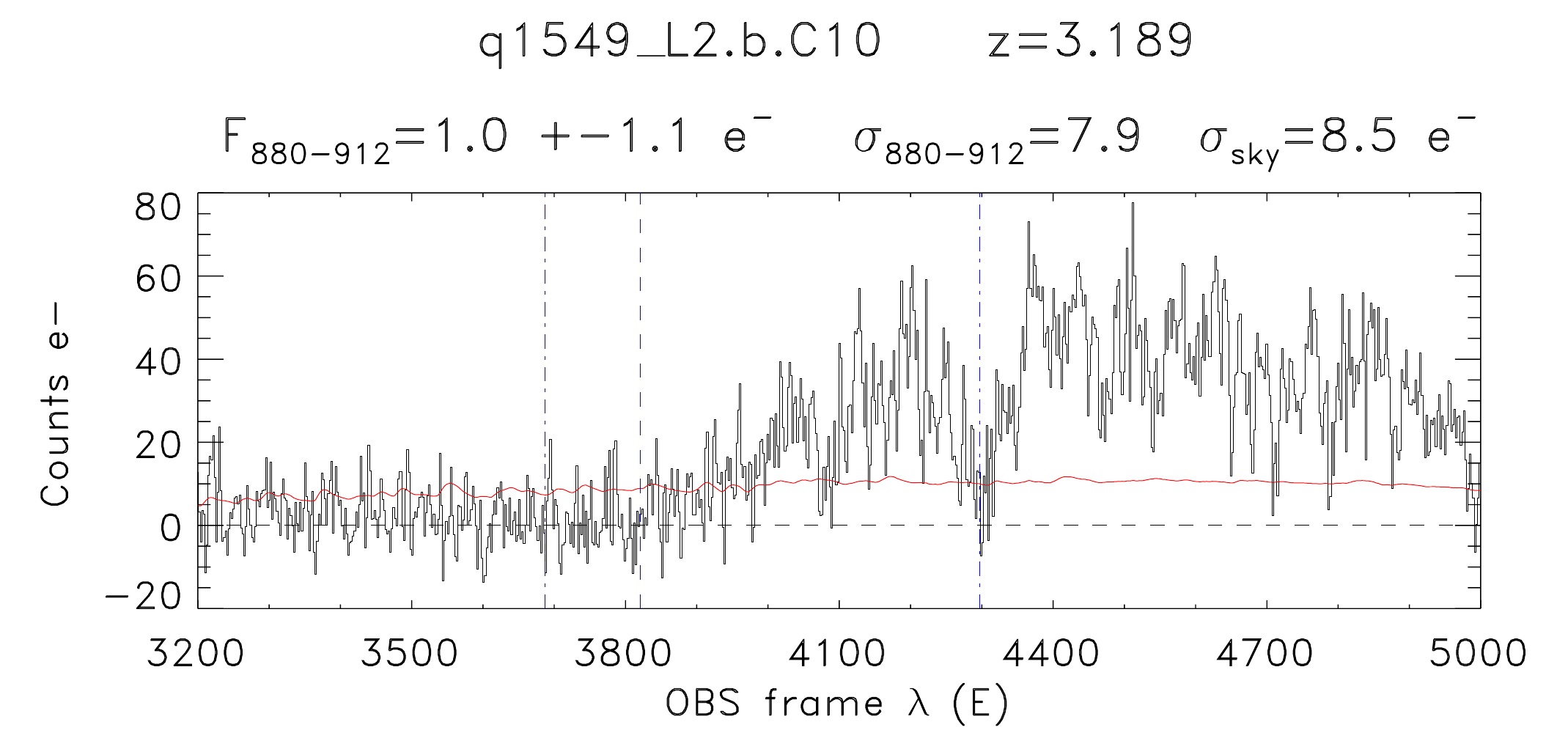}}
\centerline{\includegraphics[width=9.0cm]{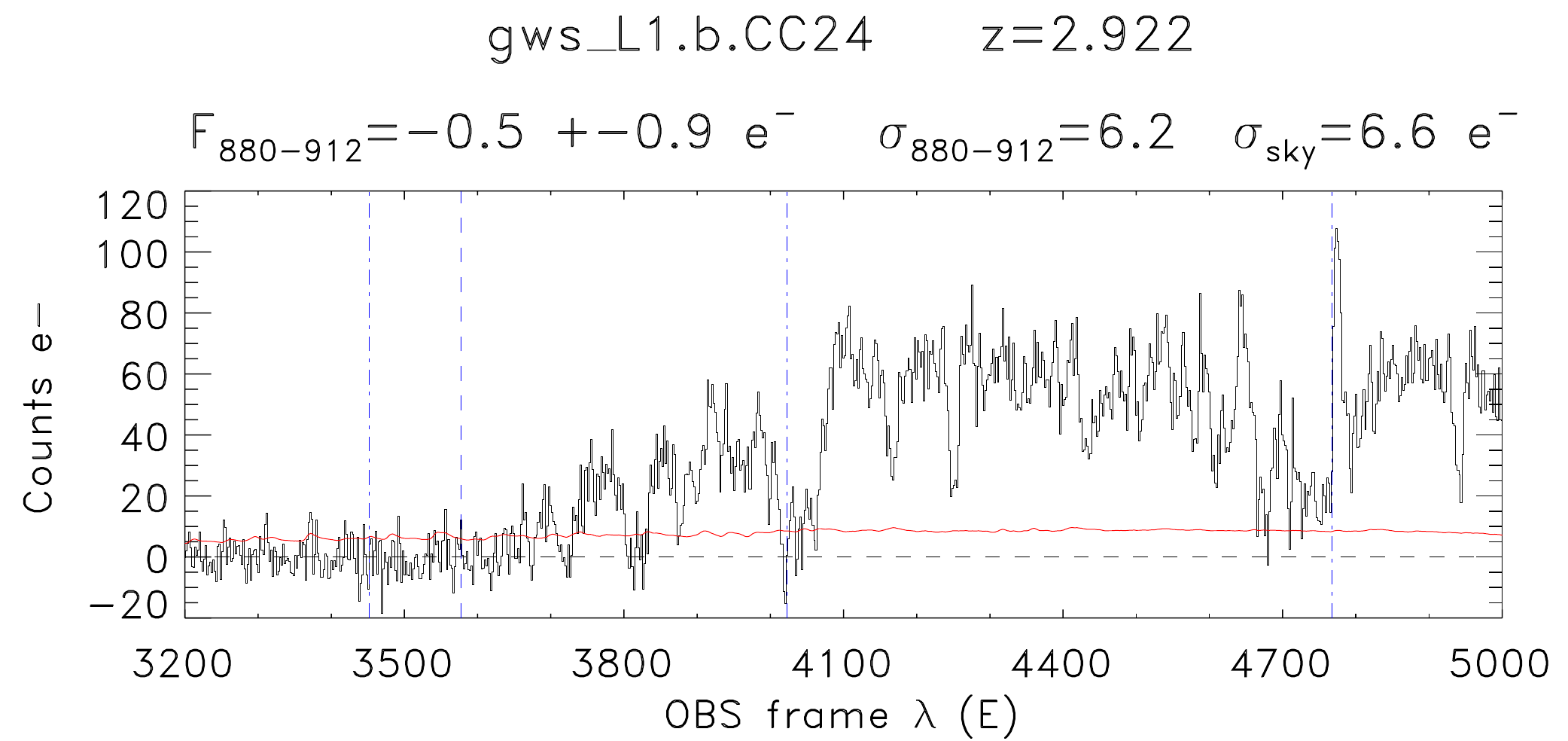}\includegraphics[width=9.0cm]{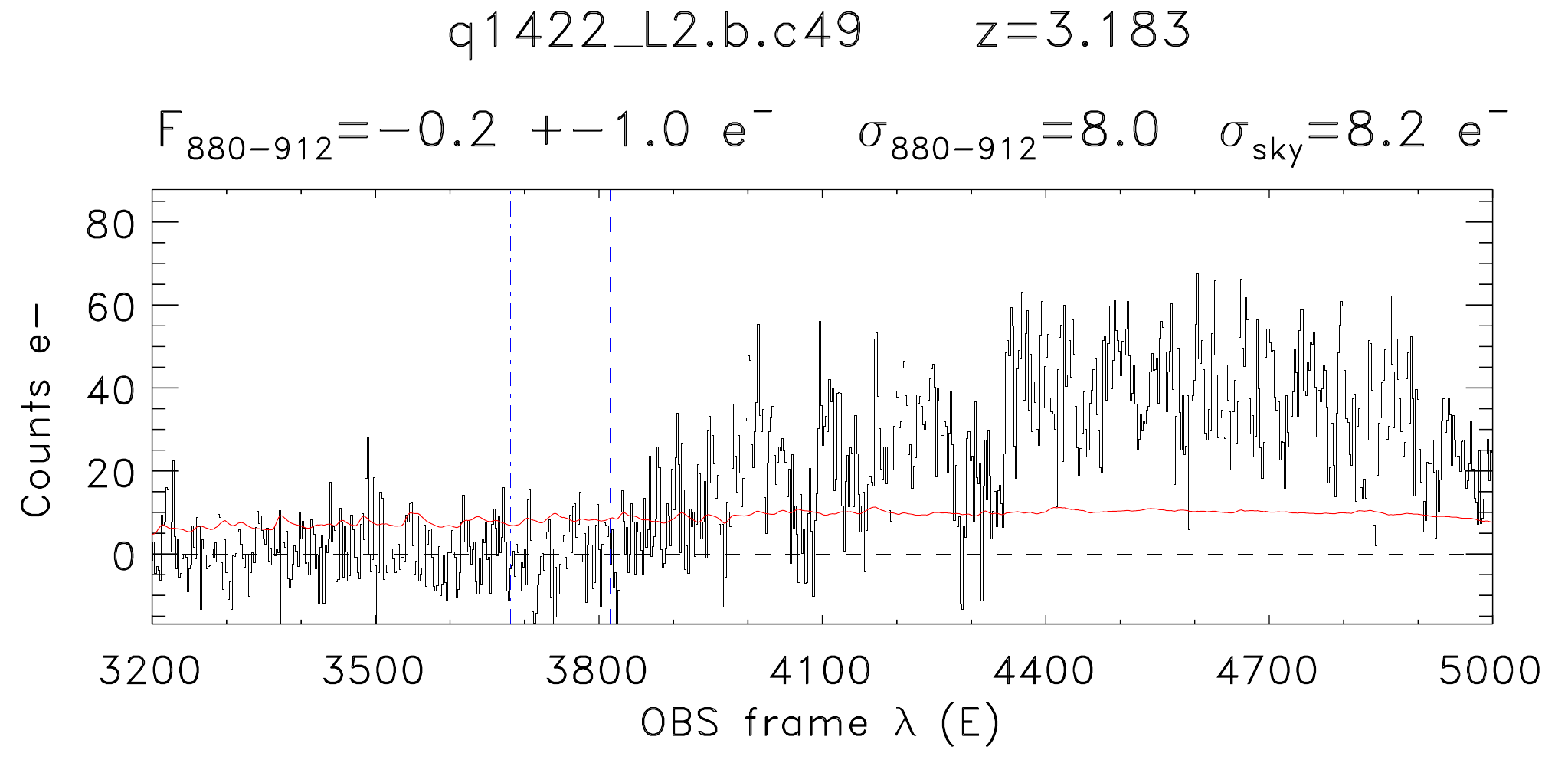}}
\centerline{\includegraphics[width=9.0cm]{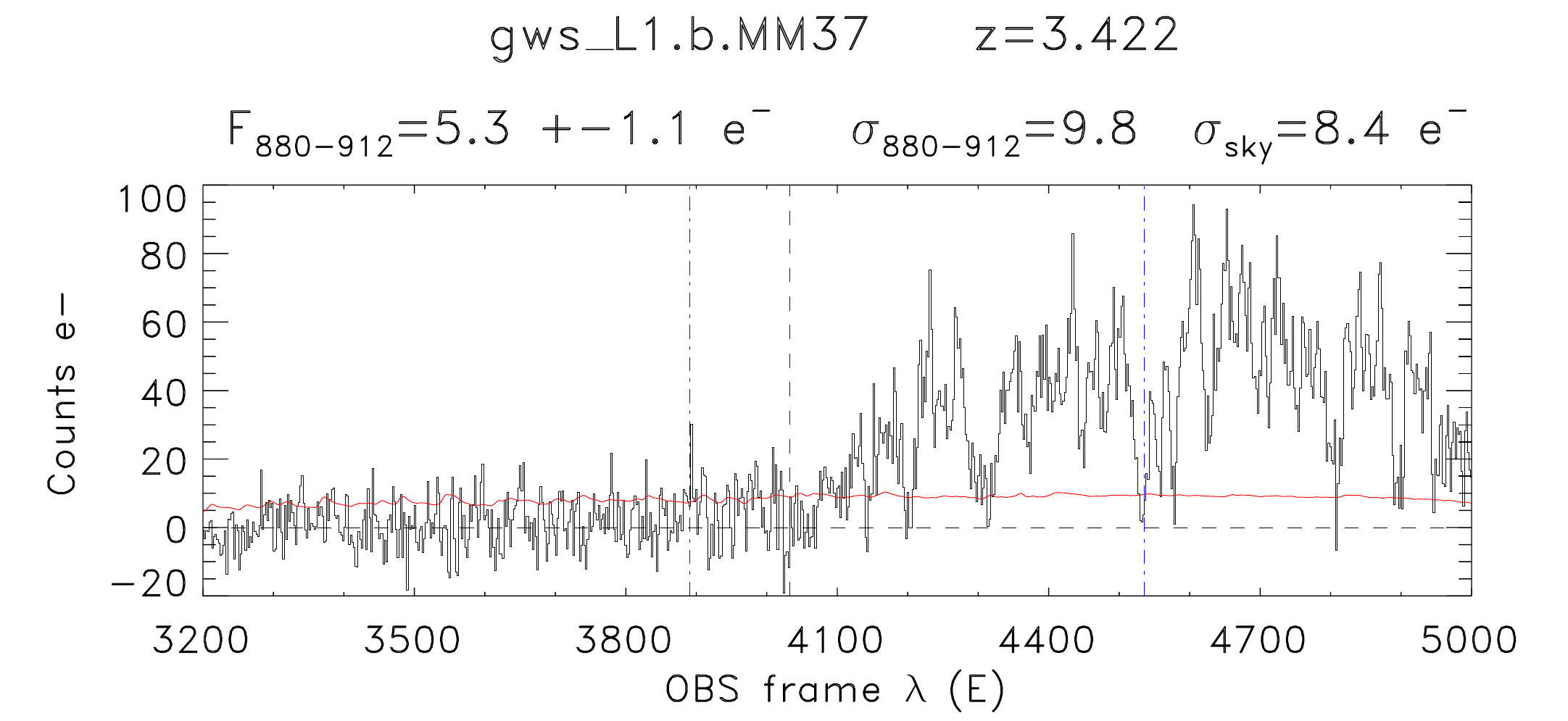}\includegraphics[width=9.0cm]{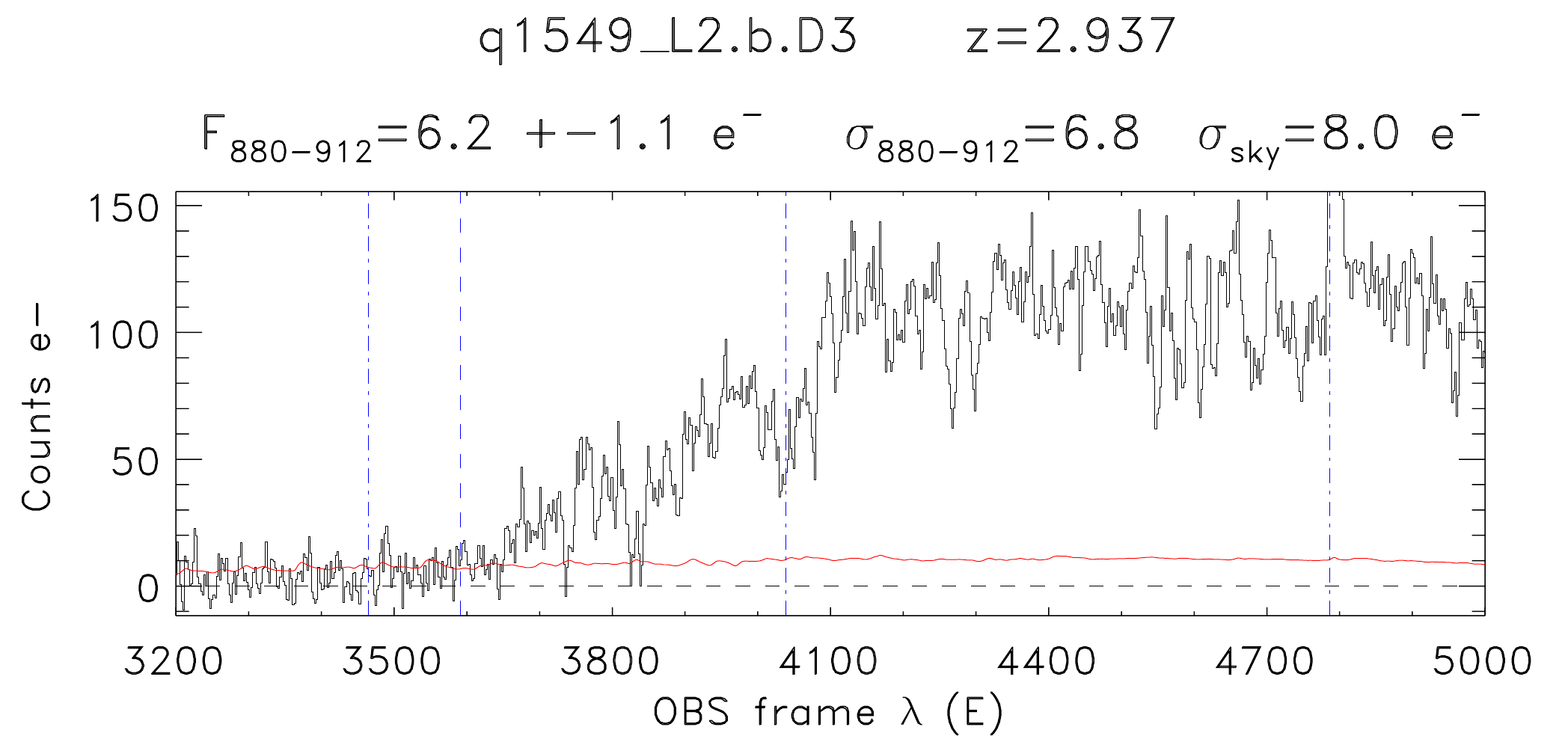}}
\caption{Example extracted LRIS-B spectra, in units of mean photoelectrons per pixel prior to re-binning and flux
calibration of the 1-D spectra. In each panel, the red curve is the 1$\sigma$ noise level per pixel expected 
from the noise model. The vertical lines in each panel mark, moving left to right, rest wavelengths 880, 912, 
1026 ($Ly\beta$), and 1216 ($Ly\alpha$).  The top left panel shows one of the two galaxies observed on
2 separate masks, after combining the independently reduced spectra in 1-D, with inverse-variance weighting. 
The annotations above each panel show the object name, redshift, mean and error in the mean of the number of photoelectrons per pixel over
the [880,912] rest wavelength interval ($F_{880-912}$), the rms per pixel over the same range ($\sigma_{880-912}$), and
the noise level per pixel expected over the same pixels based on the noise model in \S\ref{sec:noise_model}. 
   }
\label{fig:1d_counts}
\end{figure*}

In evaluating the accuracy of our noise model, we examined each extracted 1-D spectrum after wavelength 
calibration, but prior to re-sampling in the dispersion direction or any other processing of the 1-D spectra
and 1 $\sigma$ error arrays. These spectra, examples of which are shown in Figure~\ref{fig:1d_counts}, remain
in units of raw photoelectrons summed over the 10 (spatial) pixel extraction aperture of the final 2-d 
background-subtracted stack, so that the spectrum is in units of electrons per wavelength pixel (1 pix $\simeq 2.14$ \AA) 
averaged over typically 18 1800s exposures.      
Each panel in Figure~\ref{fig:1d_counts} shows the mean and standard error in the mean e$^{-}$ pix$^{-1}$ evaluated over the 
rest-frame [880,910] interval, $F_{880-912}$, the rms per pixel within that interval ($\sigma_{880-912}$), and
the expected rms per pixel given by the background pixels and noise model described in \S\ref{sec:noise_model} ($\sigma_{\rm sky}$). The top two rows
(i.e. the first 4 spectra shown) 
are formal non-detections of LyC flux, and the bottom row shows two examples of formally detected objects (Westphal/GWS-MM37 and Q1549-D3). 
These examples illustrate that the predicted and measured noise levels in the spectral regions used for LyC measurement are very
close to one another. Note that the top-left panel shows the combined spectrum of Q0933-D20 after averaging the 2 independent
1-D spectra with inverse variance weighting. 

\section{Improved IGM Opacity Models for Galaxy Sources}
\label{sec:igm_appendix}

\begin{figure}[htbp]
\centerline{\includegraphics[width=9cm]{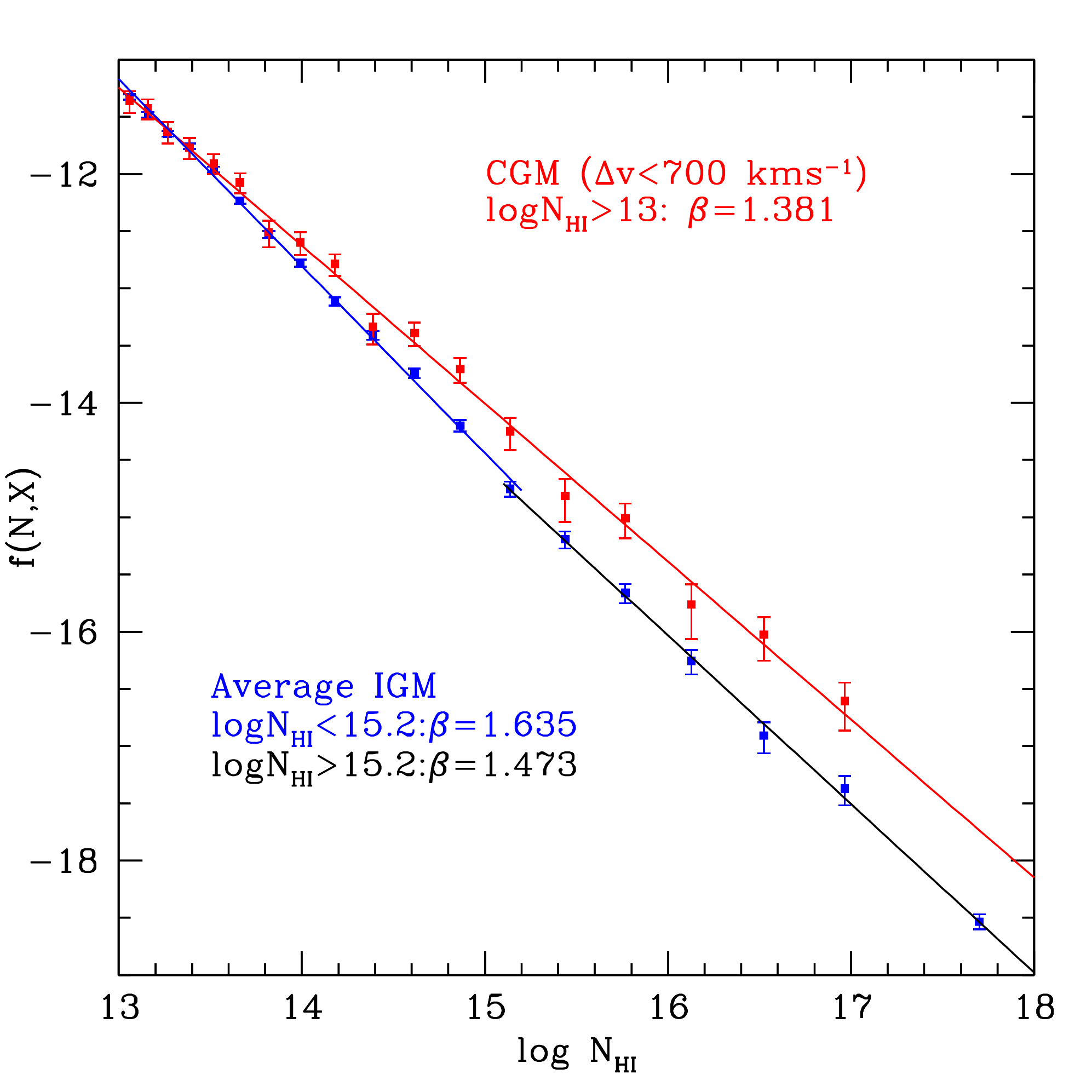}}
\caption{Distribution functions $f(\nhi,X)$ inferred from the \ion{H}{1} absorber catalog of \citet{rudie13}. 
The best-fit power law slopes $\beta$ for the ``general IGM'' (blue and black) and regions within 300 pkpc and
$\delta v < 700$ \kms\ from a galaxy in the observed galaxy sample (such regions include $\sim 7$\% of the total
effective survey volume; \citealt{rudie12a}).  The IGM has been divided into low-\nhi\ and high-\nhi\ subsamples at log~(\nhi/${\rm cm}^{-2})=15.2$ 
with separate slopes and normalizations allowed (Table~\ref{tab:fnxtab}). Note that the incidence of absorbers with log(\nhi/${\rm cm}^{-2})>16$ is $\sim 6$ times
higher in the CGM of galaxies than in the general IGM (see text for discussion.)   
\label{fig:fnx_plot}
}
\end{figure}

For the purposes of generating Monte Carlo models of the IGM opacity, we adopted a combination of 
the measured \nhi\ distribution functions $f(\nhi,X)_{\rm IGM}$ and 
$f(\nhi,X)_{\rm CGM}$ as shown in Figure~\ref{fig:fnx_plot}, both as presented by \citet{rudie13} based
on data from the Keck Baryonic Structure Survey (KBSS). 
Here $f(\nhi,X)$ is the distribution function, first introduced by \citet{carswell84}, describing the incidence of \ion{H}{1} absorption
systems per unit co-moving path length $X$, as defined by \citet{bahcall69}.  
The parameter $X$ is cosmology-dependent, but for spatially flat CDM models it is
given by  
\begin{equation}
dX = \frac{H_0}{H(z)} (1+z)^2 dz
\end{equation} 
where
\begin{equation}
H(z) = H_0\left[\Omega_{\Lambda} + \Omega_{\rm m}(1+z)^3\right]^{1/2} 
\end{equation} 
is the Hubble parameter. 

The CGM statistics (see \citealt{rudie13}) are derived based on measurements of $f(\nhi,X)$ 
for sightlines within 300~pkpc and $\pm 700$ \kms\ [$\Delta z \le 0.0023(1+z_{\rm s})$] of spectroscopically-identified
KBSS galaxies in the redshift range $2.0 \simlt z \simlt 2.8$. 
Aside from having somewhat lower redshift, the galaxies in KBSS were selected using 
very similar criteria to those of KLCS and occupy a nearly-identical range of stellar mass 
and star formation rate (see \citealt{adelberger04,steidel04}.)
Figure~\ref{fig:fnx_plot} shows that within the CGM so-defined, the incidence of \ion{H}{1} absorbers is $\sim 6$ times higher
than the IGM average at the high-\nhi\ end of the $f(\nhi,X)$ distribution; for log~\nhi$\simlt 14$ the CGM regions
asymptotically approach the statistics of the general IGM. 

In general, $f(\nhi,X)$ is redshift-independent only if the physical properties of the absorbers
do not evolve with redshift. 
Since the measurements in \citet{rudie13} are based on QSO sightlines with a path-length averaged redshift
$\langle z \rangle = 2.37$, we applied a redshift correction to the measured $f(\nhi,X)$ by expressing
the number of absorbers of column density between $\nhimin$ and $\nhimax$ as 
\begin{equation}
N_{\rm abs} = \int_{\nhimin}^{\nhimax} \int_{z1}^{z2} \nhi^{-\beta}  A(1+z)^\gamma d\nhi dz
\label{eqn:nabs}
\end{equation}
where $N_{\rm abs}$ is the number of absorbers with $\nhimin < \nhi < \nhimax$ expected over the 
redshift path interval $[z1,z2]$, $A$ is a constant chosen to match the 
observations at the effective redshift of the measured sample, $\beta$ is the slope of $f(\nhi,X)$, and
$\gamma$ is a power law exponent describing the redshift evolution of the number of absorbers per unit redshift, 
$dN_{\rm abs}/dz \propto (1+z)^{\gamma}$.    

For reasons discussed in detail by \citet{rudie13}, we
divided the full range of \nhi\ into two regimes: log~\nhi~$<15.2$ and log~\nhi~$>15.2$,  with separate $\beta$, 
$\gamma$, and normalization $A$, as indicated
in Figure~\ref{fig:fnx_plot}. The values adopted for the MC simulations are summarized in Table~\ref{tab:fnxtab}\footnote{The data points shown in Figure~\ref{fig:fnx_plot}
are identical to those presented by \cite{rudie13}; the fitted parameters $A$ and $\beta$ are slightly different due to the choice of break point  
for the two power laws describing the general IGM distribution function; in addition, for simplicity, we adopted a single power law
to describe $f(\nhi,X)$ for the CGM component. The 
resulting differences in the distribution of $t_{900}$ are small compared to the uncertainties.}. 
 
\begin{deluxetable}{lcccc}
\tabletypesize{\scriptsize}
\tablewidth{0pt}
\tablecaption{Parameters for IGM/CGM Monte Carlo Simulations}
\tablehead{
\colhead{}  & \colhead{log(\nhi/${\rm cm}^{-2}$)} & \colhead{$\beta$\tablenotemark{a}}  & \colhead{log($A$)\tablenotemark{b}} & \colhead{$\gamma$\tablenotemark{c}}} 
\startdata
IGM Low   & $12.0 - 15.2$ & $1.635$  & 9.305 & 2.50 \\
IGM High  & $15.2 - 21.0$ & $1.463$  & 7.542 & 1.00 \\
\hline \\
CGM\tablenotemark{d}   & $13.0 - 21.0$ & $1.381$ &  6.716 & 1.00 
\enddata
\tablenotetext{a}{Slope of $f(\nhi,X)$ adopted over the specified range of \nhi\ (Figure~\ref{fig:fnx_plot}.) }
\tablenotetext{b}{Normalization of the fiducial $f(\nhi,X)$ power law fits.}
\tablenotetext{c}{Power law exponent for assumed rescaling of $f(\nhi,X)$ as a function of redshift (equation~\ref{eqn:nabs}).}
\tablenotetext{d}{Used only for IGM+CGM models, for redshifts within $\left[ \Delta z= 0.0023(1+z_{\rm s})\right]$ of the source redshift.}
\label{tab:fnxtab}
\end{deluxetable}

In the present case, our main interest is a quantitative estimate of the reduction in flux in a
small wavelength range [880,910] \AA\ in the rest-frame of a galaxy source.  
The effective flux decrement in this interval includes blanketing from Lyman series absorption lines -- dominated by systems with log~(\nhi/cm$^{-2})<15.2$) -- 
and Lyman continuum opacity, dominated by higher \nhi\ systems at redshifts close to $z_{\rm s}$. 
Most of the line blanketing component is contributed by 
\lya\ and \lyb\ falling at observed wavelengths near $900(1+z_{\rm s})$ \AA. For example, at $z_{\rm s} = 3.10$, the relevant redshifts are
$z(\lya) \simeq 1.97-2.07$ and $z({\rm Ly}\beta) \simeq 2.52-2.64$.    
This implies that the measurements of \citet{rudie13} require no extrapolation in redshift to accurately estimate the line opacity
contribution to the transmission in the wavelength interval [880,910] in the rest frame for a source at $z_{\rm s} \simeq 3$.  
 \citet{rudie13} verified that the overall incidence
of the low-column density lines is consistent with $dN_{\rm abs}/dz \propto (1+z)^{2.5}$ for 
the \nhi\ range $12 < {\rm log}\nhi < 15.2$, which has been assumed in our Monte Carlo realizations. 

For Lyman continuum opacity, the most relevant absorption systems are those with $16 < {\rm log(\nhi/cm^{-2})} < 18$ (\citealt{rudie13}) and 
redshifts within our chosen LyC interval of [880,910] in the source rest-frame, i.e., 
$z_{\rm s} - \Delta z \le z \le z_{\rm s}$, where $\Delta z = 0.036(1+z_{\rm s})$. %\simeq 0.146$ for $z_{\rm s}  \simeq 3.05$.) 
The LyC opacity depends on the higher \nhi\ portion of the distribution measured by \citet{rudie13}, 
including systems with log~${\rm (\nhi/cm^{-2})}>17.2$ (Lyman limit systems), which are not well sampled in that data set.
However, \citet{rudie13} showed that the power-law extrapolation of $f(\nhi,X)$ to log~(\nhi/${\rm cm}^{-2}) > 17.2$ accurately
reproduced the best available constraints on the incidence of Lyman limit systems (LLSs) at $z \sim 2.4$, and,
assuming $\gamma = 1.0$, at $z \sim 3$ as well. 
The point representing the LLSs (at ${\rm log~(\nhi/cm^{-2}) = 17.7}$ in Figure~\ref{fig:fnx_plot} is the actual location of the data point for 
$17.2 \le {\rm log}\nhi~\le 18.2$ assuming
the extrapolated slope and the {\it observed} total redshift path density of LLSs at $z \sim 3$. 
For the MC models, LLSs were drawn from a distribution that assumes $\beta = 1.473$ 
over the full range $15.2 \le {\rm log}\nhi \le 21.0$\footnote{By number, the systems with log\nhi$>19$ 
do not contribute appreciably to the total incidence rate, so the total number of LLSs is insensitive to the assumed upper
limit on \nhi.}.

\begin{deluxetable*}{lccccccc}
\tabletypesize{\scriptsize}
\tablewidth{0pt}
\tablecaption{Model IGM Transmission Versus Redshift}
\tablehead{
  \colhead{$z_s$} & \colhead{$\langle t_{900} \rangle$\tablenotemark{a}} & \colhead{10\%}  &
\colhead{25\%} & \colhead{50\%} &   
\colhead{75\%} & \colhead{90\%} & \colhead{$\langle 1-D_B \rangle$\tablenotemark{b}}
  }
\startdata
% CGM numbers corrected on 26 May 2016
\cutinhead{IGM Only}
2.70 & $0.516\pm0.200$ &0.175 & 0.417 & 0.588 & 0.663 & 0.708 & $0.736\pm0.028$ \\
2.75 & $0.504\pm0.199$ &0.160 & 0.398 & 0.571 & 0.654 & 0.698 & $0.727\pm0.028$ \\
2.80 & $0.486\pm0.197$ &0.148 & 0.376 & 0.554 & 0.634 & 0.680 & $0.715\pm0.027$ \\
2.85 & $0.478\pm0.200$ &0.143 & 0.368 & 0.550 & 0.630 & 0.671 & $0.706\pm0.028$ \\
2.90 & $0.469\pm0.200$ &0.138 & 0.347 & 0.543 & 0.622 & 0.661 & $0.695\pm0.030$ \\
2.95 & $0.461\pm0.188$ &0.136 & 0.346 & 0.524 & 0.602 & 0.646 & $0.686\pm0.028$ \\
3.00 & $0.446\pm0.182$ &0.133 & 0.332 & 0.506 & 0.584 & 0.631 & $0.678\pm0.028$ \\
3.05 & $0.432\pm0.189$ &0.127 & 0.311 & 0.496 & 0.580 & 0.626 & $0.665\pm0.028$ \\
3.10 & $0.435\pm0.187$ &0.123 & 0.325 & 0.488 & 0.569 & 0.616 & $0.656\pm0.027$ \\
3.15 & $0.417\pm0.178$ &0.116 & 0.304 & 0.471 & 0.557 & 0.605 & $0.645\pm0.029$ \\
3.20 & $0.402\pm0.178$ &0.095 & 0.294 & 0.461 & 0.537 & 0.588 & $0.632\pm0.028$ \\
3.25 & $0.390\pm0.176$ &0.091 & 0.278 & 0.452 & 0.533 & 0.575 & $0.623\pm0.028$ \\
3.30 & $0.377\pm0.173$ &0.089 & 0.259 & 0.425 & 0.513 & 0.560 & $0.611\pm0.027$ \\
3.35 & $0.363\pm0.169$ &0.085 & 0.242 & 0.416 & 0.502 & 0.545 & $0.601\pm0.029$ \\
3.40 & $0.345\pm0.165$ &0.078 & 0.230 & 0.393 & 0.488 & 0.527 & $0.588\pm0.025$ \\
3.45 & $0.344\pm0.164$ &0.069 & 0.223 & 0.382 & 0.482 & 0.522 & $0.577\pm0.028$ \\
3.50 & $0.322\pm0.163$ &0.053 & 0.198 & 0.362 & 0.456 & 0.501 & $0.565\pm0.027$ \\
\hline
4.00 & $0.226\pm0.127$ &0.034 & 0.117 & 0.244 & 0.332 & 0.383 & $0.448\pm0.025$ \\
4.50 & $0.147\pm0.089$ &0.012 & 0.069 & 0.155 & 0.219 & 0.262 & $0.334\pm0.022$ \\
5.00 & $0.088\pm0.058$ &0.006 & 0.036 & 0.088 & 0.137 & 0.164 & $0.233\pm0.017$ \\
\cutinhead{IGM+CGM} 
      2.70  &   $0.454   \pm    0.236$  &   0.021  &   0.276  &   0.540  &   0.643  &   0.694  &  $ 0.729 \pm      0.030 $\\
      2.75  &   $0.439   \pm    0.231$  &   0.019  &   0.249  &   0.517  &   0.633  &   0.680  &  $ 0.720 \pm      0.029 $\\
      2.80  &   $0.422   \pm    0.232$  &   0.015  &   0.241  &   0.500  &   0.611  &   0.661  &  $ 0.709 \pm      0.028 $\\
      2.85  &   $0.412   \pm    0.232$  &   0.013  &   0.208  &   0.489  &   0.609  &   0.660  &  $ 0.700 \pm      0.028 $\\
      2.90  &   $0.397   \pm    0.226$  &   0.014  &   0.214  &   0.466  &   0.587  &   0.645  &  $ 0.690 \pm      0.030 $\\
      2.95  &   $0.392   \pm    0.219$  &   0.013  &   0.209  &   0.462  &   0.578  &   0.636  &  $ 0.680 \pm      0.030 $\\
      3.00  &   $0.371   \pm    0.222$  &   0.012  &   0.175  &   0.431  &   0.561  &   0.622  &  $ 0.669 \pm      0.030 $\\
      3.05  &   $0.369   \pm    0.213$  &   0.012  &   0.189  &   0.427  &   0.550  &   0.609  &  $ 0.658 \pm      0.030 $\\
      3.10  &   $0.352   \pm    0.213$  &   0.011  &   0.135  &   0.393  &   0.529  &   0.589  &  $ 0.645 \pm      0.031 $\\
      3.15  &   $0.347   \pm    0.205$  &   0.012  &   0.170  &   0.405  &   0.522  &   0.577  &  $ 0.637 \pm      0.028 $\\
      3.20  &   $0.326   \pm    0.204$  &   0.010  &   0.137  &   0.366  &   0.502  &   0.572  &  $ 0.624 \pm      0.029 $\\
      3.25  &   $0.321   \pm    0.195$  &   0.009  &   0.150  &   0.353  &   0.489  &   0.552  &  $ 0.615 \pm      0.029 $\\
      3.30  &   $0.321   \pm    0.192$  &   0.006  &   0.147  &   0.359  &   0.485  &   0.543  &  $ 0.603 \pm      0.029 $\\
      3.35  &   $0.310   \pm    0.191$  &   0.003  &   0.141  &   0.351  &   0.474  &   0.531  &  $ 0.592 \pm      0.030 $\\
      3.40  &   $0.291   \pm    0.188$  &   0.002  &   0.119  &   0.315  &   0.459  &   0.516  &  $ 0.580 \pm      0.030 $\\
      3.45  &   $0.272   \pm    0.183$  &   0.002  &   0.094  &   0.303  &   0.427  &   0.498  &  $ 0.569 \pm      0.030 $\\
      3.50  &   $0.264   \pm    0.179$  &   0.002  &   0.100  &   0.302  &   0.434  &   0.492  &  $ 0.557  \pm     0.029 $\\
\hline
4.00 & $0.196\pm0.132$ &0.001 & 0.074 & 0.204 & 0.311 & 0.365 & $0.445\pm0.025$ \\
4.50 & $0.115\pm0.091$ &0.000 & 0.026 & 0.109 & 0.191 & 0.245 & $0.330\pm0.022$ \\
5.00 & $0.068\pm0.058$ &0.000 & 0.010 & 0.064 & 0.115 & 0.152 & $0.230\pm0.017$ 
\enddata
\tablenotetext{a}{Mean and standard deviation of the IGM (or IGM+CGM) transmission over the rest-frame interval $880 \le \lambda_0 \le 910$ \AA\ for different
source redshifts}
\tablenotetext{b}{Mean and standard deviation of the transmission in the rest-frame interval $920 \le \lambda_0 \le 1015$ \AA . }
\label{tab:igm_models}
\end{deluxetable*}

Monte Carlo realizations of IGM sightlines were made, 
using the same code as in \citet{shapley06} and \citet{nestor11} 
albeit with more detailed treatment of the higher-order Lyman series lines and with normalization parameters updated as in Table~\ref{tab:fnxtab}. 
The realizations include the full range $12<$log\nhi~$<21$ 
drawn in a Poisson fashion according to the appropriate distribution function and redshift.  
Each realization for a given assumed source redshift $z_{\rm s}$ is a synthetic spectrum covering 
the observed wavelength range $3160-(1+z_{\rm s})1215.7$ \AA.   
A set of 10,000 realizations at each of 17 assumed source redshifts (2.70-3.50 in steps of 0.05) 
was performed using the ``IGM-only'' version of $f(\nhi,X)$. Table~\ref{tab:igm_models} summarizes
the results in terms of percentiles in transmission $t_{900}$, the mean unabsorbed fraction  
in the source rest-frame [880,910] interval. Also provided for each source redshift is the value of
$\langle 1-D_{\rm B} \rangle$, where $D_{\rm B}$ is the mean flux decrement from Lyman line
blanketing in the rest-frame interval [920,1015] (\citealt{oke82}). 
As shown in \S\ref{sec:igm_trans}, $\langle 1- D_{\rm B} \rangle$ is a close approximation to the
maximum transmission expected in the [880,910] LyC interval at each value of $z_{\rm s}$, for sightlines 
without significant continuum opacity from high-\nhi\ absorbers. 

As in \citet{rudie13}, a second set of Monte Carlo realizations was made by drawing absorbers from the same IGM distribution 
$f(\nhi,X)_{\rm IGM}$, {\it except} within 700 \kms [$\Delta z  < 0.0023(1+z_{\rm s})$]) of 
the source, where  
absorbers were instead drawn from the $f(\nhi,X)_{\rm CGM}$ distribution.  
This set of simulations is hereafter referred to as ``IGM+CGM''; the statistics as a function of
source redshift are summarized in the bottom half of Table~\ref{tab:igm_models}. 

\bibliographystyle{aasjournal}
\bibliography{/Users/ccs/words/refs_comb}

\end{document}